\documentclass[prb,twocolumn,showpacs,amsmath,amssymb,eqsecnum]{revtex4-1}
\usepackage{graphicx}
\usepackage{color}
\usepackage{amsmath,amssymb,bm}

\begin{document}

\title{
Current noise and Keldysh vertex function of an Anderson impurity \\ 
in the Fermi liquid regime
}

 \author{Akira Oguri}
 \affiliation{
 Department of Physics, Osaka City University, Sumiyoshi-ku, 
 Osaka 558-8585, Japan }
\affiliation{Nambu Yoichiro Institute of Theoretical and Experimental Physics, 
Osaka City University, Osaka 558-8585, Japan}

 \author{Yoshimichi Teratani}
 \affiliation{
 Department of Physics, Osaka City University, Sumiyoshi-ku, 
 Osaka 558-8585, Japan }
\affiliation{Nambu Yoichiro Institute of Theoretical and Experimental Physics, 
Osaka City University, Osaka 558-8585, Japan}

 \author{Kazuhiko Tsutsumi}
 \affiliation{
 Department of Physics, Osaka City University, Sumiyoshi-ku, 
 Osaka 558-8585, Japan }

\author{Rui Sakano}
\affiliation{
The Institute for Solid State Physics, 
the University of Tokyo, Kashiwa, Chiba 277-8581, Japan
}

\date{\today}

\begin{abstract}

We present a complete microscopic Fermi-liquid description for 
next-to-leading order transport 
through an Anderson impurity under a finite bias voltage $V$. 
It is applicable to multilevel quantum dots without particle-hole or time-reversal 
symmetry, and is constructed based on the nonequilibrium Keldysh formalism, 
taking into account the current conservation between electrons 
in the impurity levels and the conduction bands.  
Specifically, we derive the formula for the current noise generated in the steady flow
up to terms of order $(eV)^3$ at zero temperature $T=0$.
To this end, we calculate the Keldysh vertex functions 
 $\Gamma_{\sigma\sigma';\sigma'\sigma}^{ \nu_1\nu_2;\nu_3\nu_4} 
(\omega,\omega'; \omega',\omega)$,  which depend on branches     
$\nu_1, \nu_2, \nu_3$ and $\nu_4$ of the time-loop contour
and on spin degrees of freedom $\sigma$ and  $\sigma'$,  
up to linear-order terms 
with respect to  $eV$,  $T$, and frequencies $\omega$ and $\omega'$. 
The coefficients of these linear-order terms are determined by a set of the parameters, 
defined with respect to the equilibrium ground state: 
the phase shift, static susceptibilities, 
and nonlinear three-body susceptibilities of the impurity electrons.
The low-energy expressions of the vertex components are shown 
to satisfy the Ward identities with the Keldysh Green's functions 
expanded up to terms of order $\omega^2$, $(eV)^2$, and $T^2$.  
We also find that  the imaginary part of the Ward identities  
can be described in terms of the $eV$-dependent 
collision integrals for a single-quasiparticle excitation 
and that for a single quasiparticle-quasihole pair excitation. 
These collision integrals ensure  the current conservation 
of the next-to-leading order Fermi-liquid transport    
due to the quasiparticles with a finite damping rate.

\end{abstract}


\maketitle

\section{Introduction}
\label{sec:introduction}

Highly correlated low-energy quantum states 
of the Kondo systems show universal behaviors 
which can be described by the Fermi liquid theory.\cite{HewsonBook} 
It was originally developed for dilute magnetic alloys,
and has been applied later to quantum dots, for which 
the universal  Fermi-liquid behaviors 
have been observed in nonlinear current-voltage characteristics 
 \cite{GrobisGoldhaber-Gordon,ScottNatelson}   
and also in the nonequilibrium current noise.
\cite{Heiblum,Delattre2009,KobayashiKondoShot,Ferrier2016} 
Furthermore,  novel quantum systems having various kinds of internal degrees of freedom,    such as  orbitals, nuclear spins and flavors, 
bring an interesting variety to the Kondo effects, 
and have been being studied
 for carbon nanotubes,\cite{RMP-Kouwenhoven,Ferrier2016} 
 ultracold atomic gases,\cite{TakahashiColdGasKondo} 
  quark matters,\cite{QCD_Kondo} etc.

The Fermi liquid  (FL) theory for quantum impurity systems 
has been constructed based on the Kondo model\cite{NozieresFermiLiquid} 
or Anderson model.
\cite{YamadaYosida2,YamadaYosida3,YamadaYosida4,ShibaKorringa,Yoshimori} 
The ground state properties such as the residual resistivity of magnetic alloys 
and the zero-bias conductance of quantum dots can be described by 
the scattering phase shift $\delta_{\sigma}^{}$. 
The Friedel sum rule states that at zero temperature $T=0$ 
 it also corresponds to a one-point correlation function 
$\delta_{\sigma}^{}/\pi = \langle n_{d\sigma}^{} \rangle$:  
the occupation number of an impurity level 
with $\sigma$ the index for spin or the other internal degrees of freedom.
 The phase shift also determines 
the spectral weight of the impurity state at the Fermi level 
 and the energy shift of the impurity level 
due to the Coulomb interaction.

Leading order behavior of the Fermi liquid 
approaching the limit $T\to 0$ occurs, for instance,  
as a $T$-linear specific heat of impurity electrons 
 $\mathcal{C}_\mathrm{imp}^{} \propto T/T^{*}$. 
The Kondo energy scale $T^{*}$ and the Wilson ratio $R$  
together with 
the phase shift $\delta_{\sigma}^{}$   
completely describe the leading-order behaviors.
These additional parameters,  $T^*$ and $R$, 
can be expressed in terms of two-point correlation functions:  
the static susceptibilities of impurity electrons $\chi_{\sigma\sigma'}^{},$   
which can also be related to the derivative of the self-energy 
with respect to the frequency $\omega$ 
and the value of the vertex corrections at zero frequencies. 
\cite{YamadaYosida2,YamadaYosida3,YamadaYosida4}

Next leading-order behavior occurs especially in transport phenomena,     
such as the $T^2$ resistivity of magnetic alloys
\cite{YamadaYosida2,YamadaYosida3,YamadaYosida4}   
and the nonlinear  $(eV)^2$ conductance through quantum dots 
under a finite bias voltage $V$.\cite{Hershfield1,ao2001PRB,Aligia,Aligia2014,Munoz}   
In an ideal situation where quantum impurity systems 
have the particle-hole (PH) and time-reversal (TR) symmetries, 
these next leading-order terms 
 can be described in terms of the two-point correlation functions, 
mentioned above. 
This is because  in this highly symmetric case 
the quadratic  $\omega^2$,  $T^2$, and  $(eV)^2$ dependences 
occur only through the damping rate of quasiparticles.  
However, these symmetries are easily broken by external fields, 
such as a gate voltage and a magnetic field. 
When the PH or TR symmetry is broken,
the energy of a quasiparticle also shows the quadratic 
 $\omega^2$,  $T^2$, and  $(eV)^2$ dependences 
which enter through the real part of the self-energy.\cite{Yoshimori,HorvaticZlatic2}

It has recently been clarified that the energy shift of the quadratic order 
can be described in terms of three-point correlation functions.  
Extending Nozi\`{e}res' phenomenological Fermi-liquid theory,\cite{NozieresFermiLiquid}
the next-to-leading order terms for some transport coefficients 
 have been derived for the SU($N$) Kondo model by Mora, Vitushinsky {\it et al\/}
\cite{MoraSUnKondoII,Vitushinsky2008,Mora2008,MoraSUnKondoI}, 
and also for the Anderson model with\cite{FilipponeMocaWeichselbaumVonDelftMora} 
and without\cite{MoraMocaVonDelftZarand} a magnetic field 
by Filippone {\it et al\/} and  Mora {\it et al\/}, respectively.  
Correspondingly, we have presented a fully microscopic description
\cite{ao2017_1_PRL,ao2017_2_PRB,ao2017_3_PRB} 
based on the standard many-body Green's-function 
approach of Yamada-Yosida, Shiba, and Yshimori,\cite{YamadaYosida2,YamadaYosida3,YamadaYosida4,ShibaKorringa,Yoshimori}
and have shown without assuming the PH nor TR symmetry  
that  the order  $\omega^2$,  $T^2$, and $(eV)^2$ 
real parts of the self-energy for the Anderson impurity 
are determined completely by the static three-body susceptibilities  
$\chi_{\sigma\sigma'\sigma''}^{[3]}$, 
defined in Eq.\ \eqref{eq:canonical_correlation_3}. 
We have also examined behaviors of the three-body susceptibilities  
of multilevel Anderson models in a wide range of impurity-electron 
fillings,\cite{TerataniPRB2020,TerataniSakanoOguri2020,TerataniSakanoOguri2021}
using the numerical renormalization group (NRG) approach.\cite{KWW1,KWW2} 
Moreover, three-body correlations between electrons in quantum dots 
have recently been deduced successfully from nonlinear 
magnetoconductance measurements.\cite{Hata2021}

The current noise becomes one of the most important probes 
 in recent years for exploring quantum fluctuations.
The nonlinear current noise 
in the low-energy Fermi-liquid regime was studied, in early days, 
mainly in the situation where both the PH and TR symmetries are present.
 \cite{GogolinKomnikPRL,Sela2006,Golub,SelaMalecki,SakanoFujiiOguri,KarkiMoraVonDelfKiselex}
A major milestone has been achieved by Mora {\it et al\/}, without 
assuming the PH symmetry:
formula for the next-leading order terms of the current noise 
has been derived for the SU($N$) Kondo model,\cite{MoraSUnKondoII}
and for the single-orbital  Anderson model at zero magnetic field.
\cite{MoraMocaVonDelftZarand}
However, more general current-noise formulas applicable 
to the cases without the PH nor TR symmetry 
are necessary for studying nonequilibrium quantum fluctuations 
in a wide class of Kondo systems with various kinds of internal degrees of freedom.

The purpose of this paper is to present a complete microscopic description 
applicable to next-to-leading order Fermi-liquid behaviors of the current noise   
in a wide class of quantum impurity models.
To this end, we investigate the low-energy asymptotic form of the Keldysh vertex function  
$\Gamma_{\sigma\sigma';\sigma'\sigma}^{\nu_1\nu_2;\nu_3\nu_4}
(\omega,\omega';\omega'\omega)$ 
up to linear-order terms with respect to $\omega$, $\omega'$, $T$, and $(eV)$,    
for all branch-components $\nu_1,\nu_2,\nu_3,\nu_4$ of the 
Keldysh time-loop contour.
The low-energy  results of the vertex function satisfy the Ward identities 
with the Keldysh self-energy $\Sigma_{U,\sigma}^{\nu_4\nu_1}(\omega)$ 
which has been obtained up to terms of order $\omega^2$, $(eV)^2$ and $T^2$.  
It ensures the current conservation is full filled for  
the next-to-leading order transport of the quasiparticles with a finite damping rate  
and verifies the consistency in the Fermi-liquid description 
for nonlinear quantum fluctuations.\cite{Hershfield1,Hershfield2}

In this paper, we also provide two alternative derivations 
for the Ward identities for the nonequilibirum Keldysh correlation functions  
at finite bias voltage and temperature. 
The first one is based on perturbation expansion in $U$, 
and the second one is deduced from a more general 
Ward-Takahashi identity for the Anderson impurity.
We find that in the low-energy Fermi-liquid regime 
the imaginary parts of the Ward identities 
can be described in terms of 
the collision integrals,\cite{LandauPhysicalKinetics,HaugJauho} 
listed in TABLES \ref{tab:self-energy} and \ref{tab:ph-pp_propagators}: 
the fermionic collision integral $\mathcal{I}_\mathrm{K}^{}(\omega)$ describes     
the damping of a single quasiparticle, and 
the bosonic collision integrals 
$\mathcal{W}_\mathrm{K}^\mathrm{ph}(\omega)$  and 
$\mathcal{W}_\mathrm{K}^\mathrm{pp}(\omega)$ 
represent the damping of a single particle-hole pair 
 and a single particle-particle pair, respectively. 
 Furthermore,  using the Ward identities, 
we find that the causal component of the vertex function 
  $\Gamma_{\sigma\sigma';\sigma'\sigma}^{--;--}(0,0;0,0)$ 
for $\sigma\neq \sigma'$ has an $eV$-linear real part, 
as shown in TABLE \ref{tab:vertex_UD}.

The paper is organized as follows. 
In Sec.\ \ref{sec:formulation}, 
we describe the definition of 
the steady-state averages for the current and current fluctuations 
through the Anderson impurity  at finite bias voltage.
In Sec.\ \ref{sec:FL_theory_I},   
we introduce the Fermi-liquid parameters necessary 
for describing low-energy properties up to the next-leading order, 
and demonstrated how 
the asymptotic form of the retarded Green's function 
can be expressed in terms of these parameters. 
In Sec.\ \ref{sec:noise_result_first}, 
we discuss some properties of the nonlinear current noise in the Fermi-liquid regime,
leaving the derivation of the formula to the last part of this paper.
In Sec.\ \ref{sec:Keldysh_formulation}, 
we provide a perturbative derivation of the Ward identities for 
the Keldysh Green's functions at finite $eV$ and $T$.
In Sec.\ \ref{sec:nonlinear_WT} and  Appendix \ref{sec:Ward_Takahashi_derivations}, 
the nonequilibrium Ward-Takahashi identity is derived from the equation of continuity, 
and the three-point vertex functions are introduced in order to extract systematically  
the symmetrized vertex components with respect to the Keldysh time-loop branches.  
In Sec.\ \ref{sec:Collision_Fermion}, 
we show that the imaginary part of the Keldysh self-energies 
can be described by the fermionic collision integral in the Fermi-liquid regime, 
and introduce the bosonic collision integrals for the particle-hole pair
and particle-particle pair excitations.
In Sec.\ \ref{sec:current_conservation_Keldysh_selfenergy_vertex_in_FL_regime} 
and  Appendix \ref{sec:full_vertex_low_energy_form}, 
we calculate the low-energy asymptotic form of the Keldysh vertex functions, 
and show that the imaginary parts of the Ward identities  
are full filled through the relations between the fermionic and bosonic collision integrals.  
In Sec.\ \ref{sec:noise_derivation}, 
using the low-energy asymptotic form of the Keldysh vertex functions and self-energies,   
the nonlinear current-noise formula is derived up to terms order $|eV|^3$  at $T=0$.
Summary is given in Sec.\ \ref{sec:summary_IV}.

\begin{figure}[b]
 \leavevmode
\begin{minipage}{1\linewidth}
\raisebox{0.2cm}{
\includegraphics[width=0.45\linewidth]{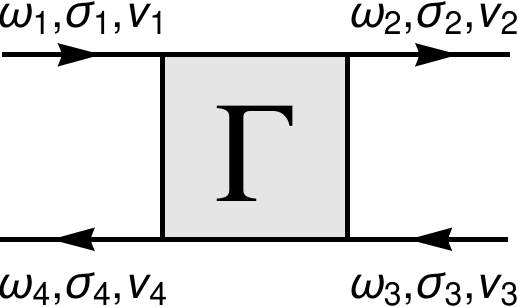}
 }
 \caption{
Vertex correction 
$\Gamma_{\sigma_1\sigma_2;\sigma_3\sigma_4}^{\nu_1\nu_2;\nu_3\nu_4}
(\omega_1,\omega_2;\omega_3\omega_4)$ between 
the electron in the levels $\sigma_i$ ($i=1,2,3,4$). 
The superscript  $\nu_i $ 
specifies  the branches of  Kedysh time-loop contour, 
for which  $\nu=-$ and $+$ represent 
the forward and return paths, respectively.
The frequencies are conserved 
such that  $\omega_1+\omega_3=\omega_2+\omega_4$. 
 }
\label{fig:Keldysh_vertex}
\end{minipage}
\end{figure}

\section{Formulation}
\label{sec:formulation}

\subsection{Anderson impurity model}

We study low-energy transport of quantum impurities 
in the Fermi-liquid regime over a wide range of electron fillings,   
using a multilevel Anderson model coupled to two different reservoirs 
on the left ($L$) and right ($R$):    
$\mathcal{H} =
\mathcal{H}_d + \mathcal{H}_c  + \mathcal{H}_\mathrm{T}$,  
\begin{subequations}
\begin{align}
\mathcal{H}_d =& \  
 \sum_{\sigma=1}^N
 \epsilon_{d\sigma}^{}\, n_{d\sigma}  
+
\frac{1}{2} \sum_{\sigma \neq \sigma'} 
U \, n_{d\sigma}^{} n_{d\sigma'}^{} \;, 
 \label{eq:H_U_spin}
\\ 
\mathcal{H}_c =& \   
\sum_{j=L,R} \sum_{\sigma=1}^N 
\int_{-D}^D  \! d\xi\,  \xi\, 
 c^{\dagger}_{\xi j \sigma} c_{\xi j \sigma}^{},
 \label{eq:Ham_cond}
\\
 \mathcal{H}_\mathrm{T} =& \   
   - \sum_{j=L,R} \sum_{\sigma=1}^N \,  v_{j}^{}
 \left( \psi_{j \sigma}^\dag d_{\sigma}^{} + 
  d_{\sigma}^{\dag} \psi_{j \sigma}^{} \right) .
 \label{eq:Ham_mix}
\end{align}
 \label{eq:Hamiltonian_general}
\!\!\!\!\! \!\!\!\! 
\end{subequations}
Here, 
 $d^{\dag}_{\sigma}$  for $\sigma=1,2,\ldots,N$ 
creates an impurity electron with energy $\epsilon _{d\sigma}$. 
The operator 
  $n_{d\sigma} = d^{\dag}_{\sigma} d^{}_{\sigma}$ represents 
the occupation number, 
 and $U$ is the Coulomb interaction between electrons in the different levels.  
For $N=2$, this Hamiltonian corresponds to the usual spin-1/2 Anderson model. 
 In this paper, we will  hereafter cite the internal $\sigma$ degrees of freedom as 
`{\it spin}'  also for $N>2$, for simplicity.
Conduction electrons  with energy $\xi$ in each of the two bands on the left and right 
are normalized such that 
$
\{ c^{\phantom{\dagger}}_{\xi j \sigma}, 
c^{\dagger}_{\xi' j'\sigma'}
\} = \delta_{jj'} \,\delta_{\sigma\sigma'}   
\delta(\xi-\xi')$ for $j=L,R$.
The linear combination of these continuous states, 
$\psi^{}_{j \sigma} \equiv  \int_{-D}^D d\xi \sqrt{\rho_c^{}} 
\, c^{\phantom{\dagger}}_{\xi j \sigma}$  with $\rho_c^{}=1/(2D)$, 
 couples to the discrete impurity level with the same $\sigma$ 
via a tunnel matrix element $v_{j}^{}$.
 It determines the width of the resonance as  
 $\Delta \equiv \Gamma_L + \Gamma_R$ with 
 $\Gamma_{j} = \pi \rho_c^{} v_{j}^2$.  
Specifically, we consider the parameter region, 
where the relevant energy scales are much smaller than the half band width $D$,
 i.e., $\max( U, \Delta, |\epsilon_{d\sigma}^{}|, |\omega|, T, |eV|)\ll D$.

In this paper, 
we study nonlinear fluctuations of the current $\widehat{J}_{j,\sigma}$ 
through a quantum dot in the low-energy Fermi-liquid regime: 
\begin{subequations}
\begin{align}
\widehat{J}_{L,\sigma} =& \     -i\,  v_L 
\left(
\psi^{\dagger}_{L\sigma} d^{}_{\sigma} 
-d^{\dagger}_{\sigma} \psi^{}_{L\sigma} \right) , 
\\
\widehat{J}_{R,\sigma} =& \  + i\, v_R 
\left(
\psi^{\dagger}_{R\sigma} d^{}_{\sigma} 
-d^{\dagger}_{\sigma} \psi^{}_{R\sigma}\right) \;. 
 \end{align}
\end{subequations}
Here,  $\widehat{J}_{L,\sigma}$ represents 
the current following from the left lead to the dot, 
and  $\widehat{J}_{R,\sigma}$ represents
  the current from  the dot to the right lead. 
These currents satisfy the conservation law     
which follows from the Heisenberg equation motion,   
$\partial  n_{d\sigma}^{} / \partial t = 
-i \, \bigl[ n_{d,\sigma},\, \mathcal{H} \bigr]$,
and  preserves the spin $\sigma$,   
 \begin{align}
\frac{\partial  n_{d\sigma}^{}}{\partial t} 
+ \widehat{J}_{R,\sigma} - \widehat{J}_{L,\sigma}\, =\, 0 \, .
\label{eq:current_conservation}
 \end{align}
These two current operators can also be 
classified in to the symmetrized part 
 \begin{align}
 \widehat{J}_{\sigma}
 \equiv    
\frac{\Gamma_L \widehat{J}_{R,\sigma} 
+\Gamma_R \widehat{J}_{L,\sigma}}{\Gamma_L+\Gamma_R} \;,
\label{eq:symmetrized_current_def}
 \end{align}
and the difference 
$\widehat{J}_{R,\sigma} -\widehat{J}_{L,\sigma}$.

\begin{figure}[b]
 \leavevmode
\begin{minipage}{1\linewidth}
\raisebox{0.2cm}{
\includegraphics[width=0.41\linewidth]{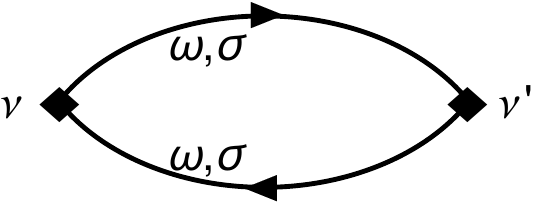}
 }
\rule{0.05\linewidth}{0cm}
\includegraphics[width=0.44\linewidth]{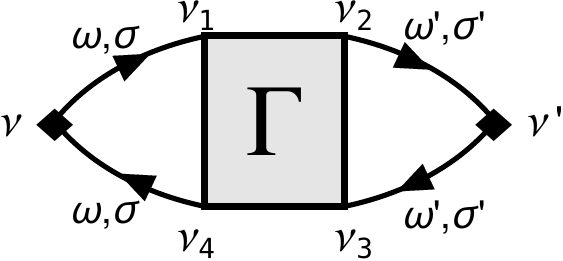}
\includegraphics[width=0.34\linewidth]{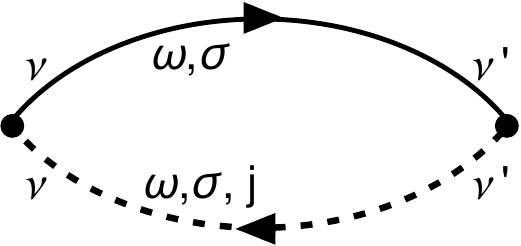}
\rule{0.1\linewidth}{0cm}
\rule{0cm}{1.8cm}
\includegraphics[width=0.34\linewidth]{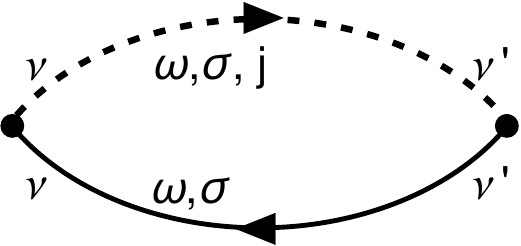}
 \caption{
Feynman diagrams for  the current-current correlation function  
 $\int_{-\infty}^{\infty} \! dt \, 
 \mathcal{K}_{\sigma'\sigma}^{\nu'\nu}(t,0)$ 
of  symmetrized current fluctuations $\widehat{J}_{\sigma}$ 
defined in Eq.\ \eqref{eq:symmetrized_current_def} 
 with  
 $\mathcal{K}_{\sigma'\sigma}^{+-}(t,0)
 \equiv  
-i \,\bigl\langle \delta \widehat{J}_{\sigma'}(t) 
\, \delta \widehat{J}_{\sigma}(0) \bigr\rangle$ and   
$\mathcal{K}_{\sigma'\sigma}^{-+}(t,0)
\equiv 
-i \,\bigl\langle \delta \widehat{J}_{\sigma}(0) 
\, \delta \widehat{J}_{\sigma'}(t)\bigr\rangle$. 
The solid line represents the Keldysh Green's function 
  $G_{\sigma}^{\nu'\nu}(\omega)$ 
of the quantum dot. 
The black diamond ({\tiny \protect \rotatebox[origin=c]{45}{$\blacksquare$}}) 
in the two diagrams on the top represents a matrix the bare current vertex 
$\bm{\lambda}_\mathrm{sym}^{\alpha}(\epsilon, \epsilon)$  for $\alpha =\nu, \nu'$
defined in Eq.\ \eqref{eq:lambda_av_U0}. 
The dashed line in the last two diagrams at the bottom  
 denotes the Green's function 
$g_{j\sigma}^{\nu'\nu}(\omega) $ of  the isolated lead on $j=L$ and $R$. 
 }
\label{fig:Kubo_Keldysh_diagram}
\end{minipage}
\end{figure}


\begin{table*}[t]
\caption{Low-energy expansion of  $dJ/dV$  up to terms of order $T^2$ and $(eV)^2$ 
for $\Gamma_L=\Gamma_R = \Delta/2$ and $\mu_L=-\mu_R =eV/2$.
}
\begin{tabular}{l} 
\hline \hline
\ \ 
$
{\displaystyle
\frac{dJ}{dV} 
}
\, =\, 
{\displaystyle
\frac{e^2}{h} \,  
\sum_{\sigma}
}\,
\Bigl[\ 
\sin^2 \delta_{\sigma}  
-  c_{T,\sigma}^{}\, \left(\pi T\right)^2
-  c_{V,\sigma}^{}\, \left(eV \right)^2 \ + \ \cdots \ 
\Bigr]\,, $
\rule{0cm}{0.65cm}
\\
\ \ 
$c_{T,\sigma}^{}\, \equiv  \,  
{\displaystyle
\frac{\pi^2}{3} 
}
\biggl[\,
 -
\biggl(\,
\chi_{\sigma\sigma}^2
+ 
2
{\displaystyle
\sum_{\sigma'(\neq \sigma)}
}
\chi_{\sigma\sigma'}^2
\,\biggr) 
\cos 2 \delta_{\sigma}
\, + 
{\displaystyle
\frac{1}{2\pi} 
}
\biggl(
  \chi_{\sigma\sigma\sigma}^{[3]}
\,+\,
{\displaystyle
\sum_{\sigma'(\neq \sigma)}
}
  \chi_{\sigma\sigma'\sigma'}^{[3]}
\biggr)
\sin 2\delta_{\sigma}\,
\biggr],$
\rule{0cm}{0.65cm}
\\
\ \ 
$
c_{V,\sigma}^{} \,\equiv   \, 
{\displaystyle
\frac{\pi^2}{4}
}
\biggl[
\,
-
\biggl(\, \chi_{\sigma\sigma}^2 +  5
{\displaystyle
\sum_{\sigma'(\neq \sigma)}
}
\chi_{\sigma\sigma'}^2
\,\biggr)
\cos 2 \delta_{\sigma} 
\,+
{\displaystyle
\frac{1}{2\pi} 
}
\biggl(
\, 
  \chi_{\sigma\sigma\sigma}^{[3]}
\,+\,
3\, 
{\displaystyle 
\sum_{\sigma'(\neq \sigma)}
}
  \chi_{\sigma\sigma'\sigma'}^{[3]}
\biggr) 
\sin 2\delta_{\sigma}
\,\biggr].
$
\rule{0cm}{0.65cm}
\\
\hline
\hline
\end{tabular}
\label{tab:CV_CT}
\end{table*}

\begin{table*}[t]
\caption{Low-bias expansion of the noise $S_\mathrm{noise}^{\mathrm{QD}}$, 
obtained at $T=0$  up to terms of order $|eV|^3$ 
for $\Gamma_L=\Gamma_R = \Delta/2$ and $\mu_L=-\mu_R =eV/2$.
}
\begin{tabular}{l} 
\hline \hline
$
{\displaystyle
S_\mathrm{noise}^{\mathrm{QD}}
}
\  =  \,  
{\displaystyle
  \frac{2e^2}{h} 
}
\ |eV| \ 
{\displaystyle
\sum_{\sigma}
}
\Bigl[\ \frac{1}{4}  \sin^2 2\delta_\sigma^{} 
\,  + \,
c_{S,\sigma}^{} 
 \, 
\left(eV\right)^2
\ + \ \cdots 
\ \Bigr] 
\,, $
\rule{0cm}{0.65cm}
\\
\rule{0.16cm}{0cm}
$c_{S,\sigma}^{} \, \equiv  \,     
{\displaystyle
\frac{\pi^2}{12} 
} \,
\Biggl[
\,  
\cos 4 \delta_{\sigma}\,
 \chi_{\sigma\sigma}^2 
+ \bigl( 2+3\cos 4 \delta_{\sigma} \bigr)
{\displaystyle
\sum_{\sigma'(\neq \sigma)}
}
\chi_{\sigma\sigma'}^2
\,+\,  
4
{\displaystyle
\sum_{\sigma' (\neq \sigma)}
}
 \cos 2\delta_{\sigma}^{} \cos 2\delta_{\sigma'}^{}
\,  \chi_{\sigma\sigma'}^2
$
\rule{0cm}{0.7cm}
\\
\rule{1.8cm}{0cm}
$
 + \, 3
{\displaystyle
\sum_{\sigma'(\neq \sigma)}
\sum_{\sigma''(\neq \sigma,\sigma')}
}
\sin 2\delta_{\sigma}^{} \,\sin 2\delta_{\sigma'}^{}
 \chi_{\sigma\sigma''}^{}  \chi_{\sigma'\sigma''}^{}
\,-\,
{\displaystyle
\frac{1}{4\pi}
} 
\biggl(
  \chi_{\sigma\sigma\sigma}^{[3]}
+
3 
{\displaystyle
\sum_{\sigma'(\neq \sigma)}
}
  \chi_{\sigma\sigma'\sigma'}^{[3]}
\biggr)  
\sin 4\delta_{\sigma}
\,\Biggr] 
$
\rule{0cm}{0.65cm}
\\
\hline
\hline
\end{tabular}
\label{tab:CS}
\end{table*}


\subsection{Average current and current fluctuations}

We consider a nonequilibrium steady state 
under a finite bias voltage  $eV \equiv \mu_L-\mu_R$, 
applied between the two leads by setting the chemical potentials 
of the left and right leads to be 
$\mu_L$ and  $\mu_R$, respectively.
The statistical density operator which describes the steady state 
in this situation can be described, 
using  the time evolution along the Keldysh time-loop contour.\cite{Caroli}

Specifically, 
the steady-state average of  the electric current 
 $J \equiv e\,\sum_{\sigma}\langle  \widehat{J}_{\sigma} \rangle$ 
 can be expressed in the following form,\cite{Hershfield1,MeirWingreen}   
\begin{align}
 J \,=  \,  
 \frac{e}{h}
\sum_{\sigma}  
\int_{-\infty}^{\infty} \!\! d\omega\,  
\bigl[\,f_L(\omega)-f_R(\omega) \,\bigr]\, 
  \mathcal{T}_{\sigma}(\omega)  \,.
\label{eq:current_formula}
\end{align}
Here,  $\mathcal{T}_{\sigma}(\omega) $ is the transmission probability 
that is determined by the retarded Green's function $G_{\sigma}^{r}(\omega)$,  
\begin{align}
\mathcal{T}_{\sigma}(\omega) \,=& \    
   - \frac{4\Gamma_L \Gamma_R}{\Gamma_L +\Gamma_R} 
 \, \mathrm{Im} \,G_{\sigma}^{r}(\omega) \,, 
\label{eq:transmissionPB}
\\ 
G_{\sigma}^{r}(\omega)
\,\equiv  & \ 
  -i 
 \int_{0}^{\infty} \! dt \, e^{i (\omega +i0^+ ) t}
 \left\langle \left\{
 d^{\phantom{\dagger}}_{\sigma}(t)\,,\, d^{\dagger}_{\sigma}(0)
\right\} \right\rangle \nonumber \\
\,= & \ 
\frac{1}{\omega -\epsilon_{d\sigma}^{} +i\Delta 
- \Sigma_{U,\sigma}^{r}(\omega)}.
\end{align}
Effects of the Coulomb repulsion on the average current $J$ 
 enters through the retarded self-energy  $\Sigma_{U,\sigma}^{r}(\omega)$ 
 which also depends on $eV$ and temperature $T$. 
In  Eq.\ \eqref{eq:current_formula}, 
 $f_{j}(\omega) \equiv f(\omega-\mu_j)$
with  $f(\omega) \equiv [e^{\omega/T}+1]^{-1}$ 
the Fermi distribution function. 
   We will choose the chemical potentials 
 $\mu_L$ and  $\mu_R$  in a way such that 
\begin{align}
\mu_L= \alpha_L eV, \qquad  
\mu_R= -\alpha_R eV,   \qquad 
\alpha_L + \alpha_R =1.
\label{eq:chemical_potentials}
\end{align}
The parameters $\alpha_L$ and $\alpha_R$ specify how the bias is applied 
relative to the Fermi level at equilibrium which is chosen to be 
the origin of one-particle energies $E_F=0$.

The main subject of this paper 
is  low-energy behavior of current noise of a quantum dot,\cite{Hershfield2}
\begin{align}
S_\mathrm{noise}^\mathrm{QD}  = \,
e^2 \sum_{\sigma\sigma'}\int_{-\infty}^{\infty} \!\!   dt  \,
\left\langle 
\delta \widehat{J}_{\sigma}(t) \, \delta \widehat{J}_{\sigma'}(0) 
+\delta \widehat{J}_{\sigma'}(0) \, \delta \widehat{J}_{\sigma}(t)
\right\rangle .
\label{eq:S_noise}
\end{align}
Here,  $\delta \widehat{J}_{\sigma}(t)  \equiv 
 \widehat{J}_\sigma (t) 
- \bigl\langle  \widehat{J}_\sigma (0)  \bigr\rangle
$ represents fluctuations of the symmetrized current.
The current-current correlation function  $S_\mathrm{noise}^\mathrm{QD}$ 
depends not only the single-quasiparticle properties 
that enter through $G_{\sigma}^{r}(\omega)$ 
but also on the collision term of two quasiparticles. 
It  is described microscopically by the Keldysh vertex corrections 
$\Gamma_{\sigma\sigma';\sigma'\sigma}^{\nu_1\nu_2;\nu_3\nu_4}
(\omega,\omega';\omega'\omega)$, illustrated in Fig.\ \ref{fig:Keldysh_vertex}.

We calculate $S_\mathrm{noise}^\mathrm{QD}$ later  
in Sec.\ \ref{sec:noise_derivation} 
up to terms of order $(eV)^3$, taking into account 
all  multiple collision processes described in Fig.\ \ref{fig:Kubo_Keldysh_diagram}.
Before giving the proof, 
the results of  the average current and the current noise 
are summarized in TABLES \ref{tab:CV_CT} and \ref{tab:CS},  respectively.  
The expansion coefficients  $c_{T,\sigma}^{}$, $c_{V,\sigma}^{}$,  
and $c_{S,\sigma}^{}$ of the next-leading order terms of 
the low-energy expansion 
can be expressed in terms of the static correlation functions 
$\chi_{\sigma\sigma'}^{}$ and $\chi_{\sigma\sigma'\sigma''}^{[3]}$, 
defined with respect to the equilibrium ground state.  
We describe the definition 
of these static correlation functions, 
which also determine the behavior of 
 the self-energy and vertex corrections,  
in the next section. 
We also present, in Sec.\ \ref{sec:noise_result_first}, some application examples of  
the transport formulas listed in TABLES \ref{tab:CV_CT} and \ref{tab:CS}.

\section{Microscopic Fermi-liquid theory}
\label{sec:FL_theory_I}

The main subject of this paper, i.e., low-energy 
behavior of the current noise, is based on the microscopic Fermi-liquid theory  
for which the roles of the three-body correlations have been clarified recently.
In this section,  we describe the latest version of FL theory  
for the retarded Green's function and the causal vertex function
 $\Gamma_{\sigma\sigma';\sigma'\sigma}^{--;--}
(\omega,\omega';\omega',\omega)$.

\subsection{Fermi-liquid parameters}
\label{subsec:FL_parameters}

Low-energy transport in the Fermi-liquid regime are 
mostly determined by  a set of the parameters 
which can be derived from the retarded Green's function 
with respect to the equilibrium ground state  
$G_{\mathrm{eq},\sigma}^{r}(\omega) 
\equiv  \left. G_{\sigma}^{r}(\omega)\right|_{T=eV=0}^{}$.   
In particular, the behavior of the retarded self-energy  
$\Sigma_{\mathrm{eq},\sigma}^{r}(\omega)$ 
at small frequencies $\omega$ plays a central role, 
\begin{align}
 \Sigma_{\mathrm{eq},\sigma}^{r}(\omega) \equiv  
\left. \Sigma_{U,\sigma}^{r}(\omega)\right|_{T=eV=0}^{}.
\end{align}
The density of states of the impurity electrons at equilibrium is given by 
\begin{align}
 \rho_{d\sigma}^{}(\omega) \,\equiv\, -\,\frac{1}{\pi} \,
\mathrm{Im}\,  G_{\mathrm{eq},\sigma}^{r}(\omega)  \;.
\label{eq:A_eq_def}
\end{align}
We will write the value at the Fermi energy $\omega=0$ 
as $\rho_{d\sigma}^{} \equiv   \rho_{d\sigma}^{}(0)$
hereafter,  suppressing the frequency argument.  
 It can also be expressed, as 
$\rho_{d\sigma}^{} = {\sin^2 \delta_{\sigma}}/{\pi \Delta}$,  
in terms of  the phase shift $\delta_{\sigma}$,  defined by 
\begin{align}
\delta_{\sigma} 
\,=\,    
\cot^{-1} \left[
\frac{\epsilon_{d\sigma}^{} 
+ \Sigma_{\mathrm{eq},\sigma}^{r}(0)}{\Delta} \right].
\label{eq:phase_shift_def}
\end{align}


At zero temperature $T=0$,  
the Friedel sum rule relates the phase shift to  the occupation number of the impurity level 
which can also be deduced from the free energy 
  $\Omega \equiv - T \log \left[\mathrm{Tr}\,e^{-\mathcal{H}/T}\right]$,   
\begin{align}
&\langle n_{d\sigma} \rangle_{\mathrm{eq}}^{} \,=\,
\frac{\partial \Omega}{\partial \epsilon_{d\sigma}^{}} 
 \,\xrightarrow{\,T\to 0\,} \,  
\frac{\delta_{\sigma}}{\pi} 
\;.
\end{align}

Some of the Fermi-liquid effects beyond 
the constant energy shift $ \Sigma_{\mathrm{eq},\sigma}^{r}(0)$ 
can be deduced from the linear susceptibilities,   
 as shown by Yamada-Yosida and  Shiba:
\cite{YamadaYosida2,YamadaYosida4,ShibaKorringa}
\begin{align}
\chi_{\sigma\sigma'}^{}\,    \equiv  &   \   
\int_0^{1/T}  \!\! d \tau \, 
\left\langle  \delta n_{d\sigma}(\tau)\,
\delta  n_{d\sigma'}\right\rangle_{\mathrm{eq}}^{}     
 \ =    \,  
- \frac{\partial^2 \Omega}
{\partial \epsilon_{d\sigma'}^{} \partial \epsilon_{d\sigma}^{}} 
\nonumber 
\\
 =  & \ 
 - \,\frac{\partial \langle n_{d\sigma} \rangle_{\mathrm{eq}}^{}}
 {\partial \epsilon_{d\sigma'}} 
\ \xrightarrow{\,T\to 0\,}    \,   
  \rho_{d\sigma}^{} \, \widetilde{\chi}_{\sigma\sigma'}^{} ,  
\label{eq:chi_org}
\rule{0cm}{0.6cm}
\\
\widetilde{\chi}_{\sigma\sigma'} 
\equiv & \   
\delta_{\sigma\sigma'}  \!+\!
\frac{\partial   \Sigma_{\mathrm{eq},\sigma}^{r}(0)}{\partial \epsilon_{d\sigma'}}
\,.
\rule{0cm}{0.75cm}
\end{align}
Here,  $\delta n_{d\sigma} \equiv   n_{d\sigma} - \langle n_{d\sigma}  
\rangle_{\mathrm{eq}}^{}$, and the reciprocal relations hold 
between the components:  
 $\chi_{\sigma\sigma'}^{}=\chi_{\sigma'\sigma}^{}$.  


For a complete description of the  FL corrections up to next-leading order, 
 the static nonlinear susceptibilities are also necessary,
\cite{MoraMocaVonDelftZarand,FilipponeMocaWeichselbaumVonDelftMora,ao2017_1_PRL,ao2017_2_PRB,ao2017_3_PRB}   
\begin{align}
\chi_{\sigma_1\sigma_2\sigma_3}^{[3]} \equiv & \  
- \!
\int_{0}^{\frac{1}{T}} \!\!\! d\tau_3 \!\! 
\int_{0}^{\frac{1}{T}} \!\!\! d\tau_2\, 
\langle T_\tau 
\delta n_{d\sigma_3} (\tau_3) \,
\delta n_{d\sigma_2} (\tau_2) \,
\delta n_{d\sigma_1}
\rangle_{\mathrm{eq}}^{} 
\nonumber \\
  = & \ 
- \,
\frac{\partial^3 \Omega }{\partial \epsilon_{d\sigma_1}\partial \epsilon_{d\sigma_2}\partial \epsilon_{d\sigma_3}} 
\,=\,  \frac{\partial \chi_{\sigma_2\sigma_3}}
{\partial \epsilon_{d\sigma_1}}\,, 
\label{eq:canonical_correlation_3}
\end{align}
where $T_\tau$ is the imaginary-time ordering operator.
These three-body correlation functions become finite 
when the PH or TR symmetry is broken. 
They contribute to 
the next-leading-order terms of the transport coefficients, 
together with  the derivative of the density of states:       
\begin{align}
\rho_{d\sigma}'  \equiv  \left.
\frac{\partial \rho_{d\sigma}^{}(\omega)}{\partial \omega} \right|_{\omega=0}^{} 
\, = \,  
\frac{\sin 2\delta_{\sigma}}{\Delta} \,
 \widetilde{\rho}_{d\sigma}^{} 
.  
\label{eq:rho_d_omega_2}
\end{align}
Here, 
$\widetilde{\rho}_{d\sigma}^{} \equiv   
 \rho_{d\sigma}^{}/z_{\sigma}^{}$  
 is the renormalized density of states for quasiparticles, with 
$z_{\sigma}^{}$ the wavefunction renormalization factor, 
\begin{align}
\frac{1}{z_{\sigma}^{}}  
\,\equiv \,      
1  -   \left.
 \frac{\partial  \Sigma_{\mathrm{eq},\sigma}^{r}(\omega)}{\partial \omega}
\right|_{\omega= 0} 
\,.
\label{eq:chi_tilde}
\end{align}

 \subsection{Fermi-liquid relations}

Why can the transport coefficients are determined by   
the static linear  and nonlinear susceptibilities at low energies 
as listed in TABLES \ref{tab:CV_CT} and \ref{tab:CS}  
 while they are originally defined 
 in terms of time-dependent correlation functions as 
  Eqs.\ \eqref{eq:current_formula} and \eqref{eq:S_noise}?
It is owing to the current conservation law 
described  in Eq.\ \eqref{eq:current_conservation}, 
from which the Ward identity 
can be deduced for various kinds of the Green's functions.
The Ward identities for the Matsubara Green's function 
have systematically been explored by Yoshimori\cite{Yoshimori} 
in the limit of $T \to 0$, 
which can be rewritten into the following form 
as a relation between the causal self-energy
$\Sigma_{\mathrm{eq},\sigma}^{--}$
and vertex function $\Gamma_{\sigma\sigma';\sigma'\sigma}^{--;--}$
of the zero-temperature formalism,
\begin{align}
\delta_{\sigma\sigma'} \frac{\partial \Sigma_{\mathrm{eq},\sigma}^{--}(\omega) }{\partial \omega} 
+ \frac{\partial \Sigma_{\mathrm{eq},\sigma}^{--}(\omega) }{\partial \epsilon_{d\sigma'}^{}} 
 = \, 
- 
\Gamma_{\sigma\sigma';\sigma'\sigma}^{--;--}(\omega,0;0,\omega) 
\,
\rho_{d\sigma'}^{} . 
\label{eq:YYY_T0_causal} 
\end{align}
At $T=0$,  the retarded self-energy is related to the causal one, 
as   
$\Sigma_{\mathrm{eq},\sigma}^{r}(\omega)
 =  
 \theta(\omega) \, \Sigma_{\mathrm{eq},\sigma}^{--}(\omega)
 +\theta(-\omega) \left\{\Sigma_{\mathrm{eq},\sigma}^{--}(\omega)\right\}^*$
with  $\theta(\omega)$ the Heaviside step function.

The diagonal  $\sigma=\sigma'$ component of the identity  
Eq.\ \eqref{eq:YYY_T0_causal} 
gives the Yamada-Yoisda relation, 
$1/z_{\sigma}^{}   =  \widetilde{\chi}_{\sigma\sigma}$, 
\cite{YamadaYosida2,Yoshimori}   
which follows from the property 
$\Gamma_{\sigma\sigma;\sigma\sigma}^{--;--}(0 , 0; 0 , 0)=0$. 
Thus, the $T$-linear specific heat of the impurity electrons 
 can be expressed  in terms of the diagonal susceptibility, as        
  $\mathcal{C}_\mathrm{imp}^{} = \frac{\pi^2}{3}
\sum_{\sigma} \chi_{\sigma\sigma}^{}\, T$.

Similarly,  at  $\omega=0$,  
 the off-diagonal $\sigma\neq \sigma'$ component 
of Eq.\ \eqref{eq:YYY_T0_causal} takes the form, 
\begin{align}
\widetilde{\chi}_{\sigma\sigma'} \, = 
 \,- \,\Gamma_{\sigma\sigma';\sigma'\sigma}^{--;--}(0, 0; 0, 0) \,
\rho_{d\sigma'}^{}\,.
\label{eq:YY2_results}
\end{align}
This shows that the susceptibilities  
determine the residual interaction between quasiparticles,\cite{HewsonRPT2001} 
\begin{align}
\widetilde{U}_{\sigma\sigma'} ^{} \equiv &  \ z_{\sigma}^{}z_{\sigma'}^{}  
 \Gamma_{\sigma\sigma';\sigma'\sigma}^{--;--}(0, 0; 0, 0) 
\, =  \,   
\frac{-\chi_{\sigma\sigma'}}{\chi_{\sigma\sigma}\,\chi_{\sigma'\sigma'}} 
\,.
\end{align}
Furthermore, the Wilson ratio $R_{\sigma\sigma'}^{}$ 
can be defined 
 as a dimensionless residual interaction for $\sigma\neq \sigma'$,
 \begin{align}
R_{\sigma\sigma'}^{}  -1
\equiv & \ 
\sqrt{\widetilde{\rho}_{d\sigma}^{}\widetilde{\rho}_{d\sigma'}^{}} 
\ \widetilde{U}_{\sigma\sigma'}^{}  
\, = \, 
 \frac{-\chi_{\sigma\sigma'}}{\sqrt{\chi_{\sigma\sigma}\,
\chi_{\sigma'\sigma'}}} \,.
 \label{eq:Wilson_ratio_general} 
\end{align}

 \subsection{Higher-order Fermi-liquid corrections}
\label{eq:second_derivative_real_self-energy}

The renormalized parameters described in the above 
are related to the first derivatives of the self-energy.  
In order to explore next-leading-order behaviors 
of the transport coefficients,  information about  
 order  $\omega^2$, $T^2$, and $(eV)^2$ terms of 
$\Sigma_{U,\sigma}^r(\omega)$ is also necessary.
\cite{FilipponeMocaWeichselbaumVonDelftMora,ao2017_1_PRL,ao2017_2_PRB}
Similarly, the low-energy expansion of 
$\Gamma_{\sigma\sigma';\sigma'\sigma}^{--;--}(\omega,\omega';\omega',\omega)$
up to  linear-order terms with respect to frequencies, $T$, and  $eV$ 
also includes alternative information. 
Here we briefly summarize the recent FL description for these high-order corrections,  
including some extensions carried out in this work.  
These corrections are also essential for our purpose  
to explore the low-energy behavior of the current noise.

\subsubsection{$\omega^2$ dependence of $\,\Sigma_{U,\sigma}^{--}$}

 The antisymmetry property of the vertex function 
 imposes a strong restriction on the two-quasiparticle 
collision processes: 
\begin{align}
&
\Gamma_{\sigma_1\sigma_2;\sigma_3\sigma_4}^{--;--}
(\omega_1, \omega_2; \omega_3, \omega_4) 
=   
-\Gamma_{\sigma_3\sigma_2;\sigma_1\sigma_4}^{--;--}
(\omega_3, \omega_2; \omega_1, \omega_4) 
\nonumber 
\\
&
\!\! 
=   
\Gamma_{\sigma_3\sigma_4;\sigma_1\sigma_2}^{--;--}
(\omega_3, \omega_4; \omega_1, \omega_2) 
= 
-\Gamma_{\sigma_1\sigma_4;\sigma_3\sigma_2}^{--;--} 
(\omega_1, \omega_4; \omega_3, \omega_2)  . 
\label{eq:vertex_antisymmetry}
\end{align}
The diagonal  $\sigma_1=\sigma_2=\sigma_3=\sigma_4$ components 
of the causal vertex  
vanish  $\Gamma_{\sigma\sigma;\sigma\sigma}^{--;--}(0,0;0,0) = 0$ 
at zero frequencies,  
and it has been used for the proof of Eq.\ \eqref{eq:chi_tilde}.\cite{Yoshimori}   
We have shown in the previous paper that the $\omega$-linear real parts of the 
diagonal vertex components also vanish at zero frequencies 
at $T=eV=0$,\cite{ao2017_2_PRB} as
\begin{align}
\left. \frac{\partial }{\partial \omega}
\mathrm{Re}\, 
\Gamma_{\sigma\sigma;\sigma\sigma}^{--;--}(\omega,0;0,\omega) 
\right|_{\omega\to 0}=\,0  
\;.
\label{eq:self_w2_N}
\end{align}
Order $\omega^2$ real part  of  the self-energy 
can be deduced from this information, 
taking a derivative of the both sides of Eq.\ \eqref{eq:YYY_T0_causal} 
with respect to $\omega$,  
\begin{align}
\left.
\frac{\partial^2}{\partial \omega^2}
\mathrm{Re}\,\Sigma_{\mathrm{eq},\sigma}^{--}(\omega)
\right|_{\omega \to 0}^{} 
\,=\, \frac{\partial^2 \Sigma_{\mathrm{eq},\sigma}^{r}(0)}
{\partial \epsilon_{d\sigma}^{2}}\,   
\ = \ 
 \frac{\partial \widetilde{\chi}_{\sigma\sigma}}{\partial \epsilon_{d\sigma}^{}} \,
\,.
\label{eq:self_w2}
\end{align}
It shows that the $\omega^2$ real part  is determined by  
the intra-level component of the three-body correlation function 
 $\chi_{\sigma\sigma\sigma}^{[3]}$.   
Physically,  this term of the self-energy induces    
 higher-order energy shifts for single-quasiparticle excitations.

It can also be deduced from  Eq.\ \eqref{eq:self_w2} 
that the diagonal  $\sigma =\sigma'$  components of the vertex 
function are pure imaginary 
up to linear order with respect to  $\omega$ and $\omega'$, 
and takes the following form at $T=eV=0$,\cite{ao2017_2_PRB}  
\begin{align}
\Gamma_{\sigma\sigma;\sigma\sigma}^{--;--}(\omega , \omega'; \omega', \omega) 
\,\rho_{d\sigma}^{2}
 =  
 i \pi 
\sum_{\sigma'(\neq \sigma)}
\chi_{\sigma\sigma'}^2
\,\bigl|\omega-\omega' \bigr| 
+ \cdots .
\label{eq:GammaUU_general_omega_dash_N}
\end{align}
Furthermore, using also the identity Eq.\  \eqref{eq:YYY_T0_causal}, 
the off-diagonal  $\sigma \neq \sigma'$ component can be deduced 
 at $T=eV=0$,  as 
\begin{align}
&
 \Gamma_{\sigma\sigma';\sigma'\sigma}^{--;--}(\omega, \omega'; \omega' ,\omega) 
\,\rho_{d\sigma}^{}\rho_{d\sigma'}^{}
\nonumber \\
&  \quad = \,    
-
\chi_{\sigma\sigma'}^{}
+ 
\rho_{d\sigma}^{}
\frac{\partial \widetilde{\chi}_{\sigma\sigma'}}
{\partial \epsilon_{d\sigma}^{}} \, \omega  
+ 
\rho_{d\sigma'}^{}
\frac{\partial \widetilde{\chi}_{\sigma'\sigma}}
{\partial \epsilon_{d\sigma'}^{}} \, \omega'   
 \nonumber \\ 
 & \qquad 
 + i \pi \,
\chi_{\sigma\sigma'}^2 
\Bigl(
\,\bigl|  \omega - \omega'\bigr| 
-
\,\bigl| \omega + \omega' \bigr| 
\Bigr)
+ \cdots. 
\label{eq:GammaUD_general_omega_dash_N}
\end{align}
Note that the imaginary parts of the vertex functions, 
which show  non-analytic $|\omega-\omega'|$ and $|\omega+\omega'|$ dependence,
 determine the damping of the two-quasiparticle collisions. 
\cite{Eliashberg,EliashbergJETP15,Yoshimori,ao2001PRB} 
In this paper, we also calculate all the Keldysh vertex components 
at finite  temperature and bias voltage up to terms of order $T$ and $eV$   
 in Sec.\ \ref{sec:current_conservation_Keldysh_selfenergy_vertex_in_FL_regime}
and  Appendix  \ref{sec:full_vertex_low_energy_form}.

\subsubsection{$T^2$ dependence of $\,\Sigma_{U,\sigma}^{--}$}

We have described in the previous paper 
that order $T^2$ term of  the retarded self-energy 
$\Sigma_{\mathrm{eq},\sigma}^r$ 
can be expressed in terms of the derivative of 
the causal vertex at $T=0$,\cite{ao2017_2_PRB} 
rewriting the proof of Yamada-Yosida\cite{YamadaYosida4}
in the following form, 
\begin{align}
&\Sigma_{\mathrm{eq},\sigma}^r(0) 
-
\left.
\Sigma_{\mathrm{eq},\sigma}^r(0) \right|_{T=0}^{} 
\, = \, \frac{(\pi  T)^2}{6}\, 
\lim_{\omega \to 0^+}
\Psi_{\sigma}^{--}(\omega)   +  \cdots \;,
\label{eq:Psi_result_T2}
\\
&\Psi_{\sigma}^{--}(\omega) 
\,\equiv  \,  
 \lim_{\omega' \to 0}
\frac{\partial}{\partial \omega'} \, 
 \sum_{\sigma'}\,
\Gamma_{\sigma \sigma';\sigma' \sigma}^{--;--}
(\omega , \omega'; \omega' , \omega) 
\rho_{d\sigma'}^{}(\omega') \;.
\label{eq:Psi_T0}
 \end{align}
The derivative with respect to $\omega'$ in the right-hand side 
can explicitly be carried out, using the results of the low-energy asymptotic form 
 Eqs.\ \eqref{eq:GammaUU_general_omega_dash_N} and 
\eqref{eq:GammaUD_general_omega_dash_N} at finite $\omega$, 
and then the coefficient for order $T^2$ term follows by taking 
the limit $\omega \to 0$, 
\begin{align}
\lim_{\omega \to 0} \Psi_{\sigma}^{--}(\omega)   =  
 \frac{1}{\rho_{d\sigma}^{}} 
\sum_{\sigma'(\neq \sigma)}
\left[
\frac{\partial \chi_{\sigma\sigma'}}{\partial \epsilon_{d\sigma'}} 
 - i \,3\pi 
\frac{1}{\rho_{d\sigma}^{}}
\chi_{\sigma\sigma'}^2
\,  \mbox{sgn}(\omega) 
\right]
\label{eq:Psi_result_+_N}
.
\end{align}
The non-analytic $\mathrm{sgn}(\omega)$ dependence, 
 appearing in the right-hand side,  
reflects the branch cut of the vertex function 
$\Gamma_{\sigma \sigma';\sigma' \sigma}^{--;--}
(\omega , \omega'; \omega' , \omega)$
along the lines  $|\omega-\omega'|=0$ and  $|\omega+\omega'|=0$ 
in the $\omega$-$\omega'$ plane.
\cite{Eliashberg,EliashbergJETP15,Yoshimori,ao2017_2_PRB}

\subsubsection{
Retarded self-energy up to order $\omega^2$, $T^2$, and $(eV)^2$
 }
\label{subsubsec:result_self_energy}

Low-energy behavior of the retarded self-energy under a finite bias voltage 
has been determined previously up to terms of order $(eV)^2$  
as well as order $\omega^2$ and $T^2$ terms, mentioned above. 
\cite{ao2017_3_PRB,TerataniSakanoOguri2020}  
We will describe alternative derivation later in Sec.\ \ref{sec:Keldysh_formulation} 
in an extended way,  using the generalized Ward identity which 
we have obtained in the present work at finite $eV$ and $T$. 
Here we present the low-energy asymptotic form of the retarded self-energy, 
 extended to multilevel quantum dots with $N>2$,
\begin{widetext}
\begin{subequations}
\begin{align}
\mathrm{Re}\, \Sigma_{U,\sigma}^r(\omega) 
\,  = & \ \ 
\Sigma_{\mathrm{eq},\sigma}^r(0)  
\,+ \bigl( 1-\widetilde{\chi}_{\sigma\sigma} \bigr)\, \omega 
\, + \frac{1}{2}\,\frac{\partial \widetilde{\chi}_{\sigma\sigma}}
{\partial \epsilon_{d\sigma}^{}}\, \omega^2 
 \, +  \frac{1}{6}\,
 \frac{1}{\rho_{d\sigma}^{}} 
\sum_{\sigma'(\neq \sigma)}
\frac{\partial \chi_{\sigma\sigma'}}{\partial \epsilon_{d\sigma'}^{}} 
\left[
\frac{3\Gamma_L \Gamma_R}{\left( \Gamma_L + \Gamma_R \right)^2} 
 \,(eV)^2 
+
 \left( \pi T\right)^2 
\right] 
\nonumber \\ 
 &  
-
\sum_{\sigma'(\neq \sigma)}
\widetilde{\chi}_{\sigma\sigma'}^{} 
\,  \alpha_\mathrm{sh}\,eV 
\,+ 
\sum_{\sigma'(\neq \sigma)}
\frac{\partial \widetilde{\chi}_{\sigma\sigma'}}{\partial \epsilon_{d\sigma}^{}} 
\,\alpha_\mathrm{sh}\,eV \, \omega
\, +  \frac{1}{2}\,
\sum_{\sigma' (\neq \sigma)} 
\sum_{\sigma'' (\neq \sigma)} 
\frac{\partial\widetilde{\chi}_{\sigma\sigma'}}{\partial\epsilon_{d\sigma''}^{}}
\, \alpha_\mathrm{sh}^2 \, (eV)^2 
\ + \,\cdots
\label{eq:self_real_ev_mag_N}
\;, \\
\mathrm{Im}\, \Sigma_{U,\sigma}^r(\omega) 
\,  = & 
\ -\,   \frac{\pi}{2}\,   
\frac{1}{\rho_{d\sigma}^{}}
\sum_{\sigma' (\neq \sigma)} \chi_{\sigma\sigma'}^2
\,
   \left[\,
    \left(\,\omega -   \alpha_\mathrm{sh}\, eV\,  \right)^2 
 + \frac{ 3\,\Gamma_L \Gamma_R}{\left( \Gamma_L + \Gamma_R \right)^2} \,(eV)^2 
 +(\pi T)^2  
    \,\right] \ + \, \cdots \;, 
\label{eq:self_imaginary_N}
\rule{0cm}{1cm}
\end{align}
 \label{eq:self_ev_mag}
\end{subequations}
\end{widetext}
and $\alpha_\mathrm{sh} \equiv 
 (\alpha_L \Gamma_L - \alpha_R \Gamma_R)/(\Gamma_L+ \Gamma_R)$. 
The parameters  $\alpha_L$ and $\alpha_R$ are defined 
in Eq.\ \eqref{eq:chemical_potentials}, 
and specify the position of the chemical potentials of 
the left  and right leads, respectively,  
with a constraint $\alpha_L+\alpha_R=1$. 
For multilevel quantum dots $N>2$, 
the nonlinear susceptibilities $\chi_{\sigma\sigma'\sigma''}^{[3]}$  
between electrons occupying three different levels  
($\sigma\neq \sigma' \neq \sigma'' \neq \sigma$)  
give finite contributions to order  $(eV)^2$ part of  
$\mathrm{Re}\, \Sigma_{U,\sigma}^r(\omega)$ 
in the case where  $\alpha_\mathrm{sh}\neq 0$. 
In contrast, 
for $\alpha_\mathrm{sh}= 0$, 
the correlations  $\chi_{\sigma\sigma'\sigma'}^{[3]}$  
between three electrons in two different levels  ($\sigma \neq \sigma'$)  
determine order $(eV)^2$ and $T^2$ terms of the real part. 
Contributions of these three-body correlation functions play 
an important role when the PH or TR symmetry is broken by external fields, 
such as  gate voltages and magnetic fields.

\section{Nonlinear current noise in the Fermi liquid regime}
\label{sec:noise_result_first}

We describe here properties of the current noise in the Fermi-liquid regime,  
leaving the complete proof later in Sec.\ \ref{sec:noise_derivation}   
as the derivation needs precise information 
about the  low-energy behavior of the Keldysh vertex corrections 
which we will shown in 
Secs.\ \ref{sec:Keldysh_formulation}--\ref{sec:current_conservation_Keldysh_selfenergy_vertex_in_FL_regime} 
and Appendix \ref{sec:full_vertex_low_energy_form}.

\subsection{Conductance and thermal noise}

We consider first the  differential conductance  $dJ/dV$ that 
can be deduced up to terms of order   $T^2$ and $(eV)^2$   
from the Landauer-type formula Eq.\ \eqref{eq:current_formula},   
using the low-energy expansion of 
$\Sigma_{U,\sigma}^{r}(\omega)$ given in Eq.\ \eqref{eq:self_ev_mag}.
Specifically,  for symmetric junctions 
with $\Gamma_L=\Gamma_R$ and $\mu_L=-\mu_R=eV/2$,  
the asymptotic form spectral function  is given by 
\begin{align}
&- \Delta \, \mathrm{Im}\,G_{\sigma}^{r}(\omega)  
\ = \ 
\sin^2 \delta_{\sigma}
+ \pi \sin 2\delta_{\sigma}\, \chi_{\sigma\sigma} 
\,\omega
\nonumber \\
&  
+  
\pi^2
\left[
\cos 2\delta_{\sigma}
\left(
\chi_{\sigma\sigma}^2
+ 
\frac{1}{2} \!
\sum_{\sigma'(\neq \sigma)}\chi_{\sigma\sigma'}^2 \! 
\right)
- 
\frac{\sin 2\delta_{\sigma}}{2\pi} \chi_{\sigma\sigma\sigma}^{[3]}
\right]  \omega^2 
\rule{0cm}{1cm}
\nonumber \\
&
+ 
\frac{\pi^2}{3} \!
\left[\,
\frac{3}{2}\cos 2 \delta_{\sigma}
\sum_{\sigma'(\neq \sigma)}\chi_{\sigma\sigma'}^2
- 
\frac{\sin 2\delta_{\sigma}}{2\pi}\,
\sum_{\sigma'(\neq \sigma)}
 \chi_{\sigma\sigma'\sigma'}^{[3]}
\,\right]
\rule{0cm}{0.9cm}
\nonumber \\
& \qquad  
\times 
\left\{\,\frac{3}{4}
\left(eV\right)^2   + \left(\pi T\right)^2
\, \right\}  
\ \  + \ \ \cdots .
\rule{0cm}{0.8cm}
\label{eq:A_including_T_eV_N_orbital}
\end{align}
 TABLE \ref{tab:CV_CT} shows the explicit expressions of 
 the coefficients $c_{T,\sigma}^{}$ and $c_{V,\sigma}^{}$ 
 for the next-leading order terms of $dJ/eV$,  
which are applicable to a wide class of quantum dots 
without the PH or TR symmetries: 
\begin{align}
\frac{dJ}{dV}  =
\frac{e^2}{h}   \sum_{\sigma=1}^N 
\left[\,
\sin^2 \delta_{\sigma}  
\,-  c_{T,\sigma}^{}\left(\pi T\right)^2
\,-  c_{V,\sigma}^{} \left(eV \right)^2  + \cdots 
\right] .
\label{eq:c_V_multi}
\end{align}

The first two terms in the right-hand side of Eq.\ \eqref{eq:c_V_multi} correspond to 
the linear conductance.   
They also determine the thermal current noise at equilibrium,   
which can be deduced from 
 the fluctuation-dissipation theorem,\cite{Hershfield2}
\begin{align}
& 
\!\!\!\!\!\!\! 
\left. S_\mathrm{noise}^{\mathrm{QD}} \right|_{V=0}\,=  \, 
4T \, \left.\frac{dJ}{dV}\right|_{V=0}^{}
\nonumber \\ 
& 
\qquad =\, 
 \frac{4e^2}{h} 
 \,T
\sum_{\sigma}
\left[
\sin^2 \delta_\sigma^{}
- c_{T,\sigma}^{}  
\left( \pi T\right)^2
 + \cdots
\right] .
\label{eq:thermal_noise}
\end{align}
We will  also describe a diagrammatic derivation 
of the equilibrium noise in the Fermi-liquid regime 
in Sec.\ \ref{subsec:thermal_noise_derivation} as a preparation 
for calculating nonlinear current noise at finite $eV$.

\subsection{Nonlinear current noise}
\label{subsec:noise_NRG_results}

We will calculate $S_\mathrm{noise}^\mathrm{QD}$  up to terms 
of order $|eV|^3$  later in Sec.\ \ref{sec:noise_derivation},  
using the diagrammatic representations 
shown in Fig.\ \ref{fig:Kubo_Keldysh_diagram}. 
In order to carry it out, 
the Keldysh Green's functions $G_{\sigma}^{\nu'\nu}(\omega)$ 
should also  be expanded up to terms of order $\omega^2$ and $(eV)^2$. 
Furthermore, all components of the Keldysh vertex function       
$\Gamma_{\sigma\sigma';\sigma'\sigma}^{\nu_1\nu_2;\nu_3\nu_4}
(\omega,\omega';\omega'\omega)$ are needed to be determined up to 
linear order terms with respect to $\omega$, $\omega'$ and $eV$. 
Leaving all these details in Secs.\ \ref{sec:Keldysh_formulation}--
\ref{sec:current_conservation_Keldysh_selfenergy_vertex_in_FL_regime}, 
we discuss here the result of the nonlinear current noise, 
is obtained for symmetric junctions with  
 $\Gamma_L=\Gamma_R$ and $\mu_L=-\mu_R= eV/2$: 
\begin{align}
\left. S_\mathrm{noise}^{\mathrm{QD}}\right|_{T=0}=   
  \frac{2e^2}{h} 
|eV| 
\sum_{\sigma}
\left[\frac{\sin^2 2\delta_\sigma^{}}{4} \,  
  + c_{S,\sigma}^{} \left(eV\right)^2 +  \cdots 
\right] .
 \end{align}
As shown in TABLE \ref{tab:CS}, 
the coefficient $c_{S,\sigma}^{}$ can be expressed 
in terms of the linear and nonlinear susceptibilities similarly to 
the other coefficients $c_{T,\sigma}^{}$ and  $c_{V,\sigma}^{}$. 
Therefore, the nonlinear Fano factor 
which is defined as the ratio of order $(eV)^3$ current noise 
 to the nonlinear current of the same order
\cite{MoraMocaVonDelftZarand,MoraSUnKondoII}   
can be expressed in the following form,
\begin{align}
F_K^{} 
\,\equiv & \ \,    
\lim_{|eV| \to 0} 
\frac{ S_\mathrm{noise}^\mathrm{QD} 
-  \frac{2e^2 |eV|}{h}\, \sum_{\sigma}\frac{\sin^2  2\delta_{\sigma}}{4} }
{- 2 |e| \, \left( J  - \frac{e^2 V }{h}\, 
\sum_{\sigma}\sin^2  \delta_{\sigma} \right)} 
\nonumber \\ 
 =  & \  \, 
\frac{ \sum_{\sigma}c_{S,\sigma}^{}}{ \sum_{\sigma}c_{V,\sigma}^{}/3}\,.
\label{eq:Fano_SUN_Lett}
\end{align}


\begin{table*}[t]
\caption{The coefficients  $C$'s for the transport coefficients given in
Eq.\ \eqref{eq:C_and_W_N2_mag_half-filling_QD}  
at finite magnetic fields $b$ for $N=2$ at half filling $\epsilon_d^{}=-\frac{U}{2}$. 
Dimensionless three-body correlation functions are defined as 
$\Theta_\mathrm{I}^{b}
\equiv 
- \frac{\sin (\pi  m_{d}) }{2\pi } 
\,\frac{\chi_{\uparrow\uparrow\uparrow}^{[3]}}{\chi_{\uparrow\uparrow}^2}$, 
and $ 
\Theta_\mathrm{II}^{b} 
\equiv  
 - \frac{\sin (\pi  m_{d}) }{2\pi} 
\frac{\chi_{\uparrow\downarrow\downarrow}^{[3]}}{\chi_{\uparrow\uparrow}^2}$.
}  
\begin{tabular}{l|l} 
\hline \hline
\ $C_{S}^{b} 
\ =  \,   
\frac{\pi^2}{192} 
\left[\, W_S^{b}  
+
\left(
\Theta_\mathrm{I}^{b}
+3 \Theta_\mathrm{II}^{b}
\right)  \, \cos (\pi  m_{d})
\,\right]$ \ \ \ \ \  \ \ 
& \ \ \ 
$
W_S^{b} 
 \,   \equiv  \, 
\cos  (2\pi m_d^{}) 
+
\bigl[\,4+5 \cos  (2\pi m_d^{}) \,\bigr] (R-1)^2
 $ 
\rule{0cm}{0.45cm}
\\
\ $
C_{V}^{b} 
\,=  \, 
\frac{\pi^2}{64}
 \,\bigl[
\,
W_{V}^{b} 
\,+ \,
\Theta_\mathrm{I}^{b} 
+
3\,
\Theta_\mathrm{II}^{b} 
\, \bigr]$ 
& \  \ \ 
$
W_{V}^{b} \,\equiv \, 
\left[ \,1+  5\left(R-1\right)^2 \,\right]\cos  (\pi m_d) 
 $ 
\rule{0cm}{0.45cm}
\\
\ $
C_{T}^{b} 
\,=  \,  
\frac{\pi^2}{48} 
\,\bigl[\,
W_{T}^{b} 
\,+\, 
\Theta_\mathrm{I}^{b} 
+
\Theta_\mathrm{II}^{b} 
\,\bigr]
 \quad$
& \ \ \  
$
W_{T}^{b} 
\ \equiv \, 
\left[\,1+  2\left(R-1\right)^2 \,\right]\cos  (\pi m_d)
$
\rule{0cm}{0.45cm}
\\
\ $
C_{\kappa,b}^\mathrm{QD} 
=  \,  
\frac{7\pi^2}{80} 
\,\bigl[\,
W_{\kappa,b}^\mathrm{QD} 
\,+\, 
\Theta_\mathrm{I}^{b} 
+
\frac{5}{21}
\Theta_\mathrm{II}^{b} 
\,\bigr]
$ 
  &  
\ \ \  
$ 
W_{\kappa,b}^\mathrm{QD} 
\, \equiv  \, 
\left[\, 1
 + \frac{6}{7}
 \left( R -1 \right)^2 \,\right]  \cos  \left(\pi m_d\right) 
$ 
\rule{0cm}{0.45cm}
\\
\hline
\hline
\end{tabular}
\label{tab:C_and_W_N2_mag_half-filling_QD}
\end{table*}


\subsubsection{Current noise in SU($N$) symmetric case}

In a special case 
where  the impurity level becomes degenerate 
$\epsilon_{d\sigma}^{} \equiv \epsilon_d^{}$ 
for all components $\sigma =1,2,\ldots,N$, 
the Hamiltonian $\mathcal{H}$  given in  Eq.\ \eqref{eq:Hamiltonian_general} 
has the  SU($N$) symmetry.
In this case,  each of the levels equally contributes to the transport, 
and thus the nonlinear Fano Factor takes the form 
$F_K^{} =   c_{S,\sigma}^{}/(c_{V,\sigma}^{}/3)$,  
with the coefficients that become independent of $\sigma$, 
\begin{align}
& \! 
c_{S,\sigma}^{}
=  \frac{\pi^2\chi_{\sigma\sigma}^2}{12}\Biggl[
\, \cos 4 \delta\,
\nonumber \\
& \qquad \quad  \ + 
\left\{
 4+5\cos 4 \delta  + 
\frac{3}{2}\bigl(1- \cos 4\delta\bigr) (N-2)
\right\}  \frac{\widetilde{K}^2}{N-1}
\nonumber \\
& \qquad \quad  \ 
-\,
 \cos 2\delta\,\Bigl\{
  \Theta_\mathrm{I}^{}
+ 3 (N-1) \Theta_\mathrm{II}^{}
\Bigr\}
\ \Biggr] ,
\label{eq:Cs_SUN}
\\
&
c_{V,\sigma}
= 
\frac{\pi^2\chi_{\sigma\sigma}^2}{4}\!
\Biggl[ \,
-\left(1+  
\frac{5\,\widetilde{K}^2}{N-1}
\right)
\cos 2 \delta 
\nonumber \\
& \qquad \qquad \qquad \ \  
+
  \Theta_\mathrm{I}^{}
 +
 3 (N-1) \Theta_\mathrm{II}^{}
\,\Biggr] .
\label{eq:Cv_SUN}
\end{align}
Here, $\widetilde{K}\equiv (N-1)(R-1)$ is a rescaled Wilson ratio: 
$R = 1-\chi_{\sigma\sigma'}^{}/\chi_{\sigma\sigma}^{}$ for $\sigma'\neq \sigma$.
The diagonal susceptibility $\chi_{\sigma\sigma}^{}$
in the right-hand side of these equations determines 
a  characteristic energy scale 
  $T^*  \equiv 1/(4\chi_{\sigma\sigma}^{})$.  
The dimensionless parameters 
 $\Theta_\mathrm{I}^{}$ and $\Theta_\mathrm{II}^{}$  
are introduced for the three-body correlation functions  
for the diagonal and off-diagonal components, respectively,  
\begin{align}
\Theta_\mathrm{I}^{} 
\,\equiv& \ 
\frac{\sin 2\delta}{2\pi}\,3
\frac{\chi_{\sigma\sigma\sigma}^{[3]}}{\chi_{\sigma\sigma}^2}\,,
\qquad \ 
  \Theta_\mathrm{II}^{} 
\,\equiv \, 
\frac{\sin 2\delta}{2\pi}\,
\frac{\chi_{\sigma\sigma'\sigma'}^{[3]} }{\chi_{\sigma\sigma}^2}\,. 
\label{eq:Theta_definition}
\end{align}
The formula for $c_{S,\sigma}^{}$ 
given in Eq.\ \eqref{eq:Cs_SUN} reproduces 
the previous result as a special case for $N=2$,
obtained by Mora, Moca, {\it et al\/} 
for the spin-1/2 Anderson model at zero magnetic field. 
\footnote{
Eq.\ (11) of  Ref.\ \onlinecite{MoraMocaVonDelftZarand}.
Their notation and our one correspond to each other such that 
$
\alpha_{\sigma}^{(1)}/\pi=    
  \chi_{\sigma\sigma}^{}$,
$
\phi_{\sigma\sigma'}^{(1)}/\pi= 
 - \chi_{\sigma\sigma'}^{}$, 
$\alpha_{\sigma}^{(2)}/\pi 
 =  
-\frac{1}{2}\, 
\chi_{\sigma\sigma\sigma}^{[3]}$, and 
$\phi_{\sigma\sigma'}^{(2)}/\pi 
=
 2 \, 
\chi_{\sigma\sigma'\sigma'}^{[3]}
$ for $\sigma'\neq \sigma$.}

The nonlinear Fano factor for the SU($N$) symmetric case is 
given by Eqs.\  \eqref{eq:Fano_SUN_Lett}--\eqref{eq:Cv_SUN}. 
It also reproduces the previous result,\cite{SakanoFujiiOguri} 
 obtained specificaly for the particle-hole symmetric case   
at which  $\delta = \pi/2$  and all the three-body correlation functions   
vanish $\chi_{\sigma\sigma'\sigma''}^{[3]} = 0$,
\begin{align}
F_K^{} 
\,\xrightarrow{\, \epsilon_d  \to -\frac{(N-1)U}{2}\,}\,
\frac{1+\frac{9\widetilde{K}^2}{N-1}}{1+\frac{5\widetilde{K}^2}{N-1}} 
\,\xrightarrow{\, U \to \infty\,}\,
 \frac{1+\frac{9}{N-1} }{1+\frac{5}{N-1}}  
\,. 
\label{eq:FK_SUN_half_filling}
\end{align}

 In the strong coupling limit $U\to \infty$, 
the occupation number  $N_d^{} \equiv \sum_\sigma \langle n_{d\sigma}^{}\rangle$ 
takes  integer values  $M=1,\,2,\,\ldots,N-1$ 
at  $\epsilon_d^{}= -(M-1/2)U$ and the phase shift 
is also locked at  $\delta = \pi M/N$.  
In this limit, 
the charge and spin susceptibilities satisfy the stationary conditions, 
especially the charge susceptibility is suppressed 
$\chi_{\sigma\sigma}^{}+(N-1)\chi_{\sigma\sigma'}^{} \to 0$, 
and thus the scaled Wilson ratio approaches  $\widetilde{K} \to 1$. 
The three body correlation functions show the property  
 $\Theta_\mathrm{I} +(N-1)\Theta_\mathrm{II} \to 0$ 
in the strong-coupling limit,
\cite{TerataniSakanoOguri2021}   
and thus  $F_K^{}$ takes the value,  
\begin{align}
F_K^{} \,  \to  \, 
 \frac{
1+\sin^2 2\delta \,
 +  \frac{9-13\sin^2 2\delta}{N-1} 
 + 2\,\Theta_\mathrm{I}^{}\,\cos 2\delta 
}{
-
\left[ \, 1+ \frac{5}{N-1}\,\right]
\cos 2\delta
-2  \, \Theta_\mathrm{I}^{}
} \;.
\end{align}
This  expression agrees with $F_K^{}$ for the SU($N$) Kondo model, 
obtained previously by Mora, Vitushinsky, {\it et al\/}.
\footnote{
  Eq.\ (51) of  Ref.\ \onlinecite{MoraSUnKondoII},  
after inserting some parenthesis for correcting minor typos}
We have calculated the three-body correlation functions of 
the SU($N$) Anderson model for $N=4,6$, 
using the NRG  in the previous paper. 
\cite{TerataniSakanoOguri2020,TerataniSakanoOguri2021}
The results show that,  for strong interactions  $U \gg \Delta$,  
the three-body correlations are characterized by  a single  parameter  
over a wide continuous filling-range  $1 \lesssim N_d^{}  \lesssim N-1$,
including the intermediate valence regions in between two 
adjacent integer-filling points  $M$ and $M + 1$ for $M=1,2,\ldots, N-2$.

\subsubsection{Current noise at finite magnetic fields}

We next provide another application of the current-noise formula 
to the case where 
the time-reversal symmetry is broken by an external magnetic field $b$.  
Specifically, we consider  the spin $1/2$ Anderson model ($N=2$) at half filling,  
taking the impurity level  to be   
 $\epsilon_{d\sigma}^{} \equiv \epsilon_d^{} - \mathrm{sgn}(\sigma)\, b$ 
with  $\epsilon_d^{}=-U/2$,  
 $\mathrm{sgn}(\uparrow)= +1$  and  $\mathrm{sgn}(\downarrow)= -1$.
In this case, the phase shift takes the form 
$\delta_{\sigma}^{} = \pi [1 + \mathrm{sgn}(\sigma)\, m_{d} ] /2$ 
with  $m_d  \equiv \langle  n_{d\uparrow} \rangle 
- \langle n_{d\downarrow}\rangle$,
and the correlation functions have the following properties:  
$\,\chi_{\uparrow\uparrow}^{}  =  \chi_{\downarrow\downarrow}^{}$,  
$\,\chi_{\uparrow\downarrow}^{}  =  \chi_{\downarrow\uparrow}^{}$,  
$\,\chi_{\downarrow\downarrow\downarrow}^{[3]} 
= -\chi_{\uparrow\uparrow\uparrow}^{[3]}$, and  
 $\,\chi_{\uparrow\uparrow\downarrow}^{[3]}
=-\chi_{\uparrow\downarrow\downarrow}^{[3]} 
$.
Therefore, 
there are  two independent three-body correlation 
functions 
$\chi_{\uparrow\uparrow\uparrow}^{[3]}$ and
$\chi_{\uparrow\downarrow\downarrow}^{[3]}$ in this case, 
which can be evaluated from the derivative of the linear susceptibilities 
 with respect to $b$, 
 \begin{align}
 \frac{\partial \chi_{\uparrow\uparrow}^{} }{\partial b}
\,=& \  -\chi_{\uparrow\uparrow \uparrow}^{[3]} 
\,+\, \chi_{\uparrow\uparrow \downarrow}^{[3]} 
\ \xrightarrow{\,\epsilon_{d} = -U/2\,} \  
 -\chi_{\uparrow\uparrow \uparrow}^{[3]} 
\,-\, \chi_{\uparrow\downarrow \downarrow}^{[3]} \,,
\nonumber \\
\frac{\partial \chi_{\uparrow\downarrow}^{} }{\partial b}
\,= & \ - \chi_{\uparrow\uparrow \downarrow}^{[3]} 
\,+\, \chi_{\uparrow\downarrow \downarrow}^{[3]} 
\ \xrightarrow{\,\epsilon_{d} = -U/2\,} \  
\,2\, \chi_{\uparrow\downarrow \downarrow}^{[3]} 
\,.
\end{align}
The transport coefficients up to the next-leading order terms 
can be described  in terms of the five FL parameters:  $m_d^{}$, 
 $T^* = 1/(4\chi_{\uparrow\uparrow}^{})$,
$R_{}^{}  = 1-\chi_{\uparrow\downarrow}^{}/ \chi_{\uparrow\uparrow}^{}$,  
and two different  three-body correlation functions, 
$\Theta_\mathrm{I}^{b}
\equiv 
- \sin (\pi  m_{d}) 
\,{\chi_{\uparrow\uparrow\uparrow}^{[3]}}/{(2\pi\chi_{\uparrow\uparrow}^2)}$,
 and 
$\Theta_\mathrm{II}^{b} 
\equiv  
 - \sin (\pi  m_{d}) \,
{\chi_{\uparrow\downarrow\downarrow}^{[3]}}
/{(2\pi \chi_{\uparrow\uparrow}^2)}$.

TABLE \ref{tab:C_and_W_N2_mag_half-filling_QD} 
shows the formulas for the coefficients of  
the low-energy expansion of the differential conductance  $dJ/dV$, 
current noise $S_\mathrm{noise}^\mathrm{QD}$, 
and thermal conductivity  $\kappa_\mathrm{QD}^{}$  for this case:
\begin{align}
&
S_\mathrm{noise}^\mathrm{QD}  
 =  
2 \,\frac{2e^2}{h}   |eV| \! 
\left[ 
\frac{\sin^2 (\pi m_d^{})}{4} 
   +  
C_S^{b}  \left( \frac{eV}{T^*}\right)^2 
\!\!  +   \cdots  
  \right] \!  ,  \! 
\nonumber 
\\
&
\!
\frac{dJ}{dV}  =   
\frac{2e^2}{h} \! 
\left[
\cos^2
 \left( \frac{\pi m_{d}}{2} \right)  \! 
- C_{T}^{b} 
\left(\frac{\pi T}{T^*}\right)^2 \!\! 
-   C_{V}^{b} 
 \left(\frac{eV}{T^*} \right)^2  \!\!  +  \cdots 
\right] \!   ,
\nonumber 
\\
&
\kappa_\mathrm{QD}^{} = 
\frac{2\pi^2 T}{3h}
  \left[
\cos^2 
 \left( \frac{\pi m_{d}}{2} \right)  
- 
 C_{\kappa,b}^\mathrm{QD} 
\left( \frac{\pi T}{T^*}\right)^2
\!+ \cdots
\right] . 
\label{eq:C_and_W_N2_mag_half-filling_QD}
\end{align}
Here, the  thermal conductivity is defined with respect to 
 the heat current  $J_Q^{}=   \kappa_\mathrm{QD}^{}\, \delta T$,        
 induced by the temperature difference  $\delta T$ between the two leads.
It is also determined by the transmission probability 
$\mathcal{T}_{\sigma}(\omega) $ defined
 in Eq.\ \eqref{eq:transmissionPB},\cite{CostiZlatic2010} as 
\begin{align}
 \kappa_\mathrm{QD}^{} \equiv  & \   
 \frac{1}{h\, T}
 \left[  
 \sum_{\sigma}
  \mathcal{L}_{2,\sigma}^{\mathrm{QD}} 
 - 
 \frac{ \left(
 \sum_{\sigma}
 \mathcal{L}_{1,\sigma}^{\mathrm{QD}}\right)^2}{\sum_{\sigma}
 \mathcal{L}_{0,\sigma}^{\mathrm{QD}}} \right]  \,, 
\\ 
\mathcal{L}_{n,\sigma}^\mathrm{QD} \, \equiv  & \ 
 \int_{-\infty}^{\infty}  
 d\omega\, 
 \omega^n \  
  \mathcal{T}_{\sigma}(\omega) \,
 \left( -
 \frac{\partial f(\omega)}{\partial \omega}
 \right)  .
 \label{eq:Ln_QD}
 \end{align}

We consider here behaviors of these coefficients 
for strong interactions $U\gg \Delta$, 
where  the system can be described by the Kondo model.
Specifically, the behavior of  $dI/dV$ in this limit 
has already been examined by Filippone {\it et al\/},
 i.e., Eq.\ (17) of Ref.\ \onlinecite{FilipponeMocaWeichselbaumVonDelftMora}, 
using a different notation.
Here we consider mainly the current $S_\mathrm{noise}^\mathrm{QD}$ and    
$\kappa_\mathrm{QD}^{}$.
In the Kondo limit, the Wilson ratio approaches  $R \to 2$ 
as the charge fluctuations are suppressed 
$\chi_{\uparrow\downarrow}^{} + \chi_{\uparrow\uparrow}^{} \to 0$. 
 Correspondingly, two different components 
of the  three-body correlation functions  take an identical value 
 $\chi_{\uparrow\downarrow \downarrow}^{[3]} \to 
\chi_{\uparrow\uparrow \uparrow}^{[3]}$, and thus  
 $\Theta_\mathrm{II}^{b} \to \Theta_\mathrm{I}^{b}$. 
Therefore, in the strong-interaction limit, 
 the transport coefficients defined in Eq.\ 
\eqref{eq:C_and_W_N2_mag_half-filling_QD}
 and TABLE \ref{tab:C_and_W_N2_mag_half-filling_QD} 
are determined by the following three FL parameters: 
 $m_d^{}$,  $\Theta_\mathrm{I}^{b}$, and  $T^*$.  
These parameters can be calculated from 
 the spin susceptibility $\chi_s^{}$ of the Kondo model 
with the Zeeman coupling  $\mathcal{H}_z^{}= -2 S_d^z\, b$  
between the impurity spin $S_d^z =(n_{d\uparrow}^{}-n_{d\downarrow}^{})/2$ 
and magnetic field $b$, 
\begin{align}
\chi_{\uparrow\uparrow}^{} 
\,\xrightarrow{U \gg \Delta\,}  & \ \ 
\frac{\chi_{s}^{}}{4}  
\, \equiv   \,  
\frac{1}{2} \frac{\partial \! \left\langle S_d^{z} \right\rangle}{\partial b}    
\,, \\
\chi_{\uparrow\uparrow \uparrow}^{[3]} 
\,\xrightarrow{U \gg \Delta\,}  & \  
-\frac{1}{8}\,  \frac{\partial \chi_s^{}}{\partial b} 
\, = \,
 -\frac{1}{4} \frac{\partial^2 \! \left\langle S_d^{z} \right\rangle}{\partial b^2}    
\;.
\end{align}
The field-dependent energy scale can be defined by
the inverse of the spin susceptibility  
 $T^* \xrightarrow{\,U\gg \Delta\,} 1/\chi_s^{}$.

\begin{figure}[t]
 \leavevmode
\centering
\includegraphics[width=0.48\linewidth]{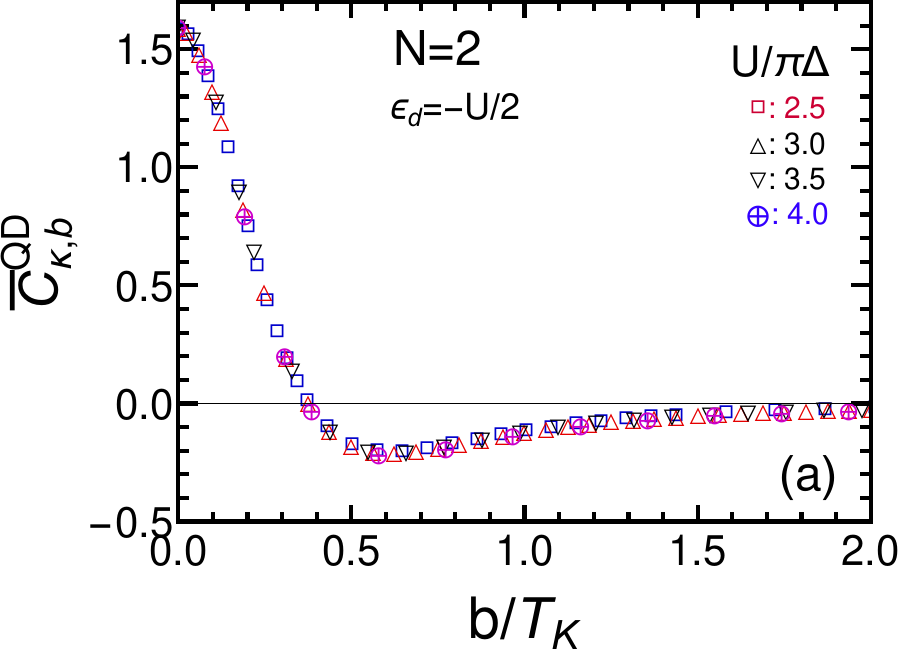}
\includegraphics[width=0.48\linewidth]{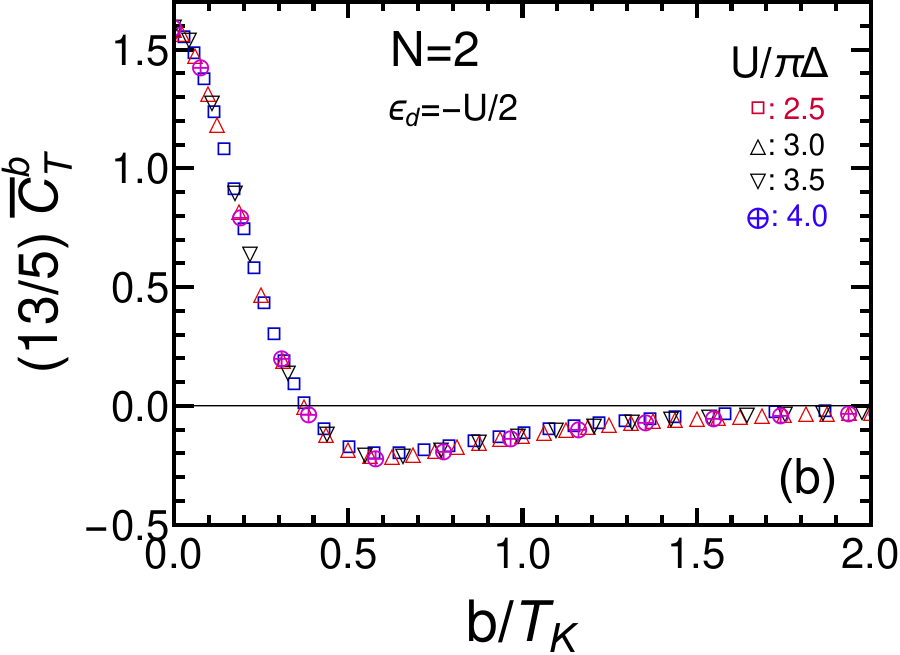}
\includegraphics[width=0.48\linewidth]{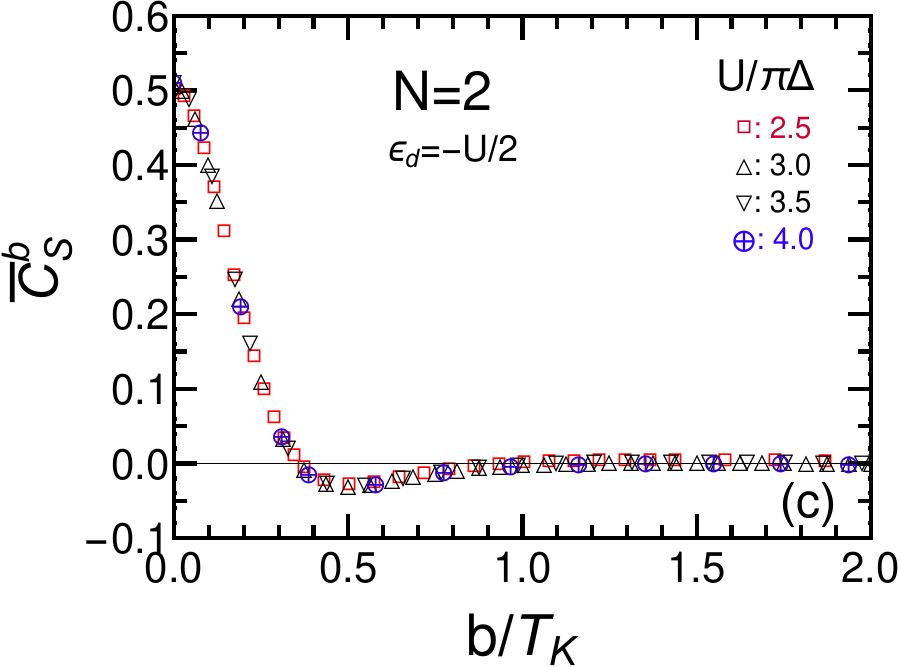}
\includegraphics[width=0.485\linewidth]{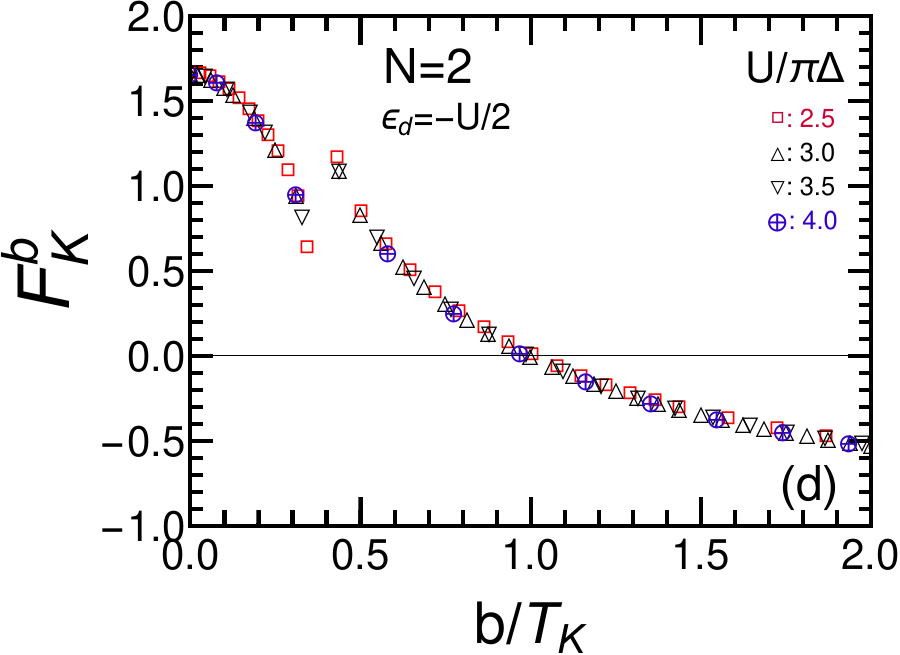}
\caption{Magnetic-field scaling of the coefficients that determine 
the next-leading-order transport of the $N=2$ Anderson impurity model.
These results are obtained at half-filling $\epsilon_d^{}=-U/2$  
for strong interactions  $U/(\pi\Delta) = 2.5, 3.0, 3.5, 4.0$, 
using the NRG approach. Figures (a)--(d) show that    
 (a):  $\overline{C}_{\kappa,b}^\mathrm{QD}  
\equiv (T_K/T^*)^2  C_{\kappa,b}^\mathrm{QD}$,
 (b): 
 $\overline{C}_{T}^{b}  
\equiv (T_K/T^*)^2  C_{T}^{b}$ multiplied by  a factor of $13/5$, 
 (c): $\overline{C}_{S}^{b} \equiv (T_K/T^*)^2  C_{S}^{b}$,  and  
 (d):  $F_K^b \equiv \overline{C}_{S}^{b}/(\overline{C}_{V}^{b}/3)$. 
The Kondo temperature is defined 
as $T_K \equiv \left. T^* \right|_{b=0}^{}$,  
at zero field with $T^* =1/(4\chi_{\uparrow\uparrow}^{})$.
}
\label{fig:spin_half_mag_kapa_T}
\end{figure}

In the Kondo limit, the nonlinear Fano factor takes 
the form\cite{TerataniSakanoOguri2021}  
\begin{align}
\frac{C_S^{b}}{C_{V}^{b}/3}  
 \,\xrightarrow{\,U\gg\Delta\,} 
\frac{4 +6\,\cos  (2\pi m_d) \,+\, 4\, \Theta_\mathrm{I}^{b}\,\cos  (\pi m_d)  
}{6\,\cos  (\pi m_d) +4\,\Theta_\mathrm{I}^{b}}
 \;.
\label{eq:Fano_2021}
\end{align}
We also find that the ratio of the $T^3$ thermal conductivity 
to the $T^2$ conductance takes a constant value  
which does not depend on magnetic field strength  $b$ or 
the Kondo coupling $J_K \propto \Delta/(\rho_c U)$,  
\begin{align}
\!\!\!\!\!\!  
\frac{C_{\kappa,b}^\mathrm{QD}}{C_{T}^{b}}
 \,\xrightarrow{\,U\gg\Delta\,} 
\  \frac{13}{5}.  
\label{eq:kappa_T_relation_QD}
\end{align}
Figure \ref{fig:spin_half_mag_kapa_T} shows 
the NRG results of these coefficients for strong interactions  $U/(\pi \Delta) \gtrsim 2.5$  
plotted against $b/T_K$. 
\footnote{NRG calculations for this case have been carried out, 
choosing the discretization parameter to be  $\Lambda=2.0$ 
and retaining typically $4000$ low-lying excited states.}
The coefficients are rescaled in a way such that 
$\overline{C}_{\kappa,b}^\mathrm{QD} 
\equiv  (T_K/T^*)^2 \,C_{\kappa,b}^\mathrm{QD}$,  
$\overline{C}_{T}^{b} \equiv (T_K/T^*)^2 \,C_{T}^{b}$, 
$\overline{C}_{S}^{b} \equiv (T_K/T^*)^2 \,C_{S}^{b}$, 
and $F_{K}^{b} \equiv C_S^{b}/(C_{V}^{b}/3)$,  
with the Kondo temperature  $T_K \equiv \left. T^*\right|_{b=0}^{}$ 
defined at zero field.  
\cite{TerataniSakanoOguri2021} 
Each of these curves shows the Kondo scaling behavior. 
We can see  that 
the universal curve for $\overline{C}_{\kappa,b}^\mathrm{QD}$ 
and that for $(15/3) \overline{C}_{T}^{b}$ are almost identical 
and overlap each other showing a good agreement with 
Eq.\ \eqref{eq:kappa_T_relation_QD}.
Note that similar relation does not hold for magnetic alloys     
for the  $T^3$ thermal conductivity 
 and the $T^2$ electrical resistivity.\cite{ao2017_3_PRB,ao2017_3_PRB_erratum_2}
Equation \eqref{eq:kappa_T_relation_QD}  
clarifies one of the FL properties 
appearing in the temperature dependence. 
The nonlinear Fano factor  
$F_K^b \equiv \overline{C}_{S}^{b}/(\overline{C}_{V}^{b}/3)$  
can be regarded as a nonequilibrium analogue. 
As the zero-points of  $\overline{C}_{S}^{b}$ and that of 
$\overline{C}_{V}^{b}$ appear at different magnetic fields,
 $F_K^b$ diverges at the zero-point of  $\overline{C}_{V}^{b}$. 

\begin{table*}[t]
\caption{Linear dependency of the Keldysh correlation functions. 
The three-point functions 
$\Phi_{\gamma,\sigma\sigma'}^{\alpha;\mu\nu}$ and 
$\Lambda_{\gamma,\sigma\sigma'}^{\alpha;\mu\nu}$ 
for $\gamma =L,R,d$ are defined in 
Eqs.\ \eqref{eq:Phi_CLR}
and \eqref{eq:Lambda_matrix}. 
Outline of the derivation  
is given in Appendix \ref{sec:linear_dependency_Keldysh}.
}
\begin{tabular}{l} 
\hline \hline
\rule{0cm}{0.5cm}
\raisebox{0.1cm}{
$
G_{\sigma}^{--} + G_{\sigma}^{++}
\,=\,G_{\sigma}^{-+} +G_{\sigma}^{+-}, 
\qquad \qquad \qquad 
\Sigma_{\sigma}^{--} + \Sigma_{\sigma}^{++}
\,=\,-\Sigma_{\sigma}^{-+} -\Sigma_{\sigma}^{+-}, 
$}
\\
\hline
$
\ 
\Phi_{\gamma,\sigma\sigma'}^{-;--}
+\Phi_{\gamma,\sigma\sigma'}^{-;++}
-\Phi_{\gamma,\sigma\sigma'}^{-;+-}
-\Phi_{\gamma,\sigma\sigma'}^{-;-+} 
\,=\,
\Phi_{\gamma,\sigma\sigma'}^{+;++}
+\Phi_{\gamma,\sigma\sigma'}^{+;--}
-\Phi_{\gamma,\sigma\sigma'}^{+;-+}
-\Phi_{\gamma,\sigma\sigma'}^{+;+-}
\,, $
\rule{0cm}{0.5cm}
\\
\raisebox{0.15cm}{
$
\Lambda_{\gamma,\sigma\sigma'}^{-;--}
+\Lambda_{\gamma,\sigma\sigma'}^{-;++}
+\Lambda_{\gamma,\sigma\sigma'}^{-;-+}
+\Lambda_{\gamma,\sigma\sigma'}^{-;+-}
\, = \, 
\Lambda_{\gamma,\sigma\sigma'}^{+;--}
+\Lambda_{\gamma,\sigma\sigma'}^{+;++}
+\Lambda_{\gamma,\sigma\sigma'}^{+;-+}
+ \Lambda_{\gamma,\sigma\sigma'}^{+;+-} 
\,,$}
\rule{0cm}{0.7cm}
\\ 
\hline
$
\Gamma_{\sigma\sigma';\sigma'\sigma}^{--;--} 
+   \Gamma_{\sigma\sigma';\sigma'\sigma}^{++;++} 
+   \Gamma_{\sigma\sigma';\sigma'\sigma}^{-+;+-}
+   \Gamma_{\sigma\sigma';\sigma'\sigma}^{+-;-+} 
 +   \Gamma_{\sigma\sigma';\sigma'\sigma}^{+-;+-} 
+   \Gamma_{\sigma\sigma';\sigma'\sigma}^{-+;-+} 
+   \Gamma_{\sigma\sigma';\sigma'\sigma}^{++;--} 
+   \Gamma_{\sigma\sigma';\sigma'\sigma}^{--;++} $
\rule{0cm}{0.45cm}
\\
\rule{0cm}{0.6cm}
\raisebox{0.1cm}{
$\, =\ 
-  \Gamma_{\sigma\sigma';\sigma'\sigma}^{+-;--} 
-  \Gamma_{\sigma\sigma';\sigma'\sigma}^{-+;--} 
-   \Gamma_{\sigma\sigma';\sigma'\sigma}^{--;+-} 
-   \Gamma_{\sigma\sigma';\sigma'\sigma}^{--;-+} 
-   \Gamma_{\sigma\sigma';\sigma'\sigma}^{-+;++} 
-   \Gamma_{\sigma\sigma';\sigma'\sigma}^{+-;++} 
-   \Gamma_{\sigma\sigma';\sigma'\sigma}^{++;-+} 
-   \Gamma_{\sigma\sigma';\sigma'\sigma}^{++;+-} \,.$
}
\\
\hline
\hline
\end{tabular}
\label{tab:Keldysh_sum_rule}
\end{table*}


\section{Keldysh Green's function for Anderson impurity}
\label{sec:Keldysh_formulation}

The current conservation is an essential requirement that 
transport theories must satisfy, and it is 
particularly important  for strongly correlated electron systems.
In this and the next section, we extend the Ward identities 
for the Keldysh Green's function which were previously obtained for  
nonequilibrium steady states 
but only for the zero-bias limit $eV \to 0$.\cite{ao2001PRB} 
Specifically,  in this section, 
we perturbatively derive the Ward identities which hold 
for finite $eV$ and can be expressed 
in terms of the Keldysh self-energy and vertex functions.  
We will also present more 
general Ward-Takahashi identity for the three-point current vertex  
which reflects directly 
the current conservation between the quantum dot and leads 
 in Sec.\  \ref{sec:nonlinear_WT}.

\subsection{General properties}

We summarize in this subsection the general properties of 
 the Keldysh Green's function of the Anderson impurity, defined by\cite{Keldysh}  
\begin{subequations}
\begin{align}
G^{--}_{\sigma} (t_1, t_2) \, \equiv&  
\, -i\, \langle \,\mbox{T}\,
d^{\phantom{\dagger}}_{\sigma}(t_1)\, d^{\dagger}_{\sigma}(t_2)
\,\rangle
\,,
\\
G^{++}_{\sigma} (t_1, t_2) \, \equiv&  
\,-i\, 
\langle \,\widetilde{\mbox{T}}\,
d^{\phantom{\dagger}}_{\sigma}(t_1)\, d^{\dagger}_{\sigma}(t_2)
\,\rangle
\,,
 \\
G^{+-}_{\sigma} (t_1, t_2) \, \equiv& 
\,-i\, 
\langle 
d^{\phantom{\dagger}}_{\sigma}(t_1)\, d^{\dagger}_{\sigma}(t_2)
\,\rangle 
\;,
\\
G^{-+}_{\sigma} (t_1, t_2) \equiv&  
\ \  i\, 
\langle 
d^{\dagger}_{\sigma}(t_2) \, d^{\phantom{\dagger}}_{\sigma}(t_1) 
\,\rangle 
\label{eq:G<}
\,,
\end{align}
\label{eq:Keldysh_Green's_function_def} \!\!\!\!\!\!\!   
\end{subequations}
where $\mbox{T}$ and $\widetilde{\mbox{T}}$ are the time-ordering and anti-time-ordering operators, respectively.
These correlation functions are defined along the Keldysh time-loop contour  
and are linearly dependent on each other, as 
 summarized in TABLE \ref{tab:Keldysh_sum_rule}.
Thus the above four Green's functions can be written in terms 
of three components: 
the  retarded  $G_{\sigma}^{r}$,  
the advanced $G_{\sigma}^{a}$, 
and the symmetrized $G_{\sigma}^\mathrm{K}$ functions, 
\begin{subequations}
\begin{align}
G_{\sigma}^{r}= & \ 
 G_{\sigma}^{--}-G_{\sigma}^{-+} ,
\label{eq:retarded_advanced_Keldysh}
\qquad \quad  
G_{\sigma}^{a}=  
 G_{\sigma}^{--}-G_{\sigma}^{+-},
\\
G_{\sigma}^\mathrm{K}\equiv  & \ 
 G_{\sigma}^{+-}+G_{\sigma}^{-+}
\,= \, 
 G_{\sigma}^{--}+G_{\sigma}^{++}\,.
\label{eq:linear_dependency_GK}
\rule{0cm}{0.3cm}
\end{align}
\end{subequations}
These  Green's functions can also be written in a matrix form,
\begin{align}
\bm{G}_{\sigma}^{}  \equiv& \    \left[ 
\begin{matrix}
G_{\sigma}^{--} & G_{\sigma}^{-+}   \cr
           G_{\sigma}^{+-} & G_{\sigma}^{++}  \cr  
\end{matrix}
                             \right]  
\label{eq:GK_def}
\\
= & \ \, 
 \frac{G_{\sigma}^{\mathrm{K}}}{2}
\left( \bm{1} + \bm{\tau}_1 \right) 
+ 
\frac{ G_{\sigma}^{r} \! + G_{\sigma}^{a}}{2} \, 
\bm{\tau}_3
- 
\frac{G_{\sigma}^{r} \! - G_{\sigma}^{a}}{2} \, 
i \bm{\tau}_2 \,. 
\rule{0cm}{0.6cm}
\nonumber 
\end{align}
Here,   $\bm{1}$ is  the $2 \times 2$ unit matrix 
and  $\bm{\tau}_j$ for $j=1,2,3$ are  
the Pauli matrices, defined by 
 \begin{align}
 \!\!\!
 \bm{\tau}_1 =
 \left[
 \begin{matrix}
 0 &   1 \cr
 1 &  0 \cr  
 \end{matrix}
 \right] , 
 \quad
 \bm{\tau}_2 =
 \left[
 \begin{matrix}
 0 &   -i \cr
 i &  \ 0  \cr  
 \end{matrix}
 \right] , 
 \quad
 \bm{\tau}_3 =
 \left[
 \begin{matrix}
 1 &  \  0 \cr
 0 &  -1 \cr  
 \end{matrix}
 \right] .
 \end{align}
Therefore, the determinant of $\bm{G}_{\sigma}^{}$ can be expressed, as 
\begin{align}
\det \left\{\bm{G}_{\sigma}^{} \right\} \,\equiv\,  
 G_{\sigma}^{--}\,G_{\sigma}^{++}-G_{\sigma}^{-+}\,G_{\sigma}^{+-} 
\,=\, 
 -\,G_{\sigma}^{r}\,G_{\sigma}^{a}, 
\label{eq:derG}
\end{align}
and this  product  $G_{\sigma}^{r}\,G_{\sigma}^{a}$  
of the retarded and advanced Green's functions 
plays an important role in the transport theory.  
We will also use later the following matrix identity 
which appears when we take the derivative of $\bm{G}_{\sigma}^{}$, 
\begin{align}
\!\!\!\!\! 
\bm{G}_{\sigma}^{}
\bigl(
\bm{1} -  \bm{\tau}_1
\bigr)
\bm{G}_{\sigma}^{}
\, = \,  
\bigl(
\bm{1} +  \bm{\tau}_1
\bigr)\, 
G_{\sigma}^{r}\,G_{\sigma}^{a}
 \,. 
\label{eq:GrGaMat}
\end{align}

The Fourier transform of the Green's function defined with respect to 
 the steady state at  $eV$ becomes a function of a single frequency $\omega$,
\begin{align}
G_{\sigma}^{\nu_1\nu_2}(\omega)= \int_{-\infty}^{\infty} 
\! dt \,e^{i\omega t} G_{\sigma}^{\nu_1\nu_2}(t,0)\,.
\end{align}
For $U=0$, it can be expressed in the following form,  
\begin{subequations}
\begin{align}
&
\!\!\!\!\!\!
G_{0\sigma}^{-+}(\omega) =  
i\, 2 \, f_\mathrm{eff}(\omega)
\,\Delta\,G_{0\sigma}^{r}(\omega)\,G_{0\sigma}^{a}(\omega) , 
\label{eq:G0_00^-+}
\\ 
&
\!\!\!\!\!\!
G_{0\sigma}^{+-}(\omega) = 
-i\, 2 \, \bigl[1-f_\mathrm{eff}(\omega)\bigr]\,\Delta\, 
        G_{0\sigma}^{r}(\omega) \,G_{0\sigma}^{a}(\omega)  , 
\label{eq:G0_00^+-}  
\\
&
\!\!\!\!\!\!
G_{0\sigma}^{--}(\omega) = 
\,G_{0\sigma}^{r}(\omega) + G_{0\sigma}^{-+}(\omega) , 
%
\label{eq:G0_00^--}
\\ 
&
\!\!\!\!\!\!
G_{0\sigma}^{++}(\omega) = 
\,-G_{0\sigma}^{a}(\omega) + G_{0\sigma}^{-+}(\omega) .
 \label{eq:G0_00^++}
\end{align}
 \label{eq:G0_Keldysh_elements} \!\!\!\!  
\end{subequations}
Here,  $ \Delta =\Gamma_L +\Gamma_R $, 
 and  the retarded  function is given by
\begin{align}
 G_{0\sigma}^{r}(\omega) \,=\,
\frac{1}{\omega-\epsilon_{d\sigma}^{}  +i (\Gamma_L+\Gamma_R)} \,. 
\end{align}
The Fourier transformed Green's functions have additional relations,   
 $G_{\sigma}^{a}(\omega) = \left\{G_{\sigma}^{r}(\omega)\right\}^*$, 
and  
  $G_{\sigma}^{++}(\omega) = -\left\{G_{\sigma}^{--}(\omega)\right\}^*$.   
Furthermore,  the greater   $G_{\sigma}^{+-}(\omega)$ 
and  lesser   $G_{\sigma}^{-+}(\omega)$ functions 
become pure imaginary.

\begin{figure}[b]
 \leavevmode
\begin{minipage}{0.84\linewidth}
 \includegraphics[width=1\linewidth]{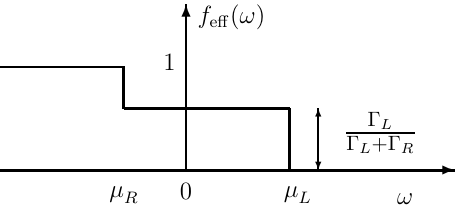}
\end{minipage}
%
%
%
%
%
%
\caption{
The nonequilibrium distribution function $f_{\mathrm{eff}}(\omega)$ 
for $\mu_L-\mu_R = eV$ and $T=0$.
The Fermi level at equilibrium 
is chosen to be the origin of single-particle energy $\omega=0$. 
}
\label{fig:distribution}
\end{figure}

The bias voltage $eV$ and temperature $T$ enter 
 the Green functions through the  distribution function 
\begin{align}
f_\mathrm{eff}(\omega) \,\equiv\,
\frac{  f_L(\omega) \,\Gamma_L 
  + f_R(\omega) \,\Gamma_R}{ 
 \Gamma_L +\Gamma_R } \, \ = \sum_{j=L,R} \frac{\Gamma_j}{\Delta}
\,f(\omega-\mu_j) .
\label{eq:f_0}
\end{align}
It shows jump discontinuities at $\omega=\mu_L$ and $\mu_R$ 
at $T=0$, as depicted in Fig.\ \ref{fig:distribution}. 
 The noninteracting Green's function can also be written in  
the following form, 
\begin{align}
\left\{\bm{G}_{0\sigma}(\omega)\right\}^{-1} 
= &   \   (\omega-\epsilon_{d\sigma}^{}) 
\, \bm{\tau}_3^{} - \bm{\Sigma}_0(\omega) \,.
\label{eq:G0_keldysh}
\end{align}
It is this noninteracting self-energy $\bm{\Sigma}_0(\omega)$ 
that brings  $f_\mathrm{eff}(\omega)$ into 
$\bm{G}_{0\sigma}(\omega)$ through  the tunneling processes, 
\begin{align}
\bm{\Sigma}_0(\omega)
\,\equiv  &  \   \ 
\bm{\tau}_3\, v_L^2\, \bm{g}_{L}^{}(\omega)\, \bm{\tau}_3
\,+ \, 
\bm{\tau}_3\, v_R^2\, \bm{g}_{R}^{}(\omega)\, \bm{\tau}_3 
\nonumber 
\\
= &\ 
-i\Delta \bigl[\,1-2f_\mathrm{eff}(\omega) \,\bigr] 
\bigl( \bm{1} -  \bm{\tau}_1^{} \bigr)
+ \Delta \bm{\tau}_2^{} \;.
\label{eq:U0_self_keldysh}
\end{align}
Here, $\bm{g}_{j}^{}(\omega)$ is the Keldysh Green's function 
for the  isolated lead on  $j =L$ and $R$,
\begin{align}
v_{j}^2 \,\bm{g}_{j}^{}(\omega) \,=  \, 
-i\Gamma_{j} \bigl[\,1-2f_{j}(\omega) \,\bigr] 
\bigl( \bm{1} +  \bm{\tau}_1^{} \bigr)
- \Gamma_{j} \bm{\tau}_2^{} \,.
\label{eq:g0_lead}
\end{align}
Note that $\bm{\Sigma}_0(\omega)$ does not depend on $\epsilon_{d\sigma}^{}$, 
and the derivative with respect to $\omega$ takes the following form, 
\begin{align}
\frac{\partial \bm{\Sigma}_0(\omega)}{\partial \omega}
\,=\, 
2i\Delta  \frac{\partial f_\mathrm{eff}(\omega)}{\partial \omega} \,
\bigl( \bm{1} -  \bm{\tau}_1^{} \bigr)\,.
\label{eq:self_U0_derivative_omega}
\end{align}

Effects of the Coulomb repulsion $U$ enter through 
the interacting self-energy,
\begin{align}
\bm{\Sigma}_{U,\sigma}^{}  =  \left[ 
\begin{matrix}
\Sigma_{U,\sigma}^{--} & \Sigma_{U,\sigma}^{-+}   \cr
           \Sigma_{U,\sigma}^{+-} & \Sigma_{U,\sigma}^{++}  \cr  
\end{matrix}
                             \right]  , 
\end{align}
with 
$\left\{\bm{G}_{\sigma}(\omega)\right\}^{-1}  = 
\left\{\bm{G}_{0\sigma}(\omega)\right\}^{-1} 
- \bm{\Sigma}_{U,\sigma}^{}(\omega)   
$.
The Dyson equation  can also be rewritten into the following form, 
using Eq.\ \eqref{eq:G0_keldysh}, 
\begin{subequations}
 \begin{align}
\left\{\bm{G}_{\sigma}(\omega)\right\}^{-1}  \,=&   \   
(\omega - \epsilon_{d\sigma}^{})\bm{\tau}_3 -
\bm{\Sigma}_{\mathrm{tot},\sigma}(\omega)\,,
\\    
\bm{\Sigma}_{\mathrm{tot},\sigma}(\omega) \,\equiv & \     
\bm{\Sigma}_0(\omega) +\bm{\Sigma}_{U,\sigma}^{}(\omega)\,.
\label{eq:matrix_self_energy_tot} 
\rule{0cm}{0.3cm}
\end{align}
\label{eq:Dyson_Keldysh} \!\!\!\!\!\!
\end{subequations}
The retarded, advanced, and symmetrized self-energies are given by  
linear combinations of the four elements of $\bm{\Sigma}_{U,\sigma}^{}$:    
\begin{subequations}
\begin{align}
& \!\!\!
\Sigma_{U,\sigma}^{r} =   
 \Sigma_{U,\sigma}^{--}+\Sigma_{U,\sigma}^{-+} ,
\label{eq:retarded_advanced_Keldysh_self}
\qquad \ \ 
\Sigma_{U,\sigma}^{a}= 
 \Sigma_{U,\sigma}^{--}+\Sigma_{U,\sigma}^{+-},
\\
& \!\!\!
\Sigma_{U,\sigma}^\mathrm{K}\,\equiv\,  
- \Sigma_{U,\sigma}^{+-}-\Sigma_{U,\sigma}^{-+}
\,=\,
 \Sigma_{U,\sigma}^{--} + \Sigma_{U,\sigma}^{++} .
\rule{0cm}{0.3cm}
\end{align}
\label{eq:self_energy_linear_dependency}
\!\!\!\!\!\!     
\end{subequations}
Note that $\Sigma_{U,\sigma}^{a}(\omega) =
\left\{\Sigma_{U,\sigma}^{r}(\omega) \right\}^*$. 
The lesser and greater self-energies, namely 
 $\Sigma_{U,\sigma}^{-+}(\omega)$ and  
$\Sigma_{U,\sigma}^{+-}(\omega)$, are 
pure imaginary. Similarly, 
 $\Sigma_{U,\sigma}^{++}(\omega) =
-\left\{\Sigma_{U,\sigma}^{--}(\omega) \right\}^*$,  
and the imaginary part of the causal self-energy 
corresponds to  $\Sigma_{U,\sigma}^\mathrm{K}(\omega)/2$,     
\begin{align}
\!\!\!\!\! 
\Sigma_{U,\sigma}^{--}(\omega)  
\, =  \, 
\frac{ 
\Sigma_{U,\sigma}^{r}(\omega)  +
\Sigma_{U,\sigma}^{a}(\omega) }{2} 
+ \frac{\Sigma_{U,\sigma}^\mathrm{K}(\omega)}{2}  .
\label{eq:causal_self_energy_in_K_component}
\end{align}
Furthermore, using Eq.\ \eqref{eq:derG},
the Green's function for the impurity electrons 
can also be expressed in the form,   
\begin{align}
&\bm{G}_{\sigma}^{}(\omega) \,= \, 
G_{\sigma}^{r}(\omega)\,G_{\sigma}^{a}(\omega)
\nonumber \\
& \qquad \quad  
 \times 
\left[ 
\begin{matrix}
\omega -\epsilon_{d\sigma}^{} 
+\Sigma_{\mathrm{tot},\sigma}^{++}(\omega) 
& 
- \Sigma_{\mathrm{tot},\sigma}^{-+}(\omega)  \cr   
\rule{0cm}{0.5cm}
-\Sigma_{\mathrm{tot},\sigma}^{+-}(\omega) & 
- \bigl( \omega -\epsilon_{d\sigma}^{} 
\bigr)
+\Sigma_{\mathrm{tot},\sigma}^{--}(\omega) 
\cr 
\end{matrix}
                                \right]   .  
\label{eq:Gdd_matrix}
\rule{0cm}{0.8cm}
\end{align}
Note that the off-diagonal Green's functions 
are proportional to the lesser or the greater self-energies.
From this property,  it can be deduced that
 the lesser and greater self-energies 
take the following form in the limit of  $eV \to 0$, 
\begin{subequations}
\begin{align}
& \!\!\!\!
\Sigma_{\mathrm{eq},\sigma}^{-+}(\omega) =     
f(\omega) 
\bigl[ \,
\Sigma_{\mathrm{eq},\sigma}^{r}(\omega)\,-\,
\Sigma_{\mathrm{eq},\sigma}^{a}(\omega)\,
\bigr] ,
\label{eq:self^{-+}_eq_Lehmann}
\\ 
&
\!\!\!\!
\Sigma_{\mathrm{eq},\sigma}^{+-}(\omega) =  
 -\bigl[1- f(\omega)\bigr]
\bigl[ \,
\Sigma_{\mathrm{eq},\sigma}^{r}(\omega)\,-\,
\Sigma_{\mathrm{eq},\sigma}^{a}(\omega)\,
\bigr].
\label{eq:self^{+-}_eq_Lehmann}
\end{align}
\label{eq:self^{-+}_self^{+-}_equilibrium}
\!\!\!\!
\end{subequations}
This is because   at equilibrium $eV=0$ 
the lesser and greater Green's functions 
can be expressed in the following form,
using the Lehmann representation,\cite{AGD} as
 \begin{subequations}
 \begin{align}
 G_{\mathrm{eq},\sigma}^{-+}(\epsilon) \,= & \      
 i\, 2\pi  f(\omega) \, \rho_{d\sigma}^{}(\omega) , 
 \label{eq:G^{-+}_eq_Lehmann}
 \\ 
 G_{\mathrm{eq},\sigma}^{+-}(\omega)\, =  &  
 -i\, 2\pi  \bigl[ 1-f(\omega)  \bigr]\, \rho_{d\sigma}^{}(\omega) .
 \label{eq:G^{+-}_eq_Lehmann}
 \end{align}
\label{eq:G_eq_Lehmann}
\!\!
 \end{subequations}
Here, $\rho_{d\sigma}^{}(\omega)$ is 
the density of states of impurity electrons,  
defined  in Eq.\ \eqref{eq:A_eq_def}.

\subsection{Nonequilibrium Ward identities} 
\label{subsec:Ward_eV_mag}

\begin{figure}[t]
 \leavevmode
\begin{minipage}{\linewidth}
\rule{0cm}{1cm}
\includegraphics[width=0.35\linewidth]{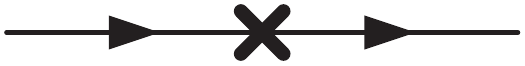}
\rule{0.1\linewidth}{0cm}
\includegraphics[width=0.37\linewidth]{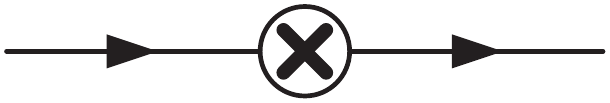}
\end{minipage}
 \caption{
Feynman diagrams for the derivatives 
$\frac{\partial}{\partial \epsilon_{d\sigma'}}
\bm{G}_{0\sigma}^{}(\omega)$ and 
$\left(
\frac{\partial}{\partial \omega}
+ \frac{\partial}{\partial \epsilon_{d\sigma}}
\right)
\bm{G}_{0\sigma}^{}(\omega)$, given in    
 Eqs.\ \eqref{eq:derivative_1_ed_only} and 
\eqref{eq:derivative_1_omega_ed}.
Here, $\times$  and $\otimes$ placed in between two propagators  
denote  the matrices 
 $\bm{\tau}_3\, \delta_{\sigma\sigma'}$ and $ -
 \frac{\partial f_\mathrm{eff}^{}(\omega)}{\partial \omega}
(-2i\Delta)\,\,\bigl( \bm{1} - \bm{\tau}_1 \bigr)$, respectively. 
}
 \label{fig:G0_derivative}
\end{figure}

We describe here a perturbative derivation of the Ward identity 
for the Anderson impurity at finite bias voltages. 
To this end, we start with the derivatives of 
the noninteracting Green's function $\bm{G}_{0\sigma}^{}$ 
with respect to external fields and a frequency.  
First of all, the derivative with respect to  $\epsilon_{d\sigma'}^{}$ is given by  
\begin{subequations}
\begin{align}
\frac{\partial \bm{G}_{0\sigma}^{}(\omega)}{\partial  \epsilon_{d\sigma'}^{}}\,  
=\, \delta_{\sigma\sigma'}\, \bm{G}_{0\sigma}^{}(\omega)\,  \bm{\tau}_3 \, 
\bm{G}_{0\sigma}^{} (\omega) .
\label{eq:derivative_1_ed_only}
\end{align}
This can be verified using Eq.\ \eqref{eq:G0_keldysh}, 
using a matrix identity 
 $\delta \bm{G} = -\bm{G} \,\{\delta \bm{G}^{-1}\}\, \bm{G}$ 
for a small variation of the Green's function  $\delta \bm{G}$.
Next one is the frequency derivative, 
which can be calculated from Eq.\ \eqref{eq:G0_Keldysh_elements},  
using the  property that $G_{0\sigma}^{r}(\omega)$ and 
$G_{0\sigma}^{a}(\omega)$ are functions 
of the difference  $\omega-\epsilon_{d\sigma}^{}$,
\begin{align}
 & 
 \!\!\!\!\!\!\!  \!\!\!\!\!\! 
\left(
\frac{\partial}{\partial \omega}
+ \frac{\partial}{\partial \epsilon_{d\sigma}^{}}
\right)^n
\bm{G}_{0\sigma}^{}(\omega) 
\nonumber \\
 =&  \ 
 - \frac{\partial^n f_\mathrm{eff}^{}(\omega) }{\partial \omega^n}
(-2i\Delta)\,
G_{0\sigma}^{r}(\omega) \,
G_{0\sigma}^{a}(\omega)\,
\left[
  \begin{matrix}
 1 & 1\cr  
 1 & 1 \cr   
 \end{matrix}
 \right]  
\nonumber \\
=&  \ 
 - \frac{\partial^n f_\mathrm{eff}^{}(\omega) }{\partial \omega^n}
(-2i\Delta)\,
\bm{G}_{0\sigma}^{}(\omega)
\,\bigl( \bm{1} - \bm{\tau}_1 \bigr) \,
\bm{G}_{0\sigma}^{}(\omega) .
\label{eq:derivative_1_omega_ed}
\end{align}
Here, we have used Eq.\ \eqref{eq:GrGaMat} to obtain  the last line. 
Similarly, as the bias voltage  enters  $\bm{G}_{0\sigma}^{}(\omega)$ 
only through $f_\mathrm{eff}^{}(\omega)$, the $eV$ derivative  
can be expressed in the form,
\begin{align}
& 
\!\!\!\!
\frac{\partial^n \bm{G}_{0\sigma}^{}(\omega) }{\partial (eV)^n}
\nonumber \\
& 
\!\!\!\!
=\,   -
 \frac{\partial^n f_\mathrm{eff}^{}(\omega)}{\partial (eV)^n}
(-2i\Delta)\,
 \bm{G}_{0\sigma}^{}(\omega)
\,\bigl( \bm{1} - \bm{\tau}_1 \bigr) \,
\bm{G}_{0\sigma}^{}(\omega).
\label{eq:derivative_1eV}
 \end{align}
\end{subequations}
These derivatives of $\bm{G}_{0\sigma}^{}(\omega)$ 
can diagrammatically be  represented as shown in Fig.\ \ref{fig:G0_derivative}. 
The Ward identities for the interacting 
self-energy $\Sigma_{U,\sigma}^{\nu\mu}(\epsilon)$ 
can be derived using the Feynman diagrammatic approach.\cite{Yoshimori}

\subsubsection{Ward identities 
for the  $\omega$ and $\epsilon_{d\sigma'}^{} derivatives$}

We consider the derivative of 
$\Sigma_{U,\sigma}^{\nu\mu}(\epsilon)$ 
with respect to $\epsilon_{d\sigma'}$. 
It can be carried out by using the chain rule 
with Eq.\ \eqref{eq:derivative_1_ed_only},  
taking the $\epsilon_{d\sigma'}$ derivative 
for the internal $\bm{G}_{0\sigma}^{}$'s 
which constitute the diagrams for the self-energy,  
\begin{align}
& \frac{\partial \Sigma_{U,\sigma}^{\nu_4\nu_1}(\epsilon)
}{\partial \epsilon_{d\sigma'}}
\nonumber \\ 
&  = \! 
\int_{-\infty}^{\infty}\! \frac{d \epsilon'}{2\pi i} 
\sum_{\nu_2\nu_3}
\Gamma_{\sigma\sigma';\sigma'\sigma}^{\nu_1\nu_2;\nu_3\nu_4}
 (\epsilon,\epsilon'; \epsilon',\epsilon)
\left\{\bm{G}_{\sigma'}^{}(\epsilon')  \bm{\tau}_3  
\bm{G}_{\sigma'}^{}(\epsilon')
\right\}_{}^{\nu_3\nu_2} \!\! .
\label{eq:Ward_Keldysh_ed}
\end{align}
Here, all contributions of the perturbation series in $U$ are included in 
the Keldysh vertex functions and the full Green's functions 
that appeared as a matrix product of the form  
$\bm{G}_{\sigma'}^{}  \bm{\tau}_3 \bm{G}_{\sigma'}^{}$. 
Note that this  identity Eq.\ \eqref{eq:Ward_Keldysh_ed}  
holds for arbitrary $eV$ and $T$.

In a similar way,  the Ward identity  for the  $\omega$ 
can also be derived,  using  Eq.\ \eqref{eq:derivative_1_omega_ed},  
\begin{widetext}
\begin{subequations}
\begin{align}
 \left(
\delta_{\sigma\sigma' }\,  \frac{\partial}{\partial \omega} 
+ \frac{\partial}{\partial \epsilon_{d\sigma'}}
\right)
\Sigma_{U,\sigma}^{\nu_4\nu_1}(\omega)
\, = & \  
-\int_{-\infty}^{\infty}\! \frac{d \omega'}{2\pi}\, 
\sum_{\nu_2\nu_3}\,
\Gamma_{\sigma\sigma';\sigma'\sigma}^{\nu_1\nu_2;\nu_3\nu_4}
 (\omega,\omega'; \omega',\omega)
\  2 \Delta
\,
G_{\sigma'}^{r}(\omega') \,G_{\sigma'}^{a}(\omega')
\,
\left(
-\frac{\partial  f_\mathrm{eff}^{} (\omega')}{\partial \omega'}
\right) 
\label{eq:Ward_NEQ_omega}
\\
\xrightarrow{\,T \to 0\,} &   \ 
- 
\sum_{\nu_2\nu_3} \sum_{j=L, R}\,
\Gamma_{\sigma\sigma';\sigma'\sigma}^{\nu_1\nu_2;\nu_3\nu_4}
 (\omega,\mu_{j}^{}; \mu_{j}^{},\omega)\,
\frac{\Delta}{\pi}\,
G_{\sigma'}^{r}(\mu_{j}^{}) \,G_{\sigma'}^{a}(\mu_{j}^{})
\,\frac{\Gamma_{j}}{\Gamma_L+\Gamma_R}
\label{eq:Ward_NEQ_omega_T0} 
\\
\xrightarrow{\,T \to 0\,,\,eV \to 0\,} &   \ 
- 
\sum_{\nu_2\nu_3}\,
\Gamma_{\sigma\sigma';\sigma'\sigma}^{\nu_1\nu_2;\nu_3\nu_4}
 (\omega,0; 0,\omega)\,
\rho_{d\sigma'}^{}
\,.
\rule{0cm}{0.7cm}
\label{eq:Ward_NEQ_omega_T0_eV0} 
\end{align}
\label{eq:Ward_NEQ_omega_all}
\end{subequations}
\end{widetext}
In order to obtain the Ward identities of this form, 
we have also used the property that 
the value of $\bm{\Sigma}_{U,\sigma}^{}(\omega)$ remains unchanged 
when the frequencies $\omega_\mathrm{in}^{}$'s  
for all the internal Green's functions along each closed loop 
 that  carries level index  $\sigma'$ 
are uniformly sifted by an arbitrary amount of $\omega_0$ in a way   
such that 
 $\omega_\mathrm{in}^{} \to \omega_\mathrm{in}^{} +\omega_0$.
\cite{Yoshimori} 
In particular, for the  diagonal  $\sigma=\sigma'$ components of 
 Eq.\ \eqref{eq:Ward_NEQ_omega},  
the internal frequencies  for the closed loops 
that carry the same level index $\sigma$ as that of 
the external one is shifted, 
choosing  $\omega_0$ to be equal to the external frequency $\omega$, 
and then the derivative $\left( {\partial}/{\partial \omega} 
+ {\partial}/{\partial \epsilon_{d\sigma}} \right)$ is carried out.

 The  off-diagonal $\sigma \neq \sigma'$ components 
of Eq.\ \eqref{eq:Ward_NEQ_omega}   
are obtained in a similar way:  shifting the frequencies 
of the closed loops carrying level index $\sigma'$
by an arbitrary amount $\omega_0$  independent of the 
external frequency  $\omega$, and then taking the 
following derivative 
\begin{align}
 \left( \frac{\partial}{\partial \omega_0} 
+ \frac {\partial}{\partial \epsilon_{d\sigma'}} \right)
\bm{\Sigma}_{U,\sigma}^{}(\omega) 
\,=\,   \frac {\partial \bm{\Sigma}_{U,\sigma}^{}(\omega) 
}{\partial \epsilon_{d\sigma'}} \,.
\label{eq:omega_0_derivative}
\end{align}
Here,  the $\omega_0$-derivative vanishes 
${\partial}\bm{\Sigma}_{U,\sigma}^{} (\omega)/{\partial \omega_0}=0$, 
and thus only  
${\partial \bm{\Sigma}_{U,\sigma}^{}(\omega) 
}/{\partial \epsilon_{d\sigma'}} $
remains in the left-hand of Eq.\ \eqref{eq:Ward_NEQ_omega}. 
For  obtaining  the right-hand side 
of  Eq.\ \eqref{eq:Ward_NEQ_omega}  for $\sigma\neq \sigma'$,  
the operator  $\left( {\partial}/{\partial \omega_0} 
+ {\partial}/{\partial \epsilon_{d\sigma'}} \right)$  
is applied to the internal Green's functions along the closed loops carrying 
the index $\sigma'$.
We have also used 
Eq.\ \eqref{eq:GrGaMat} with Eq.\ \eqref{eq:derivative_1_omega_ed} 
in order to  rewrite 
the matrix product of the full Green's function 
$\bm{G}_{\sigma'}^{} (\bm{1}- \bm{\tau}_1 )\bm{G}_{\sigma'}^{}$ 
in therms of the retarded and advanced Green's functions. 
The diagonal and off-diagonal components of Eq.\ \eqref{eq:Ward_NEQ_omega}   
are also derived from more general Ward-Takahashi identity 
that reflects the current conservation at the junctions between the dot and leads 
 in Sec.\ \ref{sec:nonlinear_WT}.

We next describe some properties of the Keldysh vertex function 
that can be deduced from the Ward identity 
 Eq.\ \eqref{eq:Ward_NEQ_omega_all}. 
In the second line that describes the $T \to 0$ limit, 
 Eq.\ \eqref{eq:Ward_NEQ_omega_T0},
the derivative of $f_\mathrm{eff}^{}(\omega')$ 
in the integrand has been replaced by a sum of  the two Dirac delta functions, 
\begin{align}
-\,\frac{\partial f_\mathrm{eff}^{}(\omega)}{\partial \omega}
\,\xrightarrow{\,T\to 0\,}\,  
\sum_{j=L,R} \, \frac{\Gamma_j^{}}{\Delta}\, \delta(\omega-\mu_j^{}) \,.
\label{eq:f_eff_derivative}
\end{align}
In particular, at  $T=eV=0$, 
 the causal component ( $\nu_4=\nu_1=-$)  
of Eq.\ \eqref{eq:Ward_NEQ_omega_T0_eV0} 
can be compared to the identity Eq.\ \eqref{eq:YYY_T0_causal} 
which has been obtained by using  
 the zero-temperature formalism.\cite{Yoshimori} 
Comparing these two results, we find that   
\begin{subequations}
\begin{align}
& 
\sum_{\nu_2\nu_3}
\Gamma_{\sigma\sigma';\sigma'\sigma}^{-\nu_2;\nu_3-}
 (\omega,0; 0,\omega) 
\,=\,
\Gamma_{\sigma\sigma';\sigma'\sigma}^{--;--}
 (\omega,0; 0,\omega) 
\,, 
\label{eq:vertex--T=ev=0}
\\ 
& 
\sum_{\nu_2\nu_3}
\Gamma_{\sigma\sigma';\sigma'\sigma}^{+\nu_2;\nu_3+}
 (\omega,0; 0,\omega) 
\,=\,
\Gamma_{\sigma\sigma';\sigma'\sigma}^{++;++}
 (\omega,0; 0,\omega) 
\label{eq:vertex++T=ev=0}
\,.
\end{align}
\label{eq:vertex--with++T=ev=0} 
\!\!\!\!\! 
\end{subequations} 
Furthermore,    at $T=eV=0$, 
the left-hand side of 
 Eq.\ \eqref{eq:Ward_NEQ_omega_T0_eV0}
can be rewritten into the following form, 
specifically for the lesser component ($\nu_4=-$, $\nu_1=+$) 
by using the property  
$\Sigma_{\mathrm{eq},\sigma}^{-+}(\omega)
=-2i f(\omega)\, \mathrm{Im}\, \Sigma_{\mathrm{eq},\sigma}^{--}(\omega)$
and $ \mathrm{Im}\, \Sigma_{\mathrm{eq},\sigma}^{--}(0) =0$ mentioned above, 
as 
\begin{align}
 & \left(
\delta_{\sigma\sigma' }\,  \frac{\partial}{\partial \omega} 
+ \frac{\partial}{\partial \epsilon_{d\sigma'}}
\right)
\Sigma_{\mathrm{eq},\sigma}^{-+}(\omega)
\nonumber \\
& =
-2i f(\omega)\, 
\mathrm{Im}\, 
 \left(
\delta_{\sigma\sigma' }\,  \frac{\partial}{\partial \omega} 
+ \frac{\partial}{\partial \epsilon_{d\sigma'}}
\right)
\Sigma_{\mathrm{eq},\sigma}^{--}(\omega)
\nonumber \\
& = 
2i f(\omega)\, 
\mathrm{Im}\, 
\Gamma_{\sigma\sigma';\sigma'\sigma}^{--;--}
 (\omega,0; 0,\omega) \rho_{d\sigma'}^{} \,.
\label{eq:lessser_causal_self_at_equilibrium}
\end{align}
Here, $\Sigma_{\mathrm{eq},\sigma}^{\mu\nu}(\omega) 
\equiv \left.\Sigma_{U,\sigma}^{\mu\nu}(\omega)\right|_{T=eV=0}$ 
and we have used  Eq.\ \eqref{eq:YYY_T0_causal} to obtain the last line.
Comparing the last line of Eq.\ 
\eqref{eq:lessser_causal_self_at_equilibrium} 
to the right-hand side of Eq.\ \eqref{eq:Ward_NEQ_omega_T0_eV0} 
for  $\nu_1=+$ and $\nu_4=-$, we find
\begin{subequations}
\begin{align}
\sum_{\nu_2\nu_3}
\Gamma_{\sigma\sigma';\sigma'\sigma}^{+\nu_2;\nu_3-}
 (\omega,0; 0,\omega) 
  = 
-2i f(\omega)\, 
\mathrm{Im}\, 
\Gamma_{\sigma\sigma';\sigma'\sigma}^{--;--}
 (\omega,0; 0,\omega) .
\label{eq:Ward_lesser_T0}
\end{align}
Similarly,  
the greater self-energy has the property 
$\Sigma_{\mathrm{eq},\sigma}^{+-}(\omega)
=-2i \bigl[1- f(\omega)\bigr]\, 
\mathrm{Im}\, \Sigma_{\mathrm{eq},\sigma}^{--}(\omega)$ 
at $T=eV=0$.  From this and the ($\nu_4=+$, $\nu_1=-$) component of 
 Eq.\ \eqref{eq:Ward_NEQ_omega_T0_eV0},  it follows that 
\begin{align}
& \!\!
\sum_{\nu_2\nu_3}
\Gamma_{\sigma\sigma';\sigma'\sigma}^{-\nu_2;\nu_3+}
 (\omega,0; 0,\omega)
\nonumber \\
& \qquad \quad 
= -2i \bigl[ 1-f(\omega) \bigr]\, 
\mathrm{Im}\, 
\Gamma_{\sigma\sigma';\sigma'\sigma}^{--;--}
 (\omega,0; 0,\omega) .
\label{eq:Ward_greater_T0}
\end{align}
\label{eq:Ward_lesser_greater_T0}
\!\!\!\!\!  
\end{subequations}

The vertex function also has another general property:     
 sixteen different components defined with respect to 
the Keldysh time-loop contour are linearly dependent,
\begin{align}
\sum_{\nu_1\nu_2\nu_3\nu_4}
\Gamma_{\sigma\sigma';\sigma'\sigma}^{\nu_1\nu_2;\nu_3\nu_4}
(\omega,\omega';\omega',\omega)
\,=\,0\,,
\label{eq:linear_dependency_Keldysh_text}
\end{align}
for which a brief explanation  
is provided in Appendix \ref{sec:linear_dependency_Keldysh}.
We can verify that both sides of Eq.\ \eqref{eq:Ward_NEQ_omega} 
 will vanish if summations over $\nu_1$ and $\nu_4$ are carried out, 
as a result of the linear dependence of the Keldysh self-energy and 
vertex function, listed in TABLE \ref{tab:Keldysh_sum_rule}.  
Furthermore, at $T=eV=0$, it can also be deduced  from 
Eqs.\ \eqref{eq:vertex--with++T=ev=0}  
\eqref{eq:Ward_lesser_greater_T0}, and  
\eqref{eq:linear_dependency_Keldysh_text}
that  
\begin{align}
& \!\!\! 
\Gamma_{\sigma\sigma';\sigma'\sigma}^{--;--} 
(\omega,0;0,\omega)
+   \Gamma_{\sigma\sigma';\sigma'\sigma}^{++;++} 
(\omega,0;0,\omega)
\nonumber \\
& \qquad \qquad \qquad \qquad 
=  \  2i \,  \mathrm{Im}  \, 
\Gamma_{\sigma\sigma';\sigma'\sigma}^{--;--} 
(\omega,0;0,\omega) \,, 
\end{align}
and 
$\Gamma_{\sigma\sigma';\sigma'\sigma}^{++;++} 
(\omega,0;0,\omega)
 = -
\left\{ 
\Gamma_{\sigma\sigma';\sigma'\sigma}^{--;--} 
(\omega,0;0,\omega)
\right\}^*$ .

\subsubsection{Ward identities for the  $eV$ derivatives}

We have described in the above the Ward identities for 
the $\omega$ and $\epsilon_{d\sigma'}^{}$ derivatives, i.e., 
Eqs.\ \eqref{eq:Ward_Keldysh_ed} and \eqref{eq:Ward_NEQ_omega}, 
respectively. 
Here we present some related Ward identities for 
the derivative of $\Sigma_{U,\sigma}^{\nu_4\nu_1}(\omega)$ 
with respect to $eV$,  which also hold at finite bias voltages. 

The bias derivative can be carried out by  
applying the differential operator $\partial/\partial (eV)$ 
to the internal Green's functions in the self-energy diagrams of all order in $U$. 
Using also the chain rule with  Eq.\ \eqref{eq:derivative_1eV}
 and the properties of  the Green's-function products 
shown in Eq.\ \eqref{eq:GrGaMat}, we obtain 
\begin{widetext}
\begin{subequations}
\begin{align}
\frac{\partial \Sigma_{U,\sigma}^{\nu_4\nu_1}(\omega)}{\partial eV} 
\, = & \  
-\int_{-\infty}^{\infty}\! \frac{d \omega'}{2\pi}\, 
\sum_{\sigma'}\sum_{\nu_2\nu_3}
\Gamma_{\sigma\sigma';\sigma'\sigma}^{\nu_1\nu_2;\nu_3\nu_4}
 (\omega,\omega'; \omega',\omega)
\, 2 \Delta \,
G_{\sigma'}^{r}(\omega') \,G_{\sigma'}^{a}(\omega')
\,
\left(
-\frac{\partial  f_\mathrm{eff}^{} (\omega')}{\partial eV}
\right)
\label{eq:Ward_NEQ_eV} 
\\
\xrightarrow{\,T \to 0\,} &   \ 
- 
\sum_{\sigma'}\sum_{\nu_2\nu_3}
\sum_{j=L, R}
\Gamma_{\sigma\sigma';\sigma'\sigma}^{\nu_1\nu_2;\nu_3\nu_4}
 (\omega, \mu_{j}^{}; \mu_{j}^{},\omega)\,
\frac{\Delta}{\pi}\,G_{\sigma'}^{r}(\mu_{j}^{}) \,G_{\sigma'}^{a}(\mu_{j}^{})
\,\left(\frac{- \Gamma_{j} \mu_j}{\Delta\, eV}\right)
\,.
\label{eq:Ward_NEQ_eV_T0} 
\rule{0cm}{0.7cm}
\end{align}
\label{eq:Ward_NEQ_eV_all} 
\end{subequations}
\end{widetext}
At $T = 0$,   the derivative $\partial f_\mathrm{eff}^{}(\omega')/\partial (eV)$ 
is replaced by the two Dirac delta functions at  $\omega' = \mu_L, \,\mu_R$, 
similarly to that for the $\omega'$ derivative,
 mentioned above for  Eq.\ \eqref{eq:f_eff_derivative}.
It also takes the following form in the zero-bias limit $eV \to 0$, 
\begin{align}
 \left. \frac{\partial f_\mathrm{eff}^{}(\omega)}{\partial eV} \right|_{eV\to 0}
 \!\!  =   -\alpha_\mathrm{sh}  \frac{\partial f(\omega)}{\partial \omega} ,
\qquad
\alpha_\mathrm{sh}  \equiv    
\frac{\alpha_L \Gamma_L - \alpha_R \Gamma_R}{\Gamma_L+ \Gamma_R}.
\end{align}
This relation  holds at finite temperatures $T\neq 0$.
Therefore, the extended Ward identities given in   
 Eqs.\ \eqref{eq:Ward_NEQ_omega} and \eqref{eq:Ward_NEQ_eV},   
reproduce the previous result  in the zero-bias limit:\cite{ao2001PRB}   
\begin{align}
\left.\frac{\partial {\bm \Sigma}_{U,\sigma}^{}(\omega)}{\partial eV}
\right|_{eV\to 0}^{}
\!\!  =\, 
 - \alpha_\mathrm{sh}  
\sum_{\sigma'}
\left(
 \delta_{\sigma\sigma'} 
\frac{\partial}{\partial \omega} 
+  \frac{\partial}{\partial \epsilon_{d\sigma'}^{}}
\right)
{\bm \Sigma}_{\mathrm{eq},\sigma}^{}(\omega)
.
\end{align}
Note that,  at $T=0$,    
 the causal component ($\nu_4=\nu_1=-$) can be calculated directly 
from Eq.\ \eqref{eq:Ward_NEQ_eV_T0}, using  Eq.\ \eqref{eq:vertex--T=ev=0}, 
as
\begin{align}
\!\! 
 \left.\frac{\partial \Sigma_{U,\sigma}^{--}(\omega)}{\partial eV}
\right|_{T \to 0 \atop  eV\to 0}^{} \! \!  = \, 
 \alpha_\mathrm{sh}  
\sum_{\sigma'}
\Gamma_{\sigma\sigma';\sigma'\sigma}^{--;--}
 (\omega,0; 0,\omega)
\,\rho_{d\sigma'}^{} .
\rule{0cm}{0.7cm}
\label{eq:self_retarded_1st_derivative_in_eV}
\end{align}

We next consider the second derivative  of $\bm{\Sigma}_\sigma(\omega)$ 
with respect to $eV$, extending the derivation given in Ref.\ \onlinecite{ao2001PRB}.  
The double derivative can be carried out, 
applying the formula Eq.\ \eqref{eq:derivative_1eV} for $n=1$ and $2$ 
to the internal Green's functions for the self-energy diagrams 
with two different first-differentiated propagators 
and one second-differentiated propagator, respectively.  
All contributions of the perturbation series in $U$ 
can be taken into account, using the Keldysh vertex function connected to  
the matrix product of the interacting propagators   
$\bm{G}_{\sigma'}^{} (\bm{1}- \bm{\tau}_1 )\bm{G}_{\sigma'}^{}$ 
that can be simplified further using  Eq.\ \eqref{eq:GrGaMat}, 
and it yields the following result:  
\begin{widetext}
\begin{align}
\!\!\!\!\! 
\left.
\frac{\partial^2  
\Sigma_{U,\sigma}^{\nu_4\nu_1}(\omega)}{
\partial (eV)^2}\right|_{eV\to 0} 
 \! = & \  \alpha_\mathrm{sh}^2 
\left(
\frac{\partial}{\partial \omega}
 + \frac{\partial}{\partial \epsilon_{d}^{}}
\right)^2 
\Sigma_{\mathrm{eq},\sigma}^{\nu_4\nu_1}(\omega) 
 \nonumber \\
  &   
- \frac{ \Gamma_L\,\Gamma_R}{ \left( \Gamma_L+ \Gamma_R \right)^2}
\int_{-\infty}^{\infty}\! \frac{d \omega'}{2\pi}\, 
\sum_{\sigma'}\sum_{\nu_2\nu_3}
\Gamma_{\sigma\sigma';\sigma'\sigma}^{\nu_1\nu_2;\nu_3\nu_4}
 (\omega,\omega'; \omega',\omega)
\, 2 \Delta \,
G_{\sigma'}^{r}(\omega') \,G_{\sigma'}^{a}(\omega')
\,
\left(
-\frac{\partial^2  f (\omega')}{\partial \omega'^2}
\right) . 
\label{eq:derivative_self_2_with_CII_new}
\end{align}
\end{widetext}
Here, we have used the property  ${\partial}/{\partial \epsilon_d^{}} 
   =  \sum_{\sigma''}  {\partial}/{\partial \epsilon_{d\sigma''}^{}}$ 
in the first term in the right-hand side.
This identity for the double derivative holds at any finite temperature,
and reproduces the previous result given for the causal component 
at  $T = 0$,\cite{,ao2017_2_PRB} 
\begin{align}
\!\!\!\! 
\left.
\frac{\partial^2  
\Sigma_{U,\sigma}^{--}(\omega)}{
\partial (eV)^2}\right|_{eV \to 0} 
\! \xrightarrow{\,T\to 0 \,} 
 & \  \ \alpha_\mathrm{sh}^2  
\left(
\frac{\partial}{\partial \omega}
+ \frac{\partial}{\partial \epsilon_{d}^{}}
\right)^2
\Sigma_{U,\sigma}^{--}(\omega)
 \nonumber \\
 & 
 + 
\frac{ \Gamma_L\,\Gamma_R}{ \left( \Gamma_L+ \Gamma_R \right)^2}
   \, \Psi_{\sigma}^{--}(\omega) .
\label{eq:derivative_self_2_with_CII_T0}
\end{align}
Here,  $\Psi_{\sigma}^{--}(\omega) $ in the right-hand side 
is the function which is defined in Eq.\ \eqref{eq:Psi_T0}, 
and it also determines the $T^2$ dependence of the self-energy. 
The bias dependence of the retarded self-energy 
shown in Eq.\  \eqref{eq:self_ev_mag}
can be deduced from these results, i.e.,   
Eqs.\ \eqref{eq:self_retarded_1st_derivative_in_eV} and 
\eqref{eq:derivative_self_2_with_CII_T0}.

\section{Ward-Takahashi identity at finite  bias voltages}
\label{sec:nonlinear_WT}

\subsection{Three-point current vertex}

In this section, 
we describe more general Ward-Takahashi identity,
 from which the Ward identities discussed in the above can naturally be deduced. 
It reflects current conservation around the quantum dot,  
and is derived nonperturbatively from the equation of motion of 
 the Keldysh three-point functions of the charge and current through 
the Anderson impurity.
We also discuss  some important properties of the three-point functions 
as a preparation for diagrammatic calculations of the current noise 
$S_\mathrm{noise}^\mathrm{QD}$. In order to carry these things out,
we extend the approach that has been used previously 
for the three-point functions in the Matsubara formalism 
to the nonequilibrium Keldysh correlation functions.
\cite{ao2001JPSJ,ao2003snote}

We consider the three-point 
correlation function of the charge fluctuation  
$\delta n_{d\sigma}^{\phantom{0}} \equiv n_{d\sigma}^{\phantom{0}} 
 - \langle n_{d\sigma}^{\phantom{0}} \rangle$ 
and that of the current fluctuations 
$\delta \widehat{J}_{j\sigma}^{\phantom{0}} 
\equiv \widehat{J}_{j\sigma}^{\phantom{0}} 
 - \langle \widehat{J}_{j\sigma}^{\phantom{0}} \rangle$ 
for  $j =L,\, R$, defined with respect to the nonequilibrium steady state,  
\begin{align} 
 \Phi_{d,\sigma\sigma'}^{\alpha;\mu\nu}(t\,; t_{1}, t_{2}) 
=&  \   -  \left \langle  T_{C} \, 
\delta n_{d\sigma'}^{\phantom{0}}(t_{}^{\alpha})\,
 d^{\phantom{\dagger}}_{ \sigma} (t_{1}^{\mu}) \, 
 d^{\dagger}_{ \sigma} (t_{2}^{\nu})                      
 \right \rangle ,
\nonumber
\\
 \Phi_{L,\sigma\sigma'}^{\alpha;\mu\nu}(t\,; t_{1}, t_{2}) 
=&  \   - \left \langle T_{C} \, \delta \widehat{J}_{L,\sigma'}(t_{}^{\alpha})\,
 d^{\phantom{\dagger}}_{\sigma} (t_{1}^{\mu}) \, 
 d^{\dagger}_{\sigma} (t_{2}^{\nu})                      
 \right \rangle ,
\nonumber
\\
\Phi_{R,\sigma\sigma'}^{\alpha;\mu\nu}(t\,; t_{1}^{}, t_{2}^{}) 
 =& \   -    \left \langle  T_{C} \, \delta \widehat{J}_{R,\sigma'}(t_{}^{\alpha})\,
 d^{\phantom{\dagger}}_{\sigma} (t_{1}^{\mu}) \, 
 d^{\dagger}_{ \sigma} (t_{2}^{\nu}) 
     \right \rangle . 
\label{eq:Phi_CLR}
\end{align} 
Here, $T_C$ is the time-ordering operator along the Keldysh time-loop contour.
The indexes $\alpha$, $\mu$, and $\nu$ specify 
at which branches of the contour, forward $-$  or backward $+$,  
the three time variables $t$, $t_1$, and $t_2$ belong, respectively. 
Therefore,  each of these three-point functions has  $2^3$ Keldysh components   
which are  linearly dependent, as shown    
 in TABLE \ref{tab:Keldysh_sum_rule} and 
 Appendix \ref{sec:linear_dependency_Keldysh}.

\begin{figure}[t] \leavevmode
\begin{minipage}{1\linewidth}
\includegraphics[width=0.9\linewidth]{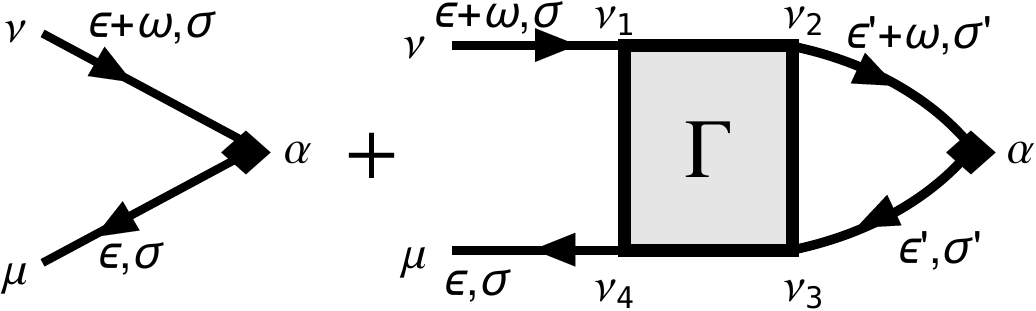}
 \caption{
Feynman diagrams for the  three-point function 
$\Phi_{\gamma,\sigma\sigma'}^{\alpha;\mu\nu}(\epsilon, \epsilon+\omega)$ 
defined in
Eq.\  \eqref{eq:3point_function_Fourier} for $\gamma=L,R,d$. 
The black diamond  ({\tiny \protect \rotatebox[origin=c]{45}{$\blacksquare$}}) 
 represents the bare current vertex 
$\bm{\lambda}_{\gamma}^{\alpha}(\epsilon, \epsilon+\omega)$ 
defined  in Eq.\ \eqref{eq:lambda_LR0}.
 }
\label{fig:3point_diagrams}
\end{minipage}
\end{figure}

In the steady state that we are considering, 
the system has the time translation symmetry. 
Therefore, these three-point functions depend only on 
two frequency variables  $\epsilon$ and $\omega$,
and the Fourier transform can be written in the following form, 
\begin{align}
& \!\!\! 
\Phi_{\gamma,\sigma\sigma'}^{\alpha;\mu\nu}(t; t_{1}, t_{2}) 
\nonumber \\
& \ \ = \int \! 
\frac{d\epsilon\,d\omega}{(2\pi)^2}
 \, 
\Phi_{\gamma,\sigma\sigma'}^{\alpha;\mu\nu}
(\epsilon,  \epsilon  + \omega)
\, e^{-i \,\epsilon (t_{1} - t)} 
\, e^{-i \,(\epsilon + \omega) (t - t_{2})}, 
\label{eq:3point_function_Fourier}
\end{align}
where  $\gamma =L,R,d$. 
We will use the matrix notation 
to express the Keldysh components, as  
$\left\{\bm{\Phi}_{\gamma,\sigma\sigma'}^{\alpha}\right\}^{\mu\nu}
= \Phi_{\gamma,\sigma\sigma'}^{\alpha;\mu\nu}$. 
Similarly, we also introduce the $2\times 2$ matrix 
for the current vertex  $\bm{\Lambda}_{\gamma,\sigma\sigma'}^{\alpha}$ 
which includes all vertex corrections due to the Coulomb interaction, as
\begin{align}
& 
  \bm{\Phi}_{\gamma,\sigma\sigma'}^{\alpha}
( \epsilon ,  \epsilon  +  \omega)
 \, = \, 
  \bm{G}_{\sigma}(\epsilon )\, 
\bm{\Lambda}_{\gamma,\sigma\sigma'}^{\alpha}
( \epsilon , \epsilon  +  \omega)
  \, \bm{G}_{\sigma}( \epsilon  +  \omega) . 
  \label{eq:Lambda_matrix} 
\end{align}
Figure \ref{fig:3point_diagrams} shows 
the Feynman diagrams for $\bm{\Phi}_{\gamma,\sigma\sigma'}^{\alpha;\mu\nu}$.
The current vertex $\bm{\Lambda}_{\gamma,\sigma\sigma'}^{\alpha}$
corresponds to the remainder parts  which are left over when 
the pair of the Green's functions with the frequency arguments 
$\epsilon+\omega$ and $\epsilon$ placed 
on the left side of each Feynman diagram are removed.
We will discuss later  
that it can be expressed in terms of the four-point 
vertex functions, as Eq.\ \eqref{eq:Bethe-Salpeter}.

\subsubsection{Relation between current noise and    
 $\,\bm{\Phi}_{R/L,\sigma\sigma'}^{\alpha}$}

It is essential for calculating the nonlinear current noise 
$S_\mathrm{noise}^\mathrm{QD}$ in the  Fermi-liquid regime 
to know the low-energy behavior of these three-point correlations.  
For instance,  the  current-current correlations 
 between  $\widehat{J}_{R,\sigma'}$ 
and $\widehat{J}_{L,\sigma}$ can be expressed 
in terms of $\bm{\Phi}_{R,\sigma\sigma'}^{\alpha}$, as
\begin{align}
S_{RL} 
\equiv& \
e^2 
\sum_{\sigma\sigma'}
\int_{-\infty}^{\infty} \!  dt\  
\nonumber \\
& \  
\times \biggl[\,
\bigl\langle \delta \widehat{J}_{R,\sigma'}(t) 
\, \delta \widehat{J}_{L,\sigma}(0) \bigr\rangle
+\bigl\langle \delta \widehat{J}_{L,\sigma}(0) 
\, \delta \widehat{J}_{R,\sigma'}(t)\bigr\rangle
\,\biggr] 
\nonumber 
\\
=& \ \frac{e^2}{\hbar} 
\sum_{\sigma\sigma'}
\sum_{\alpha=+,-}
\int_{-\infty}^{\infty}\! \frac{d\epsilon}{2\pi}\, 
\mathrm{Tr}
\Bigl[ 
\bm{\lambda}_L^{-\alpha}( \epsilon , \epsilon)\,
\bm{\Phi}_{R,\sigma\sigma'}^{\alpha}(\epsilon, \epsilon) 
\Bigr].
\end{align}
Here,   $\mathrm{Tr}$ denotes the trace that is 
taken over diagonal elements of the  $2\times 2$ Keldysh matrices. 
  $\bm{\lambda}_L^{\alpha}$ is  the bare current vertex 
accompanied by the operator $\widehat{J}_{L,\sigma}$. 
It is  illustrated in Fig.\ \ref{fig:3point_diagrams} 
 and its explicit form is given below in Eq.\ \eqref{eq:lambda_LR0}.

\subsubsection{Current vertexes  for  $U=0$:
 $\bm{\lambda}_L^{\alpha}$ and $\bm{\lambda}_R^{\alpha}$ }

In the noninteracting case $U=0$, 
the three-point correlation functions for 
the charge fluctuation  $\delta n_{d\sigma}^{}$ is given by,
\begin{align}
& \bm{\Phi}_{d,\sigma\sigma'}^{(0)\alpha}
(\epsilon, \epsilon+\omega) \,=\,  
\delta_{\sigma\sigma'}\,
\bm{G}_{\sigma}^{(0)}(\epsilon)\,\bm{\rho}_3^{\alpha}
\,\bm{G}_{\sigma}^{(0)}(\epsilon+\omega).
\label{eq:lambda_d_u0}
\end{align}
Here, the label $(0)$ in superscript represents 
that the correlation function with this label is defined for 
noninteracting electrons  $U=0$, i.e.,  
$\bm{G}_{\sigma}^{(0)} \equiv \bm{G}_{0\sigma}^{}$. 
The matrix $\bm{\rho}_{3}^{\alpha}$ 
corresponds to the projection operator 
into the forward ($-$) branch, or the backward ($+$) branch,  
of the Keldysh time-loop contour,
\begin{align}
\!\!\!\!
\bm{\rho}_{3}^{-} \equiv
\frac{\bm{1}+\bm{\tau}_3 }{2}
=\left[ 
\begin{matrix}
 1 & 0  \cr   
0 & 0 \cr  
\end{matrix}
\right] , 
\quad \  
\bm{\rho}_{3}^{+} \equiv
\frac{\bm{1}-\bm{\tau}_3}{2}
=\left[ 
\begin{matrix}
 0 & 0  \cr   
0 & 1 \cr  
\end{matrix}
\right] .
\label{eq:branch_projector}
\end{align}
The corresponding  
three-point vertex $\bm{\Lambda}_{d,\sigma\sigma'}^{(0)\alpha}$  
 can be derived from  $\bm{\Phi}_{d,\sigma\sigma'}^{(0)\alpha}$ 
using Eq.\ \eqref{eq:lambda_d_u0}, as  
  $\bm{\Lambda}_{d,\sigma\sigma'}^{(0)\alpha} = 
\delta_{\sigma\sigma'}\bm{\lambda}_{d}^{\alpha}$ 
with $\bm{\lambda}_{d}^{\alpha}\equiv \bm{\rho}_{3}^{\alpha}$.
This function  $\bm{\Lambda}_{d,\sigma\sigma'}^{(0)\alpha}$
becomes independent of the frequencies $\epsilon$ and $\omega$ 
for noninteracting electrons.

Similarly, for $U=0$, 
the three-point correlation functions 
 for current fluctuations take the following form, 
for $j =L,R$, 
\begin{align}
 & \bm{\Phi}_{j,\sigma\sigma'}^{(0)\alpha}(\epsilon, \epsilon+\omega)
 =\, 
i \, \eta_{j}^{} v_{j}^{} 
\,\delta_{\sigma\sigma'}
\Bigl[\,
\bm{G}_{d j,\sigma}^{(0)}(\epsilon)\,\bm{\rho}_3^{\alpha}
\,\bm{G}_{\sigma}^{(0)}(\epsilon+\omega)
\nonumber \\
& 
\qquad\qquad\qquad\qquad\qquad\qquad
\,-\,
\bm{G}_{\sigma}^{(0)}(\epsilon)\,\bm{\rho}_3^{\alpha}
\,\bm{G}_{j d,\sigma}^{(0)}(\epsilon+\omega)
\,\Bigr]
\nonumber \\
& \qquad \qquad \qquad \ \ 
=
\, 
\delta_{\sigma\sigma'}\,
\bm{G}_{\sigma}^{(0)}(\epsilon)\,
\bm{\lambda}_{j}^{\alpha}( \epsilon,  \epsilon  +  \omega) 
\,\bm{G}_{\sigma}^{(0)}(\epsilon+\omega) . 
\label{eq:Phi_U0}
\end{align}
Here,  $\eta_{L}^{} = -1$ and  $\eta_{R}^{} = +1$.
The inter-site Green's functions between the dot and leads 
are written as   $\bm{G}_{d j,\sigma}^{}$ and $\bm{G}_{j d,\sigma}^{}$. 
To obtain the last line of Eq.\ \eqref{eq:Phi_U0}, we have used  
the recursive properties of these inter-site Green's functions  
which hold for both interacting and noninteracting electrons,    
\begin{align}
    \bm{G}_{j d,\sigma}^{}(\epsilon) \,= & \  
-v_{j}^{} \, \bm{g}_{j}^{}(\epsilon)\,
\bm{\tau}_3\,
\bm{G}_{\sigma}^{}(\epsilon) 
    \,,         
 \nonumber \\
    \bm{G}_{d j,\sigma}^{}(\epsilon) \, = &  \  
-v_{j} \, 
\bm{G}_{\sigma}^{}(\epsilon) \,
\bm{\tau}_3\,
\bm{g}_{j}^{}(\epsilon) \,.
\end{align}
Here, $\bm{g}_{j}^{}$ is the Green's functions for the isolated lead 
on $j=L$ and $R$, defined in Eq.\ \eqref{eq:g0_lead}.
Therefore,  the bare current vertex can be written explicitly in the form, 
\begin{align}
& \bm{\lambda}_j^{\alpha}( \epsilon , \epsilon  +  \omega)
\, \equiv \, 
i\,\eta_{j}^{} v_{j}^2 
\Bigl[\,
\bm{\rho}_3^{\alpha}\,\bm{g}_{j}^{}(\epsilon+\omega)\,\bm{\tau}_3
\,-\,\bm{\tau}_3\,
\bm{g}_{j}^{}(\epsilon)\,\bm{\rho}_3^{\alpha}
\,\Bigr].
\label{eq:lambda_LR0} 
\end{align}
In particular, at $\omega=0$, 
the bare current vertexes become independent of 
 whether  $\alpha=-$ or $+$, as
\begin{subequations}
\begin{align}
\bm{\lambda}_R^{\alpha}( \epsilon , \epsilon)
\,=& \  
i\, \frac{v_R^2}{2} 
\Bigl[\,
 \bm{g}_{R}^{}(\epsilon)\,\bm{\tau}_3
\,-\,\bm{\tau}_3\,
\bm{g}_{R}^{}(\epsilon)\,\Bigr]
\nonumber \\ 
= & \  
2 \Gamma_R 
\left[
\begin{matrix}
 0 & f_R  \cr   
1-f_R  & 0 \cr  
\end{matrix}
\right]  \;, 
\\
\bm{\lambda}_L^{\alpha}( \epsilon , \epsilon)
\,= & \   
-i\, \frac{v_L^2}{2} 
\Bigl[\,
 \bm{g}_{L}^{}(\epsilon)\,\bm{\tau}_3
\,-\,\bm{\tau}_3\,
\bm{g}_{L}^{}(\epsilon)\,\Bigr]
\rule{0cm}{0.6cm}
\nonumber \\
 = & \  
-2 \Gamma_L 
\left[
\begin{matrix}
 0 & f_L  \cr   
1-f_L  & 0 \cr  
\end{matrix}
\right]  \;.
\end{align}
\label{eq:lambda_U0_w0} 
\end{subequations}
This can be verified, substituting 
the explicit form of the projection operator $\bm{\rho}_3^{\alpha}$
given in Eq.\ \eqref{eq:branch_projector}
into  the right-hand side of Eq.\ \eqref{eq:lambda_LR0}.

For later convenience, 
we introduce the following two different linear combinations 
of $\bm{\lambda}_L^{\alpha}$ and $\bm{\lambda}_R^{\alpha}$.
One is for the difference between 
 incoming and outgoing currents through the dot 
 $\widehat{J}_{R,\sigma}-\widehat{J}_{L,\sigma}$,  
and the other is for the symmetrized current 
 $ \widehat{J}_{\sigma}  \equiv    
 (\Gamma_L \widehat{J}_{R,\sigma} 
+\Gamma_R \widehat{J}_{L,\sigma})/(\Gamma_L+\Gamma_R)$
defined in Eq.\ \eqref{eq:symmetrized_current_def}, 
\begin{align}
\bm{\lambda}_J^{\alpha}( \epsilon , \epsilon+\omega)
\,\equiv & \ 
\bm{\lambda}_R^{\alpha}( \epsilon , \epsilon+\omega)
-\bm{\lambda}_L^{\alpha}( \epsilon , \epsilon+\omega)
\nonumber \\
\xrightarrow{\,\omega \to 0\,} &  \   
2 
(\Gamma_L +\Gamma_R)
\left[
\begin{matrix}
 0 & f_\mathrm{eff}^{}(\epsilon)  \cr   
1-f_\mathrm{eff}^{}(\epsilon)  & 0 \cr  
\end{matrix}
\right]  ,
\label{eq:lambda_diff_U0} 
\\
\nonumber \\
\!\!\!\!\! 
\bm{\lambda}_\mathrm{sym}^{\alpha}( \epsilon , \epsilon+\omega)
\,\equiv & \ 
\frac{
\Gamma_L \,\bm{\lambda}_R^{\alpha}( \epsilon , \epsilon+\omega)
+\Gamma_R \,\bm{\lambda}_L^{\alpha}( \epsilon , \epsilon+\omega)
}{\Gamma_L+\Gamma_R}
\nonumber \\
\xrightarrow{\,\omega \to 0\,} & \ 
\frac{-2\Gamma_L\Gamma_R}{\Gamma_L+\Gamma_R}
 \bigl[ f_L(\epsilon) -f_R(\epsilon) \bigr]
 \left[
\begin{matrix}
 0 & 1 \cr   
-1  & 0 \cr  
\end{matrix}
\right] .
\label{eq:lambda_av_U0}
%
\end{align}
Each of these bare current vertexes shows 
a pronounced frequency dependence.
While 
$\bm{\lambda}_J^{\alpha}(\epsilon,\epsilon)$ represents 
the local nonequilibrium distribution $f_\mathrm{eff}^{}(\epsilon)$, 
the symmetrized part  
 $\bm{\lambda}_\mathrm{sym}^{\alpha}(\epsilon,\epsilon)$ 
extracts the contributions of  single-particle excitations inside 
 the bias-window region through 
 $ f_L(\epsilon) -f_R(\epsilon)$.

\subsection{Symmetrization of 
$\bm{\Lambda}_{\gamma,\sigma\sigma'}^{\alpha}$ 
with respect to the forward and backward Keldysh contours
} 
\label{subsec:current_vertex_II}

We also introduce another type  symmetrization,  
defined with respect to the forward ($\alpha =-$) and 
the backward ($\alpha=+$) Keldysh branches: 
 the symmetrized (S) part  and anti-symmetrized (A) 
part of the three-point vertex functions are defined by 
\begin{subequations}
\begin{align}
\bm{\Lambda}_{p,\sigma\sigma'}^\mathrm{S}(\epsilon, \epsilon+\omega)  
\,\equiv& \,
 \bm{\Lambda}_{p,\sigma\sigma'}^{-}(\epsilon, \epsilon+\omega) 
- \bm{\Lambda}_{p,\sigma\sigma'}^{+}(\epsilon, \epsilon+\omega) \;,
 \\
\bm{\Lambda}_{p,\sigma\sigma'}^\mathrm{A}(\epsilon, \epsilon+\omega)  
\,\equiv& \,
 \bm{\Lambda}_{p,\sigma\sigma'}^{-}(\epsilon, \epsilon+\omega) 
+ \bm{\Lambda}_{p,\sigma\sigma'}^{+}(\epsilon, \epsilon+\omega) \;.
\end{align}
\!\!  \label{eq:contour_symmetrization}
\end{subequations}
Here, the label $p$ ($ =L, R, d, J, ``\mathrm{sym}"$)   
specifies the spatial components of the current and charge 
 three-point vertex functions,  including the difference $p=J$ and the average 
$p=``\mathrm{sym}"$, the bare ones of which are defined in 
Eqs.\ \eqref{eq:lambda_diff_U0} and \eqref{eq:lambda_av_U0}, 
respectively.

For interacting electrons $U=0$, 
 the symmetrized  and antisymmetrized  
three-point  vertex functions for  the impurity charge 
$\delta n_{d\sigma}^{}$ are determined by  Eqs.\  
\eqref{eq:lambda_d_u0} and \eqref{eq:branch_projector},  as 
\begin{subequations}
\begin{align}
\bm{\lambda}_d^\mathrm{S}
\equiv \bm{\rho}_{3}^{-} - \bm{\rho}_{3}^{+}=\bm{\tau}_3, \qquad
\bm{\lambda}_d^\mathrm{A}
\equiv \bm{\rho}_{3}^{-} + \bm{\rho}_{3}^{+} = \bm{1}.
\end{align}
Similarly, for the bare vertex functions for the current difference defined
 in  Eq.\ \eqref{eq:lambda_diff_U0},  
the symmeterized 
$\bm{\lambda}_J^\mathrm{S} \equiv 
\bm{\lambda}_J^\mathrm{-} -\bm{\lambda}_J^\mathrm{+}$
and anti-symmetrized 
$\bm{\lambda}_J^\mathrm{A} \equiv 
\bm{\lambda}_J^\mathrm{-} + \bm{\lambda}_J^\mathrm{+}$ 
vertexes follow  form Eq.\  \eqref{eq:lambda_LR0}, as 
\begin{align}
&\bm{\lambda}_J^\mathrm{S}( \epsilon , \epsilon  +  \omega)
\, = \, i \, 
\Bigl[\,
 \bm{\Sigma}_0(\epsilon+\omega)\,
\,-\,
\bm{\Sigma}_0(\epsilon)
\,\Bigr] 
\label{eq:current_vertex_J_S_U0}
\\
& \quad 
=  \,   
 - 2 
(\Gamma_L + \Gamma_R)\,
\Bigl[\,
 f_\mathrm{eff}^{} (\epsilon+\omega)\,
-
 f_\mathrm{eff}^{} (\epsilon)\,
\,\Bigr] 
 \left[
 \begin{matrix}
\ \  1& -1  \cr   
 -1 & \ \ 1 \cr  
 \end{matrix}
 \right], 
\nonumber 
\\
&\bm{\lambda}_J^\mathrm{A}( \epsilon , \epsilon  +  \omega)
\, = \, i \, 
\Bigl[\,
\bm{\tau}_3\, \bm{\Sigma}_0(\epsilon+\omega)\,
\,-\,
\bm{\Sigma}_0(\epsilon)\,\bm{\tau}_3
\,\Bigr] 
\label{eq:current_vertex_J_A_U0}
 \\
& \quad
\xrightarrow{\,\omega \to 0\,}   \    
4(\Gamma_L + \Gamma_R)\,
\left[
 \begin{matrix}
 \rule{0.6cm}{0cm} 0 \rule{0.6cm}{0cm}& 
 f_\mathrm{eff}^{} (\epsilon) \cr   
1- f_\mathrm{eff}^{} (\epsilon) & 0 \cr  
 \end{matrix}
 \right].
\nonumber 
\end{align}
\end{subequations}
While the symmetrized part 
  $\bm{\lambda}_J^\mathrm{S}(\epsilon,\epsilon+\omega)$
 is determined directly by the noninteracting self-energy 
$\bm{\Sigma}_0$ given in Eq.\ \eqref{eq:U0_self_keldysh},  
the antisymetrized part  $\bm{\lambda}_J^\mathrm{A}$
is determined by  the product of $\bm{\tau}_3$ and $\bm{\Sigma}_0$. 
Therefore,  at $\omega=0$,    
$\bm{\lambda}_J^\mathrm{A}$ remains finite, 
whereas $\bm{\lambda}_J^\mathrm{S}$ vanishes,  
but nevertheless its derivative has a pronounced property,  
\begin{align}
& \!\!\!\!\!\!\!\!\!\!
%
\frac{\partial}{\partial \omega}\Bigl[\,
\bm{G}_{\sigma}^{}(\epsilon)\,
\bm{\lambda}_J^\mathrm{S} (\epsilon, \epsilon  +  \omega) \, 
\bm{G}_{\sigma}^{}(\epsilon+\omega)
\,\Bigr]_{\omega \to 0}
\nonumber \\
 & \quad =\, \bm{G}_{\sigma}^{}(\epsilon)\,i\,
 \frac{\partial\bm{\Sigma}_0(\epsilon)}{\partial \epsilon}
\, \bm{G}_{\sigma}^{}(\epsilon)
 \nonumber \\
&  \quad =  \,  
 - 2 
\Delta\,
G_{\sigma}^{r}(\epsilon) \,G_{\sigma}^{a}(\epsilon)
\,
\frac{\partial  f_\mathrm{eff}^{} (\epsilon)}{\partial \epsilon}\,
\,
\left[
 \begin{matrix}
 1 & 1\cr  
 1 & 1 \cr   
 \end{matrix}
 \right] .
\label{eq:lambda_J_derivative}
\end{align}
For obtaining the last line, we have used 
 Eqs.\ \eqref{eq:GrGaMat} and \eqref{eq:self_U0_derivative_omega}.

\subsection{Ward-Takahashi identities for 
$\bm{\Phi}_{\gamma,\sigma\sigma'}^{\alpha}$ and 
$\bm{\Lambda}_{\gamma,\sigma\sigma'}^{\alpha}$ 
}
\label{subsec:WT_text}

We now come to the point  for  discussing 
 the Ward-Takahashi identity  between the Keldysh three-point functions 
 $\bm{\Phi}_{d,\sigma\sigma'}^{\alpha}$,
 $\bm{\Phi}_{R,\sigma\sigma'}^{\alpha}$, and  
 $\bm{\Phi}_{L,\sigma\sigma'}^{\alpha}$:    
\begin{align}
& i\frac{\partial }{\partial t}\, 
\bm{\Phi}_{d,\sigma\sigma'}^{\alpha}(t; t_1, t_2) 
+i \bm{\Phi}_{R,\sigma\sigma'}^{\alpha}(t; t_1, t_2) 
-i \bm{\Phi}_{L,\sigma\sigma'}^{\alpha}(t; t_1, t_2) 
\nonumber \\
& =  \
- \delta_{\sigma\sigma'}\,\delta(t-t_1) \,  
\bm{\rho}_3^{\alpha}\,\bm{\tau}_3
\,
\bm{G}_{\sigma}(t,t_2)
\nonumber \\
&  \quad \    + \, \delta_{\sigma\sigma'}\,\delta(t_2-t)\,  
\bm{G}_{\sigma}(t_1,t)
\,\bm{\tau}_3\,\bm{\rho}_3^{\alpha}
\,.
\label{eq:WT_Phi_in_time}
\end{align}
As shown in the proof given in Appendix \ref{sec:Ward_Takahashi_derivations}, 
this identity corresponds to the equation of motion of the charge component 
$\bm{\Phi}_{d,\sigma\sigma'}^{\alpha}(t;t_1,t_2)$, 
and it reflects the current conservation between the dot and leads,
described in Eq.\ \eqref{eq:current_conservation}. 
In the frequency domain,  
the Ward-Takahashi identity can be expressed in the following form,
carrying out  the Fourier transform  as Eq.\ \eqref{eq:3point_function_Fourier}, 
\begin{align}
&  \!\!\!\!
\omega \,\bm{\Phi}_{d,\sigma\sigma'}^{\alpha}(\epsilon, \epsilon+\omega) 
+i \bm{\Phi}_{R,\sigma\sigma'}^{\alpha}(\epsilon, \epsilon+\omega) 
-i \bm{\Phi}_{L,\sigma\sigma'}^{\alpha}(\epsilon, \epsilon+\omega) 
\nonumber \\
& \!\! 
 = \,  
-\,\delta_{\sigma\sigma'}\,  
\bm{\rho}_3^{\alpha}\, \bm{\tau}_3
\,
\bm{G}_{\sigma}(\epsilon+\omega)
\,+ \,\delta_{\sigma\sigma'}\,
\bm{G}_{\sigma}(\epsilon)\, \bm{\tau}_3\,\bm{\rho}_3^{\alpha}
\;.
\label{eq:WT_Phi}
\end{align}

Furthermore, it can also be rewritten into 
the relation between 
the three-point vertex functions $\bm{\Lambda}_{\gamma,\sigma\sigma'}^{\alpha}$ 
and the self-energy, using Eq.\ \eqref{eq:Lambda_matrix}, 
\begin{align}
&
\omega \bm{\Lambda}_{d,\sigma\sigma'}^{\alpha}(\epsilon, \epsilon+\omega) 
+i \bm{\Lambda}_{R,\sigma\sigma'}^{\alpha}(\epsilon, \epsilon+\omega) 
-i \bm{\Lambda}_{L,\sigma\sigma'}^{\alpha}(\epsilon, \epsilon+\omega) 
\nonumber \\
&  =\,  \delta_{\sigma\sigma'} 
\Bigl[\,
\omega\, \bm{\rho}_3^{\alpha} 
- \bm{\rho}_3^{\alpha} \,\bm{\tau}_3
\bm{\Sigma}_{\mathrm{tot},\sigma}(\epsilon+\omega)
+ \bm{\Sigma}_{\mathrm{tot},\sigma}(\epsilon)
\bm{\tau}_3\,\bm{\rho}_3^{\alpha} \,
\Bigr].
\label{eq:WT_Lambda}
\end{align}
 Here, $\bm{\Sigma}_{\mathrm{tot},\sigma}(\epsilon)
 = 
\bm{\Sigma}_0(\epsilon)
+\bm{\Sigma}_{U,\sigma}(\epsilon)
$ is the total self-energy, and 
 $\left\{ \bm{G}_{\sigma}(\epsilon) \right\}^{-1} 
= (\epsilon - \epsilon_{d\sigma}^{}) \bm{\tau}_3 -
\bm{\Sigma}_{\mathrm{tot},\sigma}(\epsilon)$,  
as shown in Eq.\ \eqref{eq:Dyson_Keldysh}. 
We will show in the following that 
 the Ward identity given in Eq.\ \eqref{eq:Ward_NEQ_omega}  
can be deduced non-perturbatively 
from the generalized one Eq.\ \eqref{eq:WT_Lambda},   
in the limit where the external frequency to be $\omega \to 0$.

\subsection{Symmetrized Ward-Takahashi identities}

The Ward-Takahashi identity Eq.\ \eqref{eq:WT_Lambda} 
can be rearranged into the symmetrized 
$\bm{\Lambda}_{p,\sigma\sigma'}^\mathrm{S} =
\bm{\Lambda}_{p,\sigma\sigma'}^{-} -
\bm{\Lambda}_{p,\sigma\sigma'}^{+}$ 
  and the  antisymmetrized 
$\bm{\Lambda}_{p,\sigma\sigma'}^\mathrm{A}=
\bm{\Lambda}_{p,\sigma\sigma'}^{-} +
\bm{\Lambda}_{p,\sigma\sigma'}^{+}
$ parts, defined in  Eq.\ \eqref{eq:contour_symmetrization}, as
\begin{subequations}
\begin{align}
&\!\!\!
\omega \,\bm{\Lambda}_{d,\sigma\sigma'}^\mathrm{S}(\epsilon, \epsilon+\omega) +i\,\bm{\Lambda}_{J,\sigma\sigma'}^\mathrm{S}(\epsilon, \epsilon+\omega) 
\nonumber \\
&   =  \,  
\delta_{\sigma\sigma'} 
\Bigl[\,
\omega\, 
\bm{\tau}_3 
-
\bm{\Sigma}_{\mathrm{tot},\sigma}(\epsilon+\omega)
+
\bm{\Sigma}_{\mathrm{tot},\sigma}(\epsilon)
\,
\Bigr], 
\label{eq:symmetrized_Lambda}
\\
& \!\!\!
\omega \,\bm{\Lambda}_{d,\sigma\sigma'}^\mathrm{A}(\epsilon, \epsilon+\omega) 
+i\,\bm{\Lambda}_{J,\sigma\sigma'}^\mathrm{A}(\epsilon, \epsilon+\omega) 
\nonumber \\
&   = \,  
\delta_{\sigma\sigma'} 
\Bigl[\,
\omega\, 
\bm{1} 
-
\bm{\tau}_3 \,
\bm{\Sigma}_{\mathrm{tot},\sigma}(\epsilon+\omega)
+
\bm{\Sigma}_{\mathrm{tot},\sigma}(\epsilon)
\,\bm{\tau}_3 
\,
\Bigr].
\label{eq:antisymmetrized_Lambda}
\end{align}
\end{subequations}
Here,
$\bm{\Lambda}_{J,\sigma\sigma'}^{q}
\equiv \bm{\Lambda}_{R,\sigma\sigma'}^{q} 
-\bm{\Lambda}_{L,\sigma\sigma'}^{q}$ 
with the label  $q$ ($=-,+, \mathrm{S}, \mathrm{A}$) 
which specifies the time-loop components.
Equation \eqref{eq:antisymmetrized_Lambda} shows that  
the antisymmetrized current vertex 
$\bm{\Lambda}_{J,\sigma\sigma'}^\mathrm{A}$
remains finite for  $\omega \to 0$, similarly to the noninteracting one  
$\bm{\lambda}_{J}^\mathrm{A}$ given in Eq.\ \eqref{eq:current_vertex_J_A_U0}, 
\begin{align}
i\,\bm{\Lambda}_{J,\sigma\sigma'}^\mathrm{A}(\epsilon, \epsilon) 
\,  = &  \ \,     
\delta_{\sigma\sigma'} 
\Bigl[\,
\bm{\Sigma}_{\mathrm{tot},\sigma}(\epsilon)\,\bm{\tau}_3 
- \bm{\tau}_3 \bm{\Sigma}_{\mathrm{tot},\sigma}(\epsilon)
\,
\Bigr]
\nonumber \\
= &  \   
2\left[
 \begin{matrix}
 \rule{0.6cm}{0cm} 0 \rule{0.5cm}{0cm}& 
-\Sigma_{\mathrm{tot},\sigma}^{-+}(\epsilon)  \cr   
 \Sigma_{\mathrm{tot},\sigma}^{+-}(\epsilon) & 0 \cr  
 \end{matrix}
 \right] 
\,\delta_{\sigma\sigma'} 
\;.
\label{eq:antisymmetrized_Lambda_w=0}
\end{align}
In contrast,    
Eq.\ \eqref{eq:symmetrized_Lambda} shows that 
 the symmetrized component vanishes  for $\omega \to 0$, 
i.e.,  $\bm{\Lambda}_{J,\sigma\sigma'}^\mathrm{S}(\epsilon, \epsilon) 
 =  0$. Nevertheless,  the $\omega$ derivative remains finite,
\begin{align}
&  
\bm{\Lambda}_{d,\sigma\sigma'}^\mathrm{S}(\epsilon, \epsilon) 
+i
\left. \frac{\partial}{\partial \omega}
\bm{\Lambda}_{J,\sigma\sigma'}^\mathrm{S}(\epsilon, \epsilon+\omega)
\right|_{\omega \to 0} 
\nonumber \\
 &  \quad \  
  =  \  
\delta_{\sigma\sigma'} 
\left[\,
\bm{\tau}_3 
-
\frac{\partial \bm{\Sigma}_{\mathrm{tot},\sigma}(\epsilon)}
{\partial \epsilon}
\,
\right].
\label{eq:WT_symmetric_w0}
\end{align}
This equation is identical to the Ward identity given in  Eq.\ \eqref{eq:Ward_NEQ_omega}. 
In order to see it more clearly, we rewrite the left-hand side of 
Eq.\ \eqref{eq:WT_symmetric_w0} further.

Contributions of the charge fluctuation term    
 $\bm{\Lambda}_{d,\sigma\sigma'}^{\mathrm{S}}(\epsilon, \epsilon)$ 
in the left-hand side of Eq.\ \eqref{eq:WT_symmetric_w0}   
can be separated into two parts described diagrammatically 
 in Fig.\ \ref{fig:3point_diagrams}. 
The  first diagram corresponds simply to 
the bare vertex $\bm{\tau}_3\,\delta_{\sigma\sigma'}$, and 
the second diagram gives contributions 
which are exactly the same as the right-hand side of 
 Eq.\ \eqref{eq:Ward_Keldysh_ed}. Therefore, 
$\bm{\Lambda}_{d,\sigma\sigma'}^{\mathrm{S}}(\epsilon, \epsilon)$ 
can be expressed in terms of the self-energy, as 
\begin{align}
\bm{\Lambda}_{d,\sigma\sigma'}^{\mathrm{S}}(\epsilon, \epsilon) 
\,= \, 
\bm{\tau}_3
\,\delta_{\sigma\sigma'}
\, + \, \frac{\partial \bm{\Sigma}_{U,\sigma}^{}(\epsilon)
}{\partial \epsilon_{d\sigma'}}
\;.
\label{eq:Lambda_d}
\end{align}
More generally, contributions of the two diagrams 
for  $\bm{\Lambda}_{p,\sigma\sigma'}^{q}(\epsilon, \epsilon+\omega)$ 
shown in Fig.\ \ref{fig:3point_diagrams} can be expressed in 
terms of the four-point Keldysh vertex functions,  
\begin{widetext}
\begin{align}
&\Lambda_{p,\sigma\sigma'}^{q:\nu\mu}
(\epsilon, \epsilon+\omega) 
\nonumber \\
& \quad 
=\, 
\delta_{\sigma\sigma'}\,
\lambda_{p}^{q:\nu\mu}
(\epsilon, \epsilon+\omega)  
+
\int_{-\infty}^{\infty}\! \frac{d \epsilon'}{2\pi i }\, 
\sum_{\mu'\nu'}
\Gamma_{\sigma\sigma';\sigma'\sigma}^{\mu\mu';\nu'\nu}
 (\epsilon+\omega,\epsilon'+\omega; \epsilon',\epsilon)
\,
\left\{\bm{G}_{\sigma'}^{}(\epsilon') \, 
\bm{\lambda}_{p}^{q}
(\epsilon', \epsilon'+\omega) \, 
\bm{G}_{\sigma'}^{}(\epsilon'+\omega)
\right\}_{}^{\nu'\mu'},
\label{eq:Bethe-Salpeter}
\end{align}
for any $q$ ($=-,+, \mathrm{S}, \mathrm{A}$)  
and any  $p$ ($= d, L, R , J, ``\mathrm{sym}"$).  
Therefore, 
the second term $i\frac{\partial}{\partial \omega}
\bm{\Lambda}_{J,\sigma\sigma'}^\mathrm{S}(\epsilon, \epsilon+\omega)$ 
in the left-hand side of Eq.\ \eqref{eq:WT_symmetric_w0} 
can  be expressed in the following form,
using Eq.\ \eqref{eq:Bethe-Salpeter}
and the properties of the bare vertex  $\bm{\lambda}_{J}^\mathrm{S}
(\epsilon, \epsilon+\omega)$ shown in 
 Eqs.\ \eqref{eq:current_vertex_J_S_U0} and \eqref{eq:lambda_J_derivative},
  as 
\begin{align}
i\!  \left.\frac{\partial}{\partial \omega}
\Lambda_{J,\sigma\sigma'}^{\mathrm{S}:\nu\mu}(\epsilon, \epsilon+\omega) 
\right|_{\omega \to 0 }
  =  \ 
- \delta_{\sigma\sigma'}
\left\{\frac{\partial \bm{\Sigma}_0 (\epsilon)}{\partial \epsilon} \right\}^{\nu\mu}  
\!  + 
\int_{-\infty}^{\infty}\! \frac{d \epsilon'}{2\pi} 
\sum_{\mu'\nu'}
\Gamma_{\sigma\sigma';\sigma'\sigma}^{\mu\mu';\nu'\nu}
 (\epsilon,\epsilon'; \epsilon',\epsilon)
\,
2 \Delta\,
G_{\sigma'}^{r}(\epsilon')\,  G_{\sigma'}^{a}(\epsilon')  
\left(-\frac{\partial  f_\mathrm{eff}^{} (\epsilon')}{\partial \epsilon'}\right)
  . 
\label{eq:LambdaS_derivative_omega}
%
\end{align}
Finally,  
substituting Eqs.\  \eqref{eq:Lambda_d} and \eqref{eq:LambdaS_derivative_omega}
into Eq.\ \eqref{eq:WT_symmetric_w0},  
we obtain the same result as Eq.\ \eqref{eq:Ward_NEQ_omega} 
from the symmetrized part of the Ward-Takahashi identity,   
\begin{align}
 \left(
\delta_{\sigma\sigma' }\,  \frac{\partial}{\partial \epsilon} 
+ \frac{\partial}{\partial \epsilon_{d\sigma'}}
\right)
\Sigma_{U,\sigma}^{\nu\mu}(\epsilon)
\, = \,  
-\int_{-\infty}^{\infty}\! \frac{d \epsilon'}{2\pi}\, 
\sum_{\mu'\nu'}\,
\Gamma_{\sigma\sigma';\sigma'\sigma}^{\mu\mu';\nu'\nu}
 (\epsilon,\epsilon'; \epsilon',\epsilon)
\  2 \Delta
\,
G_{\sigma'}^{r}(\epsilon') \,G_{\sigma'}^{a}(\epsilon')
\,
\left(
-\frac{\partial  f_\mathrm{eff}^{} (\epsilon')}{\partial \epsilon'}
\right) 
\,.
\end{align}

In addition, the antisymmetrized part of the Ward-Takahashi identity 
for  $\Lambda_{J,\sigma\sigma'}^{\mathrm{A}:\nu\mu}(\epsilon,\epsilon+\omega)$ 
also provides important information about the relation between 
the self-energy and the four-point vertex functions. 
For instance, at $\omega=0$,  Eq.\ \eqref{eq:antisymmetrized_Lambda_w=0} 
can be rewritten as an integral-form relation, 
using  Eq.\ \eqref{eq:Bethe-Salpeter} for 
 $q=\mathrm{A}$ and  $p=J$ with the property of 
the bare vertex $\lambda_{J,\sigma\sigma'}^{\mathrm{A}:\nu\mu}
(\epsilon,\epsilon+\omega)$ 
shown in Eq.\ \eqref{eq:current_vertex_J_A_U0}:   
 \begin{align}
\left[
 \begin{matrix}
 \rule{0.6cm}{0cm} 0 \rule{0.6cm}{0cm}& 
-\Sigma_{U,\sigma}^{-+}(\epsilon)  \cr   
 \Sigma_{U,\sigma}^{+-}(\epsilon) & 0 \cr  
 \end{matrix}
 \right]_{\nu\mu} 
\delta_{\sigma\sigma'} 
= 
\int_{-\infty}^{\infty}\! \frac{d \epsilon'}{2\pi}\, 
\sum_{\mu'\nu'}
\Gamma_{\sigma\sigma';\sigma'\sigma}^{\mu\mu';\nu'\nu}
 (\epsilon,\epsilon'; \epsilon',\epsilon)
\,
2\Delta\,
\left\{\bm{G}_{\sigma'}^{}(\epsilon') \, 
\left[
 \begin{matrix}
 \rule{0.6cm}{0cm} 0 \rule{0.6cm}{0cm}& 
 f_\mathrm{eff}^{} (\epsilon') \cr   
1- f_\mathrm{eff}^{} (\epsilon') & 0 \cr  
 \end{matrix}
 \right]
\bm{G}_{\sigma'}^{}(\epsilon')
\right\}_{}^{\nu'\mu'} 
.
\label{eq:generalizedWT_A_part}
\end{align}
Physically, it represents  the relation between 
the damping of a single quasiparticle in the left-hand side and 
the effects of the multiple collisions of two quasiparticles in the right-hand side.

\end{widetext}


\begin{figure}[b]
\leavevmode 
\begin{minipage}{\linewidth}
 \includegraphics[width=0.6\linewidth]{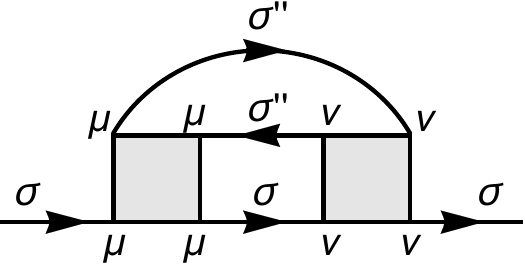}
\end{minipage}
 \caption{
Feynman diagram for $\Sigma_{\sigma}^{\nu\mu}(\omega)$ 
which yields order $\omega^2$, $(eV)^2$, $T^2$ imaginary part.
The shaded square represents 
 the scattering amplitude of quasiparticles that is given by
the full vertex correction  
$\Gamma_{\sigma,\sigma'';\sigma''\sigma}^{\nu\nu;\nu\nu}(0, 0; 0, 0)$ 
defined at $T=eV=0$. 
The inter-level $\sigma'' \neq \sigma$ components  
of the scattering amplitude is real and finite,  
whereas the intra-level ones 
identically vanish since  
 $\Gamma_{\sigma\sigma;\sigma\sigma}^{\nu\nu;\nu\nu}(0, 0; 0, 0)=0$ 
at zero frequencies 
as a result of the Pauli exclusion rule.\cite{AGD,YamadaYosida4} 
}
  \label{fig:self_im_T2}
\end{figure} 

\section{Collision integrals for the Fermi liquid at finite bias voltages}
\label{sec:Collision_Fermion}

In this section, we consider the low-energy behavior of the Keldysh self-energies  
other than the retarded $\Sigma^{r}_{U,\sigma}(\omega)$
 or advanced $\Sigma^{a}_{U,\sigma}(\omega) $ one.  
We also focus on the particle-hole and particle-particle 
excitations that determine the imaginary part 
of the Keldysh vertex functions in the Fermi-liquid regime.

\subsection{Lesser and greater self-energies in  the Fermi-liquid regime}
\label{subsec:lesser_greater_self-energy_text}

The lesser and greater self-energies,  
 $\Sigma^{-+}_{U,\sigma}(\omega)$ and $\Sigma^{+-}_{U,\sigma}(\omega)$, 
 are both pure imaginary in the frequency domain, 
and these two components  determine the imaginary part 
of the other self-energy components, 
as it can be deduced from Eqs.\ \eqref{eq:self_energy_linear_dependency} and 
\eqref{eq:causal_self_energy_in_K_component}. 
We have already discussed low-energy behavior of the retarded self-energy,  
the imaginary part of which is given by  
$2i\, \mathrm{Im}\,
\Sigma^{r}_{U,\sigma}(\omega)  = 
\Sigma^{-+}_{U,\sigma}(\omega)- \Sigma^{+-}_{U,\sigma}(\omega)$.  
However, for calculating the nonlinear current noise  $S_\mathrm{noise}^\mathrm{QD}$  
up to terms of order $(eV)^3$, 
the  low-energy asymptotic of another independent 
component  $\Sigma^\mathrm{K}_{U,\sigma}(\omega) = 
\Sigma^{-+}_{U,\sigma}(\omega) + \Sigma^{+-}_{U,\sigma}(\omega)$ 
is also necessary.

\subsubsection{Fermionic collision integrals for quasiparticles} 

The imaginary part of the self-energy $\Sigma^{\nu\mu}_{U,\sigma}(\omega)$ 
can be deduced exactly from the scattering process shown in Fig.\  \ref{fig:self_im_T2} 
up to terms of order $\omega^2$, $T^2$, and $(eV)^2$.\cite{AGD,YamadaYosida4}   
In particular, the low-energy asymptotic forms of 
the lesser and greater self-energies 
can be expressed in terms of the collision integrals of quasiparticles, 
which in the equilibrium limit $eV \to 0$ take 
the form,\cite{MorelAnderson1962} 
\begin{subequations}
\begin{align}
\mathcal{I}_\mathrm{eq}^{+-}(\omega) \equiv &  
\int\!\!d\varepsilon_1  \! \! \!\int\!\!d\varepsilon_2
 \Bigl[ 1-f(\varepsilon_1)\Bigr]
 \Bigl[ 1-f(\varepsilon_2)\Bigr]
 f(\varepsilon_1+\varepsilon_2-\omega)  
\nonumber \\
 = &  \ 
\frac{\omega^2 +  (\pi T)^2 }{2} \Bigl[\,1-f(\omega)\,\Bigr]
\,,
\\
\mathcal{I}_\mathrm{eq}^{-+}(\omega) 
\equiv & \int\!\!d\varepsilon_1 \!\! \!\int\!\!d\varepsilon_2
\, f(\varepsilon_1) \, f(\varepsilon_2)\, \Bigl[1-f(\varepsilon_1+\varepsilon_2-\omega) \Bigr]  
\nonumber \\
= & \  
\frac{\omega^2 + (\pi T)^2}{2} \,f(\omega) \,.
\end{align}
\label{eq:collision_quasiparticle_equilibrium}
\!\!\!\!\!    
\end{subequations}
Physically, $\mathcal{I}_\mathrm{eq}^{+-}(\omega)$ and 
$\mathcal{I}_\mathrm{eq}^{-+}(\omega)$ determine   
the damping rate of a single quasiparticle and a single quasihole, 
respectively. At finite $eV$, these collision integrals are extended to the following form,  replacing the Fermi function in Eq.\ \eqref{eq:collision_quasiparticle_equilibrium} 
by the nonequilibrium distribution function  $f_\mathrm{eff}(\omega)$,  as 
\begin{subequations}
\begin{align}
 & \!\!\!
\mathcal{I}^{+-}(\omega)  \equiv   
\int_{-\infty}^{\infty}\! d\omega_1 \!\int_{-\infty}^{\infty}\! d\omega_2
\nonumber \\
& \qquad \quad  \times 
\Bigl[\,1-f_\mathrm{eff}(\omega_1)\,\Bigr]
 \Bigl[\,1-f_\mathrm{eff}(\omega_2)\,\Bigr]
 \,f_\mathrm{eff}(\omega_1+\omega_2-\omega) 
\nonumber \\
& \qquad  \ \ \,     
= \sum_{j,k,\ell\atop = L,R} 
\frac{\Gamma_j\Gamma_k\Gamma_\ell}{\Delta^3}\,
\mathcal{I}_\mathrm{eq}^{+-}(\omega-\mu_j-\mu_k+\mu_\ell)\,,  
\label{eq:collision_quasiparticle_bias_+-}
\end{align}
\begin{align}
& \!\!\!\!\!  
\mathcal{I}^{-+}(\omega)  \equiv   
\int_{-\infty}^{\infty}\! d\omega_1\!\int_{-\infty}^{\infty}\! d\omega_2
\nonumber \\
&  \qquad \quad \   \times 
f_\mathrm{eff}(\omega_1)\,f_\mathrm{eff}(\omega_2)
\Bigl[\, 1- f_\mathrm{eff}(\omega_1+\omega_2-\omega)\,\Bigr] 
\nonumber \\
& \qquad \ \,    
= \sum_{j,k,\ell \atop = L,R} 
\frac{\Gamma_j\Gamma_k\Gamma_\ell}{\Delta^3}\,
\mathcal{I}_\mathrm{eq}^{-+}(\omega-\mu_j-\mu_k+\mu_\ell)\,.  
\label{eq:collision_quasiparticle_bias_-+}
\end{align}
\label{eq:collision_quasiparticle_bias}
\end{subequations}

In the Fermi-liquid regime,
 the damping rate of a quasiparticle  
is determined by the collision processes, illustrated in Fig.\ \ref{fig:self_im_T2},  
in which a single particle-hole pair with spin $\sigma''$ is virtually excited 
in the intermediate states by an incident quasiparticle with spin $\sigma$ 
via the scattering amplitude, given by the vertex functions 
 $\Gamma_{\sigma\sigma'';\sigma''\sigma}^{\nu\nu;\nu\nu}(0,0; 0,0)$ 
 at $\omega=eV=T=0$  that include all effects of multiple scatterings. 
The lesser and greater self-energies  caused by this processes 
can be expressed in terms 
of  $\mathcal{I}^{-+}(\omega)$ and $\mathcal{I}^{+-}(\omega)$,    
\begin{align}
& 
\Sigma^{-+}_{U,\sigma}(\omega)
= 
\sum_{\sigma''(\neq \sigma)}
\int\!\frac{d\omega_1}{2\pi}
\!\int\!\frac{d\omega_2}{2\pi}
\nonumber \\
 &  \qquad \qquad 
 \times
\Gamma_{\sigma''\sigma;\sigma\sigma''}^{--;--}(0,0; 0,0) \ 
\Gamma_{\sigma\sigma'';\sigma''\sigma}^{++;++}(0,0; 0,0)  
\nonumber \\
&  \qquad \qquad
 \times
\,G_{\sigma}^{-+}(\omega_1)
\,G_{\sigma''}^{-+}(\omega_2) 
\,G_{\sigma''}^{+-}(\omega_1+\omega_2-\omega) 
+ \cdots 
\nonumber \\
&=  \,   
-i \,2\pi 
\! \sum_{\sigma''(\neq \sigma)} \!  
\left|\Gamma_{\sigma\sigma'';\sigma''\sigma}^{--;--}(0,0; 0,0) \right|^2 
\rho_{d\sigma}^{} \left\{\rho_{d\sigma''}^{}\right\}^2 
  \mathcal{I}^{-+}(\omega) 
\rule{0cm}{0.6cm}
\nonumber \\
&  \quad    + \   \cdots \,,
\label{eq:lesser_asymptotic}
\end{align}
\begin{align}
 & 
\Sigma^{+-}_{U,\sigma}(\omega)
=  
\sum_{\sigma''(\neq \sigma)}
\int\!\frac{d\omega_1}{2\pi}
\!\int\!\frac{d\omega_2}{2\pi} 
 \nonumber \\
&  \qquad  \qquad
 \times
\Gamma_{\sigma\sigma'';\sigma''\sigma}^{++;++}(0,0; 0,0) \ 
\Gamma_{\sigma''\sigma;\sigma\sigma''}^{--;--}(0,0; 0,0) 
\nonumber \\
& 
 \qquad \qquad
 \times 
\,G_{\sigma}^{+-}(\omega_1)
\,G_{\sigma''}^{+-}(\omega_2) 
\,G_{\sigma''}^{-+}(\omega_1+\omega_2-\omega) 
+ \cdots 
\nonumber \\
&= \, 
i \,2\pi 
\sum_{\sigma''(\neq \sigma)}
\left|\Gamma_{\sigma\sigma'';\sigma''\sigma}^{--;--}(0,0; 0,0) \right|^2 
\rho_{d\sigma}^{} \left\{\rho_{d\sigma''}^{}\right\}^2 
 \mathcal{I}^{+-}(\omega)   
\rule{0cm}{0.6cm} 
\nonumber \\
&   \quad  + \   \cdots \,.
\label{eq:greater_asymptotic}
\end{align}
These expressions are exact up to terms of order $\omega^2$, $(eV)^2$, and $T^2$,  
and reproduce the previous results calculated  by Aligia\cite{Aligia,Aligia2014} 
using the renormalized perturbation theory.\cite{HewsonRPT2001}
In order to obtain the last lines of 
Eqs.\  \eqref{eq:lesser_asymptotic} and \eqref{eq:greater_asymptotic}, 
we have rewritten  the lesser and greater Green's functions in  the integrands 
in the following form, using Eq.\ \eqref{eq:Gdd_matrix}, 
\begin{subequations}
 \begin{align}
 G_{\sigma}^{-+}(\omega)
\,= &  \    
 - \Bigl[\, 
\Sigma_{0}^{-+}(\omega) 
+ \Sigma_{U,\sigma}^{-+}(\omega) 
\,\Bigr]
 \, G_{\sigma}^{r}(\omega)\,G_{\sigma}^{a}(\omega)\, 
\nonumber\\
 \simeq &  \   
- \Sigma_{0}^{-+}(\omega) 
 \, G_{\mathrm{eq},\sigma}^{r}(0)\,G_{\mathrm{eq},\sigma}^{a}(0)\,, 
 \\
G_{\sigma}^{+-}(\omega) 
\, = & \ 
 - \Bigl[\, 
\Sigma_{0}^{+-}(\omega) 
+ \Sigma_{U,\sigma}^{+-}(\omega) 
\,\Bigr]
 \, G_{\sigma}^{r}(\omega)\,G_{\sigma}^{a}(\omega)\, 
\rule{0cm}{0.6cm}
\nonumber
\\
\simeq&    \    
- \Sigma_{0}^{+-}(\omega) 
\, G_{\mathrm{eq},\sigma}^{r}(0)\,G_{\mathrm{eq},\sigma}^{a}(0)\,.
 \end{align}
 \label{eq:lesser_greater_lowest} 
\!\!\!\!
\end{subequations}
In the second line of each equations the total self-energy 
has been replaced by the noninteracting one:  
 $\Sigma_{0}^{-+}(\omega) = -2i \Delta f_\mathrm{eff}^{}(\omega)$,     
or $\Sigma_{0}^{+-}(\omega) = 2i \Delta [ 1-f_\mathrm{eff}^{}(\omega)]$,
given in Eq.\ \eqref{eq:U0_self_keldysh}. 
This is because the interacting self-energies 
 $\Sigma_{U,\sigma}^{-+}(\omega)$ and 
 $\Sigma_{U,\sigma}^{+-}(\omega)$ themselves 
show  $\omega^2$, $(eV)^2$, and $T^2$ dependences  
at low energies as it can be deduced from the perturbation theory in $U$, 
and they yield corrections higher than the leading-order ones. 
Furthermore, in  Eq.\ \eqref{eq:lesser_greater_lowest}, 
the retarded and advanced Green's functions 
have been replaced by the value at $\omega=eV=T=0$, i.e.,  
 $G_{\mathrm{eq},\sigma}^{r}(0)\,G_{\mathrm{eq},\sigma}^{a}(0) 
= \pi \rho_{d\sigma}^{}/\Delta$ in the last line of each equation.  

Note that the vertex correction 
that appeared in Eqs.\   
\eqref{eq:lesser_asymptotic} and \eqref{eq:greater_asymptotic} 
can be expressed in terms of  the static susceptibility $\chi_{\sigma\sigma''}^{}$ 
for different spin components $\sigma \neq \sigma''$,  
using the Fermi-liquid relations given in 
Eqs.\ \eqref{eq:chi_org} and \eqref{eq:YY2_results}, as 
\begin{align}
\chi_{\sigma\sigma''}^{}\,=\, - \,
\rho_{d\sigma}^{} \rho_{d\sigma''}^{} 
\Gamma_{\sigma\sigma'';\sigma''\sigma}^{--;--}(0,0; 0,0) \,.
\label{eq:chi_T0}
\end{align}

\subsubsection{Symmetrization of the fermionic collision integrals}

One can also deduce 
the imaginary part of the retarded self-energy  
given in Eq.\ \eqref{eq:self_imaginary_N} in another way, 
from the difference between the lesser and greater ones 
$2i\, \mathrm{Im}\, \Sigma^{r}_{U,\sigma}(\omega)  
 = \Sigma^{-+}_{U,\sigma}(\omega) - \Sigma^{+-}_{U,\sigma}(\omega)$,  
  using Eqs.\ 
\eqref{eq:collision_quasiparticle_equilibrium}--\eqref{eq:greater_asymptotic}. 
The leading-order terms in the Fermi-liquid regime are 
determined by the sum of the single-quasiparticle collision integrals 
$\mathcal{I}_{}^{+-}(\omega) +  \mathcal{I}_{}^{-+}(\omega)$,
\begin{align}
& \!\! 
\mathcal{I}_\mathrm{diff}^{}(\omega)  
\,\equiv\,
2\,\Bigl[\,
\mathcal{I}_{}^{+-}(\omega) +  
\mathcal{I}_{}^{-+}(\omega) \,\Bigr]  
 \nonumber \\
& \ 
= \Bigl( \omega - 
\alpha_\mathrm{sh}^{} eV
 \Bigr)^2
  +  
\frac{ 3\Gamma_L^{}\Gamma_R^{}}{(\Gamma_L+\Gamma_R)^2}\,(eV)^2 
+ (\pi T)^2 .
\label{eq:quasi_collision_sum}
\end{align}
Note that $\mathcal{I}_\mathrm{eq}^{+-}(\omega)+  
\mathcal{I}_\mathrm{eq}^{-+}(\omega)= \omega^2 + (\pi T)^2$; 
the Fermi distribution disappears. 

In contrast,  
$\Sigma^\mathrm{K}_{U,\sigma}(\omega)$ 
is determined by the difference 
$\mathcal{I}_{}^{+-}(\omega) -  \mathcal{I}_{}^{-+}(\omega)$ 
between the two collision terms, 
  up to terms of  order $\omega^2$, $T^2$, and $(eV)^2$, as 
\begin{align}
&\frac{\Sigma^{-+}_{U,\sigma}(\omega) 
+ \Sigma^{+-}_{U,\sigma}(\omega)}{2i}
\,=  \, 
\frac{\pi}{2 \rho_{d\sigma}} \sum_{\sigma''(\neq \sigma)} \chi_{\sigma\sigma''}^2 
\  \mathcal{I}_\mathrm{K}^{}(\omega)
\,+\, \cdots  \,.
\label{eq:self_Keldysh_compoment}
\end{align}
\begin{align}
 &\mathcal{I}_\mathrm{K}^{} (\omega) 
  \,  \equiv \,     
2\Bigl[\,\mathcal{I}_{}^{+-}(\omega)\, -\,  
\mathcal{I}_{}^{-+}(\omega)\, \Bigr]  
\nonumber 
\\
&=   
 \sum_{j,k,\ell  = L,R} 
\frac{\Gamma_j\Gamma_k\Gamma_\ell}{\Delta^3}\,
 \biggl[\,(\pi T)^2 + (\omega-\mu_j-\mu_k + \mu_\ell)^2 \,\biggr]
 \nonumber 
\\
& 
\qquad \qquad \quad 
\times 
\tanh \left(\frac{\omega-\mu_j-\mu_k + \mu_\ell}{2T}\right) \,.
\label{eq:collision_diff_general}
\end{align}
We have also shown in TABLE  \ref{tab:self-energy} 
an alternative expression, carrying out the summation over $j,k,\ell$ explicitly.
In some special cases, it takes a simplified form.
For example, at equilibrium  $eV \to 0$, it is given by 
 $\mathcal{I}_\mathrm{K}^{}(\omega) \xrightarrow{\,eV \to 0\,} 
[ \omega^2 + (\pi T)^2 ] \tanh (\frac{\omega}{2T})$.
At zero temperature, 
the hyperbolic function that comes from the Fermi function 
is replaced by the sign function, as  $1-2f(\omega)=\tanh\frac{\omega}{2T} 
\xrightarrow{\,T\to 0\,} 
\mathrm{sgn}\,  \omega$. 
Furthermore, 
for symmetric junctions $\Gamma_L=\Gamma_R=\Delta/2$ 
and symmetric bias voltages $\mu_L = -\mu_R=eV/2$,
it takes the following  form at $T=0$,
\begin{align}
\!\!\!
\mathcal{I}_\mathrm{K}^{}(\omega)
 \to &   \ 
\frac{1}{8}
\,\Biggl[ \,   
\left( \omega - \frac{3eV}{2}  \right)^2 
 \mathrm{sgn} \!  \left( \omega - \frac{3eV}{2} \right) 
\nonumber \\
& \ \ 
+3\left(  \omega -\frac{eV}{2}  \right)^2 
 \mathrm{sgn} \!  \left( \omega - \frac{eV}{2} \right) 
\nonumber \\
& \ \ 
+3\left(  \omega +\frac{eV}{2}  \right)^2 
 \mathrm{sgn} \!  \left( \omega + \frac{eV}{2} \right) 
\nonumber \\
& \ \ 
+ 
\left( \omega + \frac{3eV}{2}  \right)^2 
 \mathrm{sgn} \!  \left( \omega + \frac{3eV}{2} \right) 
 \Biggr] .
\label{eq:self_Keldysh_component_symmetric} 
\end{align}

We will show later in Sec.\  
\ref{sec:current_conservation_Keldysh_selfenergy_vertex_in_FL_regime}
that these results of the self-energies satisfy the Ward identities 
with the low-energy results of the vertex corrections 
listed in TABLES \ref{tab:vertex_UU} and \ref{tab:vertex_UD}.


\begin{table*}[t]
\caption{Fermionic collision integral 
$\mathcal{I}_\mathrm{K}^{}(\omega)$, 
 and low-energy asymptotic form of the self-energy 
at finite bias $eV = \mu_L-\mu_R$. 
The parameter $\alpha_\mathrm{sh}$ is defined such that 
 $\alpha_\mathrm{sh}\, eV  = 
 (\Gamma_L  \mu_L +  \Gamma_R \mu_R)/(\Gamma_L+ \Gamma_R)$. } 
\begin{tabular}{l} 
\hline \hline 
\rule{1.5cm}{0cm}
\rule{0cm}{0.7cm}
$\mathcal{I}_\mathrm{K}^{}(\omega)
\equiv 
\ 2 \! \displaystyle \int\!d\varepsilon_1  \! \!\int\!d\varepsilon_2\, 
\biggl\{\, \Bigl[ 1-f_\mathrm{eff}^{}(\varepsilon_1)\Bigr]
 \Bigl[ 1-f_\mathrm{eff}^{}(\varepsilon_2)\Bigr]\,
 f_\mathrm{eff}^{}(\varepsilon_1+\varepsilon_2-\omega) 
\ - \ 
 f_\mathrm{eff}^{}(\varepsilon_1) \, f_\mathrm{eff}^{}(\varepsilon_2)\, 
\Bigl[1-f_\mathrm{eff}^{}(\varepsilon_1+\varepsilon_2-\omega) \Bigr] \,\biggr\}, $
\rule{0cm}{0.7cm}
\\
\rule{2.4cm}{0cm}
 $\
\,=    \,   
\frac{\Gamma_L^{2}\Gamma_R^{}}{(\Gamma_L+\Gamma_R)^3}\  
 \Bigl[(\pi T)^2 + (\omega-\mu_L-eV)^2 \Bigr]
\tanh\frac{\omega-\mu_L-eV}{2T}
+\,   \frac{\Gamma_L^{3} +2\Gamma_L^{}\Gamma_R^{2}}
 {(\Gamma_L+\Gamma_R)^3} \ 
 \Bigl[(\pi T)^2 + (\omega-\mu_L)^2 \Bigr]
 \,\tanh\frac{\omega-\mu_L}{2T}
$
\rule{0cm}{0.7cm}
\\
\rule{2.8cm}{0cm}
$
+
 \frac{\Gamma_R^{3} +2\Gamma_L^{2}\Gamma_R^{}}
 {(\Gamma_L+\Gamma_R)^3}\ 
 \Bigl[(\pi T)^2 + (\omega-\mu_R)^2 \Bigr]
 \,\tanh\frac{\omega-\mu_R}{2T}
+\frac{\Gamma_L^{}\Gamma_R^{2}}{(\Gamma_L+\Gamma_R)^3}\  
 \Bigl[(\pi T)^2 + (\omega-\mu_R+eV)^2 \Bigr]
\tanh\frac{\omega-\mu_R+eV}{2T}  
 $.
\rule{0cm}{0.65cm}
\\
 \ \vspace{-0.35cm}
\\
\hline
$  
\Sigma^{-+}_{U,\sigma}(\omega) 
+ \Sigma^{+-}_{U,\sigma}(\omega)
\, = \,     
{\displaystyle 
i \,\frac{\pi}{\rho_{d\sigma}^{}}   
\sum_{\sigma'(\neq \sigma)} }
 \chi_{\sigma\sigma'}^2 \, \mathcal{I}_\mathrm{K}^{}(\omega)  + \cdots  $, 
\rule{0cm}{0.7cm}
\\
$
\Sigma_{U,\sigma}^{-+}(\omega) 
-\Sigma_{U,\sigma}^{+-}(\omega) 
\,  = \,
{\displaystyle 
 - \, i \, \frac{\pi}{\rho_{d\sigma}^{}} 
\sum_{\sigma'(\neq \sigma)} }
\chi_{\sigma\sigma'}^2
 \,   \biggl[\ 
\bigl( \omega - 
 \alpha_\mathrm{sh}\, eV 
\bigr)^2 
+  \frac{ 3\Gamma_L^{}\Gamma_R^{}}{(\Gamma_L+\Gamma_R)^2}\,(eV)^2 
  + (\pi T)^2 
    \ \biggr]   \ + \cdots ,
$
\rule{0cm}{0.65cm}
\\
\rule{0.15cm}{0cm}
$
 \mathrm{Re}\, \Sigma_{U,\sigma}^r(\omega) 
\,=\,
\Sigma_{\mathrm{eq},\sigma}^r(0)
\,+ \bigl( 1-\widetilde{\chi}_{\sigma\sigma} \bigr)\, \omega 
\, + 
\frac{1}{2}\,\frac{\partial \widetilde{\chi}_{\sigma\sigma}}
{\partial \epsilon_{d\sigma}^{}}\, \omega^2 
 \, +  \frac{1}{6}\,
 \frac{1}{\rho_{d\sigma}^{}} 
{\displaystyle \sum_{\sigma'(\neq \sigma)} }
\frac{\partial \chi_{\sigma\sigma'}}{\partial \epsilon_{d\sigma'}^{}} 
\left[\,
 \left( \pi T\right)^2 
+ \frac{3\Gamma_L \Gamma_R}{\left( \Gamma_L + \Gamma_R \right)^2} 
 \,(eV)^2 
\,\right] 
 $
\rule{0cm}{0.65cm}
\\
\rule{2.9cm}{0cm}
$ 
- { \displaystyle \sum_{\sigma'(\neq \sigma)} }
\widetilde{\chi}_{\sigma\sigma'}^{} 
\,  \alpha_\mathrm{sh}\,eV 
\,+ 
{\displaystyle \sum_{\sigma'(\neq \sigma)} }
\frac{\partial \widetilde{\chi}_{\sigma\sigma'}}{\partial \epsilon_{d\sigma}^{}} 
\,\alpha_\mathrm{sh}\,eV \, \omega
\, +  \frac{1}{2}\,
{\displaystyle \sum_{\sigma' (\neq \sigma)} \sum_{\sigma'' (\neq \sigma)} }
\frac{\partial\widetilde{\chi}_{\sigma\sigma'}}{\partial\epsilon_{d\sigma''}^{}}
\alpha_\mathrm{sh}^2  (eV)^2 
\ + \,\cdots \,.$
\rule{0cm}{0.5cm}
\\
\hline
\hline
\end{tabular}
\label{tab:self-energy}
\end{table*}

\begin{table*}[t]
\caption{
Bosonic collision integrals 
$\mathcal{W}_\mathrm{K}^\mathrm{ph}(\omega)$  and 
$\mathcal{W}_\mathrm{K}^\mathrm{pp}(\omega)$  
that determine low-energy behavior of the PH propagator 
$X_{\sigma\sigma'}^{\mu\nu}$ 
and PP propagator $Y_{\sigma\sigma'}^{\mu\nu}$.
}
\begin{tabular}{l} 
\hline \hline
$  \ \ 
\mathcal{W}_\mathrm{K}^\mathrm{ph}(\omega) \,\equiv  \, 
 \displaystyle \int_{-\infty}^{\infty} \!d\varepsilon \,
\biggl\{
f_\mathrm{eff}^{}(\varepsilon)
\Bigl[\,1-f_\mathrm{eff}^{}(\varepsilon + \omega) \,\Bigr]
+ f_\mathrm{eff}^{}(\varepsilon+\omega)
\Bigl[\,1-f_\mathrm{eff}^{}(\varepsilon) \,\Bigr]
\biggr\}
 $
\rule{0cm}{0.75cm}
\\
\rule{1.3cm}{0cm}
$  
=   \  
\frac{\Gamma_L^2+\Gamma_R^2}{(\Gamma_L+\Gamma_R)^2}\, 
\, \omega  \coth \frac{ \omega}{2T} 
\,+\,  \frac{\Gamma_L \Gamma_R}{(\Gamma_L+\Gamma_R)^2}
\Biggl[\, 
 \left( \omega  + eV \right) \coth \frac{\omega +eV}{2T} 
\, + \, 
\left( \omega  - eV \right) \coth \frac{\omega - eV}{2T} 
\, \Biggr].
 $
\rule{0cm}{0.65cm}
\\
$  \ \ 
\mathcal{W}_\mathrm{K}^\mathrm{pp}(\omega) \,\equiv  \, 
 \displaystyle \int_{-\infty}^{\infty} \!d\varepsilon \,
\biggl\{
\Bigl[\,1-f_\mathrm{eff}^{}(\varepsilon) \,\Bigr]
\Bigl[\,1-f_\mathrm{eff}^{}(\omega- \varepsilon) \,\Bigr]
\,+\,  
f_\mathrm{eff}^{}(\varepsilon)\,f_\mathrm{eff}^{}(\omega-\varepsilon)
\biggr\}
 $
\rule{0cm}{0.75cm}
\\
\rule{1.3cm}{0cm}
$  
=   \  
\frac{\Gamma_L^2}{(\Gamma_L+\Gamma_R)^2}\,
 \bigl(\omega-2\mu_L \bigr) \coth \frac{\omega-2\mu_L}{2T} 
\,+\,  \frac{\Gamma_R^2}{(\Gamma_L+\Gamma_R)^2}\,
\bigl( \omega-2\mu_R\bigr)   \coth \frac{\omega-2\mu_R}{2T}
\,+\,  \frac{2\Gamma_L\Gamma_R}{(\Gamma_L+\Gamma_R)^2}\,
\bigl( \omega-\mu_L -\mu_R \bigr) \coth \frac{\omega-\mu_L-\mu_R}{2T}.
 $
\rule{0cm}{0.65cm}
\\
 \ \vspace{-0.35cm}
\\
\hline
$
\quad \ 
X_{\sigma\sigma'}^{\mu\nu} (\omega)
\,\equiv \,  \displaystyle 
- \int _{-\infty}^{\infty} \!\frac{d\varepsilon}{2\pi i} \,
G_{\sigma}^{\mu\nu}(\varepsilon+\omega)
\, G_{\sigma'}^{\nu\mu}(\varepsilon)\,,
\qquad \qquad 
Y_{\sigma\sigma'}^{\mu\nu} (\omega)
\,\equiv  \,  \displaystyle 
- \int _{-\infty}^{\infty}\!\frac{d\varepsilon}{2\pi i} \,
G_{\sigma}^{\mu\nu}(\omega-\varepsilon)\,
G_{\sigma'}^{\mu\nu}(\varepsilon)\,,
$
\rule{0cm}{0.65cm}
\\
$
\mathrm{Im}\, X_{\sigma\sigma'}^{--}(\omega) 
\ = \   
\displaystyle 
\frac{1}{2i}\Bigl[\,X_{\sigma\sigma'}^{+-}(\omega)  
+X_{\sigma\sigma'}^{-+}(\omega) 
\,\Bigr] \ = \ \ 
 \pi\,  \rho_{d\sigma}^{}  \rho_{d\sigma'}^{}\, 
\mathcal{W}_\mathrm{K}^\mathrm{ph}(\omega) 
 \ + \ O\left(\omega^2, (eV)^2, T^2 \right) ,
$
\rule{0cm}{0.75cm}
\\
$
\mathrm{Im}\, Y_{\sigma\sigma'}^{--}(\omega) 
\ = \   
\displaystyle 
\frac{1}{2i}\Bigl[\,Y_{\sigma\sigma'}^{+-}(\omega)  
+Y_{\sigma\sigma'}^{-+}(\omega) 
\,\Bigr] \ = \ 
 -\, \pi\,  \rho_{d\sigma}^{}  \rho_{d\sigma'}^{}\, 
\mathcal{W}_\mathrm{K}^\mathrm{pp}(\omega) 
 \ + \ O\left(\omega^2, (eV)^2, T^2 \right) ,
$
\rule{0cm}{0.65cm}
\\
 \ \vspace{-0.35cm}
\\
\hline
\hline
\end{tabular}
\label{tab:ph-pp_propagators}
\end{table*}


\begin{figure}[b]
\leavevmode 
\begin{minipage}{\linewidth}
\includegraphics[width=0.35\linewidth]{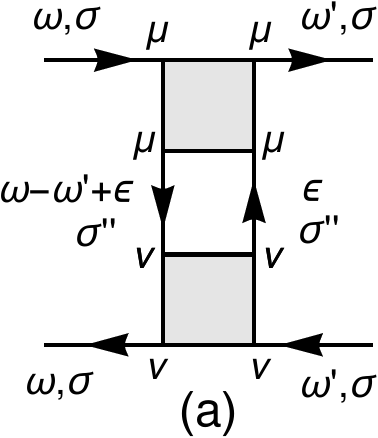}
\rule{0.1\linewidth}{0cm}
\includegraphics[width=0.35\linewidth]{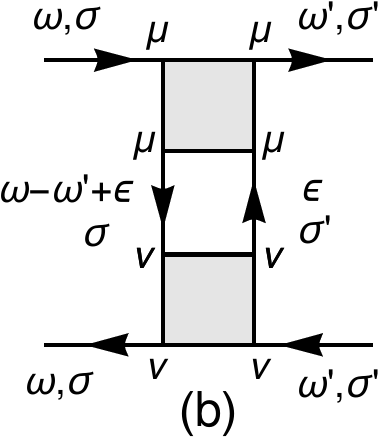}
\\
\rule{0.05\linewidth}{0cm}
\includegraphics[width=0.35\linewidth]{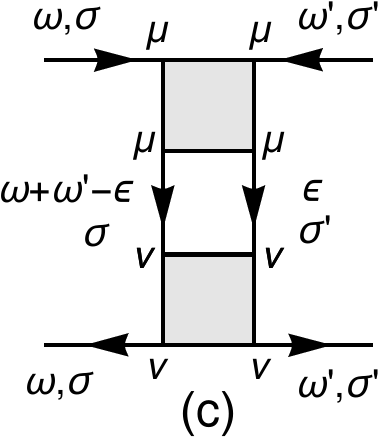}
\rule{0.05\linewidth}{0cm}
\raisebox{0.1cm}{
\includegraphics[width=0.45\linewidth]{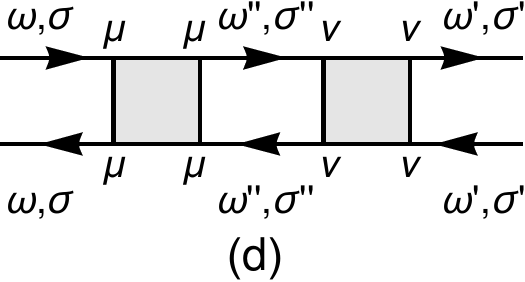}
}
\end{minipage}
 \caption{
Feynman diagrams for the Keldysh vertex functions   
$\Gamma_{\sigma\sigma';\sigma'\sigma}^{\nu\nu;\nu\nu}
(\omega, \omega'; \omega', \omega)$ 
which yield the imaginary part up to linear-order terms 
with respect to $\omega$, $\omega'$,  $eV$, and $T$.
The shaded square represents 
 the scattering amplitude of quasiparticles that is given by
the full vertex correction  
$\Gamma_{\sigma,\sigma'';\sigma''\sigma}^{\nu\nu;\nu\nu}(0, 0; 0, 0)$ 
defined at $T=eV=0$. 
The inter-level $\sigma'' \neq \sigma$  
of the scattering amplitude is real and finite,  
whereas the intra-level ones 
identically vanish since  
 $\Gamma_{\sigma\sigma;\sigma\sigma}^{\nu\nu;\nu\nu}(0, 0; 0, 0)=0$ 
at zero frequencies as a result of the Pauli exclusion rule.
}
  \label{fig:vertex_singular_general}
\end{figure} 

\subsection{Particle-hole pair and particle-particle pair excitations 
in the Fermi-liquid regime}
\label{subsec:PH_PP_propagators}

We have reexamined in the above that 
the leading-order terms of the imaginary part of the self-energies  
are determined by the three internal propagators 
which represent virtually excited quasiparticles and holes. 
Correspondingly,  the leading-order terms  
of the imaginary part of the vertex functions 
$\Gamma_{\sigma\sigma';\sigma'\sigma}^{\nu\nu;\nu\nu}
(\omega, \omega'; \omega', \omega)$
are determined by 
four types of collision processes,  illustrated 
in Fig.\ \ref{fig:vertex_singular_general}.  
The  particle-hole (PH) and particle-particle (PP) pairs 
which propagate in between the shaded squares (vertex functions)
determine  the damping rate up to terms of order  $\omega$, $\omega'$, 
$eV$, and $T$.

\subsubsection{Bosonic collision integrals for PH and PP excitations}

Leaving precise calculations of the Keldysh vertex functions 
 to Appendix \ref{sec:full_vertex_low_energy_form},  
here we consider  the low-energy behavior of the PH and PP propagators, 
 $X_{\sigma\sigma'}^{\mu\nu}(\omega)$  and
 $Y_{\sigma\sigma'}^{\mu\nu}(\omega)$, 
defined later in Eq.\ \eqref{eq:ph_pp_propagator_def}. 
The low-energy behaviors of the imaginary parts of 
the PH and PP propagators up to the linear-order terms 
with respect to  $\omega$, $eV$, and $T$ 
 can be described in terms of  the collision integrals 
 $\mathcal{W}_{}^\mathrm{ph}(\omega)$,
 $\mathcal{W}_{}^\mathrm{pp}(\omega)$, and   
 $\mathcal{W}_{}^\mathrm{hh}(\omega)$, 
defined below in Eq.\ 
\eqref{eq:define_L_omega_T_eV}--\eqref{eq:pp_collision_term}. 
In this and the next section, 
we provide a nonequilibrium Fermi-liquid description, 
using  these bosonic collision integrals.

To this end, we start with the equilibrium case  $eV = 0$, 
at which the damping of the PH and PP excitations are determined by 
a single bosonic collision integral,  
\begin{align}
\mathcal{W}_\mathrm{eq}(\omega) \, \equiv  &  
\int_{-\infty}^{\infty} \!d\varepsilon \,
f(\varepsilon) \Bigl[\,1-f(\varepsilon + \omega) \,\Bigr]
\nonumber \\
= &  \,  
\Bigl[\, 1+b(\omega)\,\Bigr]  
\lim_{D\to \infty}
\int_{-D}^{\infty} \!d\varepsilon \,
\,
\Bigl[\,f(\varepsilon)  \,-\,f(\varepsilon+\omega) \,\Bigr]
\nonumber \\
= & \   \frac{\omega}{2}\,  \left[   \coth \frac{\omega}{2T} +1 \right] . 
\label{eq:define_L_omega_T_equilibrium_appendix}
\rule{0cm}{0.6cm}
\end{align}
Here, $b(\omega)=\bigl[e^{\omega/T} -1\bigr]^{-1}$ is the Bose function.
Note that 
$\omega\,\coth (\frac{\omega}{2T}) \simeq 2T$ for $|\omega| \ll T$, 
and 
$\omega\,\coth (\frac{\omega}{2T}) 
\simeq |\omega|$ for $|\omega| \gg T$. 
Thus, 
$\mathcal{W}_\mathrm{eq}(\omega) \xrightarrow{\,T\to 0\,} 
|\omega|\, \theta(\omega)$ at $T\to 0$.

We now introduce the nonequilibrium bosonic 
collision integral  $\mathcal{W}_{}^\mathrm{ph}(\omega)$  
which determines the imaginary part of  the PH propagator at finite $eV$, 
replacing the Fermi function $f(\epsilon)$  
in Eq.\ \eqref{eq:define_L_omega_T_equilibrium_appendix}
by  $f_\mathrm{eff}(\epsilon)$,  
\begin{align}
\mathcal{W}_{}^\mathrm{ph}(\omega) \, \equiv  &  \  
\int_{-\infty}^{\infty} \!d\varepsilon \,
f_\mathrm{eff}^{}(\varepsilon)
\Bigl[\,1-f_\mathrm{eff}^{}(\varepsilon + \omega) \,\Bigr]  
\nonumber \\
=& \, \sum_{j,k = L,R} 
\frac{\Gamma_j\Gamma_k}{\Delta^2}\,
\mathcal{W}_\mathrm{eq}(\omega+\mu_j-\mu_k) . 
\label{eq:define_L_omega_T_eV}
\end{align} 
Furthermore, the time-reversal process  describes the propagation 
of a single pair which consists of a quasiparticle with frequency  $\varepsilon'$
and a quasihole with $\varepsilon' +\omega$, as 
\begin{align} 
\mathcal{W}_{}^\mathrm{ph}(-\omega) \, =   
\int_{-\infty}^{\infty} \!d\varepsilon' \,
f_\mathrm{eff}^{}(\varepsilon' + \omega )
\Bigl[\,1-f_\mathrm{eff}^{}(\varepsilon') \,\Bigr] . 
\end{align} 
Similarly, the collision integrals for the single particle-particle  
and hole-hole pairs are also given, respectively, by
\begin{align} 
\mathcal{W}^\mathrm{pp}_{}(\omega) \,\equiv &  \  
\int \! d\varepsilon\,
\Bigl[\,1-f_\mathrm{eff}^{}(\varepsilon) \,\Bigr]
\Bigl[\,1-f_\mathrm{eff}^{}(\omega-\varepsilon) \,\Bigr] , 
\nonumber \\
=& \, \sum_{j,k = L,R} 
\frac{\Gamma_j\Gamma_k}{\Delta^2}\,
\mathcal{W}_\mathrm{eq}(\omega-\mu_j-\mu_k) ,
\label{eq:pp_collision_term} 
\\
\mathcal{W}^\mathrm{hh}_{}(\omega) 
\,\equiv & \ 
\int \! d\varepsilon\,
f_\mathrm{eff}^{}(\varepsilon) \,f_\mathrm{eff}^{}(\omega-\varepsilon) 
\nonumber \\
=& \, \sum_{j,k  = L,R} 
\frac{\Gamma_j\Gamma_k}{\Delta^2}\,
\mathcal{W}_\mathrm{eq}(-\omega+\mu_j+\mu_k) .
\label{eq:hh_collision_term} 
\end{align}
We have also presented in TABLE \ref{tab:ph-pp_propagators}
more explicit forms of these bosonic collision integrals, 
carrying out the summations over $j$ and $k$.

\subsubsection{
Imaginary part of the PH and PP propagators}

We next consider the properties of   
the particle-hole and particle-particle propagators 
 $X_{\sigma\sigma'}^{\mu\nu}(\omega)$  and
 $Y_{\sigma\sigma'}^{\mu\nu}(\omega)$, defined by 
\begin{subequations}
\begin{align}
X_{\sigma\sigma'}^{\mu\nu} (\omega)
\,\equiv & \ - 
\int \!\frac{d\varepsilon}{2\pi i} \,
G_{\sigma}^{\mu\nu}(\varepsilon+\omega)
\, G_{\sigma'}^{\nu\mu}(\varepsilon),
\label{eq:ph_propagator_def}
\\
Y_{\sigma\sigma'}^{\mu\nu} (\omega)
\,\equiv & \ - 
\int \!\frac{d\varepsilon}{2\pi i} \,
G_{\sigma}^{\mu\nu}(\omega-\varepsilon)\,
G_{\sigma'}^{\mu\nu}(\varepsilon).
\label{eq:pp_propagator_def}
\end{align}
\label{eq:ph_pp_propagator_def}
\!\!\!\!\!\!
\end{subequations}
Note that the lesser and greater components 
of these propagators, i.e.,   
 $X_{\sigma\sigma'}^{-+}$,  
 $X_{\sigma\sigma'}^{+-}$,  
 $Y_{\sigma\sigma'}^{-+}$,  and 
 $Y_{\sigma\sigma'}^{+-}$, 
are  pure imaginary.  
Furthermore, the diagonal components are related to each other, 
as 
 $X_{\sigma\sigma'}^{++} (\omega) = 
- \left\{X_{\sigma\sigma'}^{--} (\omega) \right\}^*$ 
and 
 $Y_{\sigma\sigma'}^{++} (\omega) = 
- \left\{Y_{\sigma\sigma'}^{--} (\omega) \right\}^*$  
since the Green's function has the property  $G_{\sigma}^{++} (\varepsilon) = 
- \left\{G_{\sigma}^{--} (\varepsilon) \right\}^*$. 
The components of the PH and PP propagators  
are also linearly dependent, respectively,  as  
$
X_{\sigma\sigma'}^{--}
+X_{\sigma\sigma'}^{++} 
-  X_{\sigma\sigma'}^{+-} 
-X_{\sigma\sigma'}^{-+}  
 =  0 $, and 
$Y_{\sigma\sigma'}^{--} 
+Y_{\sigma\sigma'}^{++} 
-  Y_{\sigma\sigma'}^{+-} 
-Y_{\sigma\sigma'}^{-+} 
 =  0$. 
Thus,  in the frequency domain, the imaginary part of the causal 
propagator can be described in terms of  the lesser and greater components, as 
\begin{align}
\!\! \!
\mathrm{Im}\, X_{\sigma\sigma'}^{--}(\omega)
 =    
\frac{
X_{\sigma\sigma'}^\mathrm{K}(\omega)
}{2i} ,
\ \ \mathrm{Im}\, Y_{\sigma\sigma'}^{--}(\omega)
 =  
\frac{
Y_{\sigma\sigma'}^\mathrm{K}(\omega)
}{2i} .
\label{eq:ImXY--}
\end{align}
Here, 
$ 
X_{\sigma\sigma'}^\mathrm{K} \equiv  
X_{\sigma\sigma'}^{+-} +X_{\sigma\sigma'}^{-+}$ 
and 
$ 
Y_{\sigma\sigma'}^\mathrm{K} \equiv  
Y_{\sigma\sigma'}^{+-} +Y_{\sigma\sigma'}^{-+}$. 
Note that the retarded and advanced PH propagators are given  by 
$X_{\sigma\sigma'}^{r} = X_{\sigma\sigma'}^{--}- X_{\sigma\sigma'}^{-+}$ and
$X_{\sigma\sigma'}^{a} = X_{\sigma\sigma'}^{--}- X_{\sigma\sigma'}^{+-}$, 
respectively. 
The  retarded and advanced propagators for the PP excitations 
are also given by the same linear-combination forms.

The PH collision integral  $\mathcal{W}_{}^\mathrm{ph}(\omega)$, 
defined in Eq.\ \eqref{eq:define_L_omega_T_eV},  determines 
the low-energy behavior of the greater component of 
the PH propagator $X_{\sigma\sigma'}^{+-} (\omega)$ exactly 
up to  linear-order terms with respect to  $\omega$, $T$, and $eV$, 
\begin{align}
& 
\! \!  
X_{\sigma\sigma'}^{+-} (\omega)
\,\equiv    \,  
- 
\int \!\frac{d\varepsilon}{2\pi i} \,
G_{\sigma}^{+-}(\varepsilon+\omega)
\, G_{\sigma'}^{-+}(\varepsilon)
\nonumber
\\
& 
\!\!\!
\simeq \,  
 - \!\!  \int \!\frac{d\varepsilon}{2\pi i} \,
(2\pi)^2
 \rho_{d\sigma'}^{}(\varepsilon)\,
 \rho_{d\sigma}^{}(\varepsilon+\omega)
f_\mathrm{eff}^{}(\varepsilon)
\Bigl[1-f_\mathrm{eff}^{}(\varepsilon+\omega) \Bigr]
\nonumber
\\
& \!\!\!
=\,
 i \,2\pi\,  \rho_{d\sigma'}^{}\,  \rho_{d\sigma}^{}\,
\mathcal{W}_{}^\mathrm{ph}(\omega)  
\, + \, \cdots
\label{eq:X+-_low_energy}
\end{align}
To obtain the second line,  
 the lesser and greater Green's functions in the integrand has been replaced 
by the lowest-order ones, i.e.,  
$G_{\sigma}^{-+}(\omega) 
\, \simeq \,  
 -i\, 2 \pi   \rho_{d\sigma}^{} f_\mathrm{eff}^{}(\omega)$, and 
$G_{\sigma}^{+-}(\omega) 
\,\simeq\,  
 i\, 2 \pi  \rho_{d\sigma}^{} 
[ 1-f_\mathrm{eff}^{}(\omega)]$ 
in the same way as that used for Eq.\ \eqref{eq:lesser_greater_lowest}.
Correspondingly, the low-energy asymptotic form of  
the lesser PH propagator $X_{\sigma\sigma'}^{-+}(\omega)$ 
can be expressed in  terms of  $\mathcal{W}_{}^\mathrm{ph}(-\omega)$, as  
\begin{align}
& \!\! 
X_{\sigma\sigma'}^{-+}(\omega) 
\, \equiv  \, 
- \int \!\frac{d\varepsilon}{2\pi i} \,
 G_{\sigma}^{-+}(\varepsilon+\omega)\, 
G_{\sigma'}^{+-}(\varepsilon)\,
\nonumber \\
& 
= \, 
- \!\! 
\int \!\frac{d\varepsilon'}{2\pi i} \,
G_{\sigma'}^{+-}(\varepsilon'-\omega)
\, G_{\sigma}^{-+}(\varepsilon')
\ \,   = \ 
X_{\sigma'\sigma}^{+-} (-\omega) 
\nonumber \\
%
&  =  \, 
 i \,2\pi\,  \rho_{d\sigma'}^{}\,  \rho_{d\sigma}^{}\,
 \mathcal{W}_{}^\mathrm{ph}(-\omega) \,  + \, \cdots 
\label{eq:X-+_low_energy}
 \end{align}

Similarly, the low-energy behaviors of the PP propagators can be 
described in terms of the collision integrals  
 $\mathcal{W}_{}^\mathrm{pp}(\omega)$  and  
 $\mathcal{W}_{}^\mathrm{hh}(\omega)$.  
The greater $Y_{\sigma\sigma'}^{+-} (\omega)$ 
component can be calculated 
up to linear-order terms in $\omega$,  $eV$ and $T$, as 
\begin{align}
&Y_{\sigma\sigma'}^{+-} (\omega)
\, \equiv \, 
- 
\int \!\frac{d\varepsilon}{2\pi i} \,
 G_{\sigma}^{+-}(\omega-\varepsilon)
\,G_{\sigma'}^{+-}(\varepsilon)
\nonumber 
\\
& =  
 \int \!\frac{d\varepsilon}{2\pi i} \,
(2\pi)^2
 \rho_{d\sigma}^{}  \rho_{d\sigma'}^{}
\Bigl[1-f_\mathrm{eff}^{}(\varepsilon) \Bigr]
\Bigl[1-f_\mathrm{eff}^{}(\omega-\varepsilon) \Bigr] 
+\cdots 
\nonumber
\\
&= \, 
 -i 2\pi\,   \rho_{d\sigma}^{}\, \rho_{d\sigma'}^{}\, 
 \mathcal{W}_{}^\mathrm{pp}(\omega)
 \, + \,  \cdots 
.
\rule{0cm}{0.6cm}
\end{align}
The lesser component  $Y_{\sigma\sigma'}^{-+} (\omega)$ 
is determined by the HH collision integral  $\mathcal{W}_{}^\mathrm{hh}(\omega)$,  
\begin{align}
&
\!\!\!\!\!\!\!\!
\!\!\!\!\!\!\!\!
Y_{\sigma\sigma'}^{-+} (\omega)
\, \equiv   \ 
- 
\int \!\frac{d\varepsilon}{2\pi i} \,
 G_{\sigma}^{-+}(\omega-\varepsilon)
\,G_{\sigma'}^{-+}(\varepsilon)
\nonumber 
\\
& 
\!\!\!\!\!\!\!\!
\!\!\!\!\!\!\!\!
= 
 \int \!\frac{d\varepsilon}{2\pi i} \,
(2\pi)^2
 \rho_{d\sigma}^{}
 \rho_{d\sigma'}^{}\,
f_\mathrm{eff}^{}(\varepsilon) \,
f_\mathrm{eff}^{}(\omega-\varepsilon) 
+ \cdots 
\nonumber
 \\
&
\!\!\!\!\!\!\!\!
\!\!\!\!\!\!\!\!
=  \, 
 -i 2\pi\,    \rho_{d\sigma}^{}\, \rho_{d\sigma'}^{}\,
 \mathcal{W}_{}^\mathrm{hh}(\omega)
\, + \,\cdots \,.  
\rule{0cm}{0.6cm}
\end{align}

\subsubsection{Symmetrization of the bosonic collision integrals}

These lesser and greater bosonic propagators 
can be simplified further, carrying out some symmetrizations. 
For the PH pair excitations, the difference between the two 
 $X_{\sigma\sigma'}^{+-}-X_{\sigma\sigma'}^{-+}
=X_{\sigma\sigma'}^{r}-X_{\sigma\sigma'}^{a}$ gives    
 the imaginary part of the retarded propagator in the frequency domain. 
It shows the $\omega$-linear dependence 
which is determined by the collision integrals in Eqs.\ 
\eqref{eq:X+-_low_energy} and \eqref{eq:X-+_low_energy} 
at low energies,    
\begin{align}
\mathcal{W}_\mathrm{dif}^\mathrm{ph}(\omega) 
\, \equiv\, \mathcal{W}_{}^\mathrm{ph}(\omega) \, -  \,
\mathcal{W}_{}^\mathrm{ph}(-\omega) 
 \ = \ \omega
\;.
\label{eq:L_coll_diff}
\end{align} 
In the zero-bias limit  $eV \to 0$, 
the $\omega$-linear imaginary part of the retarded function 
is closely related to the Korringa-Shiba relation 
of the impurity susceptibilities,\cite{ShibaKorringa} 
 and also to  the fluctuation-dissipation theorem for the current noise.\cite{Hershfield2}

In contrast, the symmetrized PH propagator 
  $X_{\sigma\sigma'}^\mathrm{K}(\omega)$, 
described in Eq.\ \eqref{eq:ImXY--}, 
is determined by the sum of the collision integrals 
$\mathcal{W}_{}^\mathrm{ph}(\omega)$  and  
$\mathcal{W}_{}^\mathrm{ph}(-\omega)$,  as
\begin{align}
\mathcal{W}_\mathrm{K}^\mathrm{ph}(\omega) 
\equiv & \ 
\, \mathcal{W}_{}^\mathrm{ph}(\omega)  + 
\mathcal{W}_{}^\mathrm{ph}(-\omega)
\nonumber \\
= & 
\sum_{j,k =L,R} \frac{\Gamma_j\Gamma_k}{\Delta^2}
\left( \omega  +\mu_j - \mu_k \right) \,\coth \frac{\omega +\mu_j - \mu_k}{2T} . 
\label{eq:collision_intergral_PH-PP_diff}
\end{align} 
This symmetrized part  
$\mathcal{W}_\mathrm{K}^\mathrm{ph}(\omega)$ 
depends also on $eV$ and $T$. The hyperbolic function  
represents the nonequilibrium distribution,  
and it interpolates the three different  
regions  $(\omega,0,0)$,  $(0,T,0)$, and $(0,0,eV)$  
of the parameter space, consisting of $(\omega,T,eV)$.

Similarly, 
the difference between the particle-particle and hole-hole (HH) collision integrals 
is given by  
\begin{align} 
&
\!\!\!
\mathcal{W}_\mathrm{dif}^\mathrm{pp}(\omega) 
\, \equiv \,\mathcal{W}^\mathrm{pp}_{}(\omega) \,-\, \mathcal{W}^\mathrm{hh}_{}(\omega) 
\nonumber \\
&
\!\!\!
=   \, 
\omega \,-\, 2\,\frac{
\Gamma_L^2 \mu_L+\Gamma_R^2 \mu_R + \Gamma_L\Gamma_R (\mu_L+\mu_R)
}{\Delta^2}
, \rule{0cm}{0.6cm}
\label{eq:pp_dif}
\end{align} 
and the sum of the PP and HH collision integrals takes the form, 
\begin{align} 
\mathcal{W}^\mathrm{pp}_\mathrm{K}(\omega)  \equiv & \ 
\, \mathcal{W}^\mathrm{pp}_{}(\omega) 
\,+\, \mathcal{W}^\mathrm{hh}_{}(\omega) 
\nonumber \\
= &
\sum_{j,k =L,R} \frac{\Gamma_j\Gamma_k}{\Delta^2}
\left( \omega  -\mu_j - \mu_k \right) 
\,\coth \frac{\omega -\mu_j - \mu_k}{2T} .
\label{eq:collision_intergral_PH-PP_diff2}
\end{align} 

In particular,   for symmetric junctions $\Gamma_L =\Gamma_R$ and 
$\mu_L=-\mu_R=eV/2$,
 the PP and HH collision integrals coincide with the PH ones, as   
\begin{align}
\mathcal{W}_{}^\mathrm{pp}(\omega) 
\to 
\, \mathcal{W}_{}^\mathrm{ph}(\omega), 
\qquad 
\mathcal{W}_{}^\mathrm{hh}(\omega) 
\to
\, \mathcal{W}_{}^\mathrm{ph}(-\omega)  
\label{eq:ph_and_pp_symmetric_junction}
\end{align}
In this case, the difference of  these two  
takes the form $\mathcal{W}_\mathrm{dif}^\mathrm{pp}(\omega) 
\to  \omega$  which does not depend on  $eV$ and $T$. 
The symmetrized part 
$\mathcal{W}_\mathrm{K}^\mathrm{pp}(\omega)$ 
also coincides with $\mathcal{W}_\mathrm{K}^\mathrm{ph}(\omega)$, 
and takes the following form at $T=0$, 
\begin{align}
\mathcal{W}_\mathrm{K}^\mathrm{ph}(\omega) 
\to & \  
\frac{1}{4} 
\Bigl[\, 
|\omega +eV| \,+\,2|\omega| \,+\, |\omega -eV|  
\,\Bigr]
\nonumber \\
=  & \ 
\left\{
   \begin{array}{ll}
|\omega|  \,,  
&   \ \ |\omega| > |eV| \\ 
\frac{1}{2}\,|\omega| \,+\,\frac{1}{2}\,|eV|  \,, 
 & \ \   |\omega| <  |eV| 
      \rule{0cm}{0.6cm} 
\end{array} 
      \right.  
\!\! .      \rule{0cm}{0.9cm} 
\label{eq:Coll_in_bias_window_text}
\end{align}
For symmetric junctions,  
the PP and PH propagators also become identical 
at low energies, 
$Y_{\sigma\sigma'}^{+-} (\omega)
=  -X_{\sigma\sigma'}^{+-} (\omega)+\cdots$  and 
$Y_{\sigma\sigma'}^{-+} (\omega)
=  -X_{\sigma\sigma'}^{-+} (\omega)+\cdots$, 
apart from the sign.


\begin{table*}[t]
\caption{Low-energy expansion of the Keldysh vertex corrections 
for the same levels  $\sigma = \sigma'$, 
which is asymptotically exact up to linear order terms 
with respect  $eV$, $T$, and frequencies $\epsilon$ and $\epsilon'$. 
The leading-order terms of the  $\sigma = \sigma'$ components 
are pure imaginary and is determined by   
the collision $\mathcal{W}_\mathrm{K}^\mathrm{ph}(\omega)$, 
defined in Eq.\ \eqref{eq:collision_intergral_PH-PP_diff}.
The components with three identical Keldysh indexes  
such as $\Gamma_{\sigma\sigma;\sigma\sigma}^{ --;-+}$ 
and  $\Gamma_{\sigma\sigma;\sigma\sigma}^{ ++;-+}$ 
vanish in the leading order.
}
\begin{tabular}{l} 
\hline \hline
$
\Gamma_{\sigma\sigma;\sigma\sigma}^{ --;++} 
(\epsilon,\epsilon'; \epsilon',\epsilon) + 
\Gamma_{\sigma\sigma;\sigma\sigma}^{ ++;--} 
(\epsilon,\epsilon'; \epsilon',\epsilon) 
\, =  \, 
 -i\, 
{\displaystyle
\frac{2\pi}{\left\{\rho_{d\sigma}^{}\right\}^2}
\sum_{\sigma''(\neq \sigma)}
} 
\chi_{\sigma\sigma''}^{2}
\ 
\mathcal{W}_\mathrm{K}^\mathrm{ph}(\epsilon-\epsilon') 
\ + \ \cdots $
\rule{0cm}{0.65cm}
\\
$
\Gamma_{\sigma\sigma;\sigma\sigma}^{ --;++} 
(\epsilon,\epsilon'; \epsilon',\epsilon) - 
\Gamma_{\sigma\sigma;\sigma\sigma}^{ ++;--} 
(\epsilon,\epsilon'; \epsilon',\epsilon) 
\,=  \, 
 -i\, 
{\displaystyle
\frac{2\pi}{\left\{\rho_{d\sigma}^{}\right\}^2}
\sum_{\sigma''(\neq \sigma)}
} 
\chi_{\sigma\sigma''}^{2}
\, \bigl[ \, \epsilon\,-\,\epsilon' \,\bigr]  \ + \ \cdots 
$
\rule{0cm}{0.65cm}
\\
$
\Gamma_{\sigma\sigma;\sigma\sigma}^{ -+;+-} 
(\epsilon,\epsilon'; \epsilon',\epsilon) 
+\Gamma_{\sigma\sigma;\sigma\sigma}^{ +-;-+} 
(\epsilon,\epsilon'; \epsilon',\epsilon) 
\,=   \ 
 i\, 
{\displaystyle
\frac{2\pi}{\left\{\rho_{d\sigma}^{}\right\}^2}
\sum_{\sigma''(\neq \sigma)}
} 
\chi_{\sigma\sigma''}^{2}
\ 
\mathcal{W}_\mathrm{K}^\mathrm{ph}(0) 
\ + \ \cdots$
\rule{0cm}{0.65cm}
\\
$
\Gamma_{\sigma\sigma;\sigma\sigma}^{ -+;+-} 
(\epsilon,\epsilon'; \epsilon',\epsilon)  -
\Gamma_{\sigma\sigma;\sigma\sigma}^{ +-;-+} 
(\epsilon,\epsilon'; \epsilon',\epsilon) 
\,=  \, 0 \ + \ \cdots $
\rule{0cm}{0.65cm}
\\
$
\Gamma_{\sigma\sigma;\sigma\sigma}^{ -+;-+} 
(\epsilon,\epsilon'; \epsilon',\epsilon) 
\,=  \,  0 \ + \  \cdots
\,, 
\qquad \qquad  \qquad 
\Gamma_{\sigma\sigma;\sigma\sigma}^{ +-;+-} 
(\epsilon,\epsilon'; \epsilon',\epsilon) 
\,=  \,  0 \ + \  \cdots  
$ 
\rule{0cm}{0.65cm}
\\
$
\Gamma_{\sigma\sigma;\sigma\sigma}^{--;--} 
(\epsilon,\epsilon'; \epsilon',\epsilon) 
\, =  \,  
 i \,
{\displaystyle
\frac{\pi}{\left\{\rho_{d\sigma}^{}\right\}^2} \,
\sum_{\sigma''(\neq \sigma)}
}
 \chi_{\sigma\sigma''}^{2}
\ \Bigl[ \,
\mathcal{W}_\mathrm{K}^\mathrm{ph}(\epsilon-\epsilon') 
\,-\,
\mathcal{W}_\mathrm{K}^\mathrm{ph}(0) \,\Bigr]
\, + \, \cdots $, 
\rule{0.5cm}{0cm}
$\Gamma_{\sigma\sigma;\sigma\sigma}^{++;++} 
(\epsilon,\epsilon'; \epsilon',\epsilon) 
= - \left\{ \Gamma_{\sigma\sigma;\sigma\sigma}^{--;--} 
(\epsilon,\epsilon'; \epsilon',\epsilon) \right\}^*$
\rule{0cm}{0.65cm}
\\
 \ \vspace{-0.35cm}
\\
\hline
\hline
\end{tabular}
\label{tab:vertex_UU}
\end{table*}

\begin{table*}[t]
\caption{Low-energy expansion of the Keldysh vertex corrections 
for different levels $\sigma \neq \sigma'$, which 
asymptotically exact up to linear order terms 
with respect  $eV$, $T$, and frequencies $\epsilon$ and $\epsilon'$. 
The imaginary part of the $\sigma \neq \sigma'$ components 
is determined by the bosonic collision integrals 
$\mathcal{W}_\mathrm{K}^\mathrm{ph}(\omega)$,
$\mathcal{W}_\mathrm{K}^\mathrm{pp}(\omega)$,
$\mathcal{W}_\mathrm{dif}^\mathrm{pp}(\omega)$, 
and also  $\mathcal{W}_\mathrm{dif}^\mathrm{ph}(\omega)=\omega$,
defined in Eqs.\ 
\eqref{eq:L_coll_diff}--\eqref{eq:collision_intergral_PH-PP_diff2}.
The components with three identical Keldysh indexes  
such as $\Gamma_{\sigma\sigma';\sigma'\sigma}^{ --;-+}$ 
and  $\Gamma_{\sigma\sigma';\sigma'\sigma}^{ ++;-+}$ 
do not appear in the leading order at low energies. 
Note that the counter part of the causal component is given by  
 $\Gamma_{\sigma\sigma';\sigma'\sigma}^{++;++}
(\epsilon,\epsilon'; \epsilon',\epsilon)  
= - \left\{ \Gamma_{\sigma\sigma';\sigma'\sigma}^{--;--}
(\epsilon,\epsilon'; \epsilon',\epsilon) \right\}^*$, and  
$\widetilde{\chi}_{\sigma\sigma'}=\chi_{\sigma\sigma'}/\rho_{d\sigma}^{}$.
}
\begin{tabular}{l} 
\hline \hline
$
\Gamma_{\sigma\sigma';\sigma'\sigma}^{ --;++} 
(\epsilon,\epsilon'; \epsilon',\epsilon) + 
\Gamma_{\sigma\sigma';\sigma'\sigma}^{ ++;--} 
(\epsilon,\epsilon'; \epsilon',\epsilon) 
\,=  \, 
 -i\, 
{\displaystyle
\frac{2\pi}{\rho_{d\sigma}^{}\rho_{d\sigma'}^{}}
}
\ \chi_{\sigma\sigma'}^{2} \, 
\mathcal{W}_\mathrm{K}^\mathrm{ph}(\epsilon-\epsilon') 
\ + \ \cdots $
\rule{0cm}{0.65cm}
\\
$
\Gamma_{\sigma\sigma';\sigma'\sigma}^{ --;++} 
(\epsilon,\epsilon'; \epsilon',\epsilon) - 
\Gamma_{\sigma\sigma';\sigma'\sigma}^{ ++;--} 
(\epsilon,\epsilon'; \epsilon',\epsilon) 
\,=  \, 
 -i\, 
{\displaystyle
\frac{2\pi}{\rho_{d\sigma}^{}\rho_{d\sigma'}^{}}
}\ 
\chi_{\sigma\sigma'}^{2}
\, \bigl[\, \epsilon\,-\,\epsilon' \,\bigr]  \ + \ \cdots 
$
\rule{0cm}{0.65cm}
\\
$
\Gamma_{\sigma\sigma';\sigma'\sigma}^{ -+;-+} 
(\epsilon,\epsilon'; \epsilon',\epsilon) + 
\Gamma_{\sigma\sigma';\sigma'\sigma}^{ +-;+-} 
(\epsilon,\epsilon'; \epsilon',\epsilon) 
\,=  \  
 i\, 
{\displaystyle
\frac{2\pi}{\rho_{d\sigma}^{}\rho_{d\sigma'}^{}}
}
\ \chi_{\sigma\sigma'}^{2} \, \mathcal{W}^\mathrm{pp}_{K}(\epsilon+\epsilon') 
\ + \ \cdots 
$
\rule{0cm}{0.65cm}
\\
$
\Gamma_{\sigma\sigma';\sigma'\sigma}^{ -+;-+} 
(\epsilon,\epsilon'; \epsilon',\epsilon) - 
\Gamma_{\sigma\sigma';\sigma'\sigma}^{ +-;+-} 
(\epsilon,\epsilon'; \epsilon',\epsilon) 
\,=  \  
 i\, 
{\displaystyle
\frac{2\pi}{\rho_{d\sigma}^{}\rho_{d\sigma'}^{}}
}\ 
\chi_{\sigma\sigma'}^{2}\, 
\mathcal{W}_\mathrm{dif}^\mathrm{pp}(\epsilon+\epsilon') 
+ \cdots 
 $
\rule{0cm}{0.65cm}
\\
$
\Gamma_{\sigma\sigma';\sigma'\sigma}^{ -+;+-} 
(\epsilon,\epsilon'; \epsilon',\epsilon) 
+ \Gamma_{\sigma\sigma';\sigma'\sigma}^{ +-;-+} 
(\epsilon,\epsilon'; \epsilon',\epsilon) 
\,=  \   
i\, 
{\displaystyle
\frac{2\pi}{\rho_{d\sigma}^{}\rho_{d\sigma'}^{}}
}
\ 
{\displaystyle
\sum_{\sigma'' (\neq \sigma,\sigma')}
}
\chi_{\sigma\sigma''}^{} \, \chi_{\sigma''\sigma'}^{} 
\ 
\mathcal{W}_\mathrm{K}^\mathrm{ph}(0) 
 \ + \  \cdots
$ 
\rule{0cm}{0.65cm}
\\
$
\Gamma_{\sigma\sigma';\sigma'\sigma}^{ -+;+-} 
(\epsilon,\epsilon'; \epsilon',\epsilon) 
- \Gamma_{\sigma\sigma';\sigma'\sigma}^{ +-;-+} 
(\epsilon,\epsilon'; \epsilon',\epsilon) 
\,=  \ 0  \ + \  \cdots
$ 
\rule{0cm}{0.65cm}
\\
$
\mathrm{Im}\,
\Gamma_{\sigma\sigma';\sigma'\sigma}^{--;--} 
(\epsilon,\epsilon'; \epsilon',\epsilon) 
\, =  \,  
 {\displaystyle
\frac{\pi}{\rho_{d\sigma}^{}\rho_{d\sigma'}^{}} \,
}
\Biggl[\, 
 \chi_{\sigma\sigma'}^{2}
\,\biggl\{ \,
\mathcal{W}_\mathrm{K}^\mathrm{ph}(\epsilon-\epsilon') 
\,-\,
\mathcal{W}_\mathrm{K}^\mathrm{pp}(\epsilon+\epsilon') \,\biggr\}
\,-
{\displaystyle
\sum_{\sigma'' (\neq \sigma,\sigma')}
}
\chi_{\sigma\sigma''}^{} \, \chi_{\sigma''\sigma'}^{} 
\ 
\mathcal{W}_\mathrm{K}^\mathrm{ph}(0) 
\, \Biggr ]
+  \cdots $
\rule{0cm}{0.7cm}
\\
$
\mathrm{Re}\, 
\Gamma_{\sigma\sigma';\sigma'\sigma}^{--;--} 
(\epsilon,\epsilon'; \epsilon',\epsilon) \,
\,= \
{\displaystyle
\frac{1}{\rho_{d\sigma}^{}\rho_{d\sigma'}^{}} 
}
\biggl[\  
-
\chi_{\sigma\sigma'}^{}
\,+\, 
\rho_{d\sigma}^{} \frac{\partial \widetilde{\chi}_{\sigma\sigma'}}
{\partial \epsilon_{d\sigma}^{}} \  \epsilon   
\,+\, 
\rho_{d\sigma'}^{} \frac{\partial \widetilde{\chi}_{\sigma'\sigma}}
{\partial \epsilon_{d\sigma'}^{}} \  \epsilon'   
\,+\, \beta_{\sigma\sigma'}^{}\,  \alpha_\mathrm{sh} \,eV
\ \biggr] \ + \ \cdots 
$
\rule{0cm}{0.65cm}
\\
\rule{2.3cm}{0cm}
$
 \beta_{\sigma\sigma'}^{}
\,\equiv\,
-\,
\frac{\partial \rho_{d\sigma}^{} }{\partial \epsilon_{d\sigma'}} 
{\displaystyle \sum_{\sigma''(\neq \sigma)} }
\widetilde{\chi}_{\sigma\sigma''}^{}
\,-\, 
\frac{\partial \rho_{d\sigma'}^{}}{\partial \epsilon_{d\sigma}^{}}
 {\displaystyle \sum_{\sigma''(\neq \sigma')} }
\widetilde{\chi}_{\sigma'\sigma''}^{} 
\,+\,
{\displaystyle \sum_{\sigma'''(\neq \sigma, \sigma')}  }
\frac{\partial \chi_{\sigma\sigma'}^{}}{\partial \epsilon_{d\sigma'''}} 
$
\rule{0cm}{0.8cm}
\\
\vspace{-0.3cm}
\\
\hline
\hline
\end{tabular}
\label{tab:vertex_UD}
\end{table*}


\section{
Current conservation of the Keldysh vertex functions 
in the Fermi-liquid regime}
\label{sec:current_conservation_Keldysh_selfenergy_vertex_in_FL_regime}

In order to calculate the nonlinear current noise $S_\mathrm{noise}^\mathrm{QD}$, 
up to terms of order $(eV)^3$, 
we also need information about low-energy behavior of the Keldysh vertex function 
in the Fermi-liquid regime  
 $\Gamma_{\sigma\sigma';\sigma'\sigma}^{\mu\mu';\nu'\nu} 
 (\omega,\omega'; \omega',\omega)$  
up to terms of order  $\omega$, $\omega'$, and  $eV$.  
The collision processes which yield these linear-order terms 
in the imaginary part of the vertex functions are illustrated in 
Figs.\  \ref{fig:vertex_singular_general} (a)--(d).

These diagrams can be calculated in such a way that is shown 
in Appendix \ref{sec:full_vertex_low_energy_form}.  
We find that the low-energy behavior of the 
imaginary parts of the vertex functions can be described  
in terms of the collision integrals 
$\mathcal{W}_\mathrm{K}^\mathrm{ph}(\omega)$, 
and  $\mathcal{W}_\mathrm{K}^\mathrm{pp}(\omega)$,  
as shown in TABLES \ref{tab:vertex_UU} and \ref{tab:vertex_UD} 
for  $\sigma=\sigma'$ and $\sigma \neq \sigma'$, respectively. 
These results explicitly  show that all the Keldysh vertex components 
other than the causal one $\Gamma_{\sigma\sigma';\sigma'\sigma}^{--;--} 
(\omega,\omega'; \omega',\omega)$ 
and its counter part $\Gamma_{\sigma\sigma';\sigma'\sigma}^{++;++} 
(\omega,\omega'; \omega',\omega)$ are pure imaginary in the frequency domain. 
Furthermore, 
the following sum over 
 the time-loop indexes $\mu$, $\mu'$, $\nu'$, and $\nu$  vanishes,  
\begin{align}
& \Gamma_{\sigma\sigma';\sigma'\sigma}^{--;--} 
+ \Gamma_{\sigma\sigma';\sigma'\sigma}^{++;++} 
+\Gamma_{\sigma\sigma';\sigma'\sigma}^{--;++} 
+ \Gamma_{\sigma\sigma';\sigma'\sigma}^{++;--} 
+\Gamma_{\sigma\sigma';\sigma'\sigma}^{-+;+-} 
\nonumber \\ 
&
\! 
+ \Gamma_{\sigma\sigma';\sigma'\sigma}^{+-;-+} 
+\Gamma_{\sigma\sigma';\sigma'\sigma}^{-+;-+} 
+ \Gamma_{\sigma\sigma';\sigma'\sigma}^{+-;+-} 
 \,= 0  +  O\left( \omega^2, |eV|^2, T^3\right),   
\label{eq:linear_dependency_linear_order_vertex}
\end{align}
for both  $\sigma=\sigma'$ and $\sigma \neq \sigma'$. 
Note that the vertex functions with three identical time-loop indexes,   
such as  $\Gamma_{\sigma\sigma;\sigma\sigma}^{--;-+}$  and 
 $\Gamma_{\sigma\sigma;\sigma\sigma}^{-+;--}$,  
do not yield  linear-order terms 
with respect to $\omega$, $\omega'$, and $eV$.
Therefore, Eq.\ \eqref{eq:linear_dependency_linear_order_vertex} 
represents the linear dependency between the Keldysh vertex components 
at low energies 
(see also Appendix \ref{sec:linear_dependency_Keldysh}), 
\begin{align}
&\sum_{\mu\nu}\sum_{\mu'\nu'}
\Gamma_{\sigma\sigma';\sigma'\sigma}^{\mu\mu';\nu'\nu}
 (\omega,\omega'; \omega',\omega) \ = \ 0 \;. 
\label{eq:linear_dependency_Keldysh_vertex}
\end{align}
In particular, in the frequency domain, 
it describes a relationship between the imaginary parts of the vertex components.

We show in this section that the Ward identities  between 
the vertex corrections and self-energy, 
given in Eqs.\ \eqref{eq:Ward_NEQ_omega_all} and \eqref{eq:Ward_NEQ_eV_all},
can be expressed in terms of the 
collision integrals $\mathcal{I}_\mathrm{K}^{}(\omega)$,  
 $\mathcal{W}_\mathrm{K}^\mathrm{ph}(\omega)$, 
and  $\mathcal{W}_\mathrm{K}^\mathrm{pp}(\omega)$, 
which is exact  up to terms of orde  $\omega$, $\omega'$, $eV$, and $T$. 
The Ward identities, which reflect the current conservation 
through quantum dots,  also impose a strong requirement 
for the real part of causal vertex function 
$\Gamma_{\sigma\sigma';\sigma'\sigma}^{--;--} 
(\omega,\omega'; \omega',\omega)$. 
Specifically, it determines the expansion coefficient for the $eV$-linear real part,   
which contributes to the nonlinear current 
through tunnel junctions without  the inversion symmetry, i.e.,  
 $\Gamma_L\neq \Gamma_R$ or $\mu_L \neq -\mu_R$. 
\cite{TsutsumiTerataniSakanoAO2021prb}

\subsection{Current conservation and imaginary part of the Keldysh self-energies}
\label{subsec:WT_Fermi_imaginary_part}

In order to 
clarify roles of the current conservation  
in the lesser and greater self-energies,
 the product of the retarded and advanced Green's functions 
in the right-hand side of Ward identities 
  Eqs.\ \eqref{eq:Ward_NEQ_omega} and \eqref{eq:Ward_NEQ_eV}   
can be replaced by  
 $\Delta G_{\sigma'}^{r} (\omega')\,G_{\sigma'}^{a}(\omega') 
= \pi \rho_{d\sigma'}^{} + \cdots$, i.e., by the value at the Fermi level 
 $\omega'=0$ of the equilibrium ground state.  
This is because the Keldysh vertex functions, 
 $\Gamma_{\sigma\sigma';\sigma'\sigma}^{-\nu_2;\nu_3+} 
(\omega,\omega'; \omega',\omega)$ 
and
 $\Gamma_{\sigma\sigma';\sigma'\sigma}^{+\nu_2;\nu_3-}
 (\omega,\omega'; \omega',\omega)$ 
vanish at $\omega=\omega'=eV=T=0$ 
and thus the leading-order terms of the imaginary part of 
  Eqs.\ \eqref{eq:Ward_NEQ_omega} and \eqref{eq:Ward_NEQ_eV}   
for $\nu_1=-\nu_4$ 
emerge through these vertex components in the integrands.

\subsubsection{The $\omega$ derivative 
of $\Sigma_{U,\sigma}^{+-}$ and $\Sigma_{U,\sigma}^{-+}$}

We first of all  examine 
the level diagonal $\sigma=\sigma'$ part of Eq.\ \eqref{eq:Ward_NEQ_omega} 
for the greater and lesser self-energies, i.e.,  $\nu_1=-\nu_4$, 
using the low-energy asymptotic forms given 
in Eqs.\  \eqref{eq:lesser_asymptotic} and \eqref{eq:greater_asymptotic}. 
Since the quadratic dependence of the lesser and greater 
self-energies on $\omega$, $eV$, and $T$ are determined by 
the fermionic collision integrals  $\mathcal{I}_{}^{+-}(\omega)$ and  
 $\mathcal{I}_{}^{-+}(\omega)$, 
the $\omega$-linear term of the Ward identities Eq.\ \eqref{eq:Ward_NEQ_omega} 
for $\sigma=\sigma'$ 
can be extracted by taking the derivative of these functions 
with  respect to $\omega$.

For the greater self-energy $\Sigma_{U,\sigma}^{+-}(\omega)$,  
it can be calculated up to linear-order terms, 
 using Eq.\ \eqref{eq:greater_asymptotic},  
\begin{align}
 & 
\!\!\! \!\!\! \!\! 
\left(
 \frac{\partial}{\partial \omega} 
+ \frac{\partial}{\partial \epsilon_{d\sigma}}
\right)
\Sigma^{+-}_{U,\sigma}(\omega) 
\nonumber \\ 
=&  \ 
i \frac{2\pi}{\rho_{d\sigma}^{}}  
\sum_{\sigma''(\neq \sigma)} \chi_{\sigma\sigma''}^{}\, 
 \frac{\partial  \mathcal{I}^{+-}(\omega)}{\partial \omega} 
\,+\,\mathcal{O}\Bigl( \omega^2, (eV)^2 ,T^2\Bigr) \,, 
\end{align}
and the derivative of the fermionic collision integral takes the form, 
\begin{align}
\!\!\!\! \!\!
\frac{\partial \mathcal{I}_{}^{+-}(\omega)}{\partial \omega}
=&  \  \int\!d\varepsilon_1 \!\! \int\!d\varepsilon_2\,
\frac{\partial}{\partial \omega}
\Bigl[ 1-f_\mathrm{eff}(\varepsilon_1+\omega)\Bigr] 
\nonumber \\
&  \qquad  \qquad 
\times   \Bigl[ 1-f_\mathrm{eff}(\varepsilon_2)\Bigr]
 f_\mathrm{eff}(\varepsilon_1+\varepsilon_2) 
\nonumber \\
=& \ 
\int\!d\omega'\, 
 \mathcal{W}_{}^\mathrm{ph}(\omega-\omega')
 \left\{
-\frac{\partial f_\mathrm{eff}(\omega')}{\partial \omega'}
\right\}
\nonumber \\
\xrightarrow{\,T \to 0\,}&  \   
 \sum_{j,k,\ell \atop =L,R} \! 
  \frac{\Gamma_j \Gamma_k \Gamma_\ell}{\Delta^3}\,
  \mathcal{W}_\mathrm{eq}^{}(\omega+\mu_j-\mu_k-\mu_\ell).
\label{eq:Ward_greater_check_diagonal_omega}
\end{align}
Here, the bosonic one $\mathcal{W}_{}^\mathrm{ph}(\omega-\omega')$ 
appearing in the second line 
can be related to the low-energy asymptotic form 
of the vertex function  
 $\Gamma_{\sigma\sigma;\sigma\sigma}^{--;++}
 (\omega,\omega';\omega',\omega) \rho_{d\sigma}^{2}=  
-i 2\pi  \sum_{\sigma''(\neq \sigma)} \chi_{\sigma\sigma''}^2 
 \mathcal{W}_{}^\mathrm{ph}(\omega-\omega')+\cdots$, 
which is  one the results  shown in TABLE \ref{tab:vertex_UU}. 
This vertex component is determined by the collision process   
described in Fig.\ \ref{fig:vertex_singular_general} (a) 
for  $(\mu,\nu)=(-,+)$  and is calculated as Eq.\ \eqref{eq:vertex_UU_--++}.
Therefore, 
 Eq.\ \eqref{eq:Ward_greater_check_diagonal_omega} 
 shows that in the Fermi-liquid regime  
the Ward identity  Eq.\ \eqref{eq:Ward_NEQ_omega} 
for  $(\nu_4,\nu_1)=(+,-)$ 
is satisfied through this relation between  
 $\partial \mathcal{I}_{}^{+-}(\omega)/\partial \omega$ 
and $\mathcal{W}_{}^\mathrm{ph}(\omega-\omega')$,
specifically at $T \to 0$ 
the bosonic collision integral in the right-hand side 
takes the form 
$\mathcal{W}_\mathrm{eq}^{}(\varepsilon) \,\to\,
|\varepsilon| \,\theta(\varepsilon)$.  

Similarly, 
the derivative of the lesser self-energy $\Sigma_{U,\sigma}^{-+}(\omega)$  
can be described by the collision integral $\mathcal{I}_{}^{-+}(\omega)$ 
up to linear order terms with respect to  $\omega$, $eV$, and $T$, 
 using Eqs.\ \eqref{eq:lesser_asymptotic} and \eqref{eq:chi_T0},
  \begin{align}
 & 
\!\!\! \!\!\!
\left(
 \frac{\partial}{\partial \omega} 
+ \frac{\partial}{\partial \epsilon_{d\sigma}}
\right)
\Sigma^{-+}_{U,\sigma}(\omega) 
\nonumber \\ 
=&  \, 
-i \frac{2\pi}{\rho_{d\sigma}^{}}  
\sum_{\sigma''(\neq \sigma)} \chi_{\sigma\sigma''}^{}\, 
 \frac{\partial  \mathcal{I}^{-+}(\omega)}{\partial \omega} 
\,+\,\mathcal{O}\Bigl( \omega^2, (eV)^2 ,T^2\Bigr) \,, 
\end{align}
\begin{align}
\frac{\partial \mathcal{I}_{}^{-+}(\omega)}{\partial \omega}\, 
= &  \ 
\int\!d\varepsilon_1 \! \!  \int\!d\varepsilon_2\, 
\frac{\partial}{\partial \omega}\,
f_\mathrm{eff}(\varepsilon_1+\omega)\,
\nonumber \\
& \qquad \qquad 
\times
f_\mathrm{eff}(\varepsilon_2)\,
\Bigl[ 1-f_\mathrm{eff}(\varepsilon_1+\varepsilon_2) \Bigr] 
\nonumber \\
 =& \   
-\int\!d\omega' \, \mathcal{W}_{}^\mathrm{ph}(\omega'-\omega) 
 \left\{
-\frac{\partial f_\mathrm{eff}(\omega')}{\partial \omega'}
\right\}  
\nonumber \\
\xrightarrow{\,T\to 0\,}  &  \ 
-   \!\sum_{j,k,\ell \atop =L,R} \! 
  \frac{\Gamma_j \Gamma_k \Gamma_\ell}{\Delta^3}\,
  \mathcal{W}_\mathrm{eq}^{}(-\omega+\mu_k-\mu_j+\mu_\ell) .
\label{eq:Ward_lesser_check_diagonal_omega}
\end{align}
Here, the bosonic collision integral $\mathcal{W}_{}^\mathrm{ph}(\omega'-\omega)$ 
is related to the low-energy  asymptotic form of the vertex function 
$\Gamma_{\sigma\sigma;\sigma\sigma}^{++;--}
(\omega,\omega';\omega',\omega) \rho_{d\sigma}^{2}= 
-i 2\pi   \sum_{\sigma''(\neq \sigma)} \chi_{\sigma\sigma''}^2 
 \mathcal{W}_{}^\mathrm{ph}(\omega'-\omega) + \cdots$, 
which is determined by  the collision process 
described  in Fig.\ \ref{fig:vertex_singular_general} (a)  for $(\mu,\nu)=(+,-)$.
It can be calculated as  Eq.\ \eqref{eq:vertex_UU_++--} and the result    
is also  listed in TABLE \ref{tab:vertex_UU}.  
Therefore, the last line of Eq.\ \eqref{eq:Ward_lesser_check_diagonal_omega} 
also shows that  in the Fermi-liquid regime
 the Ward identity Eq.\ \eqref{eq:Ward_NEQ_omega} 
for  $\sigma=\sigma'$ and  $(\nu_4,\nu_1)=(-,+)$ is 
satisfied through this relation between 
 $\partial \mathcal{I}_{}^{-+}(\omega)/\partial \omega$ and 
$\mathcal{W}_{}^\mathrm{ph}(\omega'-\omega)$.

\subsubsection{
The  $\epsilon_{d\sigma'}^{}$ 
derivative of $\Sigma_{U,\sigma}^{+-}$ and $\Sigma_{U,\sigma}^{-+}$}

We next consider the off-diagonal $\sigma'  \neq \sigma$ components of
Eq.\ \eqref{eq:Ward_NEQ_omega}, i.e., 
the Ward identities  
for $\partial\Sigma_{U,\sigma}^{+-}(\omega)/\partial \epsilon_{d\sigma'}^{}$ and
$\partial\Sigma_{U,\sigma}^{+-}(\omega)/\partial \epsilon_{d\sigma'}^{}$ 
which show quadratic dependences on $\omega$, $eV$, and $T$
at low energies,
\begin{align}
\frac {\partial \Sigma_{U,\sigma}^{\nu,-\nu}(\omega) 
}{\partial \epsilon_{d\sigma'}} 
\ =  \ 0 
\ + \ \mathcal{O}\Bigl( \omega^2, (eV)^2 ,T^2\Bigr) \,. 
\end{align}
 It can be confirmed directly that the asymptotic forms of vertex functions 
obtained in Appendix \ref{sec:full_vertex_low_energy_form}
 satisfy the Ward identities for  $\sigma' \neq \sigma$ 
up to linear-order terms with respect to $\omega$, $eV$, and $T$; 
 substituting  the results shown in TABLE \ref{tab:vertex_UD} 
into Eq.\ \eqref{eq:Ward_NEQ_omega},  
we find that the constant and linear-order terms 
of the right-hand side vanish. 
Here we describe an alternative way to verify this consistency with the current conservation 
in terms of the collision integrals.

These Ward identities for  $\sigma' \neq \sigma$ 
have been derived  in Sec.\ \ref{sec:Keldysh_formulation}, 
using Eq.\ \eqref{eq:omega_0_derivative} which represents 
the property that the value of the self-energy is unchanged  
if the internal frequencies 
of the Green's functions along a closed-loop of level $\sigma'$
are shifted by an arbitrary amount $\omega_0$. 
The current conservation in the Fermi-liquid regime can be 
verified, applying this procedure to the collision process 
shown in Fig.\ \ref{fig:self_im_T2}. 
For the greater self-energy, it takes the form   
\begin{align}
& 
\frac {\partial \Sigma_{U,\sigma}^{+-}(\omega) 
}{\partial \epsilon_{d\sigma'}} 
\,=  \, 
 \left( \frac{\partial}{\partial \omega_0} 
+ \frac {\partial}{\partial \epsilon_{d\sigma'}} \right)
\Sigma_{U,\sigma}^{+-}(\omega) 
\nonumber \\
&= \, 
 \left( \frac{\partial}{\partial \omega_0} 
+ \frac {\partial}{\partial \epsilon_{d\sigma'}} \right) 
i \frac{2\pi \chi_{\sigma\sigma'}^{}}{\rho_{d\sigma}^{}} \,    
\mathcal{I}^{+-}(\omega)
\,+\, \cdots 
\nonumber \\
& =   \,  
i \frac{2\pi \chi_{\sigma\sigma'}^{}}{\rho_{d\sigma}^{}} \,    
\frac{\partial \mathcal{I}^{+-}(\omega)}{\partial \omega_0} 
\,+\,\mathcal{O}\Bigl( \omega^2, (eV)^2 ,T^2\Bigr) \,. 
 \label{eq:greater_WT_identity_ed_derivative}
\end{align}
Note that the fermionic collision integral  $\mathcal{I}^{+-}(\omega)$ 
can be written in the form, shifting the loop frequency by $\omega_0$ 
from the one described in  Eq.\ \eqref{eq:collision_quasiparticle_bias}, 
\begin{align}
\mathcal{I}_{}^{+-}(\omega)\, 
=&  
\int\!d\varepsilon_1 \!\! \int\!d\varepsilon_2\,
\Bigl[ 1-f_\mathrm{eff}(\varepsilon_1)\Bigr] 
\nonumber\\ 
& \times
 \Bigl[ 1-f_\mathrm{eff}(\varepsilon_2 +\omega_0)\Bigr]
f_\mathrm{eff}(\varepsilon_1+\varepsilon_2 +\omega_0-\omega) .
\end{align}
Although the derivative  ${\partial \mathcal{I}^{+-}(\omega)}/{\partial \omega_0}$ 
itself  vanishes,  it can be expressed in terms of the bosonic collision integrals, as   
\begin{align}
&\!\!\!\! \! \!
\int\!\!d\varepsilon_1 \!\! \int\!\! d\varepsilon_2
\Bigl[ 1-f_\mathrm{eff}(\varepsilon_1)\Bigr] 
\frac{\partial}{\partial \varepsilon_2} 
 \Bigl[ 1-f_\mathrm{eff}(\varepsilon_2)\Bigr]
f_\mathrm{eff}(\varepsilon_1+\varepsilon_2-\omega) 
\nonumber \\
&\!\!\!\!\!\!
=
\int\!d\omega' 
\biggl[\,
 \mathcal{W}_{}^\mathrm{ph}(\omega-\omega')
- \mathcal{W}_{}^\mathrm{pp}(\omega+\omega')
\,\biggr]
 \left\{
-\frac{\partial f_\mathrm{eff}(\omega')}{\partial \omega'}
\right\} 
\nonumber \\
&\!\!\!\! \!\!  
\xrightarrow{\,T \to 0\,} 
\sum_{j,k,\ell =L,R} \! \frac{\Gamma_j \Gamma_k \Gamma_\ell}{\Delta^3}
\biggl[\,
\,\mathcal{W}_\mathrm{eq}^{}(\omega-\mu_\ell-\mu_k+\mu_j)
\nonumber \\
& \rule{2.3cm}{0cm}
-
\,\mathcal{W}_\mathrm{eq}^{}(\omega+\mu_\ell-\mu_k-\mu_j)
\,\biggr]  \, = \, 0 \,.
\label{eq:Ward_greater_check_off-diagonal_omega}
\end{align}
Here, the bosonic 
collision integrals $\mathcal{W}_{}^\mathrm{ph}(\omega-\omega') $ and
$\mathcal{W}_{}^\mathrm{pp}(\omega+\omega') $  
that emerged in the second line
can be regarded as 
the contributions of the vertex corrections    
$\Gamma_{\sigma\sigma';\sigma'\sigma}^{--;++}
(\omega,\omega';\omega',\omega) \,
\rho_{d\sigma}^{} \rho_{d\sigma'}^{} 
=  
-i\, 2\pi   \chi_{\sigma\sigma'}^2 
 \mathcal{W}_{}^\mathrm{ph}(\omega-\omega')$, 
and  
$\Gamma_{\sigma\sigma';\sigma'\sigma}^{-+;-+}
(\omega,\omega';\omega',\omega) \,
\rho_{d\sigma}^{} \rho_{d\sigma'}^{} 
=  
i\, 2\pi   \chi_{\sigma\sigma'}^2 
\mathcal{W}_{}^\mathrm{pp}(\omega+\omega')$, 
shown in TABLE \ref{tab:vertex_UD} up to linear-order terms. 
These two vertex contributions are caused by the 
collision processes, 
 described  in Figs.\ \ref{fig:vertex_singular_general} (b) and (c) 
for $(\mu,\nu)=(-,+)$,  and are calculated 
in Eqs.\ \eqref{eq:vertex_UD_--++} and \eqref{eq:vertex_UD_-+-+}.
Therefore,  
Eqs.\  \eqref{eq:greater_WT_identity_ed_derivative}--
\eqref{eq:Ward_greater_check_off-diagonal_omega} 
show  that  the vertex function 
that is obtained in Appendix \ref{sec:full_vertex_low_energy_form} 
and the greater self-energy 
satisfy the Ward identity Eq.\ \eqref{eq:Ward_NEQ_omega} 
for $\sigma \neq \sigma'$ in the Fermi-liquid regime  
through the relation between the fermionic and bosonic collision integrals.

Similarly,  for the lesser self-energy, 
$\partial \Sigma_{U,\sigma}^{-+}(\omega)/\partial \epsilon_{d\sigma'}^{}$ 
can be expressed in terms of  
${\partial \mathcal{I}^{-+}(\omega)}/{\partial \omega_0}$ 
which takes the following form, 
\begin{align}
&\!\!\!\! \! \!
\int\!\!d\varepsilon_1 \!   \int \!d\varepsilon_2 \,
f_\mathrm{eff}(\varepsilon_1)\,
\frac{\partial}{\partial \varepsilon_2}\,  
f_\mathrm{eff}(\varepsilon_2)\,
\Bigl[ 1-f_\mathrm{eff}(\varepsilon_1+\varepsilon_2 -\omega) \Bigr] \,
\nonumber \\
&\!\!\!\! \! \!
= 
\int\!d\omega' 
\biggl[\,
 \mathcal{W}_{}^\mathrm{ph}(\omega'-\omega)
- \mathcal{W}_{}^\mathrm{hh}(\omega+\omega')
\,\biggr]
 \left\{
-\frac{\partial f_\mathrm{eff}(\omega')}{\partial \omega'}
\right\} 
\nonumber \\
&\!\!\!\! \!\!  
\xrightarrow{\,T \to 0\,}
\sum_{j,k,\ell =L,R} \frac{\Gamma_j \Gamma_k \Gamma_\ell}{\Delta^3}
\biggl[\,
\,\mathcal{W}_\mathrm{eq}^{}(-\omega+\mu_\ell-\mu_k+\mu_j)
\nonumber \\
& \rule{2.3cm}{0cm}
-
\,\mathcal{W}_\mathrm{eq}^{}(-\omega-\mu_\ell+\mu_k+\mu_j)
\,\biggr] 
\,  =  \, 0 .
\label{eq:Ward_lesser_check_off-diagonal_omega}
\end{align}
The bosonic collision integrals 
 $\mathcal{W}_{}^\mathrm{ph}(\omega'-\omega)$ and
$\mathcal{W}_{}^\mathrm{hh}(\omega'+\omega)$ 
in the second line 
correspond to the vertex functions 
$\Gamma_{\sigma\sigma';\sigma'\sigma}^{++;--}
(\omega,\omega';\omega',\omega) \,
\rho_{d\sigma}^{} \rho_{d\sigma'}^{}
=  
-i\, 2\pi   \chi_{\sigma\sigma'}^2 
 \mathcal{W}_{}^\mathrm{ph}(\omega'-\omega)$ 
and 
$\Gamma_{\sigma\sigma';\sigma'\sigma}^{+-;+-}
(\omega,\omega';\omega',\omega) \,
\rho_{d\sigma}^{} \rho_{d\sigma'}^{} 
=  
 i\,2\pi   \chi_{\sigma\sigma'}^2 
\mathcal{W}_{}^\mathrm{hh}(\omega+\omega')$ 
that emerge in the right-hand side of Ward identity Eq.\ \eqref{eq:Ward_NEQ_omega}.  
These two vertex components 
represent the contributions 
of the collision processes, described in Figs.\ \ref{fig:vertex_singular_general} (b) and (c), 
 and are calculated in Eqs.\ \eqref{eq:vertex_UD_++--} and \eqref{eq:vertex_UD_+-+-}.

\subsubsection{
The $eV$ derivative of $\Sigma_{U,\sigma}^{+-}$ and $\Sigma_{U,\sigma}^{-+}$}

We show here  that the low-energy asymptotic forms of the vertex functions, 
given in TABLES \ref{tab:vertex_UU} and \ref{tab:vertex_UD}, 
satisfy the Ward identity Eq.\ \eqref{eq:Ward_NEQ_eV_all} 
with 
$\partial \Sigma_{U,\sigma}^{+-}(\omega)/\partial (eV)$ and
$\partial \Sigma_{U,\sigma}^{-+}(\omega)/\partial (eV)$. 
Since the quadratic  $\omega^2$, $(eV)^2$, and $T^2$ dependences 
of $\Sigma_{U,\sigma}^{+-}(\omega)$ and $\Sigma_{U,\sigma}^{-+}(\omega)$ 
are described  by 
the fermionic collision integrals  $\mathcal{I}_{}^{+-}(\omega)$ and  
 $\mathcal{I}_{}^{-+}(\omega)$,
the linear order terms of these derivatives can be calculated by using   
Eqs.\ \eqref{eq:lesser_asymptotic} and \eqref{eq:greater_asymptotic},
\begin{align}
\!\! \frac{\partial \Sigma^{+-}_{U,\sigma}(\omega) }{\partial eV} 
 = & \,   
i\,\frac{2\pi}{\rho_{d\sigma}^{}}   
\sum_{\sigma'(\neq \sigma)}
 \chi_{\sigma\sigma'}^2 \,
\frac{\partial \mathcal{I}^\mathrm{+-}_{}(\omega)}{\partial eV}    \,+ \cdots ,
\\
\!\! 
\frac{\partial \Sigma^{-+}_{U,\sigma}(\omega)  }{\partial eV} 
 = & \,  
-i\,\frac{2\pi}{\rho_{d\sigma}^{}}   
\sum_{\sigma'(\neq \sigma)}
 \chi_{\sigma\sigma'}^2 \,
\frac{\partial \mathcal{I}^\mathrm{-+}_{}(\omega)}{\partial eV}    \,+ \cdots .
\end{align}

For the greater self-energy,  $\partial \mathcal{I}_{}^{+-}(\omega)/\partial (eV)$ 
 can be expressed in the following form, 
 taking the derivative of the three nonequilibrium distribution 
functions $f_\mathrm{eff}$  
in the integrand in Eq. \eqref{eq:collision_quasiparticle_bias}, 
\begin{align}
& \frac{\partial \mathcal{I}_{}^{+-}(\omega)}{\partial eV}\, 
 = 
 \int\!d\omega'\, 
\biggl[\,
2 \mathcal{W}_{}^\mathrm{ph}(\omega-\omega')
-  \mathcal{W}_{}^\mathrm{pp}(\omega+\omega')
\,\biggr]
\nonumber \\
& \rule{5.25cm}{0cm}
 \times \left\{
-\frac{\partial f_\mathrm{eff}(\omega')}{\partial eV}
\right\}
\nonumber \\
&\xrightarrow{\,T \to 0\,} 
\sum_{j,k,\ell \atop =L,R} 
\frac{\Gamma_j\Gamma_k \Gamma_\ell}{\Delta^3}
\left(\frac{\mu_\ell -\mu_j -\mu_k}{eV}\right)
\nonumber \\
& \qquad \qquad \qquad \qquad  \quad
\times  \mathcal{W}_\mathrm{eq}^{}(\omega+\mu_\ell-\mu_j-\mu_k).
\label{eq:Ward_greater_check_eV}
\end{align}
Here,  two $\mathcal{W}_{}^\mathrm{ph}$'s 
and $\mathcal{W}_{}^\mathrm{pp}$  in the integrand 
 correspond to the vertex contributions 
$\Gamma_{\sigma\sigma;\sigma\sigma}^{--;++}
(\omega,\omega';\omega',\omega)$,
$\Gamma_{\sigma\sigma';\sigma'\sigma}^{--;++}
(\omega,\omega';\omega',\omega)$,  and 
$\Gamma_{\sigma\sigma';\sigma'\sigma}^{-+;-+}
(\omega,\omega';\omega',\omega)$ 
up to linear-order terms,  
shown  in TABLES \ref{tab:vertex_UU} and \ref{tab:vertex_UD}. 
These vertex contributions are caused by the collision processes 
described in Figs.\ \ref{fig:vertex_singular_general} (a), (b), and (c) 
for  $(\mu,\nu)=(-,+)$, 
and are calculated in Appendix   \ref{sec:full_vertex_low_energy_form}. 
Furthermore, the last line of Eq.\ \eqref{eq:Ward_greater_check_eV}, 
in which  $\mathcal{W}_\mathrm{eq}^{}(\varepsilon) \xrightarrow{\,T\to 0\,} 
|\varepsilon| \,\theta(\varepsilon)$, 
explicitly shows that these contributions coincide with 
the that in the right-hand side of the Ward identity Eq.\ \eqref{eq:Ward_NEQ_eV_T0}.

Similarly, the low-energy behavior of 
 $\partial \Sigma_{U,\sigma}^{-+}(\omega)/\partial eV$  
is determined by the derivative of  $\mathcal{I}_{}^{-+}(\omega)$, 
\begin{align}
&\frac{\partial \mathcal{I}_{}^{-+}(\omega)}{\partial eV}\, 
 = \,
-\int\!d\omega' \, 
\biggl[\,
2 \mathcal{W}_{}^\mathrm{ph}(\omega'-\omega) 
-\mathcal{W}_{}^\mathrm{hh}(\omega+\omega') 
\,\biggr] 
\nonumber \\
&  \rule{5.65cm}{0cm}
\times 
\left\{
-\frac{\partial f_\mathrm{eff}(\omega')}{\partial eV}
\right\} 
\nonumber \\
&\xrightarrow{\,T \to 0\,} 
-
\sum_{j,k,\ell \atop =L,R} 
\frac{\Gamma_j\Gamma_k \Gamma_\ell}{\Delta^3}
\left(\frac{\mu_\ell -\mu_j -\mu_k}{eV}\right)
\nonumber \\
& \qquad \qquad \qquad \qquad \quad  
\times \mathcal{W}_\mathrm{eq}^{}(-\mu_\ell +\mu_j +\mu_k -\omega) 
 .
\label{eq:Ward_lesser_check_eV}
\end{align}
In this case,  two $\mathcal{W}_{}^\mathrm{ph}$'s 
and one $\mathcal{W}_{}^\mathrm{hh}$  
in the integrand correspond to vertex contributions 
$\Gamma_{\sigma\sigma;\sigma\sigma}^{++;--}
(\omega,\omega';\omega',\omega)$,
$\Gamma_{\sigma\sigma';\sigma'\sigma}^{++;--}
(\omega,\omega';\omega',\omega)$,  and 
$\Gamma_{\sigma\sigma';\sigma'\sigma}^{+-;+-}
(\omega,\omega';\omega',\omega)$ 
up to linear-order terms,  
shown  in TABLES \ref{tab:vertex_UU} and \ref{tab:vertex_UD}. 
These vertex components are caused by the collision processes 
described in Figs.\ \ref{fig:vertex_singular_general} (a), (b), and (c) 
for $(\mu,\nu)=(+,-)$, and  are
 calculated in Appendix  \ref{sec:full_vertex_low_energy_form}. 
In particular, the last line 
of Eq.\ \eqref{eq:Ward_lesser_check_eV}
shows that  at $T=0$ these contributions coincide with  
that in the right-hand side of the Ward identity Eq.\ \eqref{eq:Ward_NEQ_eV_T0}.

\subsection{Real part of the vertex function at finite $eV$}
\label{sec:real_vertex_eV_liniear}

We next show that the asymptotic form of the real part of the vertex function 
at finite bias voltages 
 can also be deduced from  the Ward identity Eq.\ \eqref{eq:Ward_NEQ_omega_T0}. 
That is the explicit expression of the  $eV$-linear real part of 
$\Gamma_{\sigma\sigma';\sigma'\sigma}^{--;--} (0,0; 0,0)$,  
which has not been taken into account 
 in Eq.\ \eqref{eq:GammaUD_general_omega_dash_N}. 
To be specific, we consider nonequilibrium behaviors at $T=0$ 
in this subsection.

The Keldysh vertex components other than the causal one and its counter part 
are pure imaginary, as shown explicitly 
in TABLES \ref{tab:vertex_UU} and \ref{tab:vertex_UD},  
 up terms of order $\omega$, $\omega'$, and $eV$. 
Thus, only  
$\Gamma_{\sigma\sigma';\sigma'\sigma}^{--;--}
(\omega,\omega';\omega',\omega)$ and 
$\Gamma_{\sigma\sigma';\sigma'\sigma}^{++;++}
(\omega,\omega';\omega',\omega)$ have the real parts 
in the Fermi-liquid regime, which 
can be related to the real part of the 
retarded self-energy which is identical to the real part 
of the causal one  
$\mathrm{Re}\,\Sigma_{U,\sigma}^{r}(\omega) 
\equiv \mathrm{Re}\, \Sigma_{U,\sigma}^{--}(\omega)$, 
using the Ward identity Eq.\ \eqref{eq:Ward_NEQ_omega_T0}: 
 \begin{align}
& \!\!\!\!\!\!\!\!  
\left(
\delta_{\sigma\sigma'}\,  \frac{\partial}{\partial \omega} 
+ \frac{\partial}{\partial \epsilon_{d\sigma'}}
\right)
\mathrm{Re}\,\Sigma_{U,\sigma}^{r}(\omega)
\nonumber \\
&
\ \ 
=  \ 
- 
 \sum_{j=L, R}\,
\mathrm{Re}\,
\Gamma_{\sigma\sigma';\sigma'\sigma}^{--;--}
 (\omega,\mu_{j}^{}; \mu_{j}^{},\omega)\,
A_{d\sigma'}^{(1)}(\mu_{j}^{})
\,\frac{\Gamma_{j}}{\Delta} 
\nonumber \\
& 
\qquad +\, O\left(\omega^2,(eV)^2\right)
 \,.
\label{eq:Ward_real_check_omega}
\end{align}
Here, the product of the two Green's that appeared 
in the right-hand side of  Eq.\ \eqref{eq:Ward_NEQ_omega_T0} 
has been rewritten into the form  
 $\Delta G_{\sigma}^{r}(\omega)\,G_{\sigma}^{a}(\omega) /\pi 
 =A_{d\sigma}^{(1)}(\omega) + O(\omega^2, (eV)^2)$, 
using the spectral function  $A_{d\sigma}^{(1)}(\omega)$ 
which takes into account the self-energy corrections 
up to linear-order terms in $\omega$ and $eV$:  
\begin{align}
& \!\!\!
A_{d\sigma}^{(1)}(\omega) 
\,\equiv \,\frac{1}{\pi}\, 
\frac{\Delta}{
\Bigl(\widetilde{\chi}_{\sigma\sigma}\,\omega   - \epsilon_{d\sigma}^{*} 
+ \! 
\displaystyle \sum_{\sigma''(\neq \sigma)}
\widetilde{\chi}_{\sigma\sigma''}^{} \,  \alpha_\mathrm{sh}
\,eV \Bigr)^2+\Delta^2} 
\nonumber\\
 & =\, 
\rho_{d\sigma}^{} + 
\rho_{d\sigma}'
\left[\,\omega \,+ 
 \sum_{\sigma''(\neq \sigma)}
\frac{\widetilde{\chi}_{\sigma\sigma''}^{}}{\widetilde{\chi}_{\sigma\sigma}^{}} 
\,\alpha_\mathrm{sh} \,eV  \, \right] + \cdots .  
\label{eq:A_1_effective}
\end{align}
Here, $\rho_{d\sigma}'$ is the derivative of the equilibrium density of state, 
defined in Eq.\ \eqref{eq:rho_d_omega_2}. 
The spectral function $A_{d\sigma}^{(1)}(\omega)$ 
includes the order $eV$ energy shift emerging for 
asymmetric junctions with $\alpha_\mathrm{sh}^{}\neq 0$   
in addition to the equilibrium energy shift 
$\epsilon_{d\sigma}^{*} \equiv \epsilon_{d\sigma}^{}
+\Sigma_{\mathrm{eq},\sigma}^{r}(0)$.

As we already know  the low-energy asymptotic form of the 
retarded self-energy as shown in  Eq.\ \eqref{eq:self_real_ev_mag_N},
the derivatives  in the left-hand side of Eq.\ \eqref{eq:Ward_real_check_omega} 
can be calculated explicitly up to terms of order $\omega$ and $eV$:
 it vanishes for  $\sigma=\sigma'$, 
\begin{align}
& \!\!\!\!\!\! 
\left(
 \frac{\partial}{\partial \omega} 
+ \frac{\partial}{\partial \epsilon_{d\sigma}}
\right)
\mathrm{Re}\,\Sigma_{U,\sigma}^{r}(\omega)
\,=\, 0 \,+\, O\left(\omega^2,(eV)^2\right) \,,
\end{align}
and it takes the following form  for $\sigma \neq \sigma'$,
\begin{align}
&\!\!  
\frac{\partial}{\partial \epsilon_{d\sigma'}}
\mathrm{Re}\,\Sigma_{U,\sigma}^{r}(\omega) 
\nonumber \\ 
&\!\!\!\!  
=\,
\widetilde{\chi}_{\sigma\sigma'}  
-\frac{\partial\widetilde{\chi}_{\sigma\sigma}}{\partial \epsilon_{d\sigma'}}\, \omega
- \!
\sum_{\sigma''(\neq \sigma)}
\frac{\partial \widetilde{\chi}_{\sigma\sigma''}^{}}{\partial \epsilon_{d\sigma'}} 
\,  \alpha_\mathrm{sh}\,eV 
+ \cdots.
\label{eq:derivative_ReSelfRetarded_omega_general_text}
\end{align}

Therefore, 
 $\mathrm{Re}\, \Gamma_{\sigma\sigma;\sigma\sigma}^{--;--}
(\omega,\omega';\omega',\omega)$ 
  in right-hand side of Eq.\ \eqref{eq:Ward_real_check_omega} 
can be deduced from these expressions using also Eq.\ \eqref{eq:A_1_effective}. 
As shown in Eq.\ \eqref{eq:GammaUU_general_omega_dash_N},  
the real part of causal vertex function for $\sigma=\sigma'$, 
 does not have a constant and linear-order terms 
with respect $\omega$ and $\omega'$ at $eV=0$. 
This does not change up to order $eV$,  
\begin{align}
&\!\!\!\!
\mathrm{Re}\, \Gamma_{\sigma\sigma;\sigma\sigma}^{--;--}(\omega, \omega'; \omega' ,\omega) 
= 0 + O\left(\omega^2, \omega'^2, (eV)^2\right) .
\label{eq:GammaUU_general_omega_dash_N_II}
\end{align}
We also find that 
the real part of the causal vertex function  for $\sigma \neq \sigma'$ 
captures the  $eV$-linear term when the inversion symmetry 
is broken in a way such that $\alpha_\mathrm{sh}^{}\neq 0$, 
\begin{align}
&\!\!
\mathrm{Re}\, \Gamma_{\sigma\sigma';\sigma'\sigma}^{--;--}(\omega, \omega'; \omega' ,\omega) 
\,\rho_{d\sigma}^{}\rho_{d\sigma'}^{}
\nonumber \\
&=  \,       
-
\chi_{\sigma\sigma'}^{}
+ 
\rho_{d\sigma}^{}
\frac{\partial \widetilde{\chi}_{\sigma\sigma'}}
{\partial \epsilon_{d\sigma}^{}} \, \omega  
+ 
\rho_{d\sigma'}^{}
\frac{\partial \widetilde{\chi}_{\sigma'\sigma}}
{\partial \epsilon_{d\sigma'}^{}} \, \omega'   
+ \beta_{\sigma\sigma'}^{}\,  \alpha_\mathrm{sh} \,eV
\nonumber \\
& \quad \,+\, O\left(\omega^2, \,\omega'^2,(eV)^2\right)\,
\rule{0cm}{0.6cm}
\label{eq:GammaUD_general_omega_dash_N_II}
 \\
&\beta_{\sigma\sigma'}^{} 
\,\equiv \,
 -\,
\frac{\partial \rho_{d\sigma}^{} }{\partial \epsilon_{d\sigma'}} 
\sum_{\sigma''(\neq \sigma)}
\widetilde{\chi}_{\sigma\sigma''}^{}
\,-\, 
\frac{\partial \rho_{d\sigma'}^{}}{\partial \epsilon_{d\sigma}^{}}
 \sum_{\sigma''(\neq \sigma')}
\widetilde{\chi}_{\sigma'\sigma''}^{} 
\rule{0cm}{0.7cm}
\nonumber \\
& \qquad \quad 
+ 
\sum_{\sigma'''(\neq \sigma, \sigma')}
\frac{\partial \chi_{\sigma\sigma'}^{}}{\partial \epsilon_{d\sigma'''}} \;.
\label{eq:GammaUD_general_omega_dash_N_II_beta}
\end{align}

We have also confirmed that  these results for the real 
parts of the vertex functions  Eqs.\   
\eqref{eq:GammaUU_general_omega_dash_N_II} and 
\eqref{eq:GammaUD_general_omega_dash_N_II}
satisfy the Ward identity for the  $eV$ derivative 
given in  Eq.\ \eqref{eq:Ward_NEQ_eV_T0}, 
which  takes the following form for the real part,  
 \begin{align}
 &
\frac{\partial}{\partial eV} \, 
\mathrm{Re}\,\Sigma_{U,\sigma}^{r}(\omega)
\nonumber \\ 
&
=    
- 
\sum_{\sigma'}\! 
\sum_{j=L, R}\! 
\mathrm{Re}\,
\Gamma_{\sigma\sigma';\sigma'\sigma}^{--;--}
 (\omega,\mu_{j}^{}; \mu_{j}^{},\omega) 
A_{d\sigma'}^{(1)}(\mu_{j}^{})
\left(\frac{- \Gamma_{j} \mu_j}{\Delta\, eV}\right)
\nonumber \\
& \quad \,+\, O\left(\omega^2,(eV)^2\right)
\,.
\rule{0cm}{0.5cm}
\label{eq:Ward_real_check_eV}
\end{align}
Namely, the right-hand side of this identity   Eq.\ \eqref{eq:Ward_real_check_eV} 
coincides with  the derivative of 
 $\mathrm{Re}\, \Sigma_{U,\sigma}^r(\omega)$ on the left-hand side 
that can be calculated directly using 
 Eq.\ \eqref{eq:self_real_ev_mag_N},
\begin{align}
&\!\!
\frac{\partial \,\mathrm{Re}\, \Sigma_{U,\sigma}^r(\omega) }{\partial eV}
 =    -
\sum_{\sigma''(\neq \sigma)}
\left[
\widetilde{\chi}_{\sigma\sigma''}^{} 
 - \frac{\partial \widetilde{\chi}_{\sigma\sigma''}}{\partial \epsilon_{d\sigma}^{}} 
\, \omega  \, \right] 
\,\alpha_\mathrm{sh} 
\nonumber \\
&  \qquad \qquad \quad 
+  
\frac{\Gamma_L \Gamma_R}{\Delta^2} 
\sum_{\sigma''(\neq \sigma)} 
 \frac{1}{\rho_{d\sigma}}
 \frac{\partial \chi_{\sigma\sigma''}}{\partial \epsilon_{d\sigma''}} 
\,   eV 
\nonumber \\
 &  \qquad   \qquad  \quad 
+ 
\sum_{\sigma' (\neq \sigma)} 
\sum_{\sigma'' (\neq \sigma)} 
\frac{\partial\widetilde{\chi}_{\sigma\sigma'}}{\partial\epsilon_{d\sigma''}^{}}
\,\alpha_\mathrm{sh}^2\,eV 
+ \cdots .
\end{align}

\section{
Current noise formula in the Fermi-liquid regime}
\label{sec:noise_derivation}

We are now at the stage of being able to calculate 
the current noise $S_\mathrm{noise}^\mathrm{QD}$ 
at finite bias voltages up to order $(eV)^3$. 
It is defined in terms of  the current-current correlation function 
  $\mathcal{K}_{\sigma'\sigma}^{\alpha'\alpha}$, 
\begin{align}
\!\!\!\! 
S_\mathrm{noise}^\mathrm{QD}\,= \, 
 e^2 \!
\int_{-\infty}^{\infty} \!  dt   
\sum_{\sigma\sigma'}
i \Bigl[\,
\mathcal{K}_{\sigma'\sigma}^{+-}(t,0)
+\mathcal{K}_{\sigma'\sigma}^{-+}(t,0)
\,\Bigr] . 
\label{eq:S_noise_text}
\end{align}
Here,
$\mathcal{K}_{\sigma'\sigma}^{+-}(t,0)
 \equiv  
-i \,\bigl\langle \delta \widehat{J}_{\sigma'}(t) 
\, \delta \widehat{J}_{\sigma}(0) \bigr\rangle$,   
$\mathcal{K}_{\sigma'\sigma}^{-+}(t,0)
\equiv 
-i \,\bigl\langle \delta \widehat{J}_{\sigma}(0) 
\, \delta \widehat{J}_{\sigma'}(t)\bigr\rangle$.
The symmetrized current operator  $\widehat{J}_{\sigma}$ is  defined  in 
Eq.\ \eqref{eq:symmetrized_current_def}. 
In the Keldysh-Feynman diagrammatic representation, 
this operator can be treated as a matrix current vertex 
  $\bm{\lambda}_\mathrm{sym}^{\alpha}(\epsilon,\epsilon+\omega)$  
defined  in Eq.\ \eqref{eq:lambda_av_U0} in the frequency domain,  and  
in  Fig.\ \ref{fig:Kubo_Keldysh_diagram} 
it is  illustrated  as a black diamond 
 (${\tiny \protect \rotatebox[origin=c]{45}{$\blacksquare$}}$). 
In particular, at $\omega=0$,   
 the current vertex for the constant noise 
can be expressed in the following form 
using the Pauli matrix $\bm{\tau}_2$,     
\begin{align}
\bm{\lambda}_\mathrm{sym}^{\alpha}(\epsilon , \epsilon)
\,=\, 
\frac{-2\Gamma_L\Gamma_R \Bigl[\,f_L(\epsilon) -f_R(\epsilon) \,\Bigr]}
{\Gamma_L+\Gamma_R}
\   i \bm{\tau}_2 
\;,
\end{align}
and becomes independent of whether  $\alpha=-$ or $+$. 
At  $T=0$, this current vertex 
$\bm{\lambda}_\mathrm{sym}^{\alpha}(\epsilon , \epsilon)$ 
takes a finite value just in the bias-window region $\mu_R \leq \epsilon \leq \mu_L$,
and it identically vanishes at equilibrium $eV=0$.

We  calculate $S_\mathrm{noise}^\mathrm{QD}$ 
separating it  into four parts,
\begin{align}
 S_\mathrm{noise}^\mathrm{QD} 
\,=& \  
 \left(
\frac{\Gamma_L}{\Gamma_L+\Gamma_R}
\right)^2 \! \delta S_{RR}
+ \! \left(
\frac{\Gamma_R}{\Gamma_L+\Gamma_R}
\right)^2 \! \delta S_{LL}
\nonumber \\
 &  + \, S_\mathrm{sym}^\mathrm{qp} \,+\, S_\mathrm{sym}^\mathrm{coll} \,.
\label{eq:noise_4parts}
\rule{0cm}{0.5cm}
\end{align}
The first two terms $\delta S_{RR}$ and  $\delta S_{LL}$ 
represent the contributions of the processes, 
 illustrated in the bottom row of Fig.\ \ref{fig:Kubo_Keldysh_diagram},  
and can be expressed in the form,
\begin{align}
\delta S_{jj} 
=& \ \frac{e^2}{\hbar} 
\,i\,4\Gamma_j^{} 
\sum_{\sigma}
\int_{-\infty}^{\infty}\!\!  \frac{d\epsilon}{2\pi}\,
 \Bigl[\,
   f_j\,G_{\sigma}^{+-}
-  \left(1-f_j \right)\,G_{\sigma}^{-+} 
\,\Bigr] ,
\label{eq:Noise_dS_part}
\end{align}
for $j=L,R$. 
In the Feynman diagrams for these processes,  the dashed line 
represents the bare conduction-electron Green's function 
 $\bm{g}_{j}^{}$   
of the isolated leads on $j=L$ and $R$, 
defined in Eq.\ \eqref{eq:g0_lead}.
The solid line represents the full impurity-electron Green's function
 $\bm{G}_{\sigma}^{}$.   
The contributions of  $\delta S_{RR}$ and  $\delta S_{LL}$ 
on the total noise $S_\mathrm{noise}^\mathrm{QD}$ 
can also be expressed in the following form, 
rewriting the lesser and greater Green's functions 
in Eq.\ \eqref{eq:Noise_dS_part} by using  Eq.\ \eqref{eq:Gdd_matrix}, 
\begin{widetext}
\begin{align}
 & 
\!\!\!\!\! \!\!\!\!\! 
\left(
\frac{\Gamma_L}{\Gamma_L+\Gamma_R}
\right)^2 \delta S_{RR}^{}
+ \left(
\frac{\Gamma_R}{\Gamma_L+\Gamma_R}
\right)^2 \delta S_{LL}^{}
\nonumber \\
&  \rule{0cm}{0.75cm}
= \,   
\frac{2e^2}{h} 
4\Gamma_L^{}\Gamma_R^{}
\sum_{\sigma}
\int_{-\infty}^{\infty}\!\!  d\epsilon\,
G_{\sigma}^{r}G_{\sigma}^{a}
\Biggl[\ 
f_R(1-f_R) +f_L(1-f_L)
\,+\, \left\{1 -\frac{2\Gamma_L\Gamma_R}{(\Gamma_R +\Gamma_L )^2}
\right\}  (f_L-f_R)^2
 \nonumber \\
& 
\qquad \qquad \qquad \qquad \qquad \qquad \qquad \ 
+ 
 \frac{\Gamma_R f_L + \Gamma_L f_R }{(\Gamma_L+\Gamma_R)^2}
\,\frac{\Sigma_{U,\sigma}^{+-}}{2i}
\  - \  \frac{\Gamma_R (1-f_L) + \Gamma_L (1-f_R) }{(\Gamma_L+\Gamma_R)^2}
\,\frac{\Sigma_{U,\sigma}^{-+}}{2i}
\ \Biggr]\; .
\label{eq:weighted_sum_delSRR_delSLL}
\end{align}
The remaining two parts,  
 $S_\mathrm{sym}^\mathrm{qp}$ and $S_\mathrm{sym}^\mathrm{coll}$ 
of  Eq.\ \eqref{eq:noise_4parts}, 
represent the contributions of the diagrams shown 
in the top row  of Fig.\ \ref{fig:Kubo_Keldysh_diagram}, 
i.e., the bubble diagram on the left 
and the one with vertex corrections on the right, 
\begin{align}
 S_\mathrm{sym}^\mathrm{qp}
 \,= & \       \frac{e^2}{\hbar} 
\sum_{\alpha=+,-} \sum_{\sigma}
\int_{-\infty}^{\infty}\! \frac{d\epsilon}{2\pi}\, 
\mathrm{Tr}
\Bigl[ \,
\bm{\lambda}_\mathrm{sym}^{\alpha}( \epsilon , \epsilon)
\bm{G}_{\sigma}^{}(\epsilon)
\bm{\lambda}_\mathrm{sym}^{-\alpha}( \epsilon , \epsilon)
\bm{G}_{\sigma}^{}(\epsilon)
\Bigr] 
\nonumber \\
=  &  \     
\frac{2e^2}{h} 
\left[
\frac{2\,\Gamma_L\Gamma_R}{\Gamma_L+\Gamma_R}\right]^2
\sum_{\sigma}
\int_{-\infty}^{\infty}\! d\epsilon\, 
\bigl[ f_L (\epsilon)-f_R(\epsilon) \bigr]^2
\left[\bigl\{G_{\sigma}^{r}(\epsilon) -G_{\sigma}^{a}(\epsilon)\bigr\}^2
 +2G_{\sigma}^{r}(\epsilon)\, G_{\sigma}^{a}(\epsilon)\right] \,, 
\label{eq:Noise_bubble}
\\
 S_\mathrm{sym}^\mathrm{coll} \, =  & \   
\frac{e^2}{\hbar}
\left[\frac{2\,\Gamma_L\Gamma_R}{\Gamma_L+\Gamma_R}\right]^2
\sum_{\alpha=+,-}
\int_{-\infty}^{\infty} \! \frac{d\epsilon}{2\pi}
\int_{-\infty}^{\infty} \! \frac{d\epsilon'}{2\pi}\, 
\bigl[ f_L(\epsilon) -f_R(\epsilon) \bigr]
\bigl[ f_L(\epsilon') -f_R(\epsilon') \bigr]
\, \mathcal{Q}(\epsilon,\epsilon') \;,
\label{eq:Noise_vertex}
\rule{0cm}{0.8cm}
\\
 \mathcal{Q}(\epsilon,\epsilon') 
\, \equiv  & \,   
(- i )
\sum_{\sigma\sigma'}
\sum_{\mu\nu \atop \mu'\nu'}
\Gamma_{\sigma\sigma';\sigma'\sigma}^{\mu\mu';\nu'\nu}
 (\epsilon,\epsilon'; \epsilon',\epsilon)
\, \bigl\{\bm{G}_{\sigma}^{}(\epsilon)
\, i \bm{\tau}_2 
\bm{G}_{\sigma}^{}(\epsilon)
\bigr\}_{\mu\nu}
\bigl\{\bm{G}_{\sigma'}^{}(\epsilon')
\, i \bm{\tau}_2 
\bm{G}_{\sigma'}^{}(\epsilon')
\bigr\}_{\nu'\mu'} \;.
\label{eq:Q_dif}
\rule{0cm}{0.6cm}
\end{align}
\end{widetext}
In order to obtain Eq.\ \eqref{eq:Noise_bubble}, 
we have decomposed the product of the Green's function into the three parts,  
using the basis set, described in Eq.\ \eqref{eq:GK_def},  as 
\begin{align}
& \!\!\!
\bm{G}_{\sigma}^{}
\, i \bm{\tau}_2 
\bm{G}_{\sigma}^{}
\,=   \,   
a_{1\sigma}^{} \left( \bm{1} + \bm{\tau}_1 \right)
\,+\, a_{3\sigma}^{} \bm{\tau}_3 
\,-\, a_{2\sigma}^{}  i \bm{\tau}_2 \,. 
\label{eq:bare_current_vertex_decomposition}
\end{align}
The coefficients can be determined, for instance, in a way such that   
$\mathrm{Tr}
\Bigl[ 
i\bm{\tau}_2
\bm{G}_{\sigma}^{}
i\bm{\tau}_2
\bm{G}_{\sigma}^{}
\Bigr] =   2 a_{2\sigma}^{}$, as 
\begin{subequations}
\begin{align}
 a_{1\sigma}^{}= & \ 
\frac{  \left(G_{\sigma}^{+-} + G_{\sigma}^{-+}\right) 
 \left(G_{\sigma}^{r}-G_{\sigma}^{a} \right)}{2} 
,
&
\\
 a_{3\sigma}^{}=& \  
 \frac{\left\{G_{\sigma}^{r}\right\}^2-\left\{G_{\sigma}^{a}\right\}^2}{2},
\\
 a_{2\sigma}^{}= &  \ 
\frac{\left\{G_{\sigma}^{r}\right\}^2+\left\{G_{\sigma}^{a}\right\}^2}{2}.
\end{align}
\label{eq:a_coefficients} \!\!\!\!\! 
\end{subequations}

The free-quasiparticle part, i.e.,  $S_\mathrm{sym}^\mathrm{qp}$ 
of Eq.\ \eqref{eq:Noise_bubble},  
vanishes at equilibrium $eV=0$ and 
shows an  $|eV|$-linear dependence at small bias voltages 
since the distribution function $(f_L -f_R)^2$ 
restricts the region of the integration to the  
 inside of the bias window $\mu_R \leq \epsilon \leq \mu_L$.

The fourth part, $S_\mathrm{sym}^\mathrm{coll}$ of  Eq.\ \eqref{eq:Noise_vertex},  
represents the contributions of collisions between quasiparticles,   
which enter through the vertex functions in 
the kernel  $\mathcal{Q}(\epsilon,\epsilon')$ 
defined as  Eq.\ \eqref{eq:Q_dif}. 
The domains of integrations with respect to $\epsilon$ and  $\epsilon'$ 
are also restricted by the distribution functions $f_L (\epsilon)-f_R(\epsilon)$ and 
 $f_L (\epsilon')-f_R(\epsilon')$. 
 Therefore,  this part $S_\mathrm{sym}^\mathrm{coll}$ 
 does not show an $|eV|$-linear dependence, 
and the restrictions due to the bias window themselves  
cause a contribution of order $|eV|^2$ at small bias voltages.

\subsection{Thermal noise at equilibrium $eV=0$}
\label{subsec:thermal_noise_derivation}

Here we briefly show how the previous result for the thermal noise at equilibrium  
can be reproduced in our formulation.  
At $eV=0$, the contributions of  $S_\mathrm{sym}^\mathrm{qp}$ 
and $S_\mathrm{sym}^\mathrm{coll}$  
vanish among the four parts of the current noise, 
described in Eq.\ \eqref{eq:noise_4parts}.  
Furthermore,  the lesser and greater Green's functions
that determine  $\delta S_{RR}$ and  $\delta S_{LL}$   
through Eq.\ \eqref{eq:Noise_dS_part}
can be written in terms of 
the retarded Green's function $(-1/\pi)\, \mathrm{Im}\,G_{\mathrm{eq},\sigma}^{r}$,
using  Eq.\  \eqref{eq:G_eq_Lehmann}. 
Therefore, the well-known thermal noise formula  
is derived from Eqs.\ \eqref{eq:noise_4parts} and  \eqref{eq:Noise_dS_part},
 as   
\begin{align}
 S_\mathrm{noise}^\mathrm{QD} 
 & \xrightarrow{\,eV \to 0 \,}    
 \left(
\frac{\Gamma_L}{\Gamma_L+\Gamma_R}
\right)^2 \! \delta S_{RR}
+ \! \left(
\frac{\Gamma_R}{\Gamma_L+\Gamma_R}
\right)^2 \! \delta S_{LL} 
\nonumber \\
&= \, 4T \  \frac{e^2}{h}
\sum_{\sigma}
\int_{-\infty}^{\infty} \! d\epsilon\, 
\left(-\frac{\partial f(\epsilon)}{\partial \epsilon}\right)  
\,  \mathcal{T}_{\mathrm{eq},\sigma}^{}(\epsilon) 
\,.
\label{eq:thermal_noise_2}
\rule{0cm}{0.4cm}
\end{align}
Here,  
$\mathcal{T}_{\mathrm{eq},\sigma}^{}(\epsilon) 
\equiv  \mathcal{T}_{\sigma}(\epsilon)\big|_{eV=0}$   
is the transmission probability,  defined in Eq.\ \eqref{eq:transmissionPB}. 
We have also used the relation 
 $T (-\partial f/ \partial \epsilon)  =f(1-f)$  of the Fermi function part. 
In the low-temperature Fermi-liquid regime, the thermal noise can be deduced 
up to terms of order $T^3$ 
by using the low-temperature expansion of the linear conductance, 
given in Eq.\ \eqref{eq:thermal_noise}.

\subsection{The $|eV|$-linear current noise at $T=0$}

We next consider the $|eV|$-linear contributions of the 
current noise  $S_\mathrm{noise}^\mathrm{QD}$ 
that  can be determined by the first three terms 
of Eq.\ \eqref{eq:noise_4parts}.  
The order $|eV|$ contributions of  
 $S_\mathrm{sym}^\mathrm{qp}$ can be calculated, 
taking  the frequency argument for the 
Green's functions in Eq.\ \eqref{eq:Noise_bubble}
to be $\epsilon=0$ 
as the linear dependence is determined by the width of the bias window.
 For the weighted sum of  $\delta S_{RR}$ and $\delta S_{LL}$ in 
Eq.\  \eqref{eq:weighted_sum_delSRR_delSLL}, 
the lesser or greater self-energy that appeared in the right-hand side   
does not yield order  $|eV|$ contributions.
This is because these two self-energies $\Sigma_{U,\sigma}^{-+}(\epsilon)$ 
and $\Sigma_{U,\sigma}^{+-}(\epsilon)$ in the integrand  
show the $\epsilon^2$ and  $|eV|^2$ behaviors,  
as described in Eqs.\ \eqref{eq:lesser_asymptotic} and \eqref{eq:greater_asymptotic},   
and yield high-order terms which we consider later in the next subsection. 
Therefore, the sum of these contributions determine 
the linear current noise  at $T=0$, as  
\begin{align}
& 
\!\!
S_\mathrm{noise}^\mathrm{QD} 
 \,\xrightarrow{\,T \to 0 \,} \,   
\left(
\frac{\Gamma_L}{\Gamma_L+\Gamma_R}
\right)^2 \delta S_{RR}^{}
+ \left(
\frac{\Gamma_R}{\Gamma_L+\Gamma_R}
\right)^2 \delta S_{LL}^{}
\nonumber \\
& \qquad \qquad \quad \ \   + 
S_\mathrm{sym}^\mathrm{qp} 
 \ + \  O\left(|eV|^2\right)
\rule{0cm}{0.6cm}
\nonumber \\
& 
\!\!\!
= 
\frac{2e^2}{h} \,  \bigl|eV\bigr|\,
\sum_{\sigma}
 \mathcal{T}_{\mathrm{eq},\sigma}^{}(0)
\Bigl[\,  1-
\mathcal{T}_{\mathrm{eq},\sigma}^{}(0)
\,\Bigr] + \cdots
.
\label{eq:Noise_linear}
\end{align}
It reproduces the well-established linear-noise formula,   
which is determined by the transmission probability 
at the Fermi level 
 $\mathcal{T}_{\mathrm{eq},\sigma}^{}(0)=\sin^2 \delta_{\sigma}^{}$.

\subsection{Order $|eV|^3$ terms of 
 current noise $S_\mathrm{noise}^\mathrm{QD}$}

We calculate order $|eV|^3$ terms of 
the current noise $S_\mathrm{noise}^\mathrm{QD}$
in the rest of this section, 
specifically for symmetric junctions $\Gamma_L=\Gamma_R$  
and  $\mu_L = -\mu_R=eV/2$, at  $T = 0$. 
To this end, we consider first of all  the contributions of 
 the first three parts of Eq.\ \eqref{eq:noise_4parts}, 
for which effects of the Coulomb interaction enter only through  
the Green's function $G_{\sigma}^{\mu\nu}(\epsilon)$, or  
 the self-energy $\Sigma_{U,\sigma}^{\mu\nu}(\epsilon)$, 
as shown in Eqs.\  \eqref{eq:weighted_sum_delSRR_delSLL} and  
\eqref{eq:Noise_bubble}. 
Then, we calculate 
the remaining part $S_\mathrm{sym}^\mathrm{coll}$,
which includes effects  of vertex corrections 
as shown in Eq.\ \eqref{eq:Noise_vertex}, 
 later in this subsection.

\subsubsection{Contributions of bubble diagrams}

The free quasiparticle contributions without 
vertex corrections are described by the first three 
  parts of Eq.\ \eqref{eq:noise_4parts}, as mentioned.    
The sum of these contributions takes 
a simplified form for symmetric junctions $\Gamma_L=\Gamma_R$ 
and  $\mu_L = -\mu_R=eV/2$, at  $T = 0$, 
\begin{align}
& 
\left(
\frac{\Gamma_L}{\Gamma_L+\Gamma_R}
\right)^2 \delta S_{RR}^{}
\,+\, \left(
\frac{\Gamma_R}{\Gamma_L+\Gamma_R}
\right)^2 \delta S_{LL}^{}
 \, +\,  S_\mathrm{sym}^\mathrm{qp}  
\nonumber \\
& 
\!\!
= \frac{2e^2}{h} 
\sum_{\sigma}
\int_{-\infty}^{\infty}\! \! d\epsilon\, 
 (f_L-f_R)^2
\Delta^2
\! \left[
\left(
\frac{G_{\sigma}^{r}-G_{\sigma}^{a}}{2}\right)^2
+ G_{\sigma}^{r}G_{\sigma}^{a}
\right]
\rule{0cm}{0.8cm}
\nonumber \\
& 
 +
 \frac{2e^2}{h} 
\sum_{\sigma}
\int_{-\infty}^{\infty}\! d\epsilon\ 
\Delta\,
G_{\sigma}^{r}G_{\sigma}^{a}
\rule{0cm}{0.6cm}
\nonumber \\
& 
\qquad  \ \ 
\times 
\left[\,
 \frac{f_L +f_R}{4i} \,\Sigma_{U,\sigma}^{+-}
- \frac{(1-f_L) + (1-f_R) }{4i} \,\Sigma_{U,\sigma}^{-+}
\,\right] .
\end{align}
In order to extract order $|eV|^3$ terms from this equation,  
the lesser and greater self-energies in the last line 
 can be replaced by the low-energy asymptotic forms, 
given in Eqs.\ \eqref{eq:lesser_asymptotic}
and \eqref{eq:greater_asymptotic} 
which exactly describe the $\epsilon^2$ and $(eV)^2$ dependences of 
$\Sigma_{U,\sigma}^{-+}(\epsilon)$ and
  $\Sigma_{U,\sigma}^{+-}(\epsilon)$.  
Thus, it can be rewritten into the form, 
\begin{align}
& 
\left(
\frac{\Gamma_L}{\Gamma_L+\Gamma_R}
\right)^2 \delta S_{RR}^{}
+ \left(
\frac{\Gamma_R}{\Gamma_L+\Gamma_R}
\right)^2 \delta S_{LL}^{}
\,  +\,  S_\mathrm{sym}^\mathrm{qp}  
\nonumber \\
& 
\!\!
= \frac{2e^2}{h} 
\sum_{\sigma}
\int_{-\frac{|eV|}{2}}^{\frac{|eV|}{2}}\!\!   d\epsilon\, 
\Delta^2
\! \left[
\left\{
\frac{G_{\sigma}^{r}(\epsilon)-G_{\sigma}^{a}(\epsilon)}{2}\right\}^2
\!
+ G_{\sigma}^{r}(\epsilon)\,G_{\sigma}^{a}(\epsilon)
\right]
\rule{0cm}{0.8cm}
\nonumber \\
& \ \ 
 +
 \frac{2e^2}{h} 
\sum_{\sigma}
\frac{\pi \Delta G_{\sigma}^{r}(0)G_{\sigma}^{a}(0)}{\rho_{d\sigma}^{}} 
\sum_{\sigma'' (\neq \sigma)} \chi_{\sigma\sigma''}^{2}
\rule{0cm}{0.6cm}
\nonumber \\
& \ \ 
\times 
\int_{-\infty}^{\infty}\! d\epsilon\, 
\left[\,  \frac{f_L +f_R}{2} \,\mathcal{I}^{+-}
+\frac{(1-f_L) + (1-f_R)}{2} \,\mathcal{I}^{-+}
\,\right]  
\nonumber 
\\
& 
\ \ + \cdots \,.
\rule{0cm}{0.6cm}
\label{eq:noise_bubble_integrals}
\end{align}
The contributions of the first integral in Eq.\ \eqref{eq:noise_bubble_integrals} 
can be calculated, expanding 
  $G_{\sigma}^{r}(\epsilon)$ and  $G_{\sigma}^{r}(\epsilon)$ 
in the integrand up to order $\epsilon^2$ and $|eV|^2$ 
using the asymptotic form of the self-energy summarized in TABLE \ref{tab:self-energy}.  
Note that 
 $\Delta G_{\sigma}^{r}(0)G_{\sigma}^{a}(0) =  \pi \rho_{d\sigma}^{}$,   
and the low-energy expansion of $G_{\sigma}^{r}(\epsilon)-G_{\sigma}^{a}(\epsilon)$  
for symmetric junctions 
is  given in Eq.\ \eqref{eq:A_including_T_eV_N_orbital}. 
The product of the retarded and advanced Green's functions 
 $G_{\sigma}^{r}(\epsilon)\,G_{\sigma}^{a}(\epsilon)$ 
can also be expanded in a similar way.
The second integral in Eq.\ \eqref{eq:noise_bubble_integrals}
 can be carried out,  using the explicit form of the collision integrals 
$\mathcal{I}^{-+}(\epsilon)$ and $\mathcal{I}^{+-}(\epsilon)$ 
given in Eq.\ \eqref{eq:collision_quasiparticle_bias},  as 
\begin{align}
&
\int_{-\infty}^{\infty}\! d\epsilon\, 
\left[\,  \frac{f_L +f_R}{2} \,\mathcal{I}^{+-}
+\frac{(1-f_L) + (1-f_R) }{2} \,\mathcal{I}^{-+}
\,\right]  
\nonumber \\
=& \ 
\int_{-\frac{3|eV|}{2}}^{\frac{-|eV|}{2}}\! d\epsilon\, 
\frac{\left(\epsilon+\frac{3|eV|}{2}\right)^2}{32} 
+ \int_{-\frac{3|eV|}{2}}^{\frac{|eV|}{2}}\! d\epsilon\, 
\frac{\left(\epsilon+\frac{3|eV|}{2}\right)^2}{32} 
\nonumber \\
&\ 
+ 3 \int_{-\frac{|eV|}{2}}^{\frac{|eV|}{2}}\! d\epsilon\, 
\frac{\left(\epsilon+\frac{|eV|}{2}\right)^2}{32} 
+ 3 \int_{-\frac{|eV|}{2}}^{\frac{|eV|}{2}}\! d\epsilon\, 
\frac{\left(\epsilon-\frac{|eV|}{2}\right)^2}{32} 
\nonumber \\
&\ 
+ \int_{-\frac{|eV|}{2}}^{\frac{3|eV|}{2}}\! d\epsilon\, 
\frac{\left(\epsilon-\frac{3|eV|}{2}\right)^2}{32} 
+ 
\int_{\frac{|eV|}{2}}^{\frac{3|eV|}{2}}\! d\epsilon\, 
\frac{\left(\epsilon-\frac{3|eV|}{2}\right)^2}{32} 
\nonumber \\
  =&  \quad \frac{|eV|^3}{4} \,.
\end{align}

Therefore, the sum of the first three parts of Eq.\ \eqref{eq:noise_4parts}  
can be expressed in the following form, 
which is exact up to  terms of order $|eV|^3$ for symmetric junctions at $T=0$,  
\begin{align}
& \!\! 
S_\mathrm{sym}^\mathrm{qp}
+\left(
\frac{\Gamma_L}{\Gamma_L+\Gamma_R}
\right)^2 \delta S_{RR}
+\left(
\frac{\Gamma_R}{\Gamma_L+\Gamma_R}
\right)^2 \delta S_{LL}
\nonumber \\
&=\, 
\frac{2e^2}{h} \,  |eV| 
\sum_{\sigma}
\frac{\sin^2 2\delta_{\sigma}^{}}{4} 
\nonumber \\
& \ \ 
-  \frac{2e^2}{h} \,  \frac{|eV|^3}{3}
\sum_{\sigma}
\Biggl[\,
c_{V,\sigma}^{} \cos 2\delta_{\sigma}^{} 
 +  \frac{\pi^2}{4} \sin^2 2\delta_{\sigma}\, \chi_{\sigma\sigma}^2 
\nonumber \\
& \qquad \qquad \qquad  \qquad  \ \  
+ \frac{\pi^2}{2} \sum_{\sigma'(\neq \sigma)}\chi_{\sigma\sigma'}^2  \,
\Biggr]
   \ + \    \cdots \;.
\label{eq:noise_bubble_conclusion}
\end{align}
Here, $c_{V,\sigma}^{}$  is the coefficient for the 
order $(eV)^2$ term of the differential conductance $dJ/dV$, 
given in Eq.\  \eqref{eq:c_V_multi} and TABLE \ref{tab:CV_CT}.

\subsubsection{Contributions of vertex corrections $S_\mathrm{sym}^\mathrm{coll}$ }

The remaining term, 
$S_\mathrm{sym}^\mathrm{coll}$ of Eq.\ \eqref{eq:noise_4parts},  
represents  the contributions of the two-quasiparticle collision processes, 
which corresponds to the second diagram 
in the first row of Fig.\ \ref{fig:Kubo_Keldysh_diagram}.
Equation \eqref{eq:Noise_vertex} takes a simplified form, 
for symmetric junctions  $\Gamma_L=\Gamma_R = \Delta/2$ and 
$\mu_L = -\mu_R =eV/2$, especially at $T=0$, 
\begin{align}
S_\mathrm{sym}^\mathrm{coll}  
\,= \,   
\frac{2e^2}{h}
\frac{\Delta^2}{8\pi}
\int_{-\frac{|eV|}{2}}^{\frac{|eV|}{2}}  \! d\epsilon
\! \int_{-\frac{|eV|}{2}}^{\frac{|eV|}{2}} \! d\epsilon' \, 
\mathcal{Q}(\epsilon,\epsilon').
\label{eq:noise_vertex_integral}
\end{align}
The domains of the integrals with respect to $\epsilon$ and $\epsilon'$  
are restricted to the inside of the bias window. Therefore, 
in order to calculate $S_\mathrm{sym}^\mathrm{coll}$ up to order $|eV|^3$, 
the kernel $\mathcal{Q}(\epsilon,\epsilon')$ should be expanded  
up to linear order  in $\epsilon$, $\epsilon'$, and $|eV|$. 
To this end, we rewrite   $\mathcal{Q}(\epsilon,\epsilon')$ 
into the following form 
by substituting  Eq.\ \eqref{eq:bare_current_vertex_decomposition} 
into  Eq.\ \eqref{eq:Q_dif}, 
\begin{align}
\! 
\mathcal{Q}(\epsilon,\epsilon') 
= 
\sum_{\sigma\sigma'} 
\sum_{l=1}^3\sum_{m=1}^3 
a_{l\sigma}^{}(\epsilon)\, a_{m\sigma'}^{}(\epsilon')\, 
\overline{\mathcal{Q}}_{\sigma\sigma'}^{lm}(\epsilon,\epsilon')\,.
\label{eq:Q_9parts}
\end{align}
Here,  $\overline{\mathcal{Q}}_{\sigma\sigma'}^{lm}(\epsilon,\epsilon')$ 
corresponds to the vertex contributions,  for which the summations over 
 the Keldysh-branch indexes  ($\mu,\mu';\nu',\nu$) in Eq.\ \eqref{eq:Q_dif} 
have been carried out with matrix set  $\bm{1} +\bm{\tau}_1$,
$\bm{\tau}_3$, and $i\bm{\tau}_2$.
Specifically, 
for symmetric junctions $\Gamma_L=\Gamma_R = \Delta/2$ and 
$\mu_L = -\mu_R =eV/2$,  only the two components   
$\overline{\mathcal{Q}}_{\sigma\sigma'}^{33}(\epsilon,\epsilon')$ and 
 $\overline{\mathcal{Q}}_{\sigma\sigma'}^{22}(\epsilon,\epsilon')$ 
among nine possible ($\ell$, $m$) configurations   
 have linear-order terms with respect to $\epsilon$, $\epsilon'$ and $|eV|$,  
and thus  Eq.\ \eqref{eq:Q_9parts} takes a simplified form, 
\begin{align}
\!\!
\mathcal{Q}(\epsilon,\epsilon') 
= &
\sum_{\sigma\sigma'} 
\biggl[\,
a_{3\sigma}^{}(\epsilon) \,a_{3\sigma'}^{}(\epsilon')\, 
\overline{\mathcal{Q}}_{\sigma\sigma'}^{33}(\epsilon,\epsilon')
\nonumber 
\\
& \quad \ \   +
a_{2\sigma}^{}(\epsilon) \,a_{2\sigma'}^{}(\epsilon')\, 
\overline{\mathcal{Q}}_{\sigma\sigma'}^{22}(\epsilon,\epsilon') 
\biggr]  +  \cdots .
\label{eq:Q_9parts_symmetric_bias}
\end{align} 

One of the reasons for this is that  the coefficient $a_{1\sigma}^{}(\epsilon)$, 
defined in Eq.\ \eqref{eq:a_coefficients},  
 vanishes identically inside the bias-window region  $|\epsilon| \leq |eV|/2$. 
It follows from the low-energy properties of 
the lesser and greater Green's functions,
described in  Eq.\  \eqref{eq:lesser_greater_lowest},  
\begin{align}
G_{\sigma}^{+-}(\epsilon) + G_{\sigma}^{-+}(\epsilon) \,\simeq \, 
-2i \Delta \,\bigl[ 1 -2 f_\mathrm{eff}^{}(\epsilon) \bigr]\, 
  G_{\sigma}^{r}(\epsilon)\,G_{\sigma}^{a}(\epsilon)\,. 
\end{align}
Thus, the components which are  accompanied by $a_{1\sigma}^{}(\epsilon)$ or 
 $a_{1\sigma'}^{}(\epsilon')$ vanish  
 in the right-hand side of Eq.\ \eqref{eq:Q_9parts} 
since $1-2f_\mathrm{eff}^{}(\epsilon) \equiv 0$ 
 at $|\epsilon|<|eV|/2$  for symmetric junctions.  
Another reason is that  two other components   
$\overline{\mathcal{Q}}_{\sigma\sigma'}^{32}(\epsilon,\epsilon')$ and 
$\overline{\mathcal{Q}}_{\sigma\sigma'}^{23}(\epsilon,\epsilon')$ 
do not  have linear-order terms in $\epsilon$, $\epsilon'$ and $|eV|$:  
\begin{align}
 \overline{\mathcal{Q}}_{\sigma\sigma'}^{32}(\epsilon,\epsilon')
&= i \sum_{\mu\nu}
\sum_{\mu'\nu'}
\left\{\bm{\tau}_3\right\} _{\mu\nu}
\Gamma_{\sigma\sigma';\sigma'\sigma}^{\mu\mu';\nu'\nu}
 (\epsilon,\epsilon'; \epsilon',\epsilon)
\left\{i\bm{\tau}_2\right\} _{\nu'\mu'}
\nonumber \\
 & =  
i\left[\, \Gamma_{\sigma\sigma';\sigma'\sigma}^{-+;--}
+\Gamma_{\sigma\sigma';\sigma'\sigma}^{+-;++}
-\Gamma_{\sigma\sigma';\sigma'\sigma}^{--;+-}
-\Gamma_{\sigma\sigma';\sigma'\sigma}^{++;-+} \,\right]
\nonumber \\
&  =   \ 0 
 \ + \ O(\epsilon^2,  \epsilon'^2, |eV|^2)
  \rule{0cm}{0.6cm}
\;, 
\end{align}
and also $\overline{\mathcal{Q}}_{\sigma\sigma'}^{23}(\epsilon,\epsilon')=0 
  +  O(\epsilon^2,  \epsilon'^2, |eV|^2)$. 
The is because 
the Keldysh vertex components with three identical branch indexes,   
such as  $\Gamma_{\sigma\sigma;\sigma\sigma}^{+-;++}$  and 
 $\Gamma_{\sigma\sigma;\sigma\sigma}^{--;+-}$, 
do not have a constant and linear order terms 
with respect to $\epsilon$, $\epsilon'$, and $eV$.

The two components  
 $\overline{\mathcal{Q}}_{\sigma\sigma'}^{33}(\epsilon,\epsilon')$   
and 
 $\overline{\mathcal{Q}}_{\sigma\sigma'}^{22}(\epsilon,\epsilon')$  
in Eq.\ \eqref{eq:Q_9parts_symmetric_bias} 
 can be calculated, using  the low-energy expansion of the vertex functions,  
 given in TABLES \ref{tab:vertex_UU} and \ref{tab:vertex_UD}.   
These components for $\sigma=\sigma'$ can be expressed in the following form  
at  $|\epsilon|<|eV|/2$ and $|\epsilon'|<|eV|/2$,  
\begin{align}
 \overline{\mathcal{Q}}_{\sigma\sigma}^{33}(\epsilon,\epsilon')
&= -i \sum_{\mu\nu}
\sum_{\mu'\nu'}
\left\{\bm{\tau}_3\right\} _{\mu\nu}
\Gamma_{\sigma\sigma;\sigma\sigma}^{\mu\mu';\nu'\nu}
 (\epsilon,\epsilon'; \epsilon',\epsilon)
\left\{\bm{\tau}_3\right\} _{\nu'\mu'}
\nonumber \\
 & \!\! =  
-i\left[\,\Gamma_{\sigma\sigma;\sigma\sigma}^{--;--}
+\Gamma_{\sigma\sigma;\sigma\sigma}^{++;++}
-\Gamma_{\sigma\sigma;\sigma\sigma}^{-+;+-}
-\Gamma_{\sigma\sigma;\sigma\sigma}^{+-;-+}  \,\right]
\nonumber \\
& \!\!  =   \frac{\pi}{\rho_{d\sigma}^2} \sum_{\sigma''(\neq \sigma)}
 \chi_{\sigma\sigma''}^{2} 
\Bigl [\,
|\epsilon-\epsilon'| - |eV|
\, \Bigr ]
 \ + \ \cdots
\;, 
\label{eq:Q33_UU}
\end{align}
and 
\begin{align}
 \overline{\mathcal{Q}}_{\sigma\sigma}^{22}(\epsilon,\epsilon')
&= -i \sum_{\mu\nu}
\sum_{\mu'\nu'}
\left\{i\bm{\tau}_2\right\} _{\mu\nu}
\Gamma_{\sigma\sigma;\sigma\sigma}^{\mu\mu';\nu'\nu}
 (\epsilon,\epsilon'; \epsilon',\epsilon)
\left\{i\bm{\tau}_2\right\} _{\nu'\mu'}
\nonumber \\
 & \!\! =  
-i\left[\, \Gamma_{\sigma\sigma;\sigma\sigma}^{-+;-+}
+\Gamma_{\sigma\sigma;\sigma\sigma}^{+-;+-}
-\Gamma_{\sigma\sigma;\sigma\sigma}^{--;++}
-\Gamma_{\sigma\sigma;\sigma\sigma}^{++;--} \,\right]
\nonumber \\
& \!\!  =   \frac{\pi}{\rho_{d\sigma}^2} \sum_{\sigma''(\neq \sigma)}
 \chi_{\sigma\sigma''}^{2} 
\Bigl[\,
|\epsilon-\epsilon'| + |eV|
\, \Bigr]
 \ + \ \cdots
\;. 
\label{eq:Q22_UU}
\end{align}
Similarly, at  $|\epsilon|<|eV|/2$ and $|\epsilon'|<|eV|/2$,  
the components for $\sigma\neq\sigma'$ are given by 
\begin{align}
 \overline{\mathcal{Q}}_{\sigma\sigma'}^{33}(\epsilon,\epsilon')
&= -i \sum_{\mu\nu}
\sum_{\mu'\nu'}
\left\{\bm{\tau}_3\right\} _{\mu\nu}
\Gamma_{\sigma\sigma';\sigma'\sigma}^{\mu\mu';\nu'\nu}
 (\epsilon,\epsilon'; \epsilon',\epsilon)
\left\{\bm{\tau}_3\right\} _{\nu'\mu'}
\nonumber \\
 & \!\!\! 
=  
-i\left[\, \Gamma_{\sigma\sigma';\sigma'\sigma}^{--;--}
+\Gamma_{\sigma\sigma';\sigma'\sigma}^{++;++}
-\Gamma_{\sigma\sigma';\sigma'\sigma}^{-+;+-}
-\Gamma_{\sigma\sigma';\sigma'\sigma}^{+-;-+} \,\right]
\nonumber \\
&  \!\!\! 
=   \frac{\pi}{\rho_{d\sigma}^{}\rho_{d\sigma'}^{}} 
\biggl [\,
\chi_{\sigma\sigma'}^{2}
\Bigl(\,
 |\epsilon-\epsilon'| \, - \,   |\epsilon+\epsilon'|
\,\Bigr) 
\nonumber \\
& \qquad \qquad \ \ 
 - 2 |eV|\sum_{\sigma_3(\neq \sigma,\sigma')} 
\chi_{\sigma\sigma_3}^{}\chi_{\sigma_3\sigma'}^{}
\, \biggr]
  +  \cdots , 
\label{eq:Q33_UD}
\end{align}
and 
\begin{align}
 \overline{\mathcal{Q}}_{\sigma\sigma'}^{22}(\epsilon,\epsilon')
&= -i \sum_{\mu\nu}
\sum_{\mu'\nu'}
\left\{i\bm{\tau}_2\right\} _{\mu\nu}
\Gamma_{\sigma\sigma';\sigma'\sigma}^{\mu\mu';\nu'\nu}
 (\epsilon,\epsilon'; \epsilon',\epsilon)
\left\{i\bm{\tau}_2\right\} _{\nu'\mu'}
\nonumber \\
 & \!\!\! 
=  
-i\left[\, \Gamma_{\sigma\sigma';\sigma'\sigma}^{-+;-+}
+\Gamma_{\sigma\sigma';\sigma'\sigma}^{+-;+-}
-\Gamma_{\sigma\sigma';\sigma'\sigma}^{--;++}
-\Gamma_{\sigma\sigma';\sigma'\sigma}^{++;--} \,\right]
\nonumber \\
& \!\!\!  
=  \frac{\pi}{\rho_{d\sigma}^{}\rho_{d\sigma'}^{}} 
\chi_{\sigma\sigma'}^{2}
\Bigl [\,
 |\epsilon-\epsilon'| \, + \,   |\epsilon+\epsilon'|
\,+ 2 |eV|\, \Bigr ]
\nonumber \\
& \  \ + \ \cdots .
\label{eq:Q22_UD}
\end{align}
In order to obtain these expressions,
 we have used the properties of  the bosonic collision integrals, 
 described in Eqs.\ \eqref{eq:ph_and_pp_symmetric_junction} and 
 \eqref{eq:Coll_in_bias_window_text}.  
Note that 
 $a_{3\sigma}^{}(\epsilon)$ and  $a_{2\sigma}^{}(\epsilon)$ 
 in Eq.\ \eqref{eq:Q_9parts_symmetric_bias}  
can be replaced by their zero-frequency values 
\begin{align}
a_{3\sigma}^{}(0)  
 = 
i \, \frac{\pi \rho_{d\sigma}^{}}{\Delta}\,
\sin 2\delta_{\sigma}^{} , 
\qquad 
a_{2\sigma}^{}(0) 
 =  \ 
 \frac{\pi \rho_{d\sigma}^{}}{\Delta} 
\cos 2 \delta_{\sigma}^{}  ,
\end{align}
since their $\epsilon$ dependence yields the corrections higher 
than the order $|eV|^3$ ones.

We can now calculate 
the vertex contributions $S_\mathrm{sym}^\mathrm{coll}$ 
by substituting 
Eqs.\ \eqref{eq:Q33_UU} and \eqref{eq:Q22_UD}
 into Eq.\ \eqref{eq:Q_9parts_symmetric_bias}, 
and then carrying out the  double integral 
in Eq.\ \eqref{eq:noise_vertex_integral} that is given by  
\begin{align}
\int_{-\frac{|eV|}{2}}^{\frac{|eV|}{2}}\! d\epsilon\! 
\int_{-\frac{|eV|}{2}}^{\frac{|eV|}{2}}\! d\epsilon'\,
 \left|\epsilon \pm \epsilon' \right|
\ =  & \ \frac{1}{3}\, |eV|^3 \;.
\end{align}
Consequently, the vertex contributions to 
the nonlinear current noise  $S_\mathrm{sym}^\mathrm{coll}$
are obtained exactly up to order $|eV|^3$ at $T=0$,  
\begin{align}
&
S_\mathrm{sym}^\mathrm{coll}  \,  =  \  
 \frac{2e^2}{h} \frac{\pi^2 |eV|^3}{12}
\sum_{\sigma} 
\sum_{\sigma'(\neq \sigma)} 
\nonumber \\
& \quad  \ \times 
\Biggl[\, 
\Bigl(\sin^2 2\delta_{\sigma}^{}\, + 2  \cos^2 2\delta_{\sigma}^{} \Bigr)
\, \chi_{\sigma\sigma'}^{2} 
\nonumber \\
&  \qquad  \quad 
+ 4\cos 2\delta_{\sigma}^{}\cos 2\delta_{\sigma'}^{} 
\, \chi_{\sigma\sigma'}^{2} 
\nonumber \\
&   \qquad \quad 
+ 3 \sin 2\delta_{\sigma}^{}\sin 2\delta_{\sigma'}^{} 
\!\sum_{\sigma''(\neq \sigma,\sigma')} \!
\chi_{\sigma\sigma''}^{}\chi_{\sigma''\sigma'}^{} \Biggr]
 +  \cdots . 
\label{eq:noise_vertex_conclusion}
\end{align}

\subsection{Total current noise $S_\mathrm{noise}^{\mathrm{QD}}$ 
up to order $|eV|^3$}

We have calculated the nonlinear current noise in the above,
separating it into four parts as shown in Eq.\ \eqref{eq:noise_4parts}. 
The contributions of the first three parts 
which are described by the bubble diagrams  
are given in  Eq.\  \eqref{eq:noise_bubble_conclusion}, 
and the remaining part that represents  contributions of the vertex corrections  
are given in Eq.\ \eqref{eq:noise_vertex_conclusion}.
Adding all contributions, 
we obtain the following formula for the nonlinear current noise  
that is exact  up to order $|eV|^3$ at  $T=0$ 
 for symmetric junctions  $\Gamma_L=\Gamma_R$ and $\mu_L=-\mu_R= eV/2$,
\begin{align}
S_\mathrm{noise}^\mathrm{QD}
\,= &  \  
  \frac{2e^2}{h} \,
|eV| 
\sum_{\sigma}
\left[\,
\frac{\sin^2 2\delta_\sigma^{}}{4} \,  
\,  + \,
c_{S,\sigma}^{} 
 \, 
\left(eV\right)^2
\ + \ \cdots 
\,
\right] .
 \end{align}
The coefficient $c_{S,\sigma}^{} $ is given by 
 \begin{align}
\!\! 
c_{S,\sigma}^{} = & \    
\frac{\pi^2}{12} 
\Biggl[
\,  
\cos 4 \delta_{\sigma}\,
 \chi_{\sigma\sigma}^2 
+ \bigl( 2+3\cos 4 \delta_{\sigma} \bigr)
\sum_{\sigma'(\neq \sigma)}\chi_{\sigma\sigma'}^2
\nonumber 
\\
& 
\quad \ \  +\,  
4\sum_{\sigma' (\neq \sigma)} \cos 2\delta_{\sigma}^{} \cos 2\delta_{\sigma'}^{}
\,  \chi_{\sigma\sigma'}^2
\nonumber \\
& 
\quad \ \   +\,3 \sum_{\sigma'(\neq \sigma)}
\sum_{\sigma''(\neq \sigma,\sigma')}
\sin 2\delta_{\sigma}^{} \,\sin 2\delta_{\sigma'}^{}
 \chi_{\sigma\sigma''}^{}  \chi_{\sigma'\sigma''}^{}
\nonumber \\
& \quad \ \   
-\,
\biggl(
  \chi_{\sigma\sigma\sigma}^{[3]}
+
3 \sum_{\sigma'(\neq \sigma)}
  \chi_{\sigma\sigma'\sigma'}^{[3]}
\biggr) 
\frac{\sin 4\delta_{\sigma}}{4\pi} 
\,\Biggr] .
\end{align}
This is one of the most important results of this paper, 
 and is also described in Sec.\ \ref{sec:noise_result_first}.

\section{Summary}
\label{sec:summary_IV}

In summary, we have investigated thoroughly low-energy behaviors   
of the Keldysh vertex functions 
  $\Gamma_{\sigma\sigma';\sigma'\sigma}^{\nu_1\nu_2;\nu_3\nu_4}
(\omega,\omega';\omega',\omega)$ 
of the Anderson impurity model 
in a nonequilibrium steady state under a finite bias voltage $eV$. 
This information  is essential to determine the next-to-leading order transport 
of the nonlinear current noise  through quantum dots 
in the Fermi-liquid regime.

We have provided 
a Feynman-diagrammatic derivation of  the Ward identities, 
which describes the relations between the Keldysh vertex functions 
and the self-energies $\Sigma_{U,\sigma}^{\nu_4\nu_1}(\omega)$. 
In the perturbation theory in $U$, 
effects of the bias voltage $eV$ and temperature $T$ enter 
through the nonequlibirum distribution function $f_\mathrm{eff}(\omega)$
that is contained in the noninteracting 
propagators  $G_{0\sigma}^{\nu_4\nu_1}(\omega)$, 
given in  Eq.\  \eqref{eq:G0_Keldysh_elements}. 
It appears in the right-hand side of the Ward identities 
Eqs.\ \eqref{eq:Ward_NEQ_omega_all} and \eqref{eq:Ward_NEQ_eV_all}, 
and each of the derivatives  $\partial f_\mathrm{eff}(\omega)/\partial \omega$ 
and $\partial f_\mathrm{eff}(\omega)/\partial (eV)$ 
evolves into the two Dirac delta functions  at $\omega = \mu_L$ and $\mu_R$   
 in the limit  $T\to 0$. 

We have also verified that these relations can be derived from more general 
Ward-Takahashi identities for the three-point Keldysh correlations functions   
$\bm{\Phi}_{\gamma,\sigma\sigma'}^{\alpha}(\epsilon,\epsilon+\omega)$,  
or the corresponding three-point vertex functions 
 $\bm{\Lambda}_{\gamma,\sigma\sigma'}^{\alpha}(\epsilon,\epsilon+\omega)$ 
for $\gamma=L,R,d$  and  $\alpha= +,-$, 
given in Eqs.\ \eqref{eq:WT_Phi} and \eqref{eq:WT_Lambda}. 
These equations explicitly show that 
the Ward identity Eq.\ \eqref{eq:Ward_NEQ_omega_all} 
reflects the local current conservation between the impurity site and the conduction bands 
 for each $\sigma$ component.  

Thus, the current conservation plays an essential role 
to the next-to-leading order transport in the Fermi-liquid regime 
since a quasiparticle shows the damping  
of order $\omega^2$, $(eV)^2$ and $T^2$.
It is determined the scattering processes illustrated in Fig.\ \ref{fig:self_im_T2}, 
and its precise dependences on $\omega$, $eV$ and $T$ can be described 
in terms of the fermionic collision integrals  
$\mathcal{I}^{+-}(\omega)$  and  $\mathcal{I}^{-+}(\omega)$, 
defined   in Sec.\ \ref{sec:Collision_Fermion}.
Correspondingly, the imaginary part of the vertex functions 
are determined by the scattering processes illustrated in 
Fig.\ \ref{fig:vertex_singular_general} at low energies 
up to linear-order terms with respect to $\omega$, $eV$ and $T$;  
the results are summarized in TABLES \ref{tab:vertex_UU} and \ref{tab:vertex_UD}.   
The imaginary part of the vertex functions 
can be expressed in terms of the bosonic collision integrals 
 $\mathcal{W}_{}^\mathrm{ph}(\omega)$,
 $\mathcal{W}_{}^\mathrm{pp}(\omega)$, and   
 $\mathcal{W}_{}^\mathrm{hh}(\omega)$    
 which represent the single particle-hole,  particle-particle, 
and hole-hole propagating processes. 


The imaginary parts of the Ward identities are full filled in the Fermi-liquid regime 
through the relations between these bosonic collision integrals 
and the fermionic ones, 
as shown in 
Sec.\ \ref{sec:current_conservation_Keldysh_selfenergy_vertex_in_FL_regime}. 
Furthermore,  from the Ward identities, 
we have also deduced the $eV$-linear real part  
which emerges for the causal  vertex component, i.e.,   
  $\mathrm{Re}\,\Gamma_{\sigma\sigma';\sigma'\sigma}^{--;--}
(\omega,\omega';\omega',\omega)$ for $\sigma\neq \sigma'$. 
This real part  becomes finite 
 when the inversion symmetry 
of the tunnel junction is broken in a way such that 
$\Gamma_L^{} \mu_L^{}+ \Gamma_R^{} \mu_R^{} \neq 0$, 
and its value is determined by 
the  nonlinear three-body susceptibilities 
and the derivative of the density of states.

Using these exact low-energy asymptotic forms  
of the Keldysh vertex functions and Green's functions, 
we have calculated the current noise $S_\mathrm{noise}^\mathrm{QD}$ 
  up to terms of order $|eV|^3$, 
and have applied it to some typical cases: 
the SU($N$) Anderson model, 
and the $S=1/2$ Anderson model in a magnetic field $b$. 
Our formula is applicable to a wide class of quantum dots 
without particle-hole or time-reversal symmetry, 
for any value of the impurity-electron filling $\langle n_{d\sigma}^{}\rangle$ 
which varies with parameters  $U$, $\epsilon_{d\sigma}^{}$, and $\Delta$. 
In this paper,  the order $|eV|^3$ current noise has been obtained  
specifically for symmetric tunnel junctions satisfying 
the conditions  $\Gamma_L=\Gamma_R= \Delta/2$ and $\mu_L = -\mu_R =eV/2$, 
 just for simplicity. 
Nevertheless, calculations carried out 
in Sec.\ \ref{sec:noise_derivation} can be extended straightforwardly 
to quantum dots with no such junction symmetries, and it is left for a future work.
The necessary information for carrying it out, namely 
 the low-energy asymptotic form of the Keldysh vertex functions
 in the Fermi-liquid regime,  has already been obtained in this paper 
without assuming these symmetries for tunnel junctions. 


\begin{acknowledgments}
We would like to thank K.\ Kobayashi, T.\ Hata,   M.\ Ferrier, 
and A.\ C.\ Hewson for valuable discussions.  
This work was supported by 
JSPS KAKENHI Grant Numbers  JP18J10205, JP18K03495, 
JP21K03415,  JST CREST Grant No.\ JPMJCR1876, 
and the Sasakawa Scientific Research Grant from the Japan Science Society No.\ 2021-2009.
\end{acknowledgments}


\appendix

\section{Linear dependency of the Keldysh correlation functions}
\label{sec:linear_dependency_Keldysh}

We describe here briefly the linear dependence among 
 the components of the Keldysh correlation functions. 
To this end, we quickly look back the situation for the 
single-particle Green's function. 

By definition given in Eq.\ \eqref{eq:Keldysh_Green's_function_def}, 
the components of $G_{\sigma}^{\mu\nu}$ are linearly dependent,  as 
\begin{align}
\!\!\!\!  
\sum_{\mu\nu}
\mathrm{sgn}(\mu\nu)\,  G_{\sigma}^{\mu\nu}
 =
 G^{--}_{\sigma} \! +  G^{++}_{\sigma}\!  
- G^{+-}_{\sigma}\!   -  G^{-+}_{\sigma} 
 =  0.
\label{eq:keldysh_sum_G_appendix}
\end{align}
Corresponding relation between the self-energy components can be 
deduced from this relations. It can be carried out, substituting 
 the expressions of $G_{\sigma}^{\mu\nu}$  
in the right-hand side of the Dyson equation, 
 $\bm{G}_{\sigma}= \bm{G}_{0\sigma} 
+\bm{G}_{0\sigma}\bm{\Sigma}_{U,\sigma}\,\bm{G}_{\sigma}$,  
into  Eq.\ \eqref{eq:keldysh_sum_G_appendix},   
and using also Eq.\ \eqref{eq:GK_def}
 to rewrite  $G^{\mu\nu}_{\sigma}$ further 
in terms of  $G^{r}_{\sigma}$,  $G^{a}_{\sigma}$, and $G^\mathrm{K}_{\sigma}$,
 as 
\begin{align}
\sum_{\mu\nu}
\mathrm{sgn}(\mu\nu)\,  G_{0\sigma}^{\mu\nu}\, +\, 
G_{0\sigma}^a \,G_{\sigma}^r \sum_{\mu'\nu'} 
\Sigma_{U,\sigma}^{\mu'\nu'} 
  \,=\, 0\, .
\label{eq:keldysh_sum_G_appendix_2}
\end{align}
The first term in the left-hand side vanishes 
because the noninteracting Green's function  $G_{0\sigma}^{\mu\nu}$ 
also satisfies the same linear-dependent relation 
as Eq.\ \eqref{eq:keldysh_sum_G_appendix}.  
Furthermore, as the product  $G^{a}_{0\sigma} G^{r}_{\sigma}$ 
in the left-hand side of Eq.\ \eqref{eq:keldysh_sum_G_appendix_2} 
does not identically vanish, it follows that 
\begin{align}
 \Sigma_{U,\sigma}^{--} +  \Sigma_{U,\sigma}^{++} 
+ \Sigma_{U,\sigma}^{+-} + \Sigma_{U,\sigma}^{-+} \, =\,  0\,. 
\end{align}
In the next two subsections, we examine  the linear dependency 
of  $\Lambda_{\gamma,\sigma\sigma'}^{\alpha;\mu\nu}$ 
and $\Gamma_{\sigma_1\sigma_2;\sigma_3\sigma_4}^{\nu_1\nu_2;\nu_3\nu_4}$,
following along the similar line.

\subsection{Linear dependence  
of $\Lambda_{\gamma,\sigma\sigma'}^{\alpha;\mu\nu}$ }

The  three-point function $\Phi_{\gamma,\sigma\sigma'}^{\alpha;\mu\nu}$
for  $\gamma =L, R, d$  are also linearly dependent  
as it can be verified directly from the definition given in Eq.\ \eqref{eq:Phi_CLR},   
\begin{align}
&
\left(
\Phi_{\gamma,\sigma\sigma'}^{-;--}
+\Phi_{\gamma,\sigma\sigma'}^{-;++}
-\Phi_{\gamma,\sigma\sigma'}^{-;+-}
-\Phi_{\gamma,\sigma\sigma'}^{-;-+} 
\right)
\nonumber \\
& \quad 
-
\left(
\Phi_{\gamma,\sigma\sigma'}^{+;++}
+\Phi_{\gamma,\sigma\sigma'}^{+;--}
-\Phi_{\gamma,\sigma\sigma'}^{+;-+}
-\Phi_{\gamma,\sigma\sigma'}^{+;+-}
\right) \, = \, 0 .
\label{eq:keldysh_sum_Phi_appendix}
\end{align}
Therefore,  the components of the vertex  
$\Lambda_{\gamma,\sigma\sigma'}^{\alpha;\mu\nu}
(\epsilon,\epsilon+\omega)$, 
defined in Eq.\ \eqref{eq:Lambda_matrix} such that 
 $\bm{\Phi}_{\gamma,\sigma\sigma'}^{\alpha}
= \bm{G}_{\sigma}^{}
\bm{\Lambda}_{\gamma,\sigma\sigma'}^{\alpha}
\bm{G}_{\sigma}^{}$ in the frequency domain,  
are also linearly dependent.
The left-hand side of Eq.\ \eqref{eq:keldysh_sum_Phi_appendix}  
 can be rewritten  in terms 
of  $\Lambda_{\gamma,\sigma\sigma'}^{\alpha;\mu\nu}$, 
  $G^{r}_{\sigma}$,  $G^{a}_{\sigma}$, and $G^\mathrm{K}_{\sigma}$,
using also Eq.\ \eqref{eq:GK_def},  as 
\begin{align}
&
\!\!\! 
  G_{\sigma}^{a}(\epsilon )\, 
  \, G_{\sigma}^{r}( \epsilon  +  \omega)  
\biggl[\,
\Lambda_{\gamma,\sigma\sigma'}^{-;--}
+\Lambda_{\gamma,\sigma\sigma'}^{-;++}
+\Lambda_{\gamma,\sigma\sigma'}^{-;-+}
+\Lambda_{\gamma,\sigma\sigma'}^{-;+-}
\nonumber \\
& \qquad \   
-\Lambda_{\gamma,\sigma\sigma'}^{+;--}
-\Lambda_{\gamma,\sigma\sigma'}^{+;++}
-\Lambda_{\gamma,\sigma\sigma'}^{+;-+}
-\Lambda_{\gamma,\sigma\sigma'}^{+;+-}
\,\biggr]  =  0.
\end{align}
Thus,  from this, it follows that 
\begin{align}
 \sum_{\alpha=\pm} \sum_{\mu\nu}\, 
\mathrm{sgn}(\alpha) 
\, \Lambda_{\gamma,\sigma\sigma'}^{\alpha;\mu\nu}(\epsilon,\epsilon+\omega)
\  = \  0\,.
\end{align}

\subsection{Linear dependence  
of $\Gamma_{\sigma_1\sigma_2;\sigma_3\sigma_4}^{\nu_1\nu_2;\nu_3\nu_4}$}
\label{subsec:GII}

We next consider the four-point correlation function, defined by 
\begin{align}
& \!\!
\mathcal{G}_{\sigma_4,\sigma_2;\sigma_3,\sigma_1}^{\mu_4\mu_2;\mu_3\mu_1}
(t_{4}^{},t_{2}^{};t_{3}^{},t_{1}^{})  
\nonumber \\
& \ \equiv\,  
  i  
\,
\left \langle T_C^{}\,d_{\sigma_4}^{}(t_{4}^{\mu_4})
\,d_{\sigma_2}^{}(t_{2}^{\mu_2})
\,d_{\sigma_3}^{\dagger}(t_{3}^{\mu_3})\,
d_{\sigma_1}^{\dagger}(t_{1}^{\mu_1}) \right \rangle . 
\label{eq:GII_4point}
\end{align}
It can be expressed  
 in terms of the four-point vertex corrections, as 
\begin{align}
&\mathcal{G}_{\sigma_4,\sigma_2,\sigma_3,\sigma_1}^{\mu_4\mu_2;\mu_3\mu_1}(t_4,t_2;t_3,t_1)  
\nonumber \\ 
&=  \, 
 -\,i  \,
G_{\sigma_1}^{\mu_4\mu_1}(t_4,t_1)\,
G_{\sigma_2}^{\mu_2\mu_3} (t_2,t_3)\,
\delta_{\sigma_4\sigma_1}^{} \delta_{\sigma_2\sigma_3}^{}
\nonumber \\
& \quad 
+ \,i  \, 
G_{\sigma_4}^{\mu_4\mu_3}(t_4,t_3)\,
G_{\sigma_1}^{\mu_2\mu_1}(t_2,t_1)\,
\delta_{\sigma_2\sigma_1}^{}\delta_{\sigma_4\sigma_3}^{}
\nonumber \\
& \quad  
 + 
\int 
\prod_{k=1}^4 dt_{k}' 
\sum_{\nu_1\nu_2 \atop \nu_3\nu_4}
\Gamma_{\sigma_1\sigma_2;\sigma_3\sigma_4}^{\nu_1\nu_2;\nu_3\nu_4}(t_1',t_2';t_3', t_4') 
\nonumber \\
& \quad \  \times 
G_{\sigma_4}^{\nu_1\mu_1}(t_1',t_1)
G_{\sigma_2}^{\mu_2\nu_2}(t_2,t_2')
G_{\sigma_3}^{\nu_3\mu_3}(t_3',t_3)
G_{\sigma_1}^{\mu_4\nu_4}(t_4,t_4').
\label{eq:BS_equation_4point}
 \end{align}
By definition,  sixteen Keldysh components of 
$\mathcal{G}_{\sigma_4,\sigma_2;\sigma_3,\sigma_1}
^{\mu_4\mu_2;\mu_3\mu_1}$ are linear dependent, 
and the relation among them is given by  
\begin{align}
&  \!\! 
\sum_{\mu_4\mu_2 \atop \mu_3 \mu_1} 
\mathrm{sgn}(\mu_4\mu_2\mu_3\mu_1)\  
\mathcal{G}_{\sigma_4,\sigma_2;\sigma_3,\sigma_1}^{\mu_4\mu_2;\mu_3\mu_1}
\nonumber  \\
&  
\equiv\    \mathcal{G}_{\sigma_4,\sigma_2;\sigma_3,\sigma_1}^{--;--} 
+   \mathcal{G}_{\sigma_4,\sigma_2;\sigma_3,\sigma_1}^{++;++} 
+   \mathcal{G}_{\sigma_4,\sigma_2;\sigma_3,\sigma_1}^{-+;+-}
+   \mathcal{G}_{\sigma_4,\sigma_2;\sigma_3,\sigma_1}^{+-;-+} 
\nonumber \\
& \quad 
 +   \mathcal{G}_{\sigma_4,\sigma_2;\sigma_3,\sigma_1}^{+-;+-} 
+   \mathcal{G}_{\sigma_4,\sigma_2;\sigma_3,\sigma_1}^{-+;-+} 
+   \mathcal{G}_{\sigma_4,\sigma_2;\sigma_3,\sigma_1}^{++;--} 
+   \mathcal{G}_{\sigma_4,\sigma_2;\sigma_3,\sigma_1}^{--;++} 
\nonumber \\
& \quad
-   \mathcal{G}_{\sigma_4,\sigma_2;\sigma_3,\sigma_1}^{+-;--} 
-   \mathcal{G}_{\sigma_4,\sigma_2;\sigma_3,\sigma_1}^{-+;--} 
-   \mathcal{G}_{\sigma_4,\sigma_2;\sigma_3,\sigma_1}^{--;+-} 
-   \mathcal{G}_{\sigma_4,\sigma_2;\sigma_3,\sigma_1}^{--;-+} 
\nonumber \\
& \quad  
-   \mathcal{G}_{\sigma_4,\sigma_2;\sigma_3,\sigma_1}^{-+;++} 
-   \mathcal{G}_{\sigma_4,\sigma_2;\sigma_3,\sigma_1}^{+-;++} 
-   \mathcal{G}_{\sigma_4,\sigma_2;\sigma_3,\sigma_1}^{++;-+} 
-   \mathcal{G}_{\sigma_4,\sigma_2;\sigma_3,\sigma_1}^{++;+-}
\nonumber \\
& = \ 0 \, .
\label{eq:keldysh_sum_GII_appendix}
\end{align}

This relation can be rewritten in terms of 
$\Gamma_{\sigma_1\sigma_2;\sigma_3\sigma_4}^{\nu_1\nu_2;\nu_3\nu_4}$, 
  $G^{r}_{\sigma}$,  $G^{a}_{\sigma}$, and $G^\mathrm{K}_{\sigma}$, by 
substituting  Eq.\ \eqref{eq:BS_equation_4point} 
into the left-hand side of  Eq.\ \eqref{eq:keldysh_sum_GII_appendix}, 
and using  Eq.\ \eqref{eq:GK_def}, as
\begin{align}
&  \!\!\!\!\! \!
\sum_{\mu_4\mu_2 \atop \mu_3 \mu_1} 
\mathrm{sgn}(\mu_4\mu_2\mu_3\mu_1)\  
\mathcal{G}_{\sigma_4,\sigma_2;\sigma_3,\sigma_1}^{\mu_4\mu_2;\mu_3\mu_1}
(t_4,t_2;t_3, t_1) 
\nonumber \\
& 
\!\!\!   = 
\int \prod_{k=1}^4  dt_k'
\sum_{\nu_1\nu_2 \atop \nu_3\nu_4}
\Gamma_{\sigma_1\sigma_2;\sigma_3\sigma_4}^{\nu_1\nu_2;\nu_3\nu_4}
(t_1',t_2';t_3', t_4') 
\nonumber \\
& \quad \times 
G_{\sigma_1}^r(t_1',t_1)
G_{\sigma_2}^a(t_2,t_2')
G_{\sigma_3}^r(t_3',t_3)
G_{\sigma_4}^a(t_4,t_4') 
\, =  \, 0 .
\label{eq:Keldysh_sum_rule_appendix_rewrite}
\end{align}
Note that the contributions of the disconnected parts, 
which correspond to the first two terms in 
the right-hand side of  Eq.\ \eqref{eq:BS_equation_4point},  
vanish 
\begin{align}
\sum_{\mu_1\mu_2 \atop \mu_3\mu_4}
\mathrm{sgn}(\mu_4\mu_2\mu_3\mu_1)\  
G_{\sigma_1}^{\mu_4\mu_1}
\,G_{\sigma_2}^{\mu_2\mu_3} 
\,=\, 0.
\end{align}
This can be verified, using  
 Eq.\ \eqref{eq:keldysh_sum_G_appendix}, 
 or directly from Eq.\ \eqref{eq:GK_def}. 

The relation obtained in Eq.\  \eqref{eq:Keldysh_sum_rule_appendix_rewrite}
 holds for arbitrary $t_1$, $t_2$, $t_3$, and  $t_4$,  
and thus  
\begin{align}
\sum_{\nu_1\nu_2 \atop \nu_3\nu_4}
\Gamma_{\sigma_1\sigma_2;\sigma_3\sigma_4}^{\nu_1\nu_2;\nu_3\nu_4}
(t_1',t_2';t_3', t_4') 
 \  =  \ 0 .
\label{eq:Keldysh_sum_rule_vertex_time}
\end{align}
The linear dependence among  the Fourier transformed functions  
$\Gamma_{\sigma_1\sigma_2;\sigma_3\sigma_4}^{\nu_1\nu_2;\nu_3\nu_4}(\omega_1,\omega_2;\omega_3, \omega_4)$ can also be expressed 
in the same form as Eq.\ \eqref{eq:Keldysh_sum_rule_vertex_time}, 
  in the frequency domain defined by  
\begin{align}
& \! 
\int \prod_{k=1}^4  dt_k\,
e^{i (\omega_4 t_4+\omega_2 t_2   -\omega_3 t_3- \omega_1 t_1 )}\,
\Gamma_{\sigma_1\sigma_2;\sigma_3\sigma_4}^{\nu_1\nu_2;\nu_3\nu_4}(t_1,t_2;t_3, t_4) 
\nonumber \\
& =\  2\pi\,  \delta(\omega_1+\omega_3 -\omega_2 -\omega_4) 
\ 
\Gamma_{\sigma_1\sigma_2;\sigma_3\sigma_4}^{\nu_1\nu_2;\nu_3\nu_4}(\omega_1,\omega_2;\omega_3, \omega_4)  .
\end{align}

\section{
Derivation of Ward-Takahashi identity for finite $eV$ and $T$
}
\label{sec:Ward_Takahashi_derivations}

In this Appendix, we provide a derivation of the Ward-Takahashi identity 
for the Keldysh three-point functions defined  in Sec.\ \ref{sec:nonlinear_WT}:  
\begin{align}
& i\frac{\partial }{\partial t}\, 
\bm{\Phi}_{d,\sigma\sigma'}^{\alpha}(t; t_1, t_2) 
+i\,\bm{\Phi}_{R,\sigma\sigma'}^{\alpha}(t; t_1, t_2) 
-i\,\bm{\Phi}_{L,\sigma\sigma'}^{\alpha}(t; t_1, t_2) 
\nonumber \\
& =  \
- \delta_{\sigma\sigma'}\,\delta(t-t_1) \,  
\bm{\rho}_3^{\alpha}\,\bm{\tau}_3
\,
\bm{G}_{\sigma}(t,t_2)
\nonumber \\
&  \quad \    + \, \delta_{\sigma\sigma'}\,\delta(t_2-t)\,  
\bm{G}_{\sigma}(t_1,t)
\,\bm{\tau}_3\,\bm{\rho}_3^{\alpha}
\,.
\label{eq:WT_Phi_in_time_II}
\end{align}
This identity, 
which is also given Eq.\ \eqref{eq:WT_Phi_in_time},  
reflects the conservation of the current flowing between the dot and leads,   
\begin{align}
 \frac{\partial }{\partial t}\,\delta n_{d,\sigma} (t) \,+
\,\delta J_{R,\sigma} \,-\,  \,\delta J_{L,\sigma} \,=\, 0 \,.
\label{eq:EOM_deltaNd}
\end{align}
It can be deduced from the equation of motion  for  
$\bm{\Phi}_{d,\sigma\sigma'}^{\alpha}$ along  
 the forward  $\alpha=-$ and backward  $\alpha=+$ branches of  
the Keldysh  time-loop contour. 
In the following two subsections,   
Eq.\ \eqref{eq:WT_Phi_in_time_II} 
is derived separately for these two branches as 
Eqs.\ \eqref{eq:WTmatrix-} and \eqref{eq:WTmatrix+}. 
Note that  $\bm{\rho}_3^{\alpha}$, defined in Eq.\ \eqref{eq:branch_projector}, 
is the projection operator into the subspace of the $\alpha$ branch, 
and it has the properties  $\bm{\rho}_3^{\alpha} \bm{\tau}_3 
= \bm{\tau}_3 \bm{\rho}_3^{\alpha} = \eta_{\alpha }\bm{\rho}_3^{\alpha}$ 
with  $\eta_- =+1$ and $\eta_+ =-1$.

\subsection{
\!\!\!\!\! 
Forward-branch  
time evolution of  $\bm{\Phi}_{d,\sigma\sigma'}^{-}(t; t_1, t_2)$
} 

We consider first  the forward branch $\alpha=-$,   
 and show that the  equation of motion 
of $\bigl\{\bm{\Phi}_{d,\sigma\sigma'}^{-}\bigr\}^{\mu\nu} 
= \Phi_{d,\sigma\sigma'}^{-;\mu\nu}$ 
can be expressed in the following form,  
\begin{align}
& 
\frac{\partial }{\partial t}\, 
\bm{\Phi}_{d,\sigma\sigma'}^{-}(t; t_1, t_2) 
+ \bm{\Phi}_{R,\sigma\sigma'}^{-}(t; t_1, t_2) 
- \bm{\Phi}_{L,\sigma\sigma'}^{-}(t; t_1, t_2) 
\nonumber \\
 =&   \ 
 \delta_{\sigma\sigma'}
\Bigl[\,
\delta(t-t_1) 
\bm{\rho}_3^-
\, i\bm{G}_{\sigma}(t,t_2)
\, - 
\delta(t_2-t)
\, i \bm{G}_{\sigma}(t_1,t) \bm{\rho}_3^- 
\,\Bigr].
\rule{0cm}{0.5cm}
\label{eq:WTmatrix-}
\end{align}

One of the four matrix elements of $\bm{\Phi}_{d,\sigma\sigma'}^{-}$ 
is the causal function, i.e., 
  $\Phi_{d,\sigma\sigma'}^{-;\mu\nu}$ for $\mu=\nu=-$.  
The time derivative of this function can be calculated,  
following the standard approach  for the causal three-point functions,
described, for instance, in Ref.\ \onlinecite{SchriefferBook}, 
\begin{align}
&
\frac{\partial }{\partial t}\, 
\Phi_{d,\sigma\sigma'}^{-;--}(t; t_1, t_2) 
+ \Phi_{R,\sigma\sigma'}^{-;--}(t; t_1, t_2) 
- \Phi_{L,\sigma\sigma'}^{-;--}(t; t_1, t_2) 
\nonumber \\
 = &   \ 
-\delta(t-t_1) \,
 \left \langle T \left[\, \delta n_{d,\sigma'} (t_1)\,,\, 
d_{\sigma}^{}(t_1) \,\right]   
d_{\sigma}^{\dagger} (t_2) 
            \, \right \rangle 
\nonumber \\
& \ \,   
- \, \delta(t_2-t)\,  
 \left \langle T \,   d_{\sigma}^{}(t_1) 
\left[\, \delta n_{d,\sigma'}(t_2)\,,\, 
d_{\sigma}^{\dagger}(t_2)\,\right]\, \right \rangle
\nonumber \\
 =&  
\ \, \delta_{\sigma\sigma'}
\Bigl[\,
\,\delta(t-t_1) \,  i G_{\sigma}^{--}(t,t_2)
- \,
\delta(t_2-t)\, i G_{\sigma}^{--}(t_1,t)
\,\Bigr].
\rule{0cm}{0.5cm}
\label{eq:WTmatrix-;--}
\end{align}
Here, we have used the current conservation law Eq.\ \eqref{eq:EOM_deltaNd} 
and the {equal-time\/} commutation relations     
$\bigl[ n_{d\sigma'}^{},\, d_{\sigma}^{} \bigr]
= - \delta_{\sigma\sigma'}\,d_{\sigma}^{}$, and 
$\bigl[ n_{d\sigma'}^{},\, d_{\sigma}^{\dagger} \bigr]
= \delta_{\sigma\sigma'}\,d_{\sigma}^{\dagger}$. 
Note that the Dirac delta functions arise from the derivatives 
of the Heaviside step functions,  $\theta(t)$'s,  that 
specify the time-ordering of $t_1$, $t_2$, and $t$.

We next consider the component with  $\mu=\nu=+$,  
for which the time ordering is much simpler, 
\begin{align}
-\Phi_{d,\sigma\sigma'}^{-;++}(t; t_1, t_2)  =&  
-\theta(t_1-t_2)
 \left \langle  
d_{\sigma}^{\dagger} (t_2) \,d_{\sigma}^{}(t_1) \,
\delta n_{d,\sigma'} (t)   \right \rangle 
\nonumber \\
& 
+ 
\theta(t_2-t_1)
 \left \langle  
d_{\sigma}^{}(t_1) \,d_{\sigma}^{\dagger} (t_2) \,
\delta n_{d,\sigma'} (t)  \right \rangle .
\end{align}
Since the  $t$-dependence  enters only through 
the charge fluctuation operator  $\delta n_{d,\sigma'} (t)$ 
in the right-hand side,
the equation of motion of this component   
is determined simply by $\partial \delta n_{d,\sigma'} (t)/\partial t$ 
that satisfies Eq.\  \eqref{eq:EOM_deltaNd}. Therefore,  
\begin{align}
&
\frac{\partial }{\partial t}
\Phi_{d,\sigma\sigma'}^{-;++}(t; t_1, t_2) 
+\Phi_{R,\sigma\sigma'}^{-;++}(t; t_1, t_2) 
-\Phi_{L,\sigma\sigma'}^{-;++}(t; t_1, t_2) 
\nonumber \\
& \!\!  =  \  0 \, .
\label{eq:WTmatrix-;++}
\end{align}

 The other component with  $\mu=+$ and $\nu=-$ 
can also be expressed, using the step function for time-ordering, as 
\begin{align}
-\Phi_{d,\sigma\sigma'}^{-;+-}(t; t_1, t_2) \,  = &  
\ \theta(t-t_2)
 \left \langle  
d_{\sigma}^{}(t_1) \,
\delta n_{d,\sigma'} (t) \,d_{\sigma}^{\dagger} (t_2)  \right \rangle 
\nonumber \\
& 
\! 
+  \theta(t_2-t)
 \left \langle  d_{\sigma}^{}(t_1) \,d_{\sigma}^{\dagger} (t_2) \,
\delta n_{d,\sigma'} (t)  \right \rangle  . 
\end{align}
Thus, the time-derivative of this function is given by 
\begin{align}
& \frac{\partial }{\partial t}\, 
\Phi_{d,\sigma\sigma'}^{-;+-}(t; t_1, t_2) 
+ \Phi_{R,\sigma\sigma'}^{-;+-}(t; t_1, t_2) 
- \Phi_{L,\sigma\sigma'}^{-;+-}(t; t_1, t_2) 
 \nonumber \\
= & \   
-\delta(t-t_2)
 \left \langle  \,
d_{\sigma}^{}(t_1) \,
\left[\,
\delta n_{d,\sigma'} (t_2)\,,\,
d_{\sigma}^{\dagger} (t_2) \,\right]\, \right \rangle 
\rule{0cm}{0.4cm}
 \nonumber \\ 
= & \   
-\delta_{\sigma\sigma'}\,
\delta(t-t_2)\,iG_{\sigma}^{+-}(t_1,t) \,.
\label{eq:WTmatrix-;+-}
\end{align}

Similarly, the time-ordering of the last one of four forward-branch components, i.e., 
 the one with  $\mu=-$ and $\nu=+$,  can be expressed in the form,
\begin{align}
- \Phi_{d,\sigma\sigma'}^{-;-+}(t; t_1, t_2)  = &  
-\theta(t-t_1)
 \left \langle  \,
d_{\sigma}^{\dagger} (t_2) \,
\delta n_{d,\sigma'} (t)\,
d_{\sigma}^{}(t_1) \,
 \, \right \rangle 
\nonumber \\
& 
- \, 
\theta(t_1-t)
 \left \langle  \,
d_{\sigma}^{\dagger} (t_2) \,
d_{\sigma}^{}(t_1) \,
\delta n_{d,\sigma'} (t)
            \, \right \rangle .
\end{align}
Taking the derivative with respect to $t$, we obtain
\begin{align}
&\frac{\partial }{\partial t}\, 
\Phi_{d,\sigma\sigma'}^{-;-+}(t; t_1, t_2) 
+ \Phi_{R,\sigma\sigma'}^{-;-+}(t; t_1, t_2) 
- \Phi_{L,\sigma\sigma'}^{-;-+}(t; t_1, t_2) 
 \nonumber \\
 = & \   
\delta(t-t_1) \left \langle  \,d_{\sigma}^{\dagger} (t_2)\, 
\left[\,\delta n_{d,\sigma'} (t)\,,\,
d_{\sigma}^{}(t_1) 
\,\right]
\,
 \right \rangle 
\nonumber \\ 
 = & \   
\delta_{\sigma\sigma'}\,
\delta(t-t_1)
\,iG_{\sigma}^{-+}(t,t_2) \,.
\label{eq:WTmatrix-;-+}
\end{align}
The four equations 
\eqref{eq:WTmatrix-;--},
\eqref{eq:WTmatrix-;++},
\eqref{eq:WTmatrix-;+-}, and 
\eqref{eq:WTmatrix-;-+} can be expressed as one matrix equation, 
as described in  Eq.\ \eqref{eq:WTmatrix-}.


\subsection{
\!\!\!\!\! 
Backward-branch time evolution of   
$\bm{\Phi}_{d,\sigma\sigma'}^{+}(t; t_1, t_2)$}

Time evolution of  the three-point function 
$\Phi_{d,\sigma\sigma'}^{\alpha;\mu\nu}$ along the backward branch $\alpha=+$  
 can  also be deduced from the equation of continuity Eq.\ \eqref{eq:EOM_deltaNd}, 
in a similar way that is described in the above to obtain Eq.\  \eqref{eq:WTmatrix-}.   
The results can be written in the form, 
\begin{align}
&
\frac{\partial }{\partial t}\, 
\bm{\Phi}_{d,\sigma\sigma'}^{+}(t; t_1, t_2) 
+\bm{\Phi}_{R,\sigma\sigma'}^{+}(t; t_1, t_2) 
-\bm{\Phi}_{L,\sigma\sigma'}^{+}(t; t_1, t_2) 
\nonumber \\
 =&   \ \delta_{\sigma\sigma'}
\Bigl[
-\delta(t-t_1) \bm{\rho}_3^+
\,i\bm{G}_{\sigma}(t,t_2)
+ \delta(t_2-t)\, i 
\bm{G}_{\sigma}(t_1,t)  \bm{\rho}_3^+ \Bigr]. 
\label{eq:WTmatrix+}
\end{align}
Here we provide the derivation of each element
of this matrix equation  in the following.

The three-point function for $\mu=\nu=+$ 
corresponds to the counterpart of the causal function 
described in Eq.\ \eqref{eq:WTmatrix-;--}. 
Therefore, the time derivative of this function can also be calculated,  
following the same standard approach,\cite{SchriefferBook}
as 
\begin{align}
&
\frac{\partial }{\partial t}\, 
\Phi_{d,\sigma\sigma'}^{+;++}(t; t_1, t_2) 
+\Phi_{R,\sigma\sigma'}^{+;++}(t; t_1, t_2) 
-\Phi_{L,\sigma\sigma'}^{+;++}(t; t_1, t_2) 
\nonumber \\ 
=  & \ \   
\delta(t-t_1) \,
 \left \langle \widetilde{T} \left[\, \delta n_{d,\sigma'} (t_1)\,,\, 
d_{\sigma}^{}(t_1) \,\right]   
d_{\sigma}^{\dagger} (t_2) 
            \, \right \rangle 
\nonumber \\
&  
+ \, \delta(t_2-t)\,  
 \left \langle \widetilde{T} \,   d_{\sigma}^{}(t_1) 
\left[\, \delta n_{d,\sigma'}(t_2)\,,\, 
d_{\sigma}^{\dagger}(t_2)\,\right]\, \right \rangle
\nonumber \\
= & \ 
\delta_{\sigma\sigma'}\Bigl[\,
-\delta(t-t_1) \,  i G_{\sigma}^{++}(t,t_2)
+  \delta(t_2-t)\, i G_{\sigma}^{++}(t_1,t) 
\,\Bigr].
\label{eq:WTmatrix+;++}
\end{align}

For the component of $\mu=\nu=-$,
the time-ordering can explicitly be written in the form, 
\begin{align}
-\Phi_{d,\sigma\sigma'}^{+;--}(t; t_1, t_2)  = & \   
\theta(t_1-t_2)
 \left \langle  
\delta n_{d,\sigma'} (t)\,
d_{\sigma}^{}(t_1) \,d_{\sigma}^{\dagger} (t_2)  \right \rangle  
\nonumber \\ 
& \! 
-\theta(t_2-t_1)
 \left \langle  
\delta n_{d,\sigma'} (t) \,
d_{\sigma}^{\dagger} (t_2) \,d_{\sigma}^{}(t_1)  \right \rangle .
\end{align}
Since the step functions do not depend on $t$  in this case, 
the equation of motion does not  have a source term,    
\begin{align}
&\frac{\partial }{\partial t}
\Phi_{d,\sigma\sigma'}^{+;--}(t; t_1, t_2) 
+\Phi_{R,\sigma\sigma'}^{+;--}(t; t_1, t_2) 
-\Phi_{L,\sigma\sigma'}^{+;--}(t; t_1, t_2) 
\nonumber \\
& \!\!  = \  0 \,.
\label{eq:WTmatrix+;--}
\end{align}

In contrast, the time-ordering of the component with $\mu=-$ and $\nu=+$ 
depends on $t$, 
\begin{align}
-\Phi_{d,\sigma\sigma'}^{+;-+}(t; t_1, t_2)  = & \ 
-\theta(t-t_2)
 \left \langle  d_{\sigma}^{\dagger} (t_2) \,
\delta n_{d,\sigma'} (t)\, d_{\sigma}^{}(t_1)  \right \rangle 
\nonumber \\
& 
- \, 
\theta(t_2-t)
 \left \langle  \delta n_{d,\sigma'} (t)\,
d_{\sigma}^{\dagger} (t_2) \, d_{\sigma}^{}(t_1)  \right \rangle .
\end{align}
Thus, the equation of motion has a source term in the right-hand side, as  
\begin{align}
& \frac{\partial }{\partial t}\, 
\Phi_{d,\sigma\sigma'}^{+;-+}(t; t_1, t_2) 
+ \Phi_{R,\sigma\sigma'}^{+;-+}(t; t_1, t_2) 
- \Phi_{L,\sigma\sigma'}^{+;-+}(t; t_1, t_2) 
 \nonumber \\
= & \   
\delta(t-t_2)
 \left \langle  \,
\left[\,
d_{\sigma}^{\dagger} (t)\, 
\,,\,
\delta n_{d,\sigma'} (t)
\,\right]
d_{\sigma}^{}(t_1) 
\,
 \right \rangle 
\rule{0cm}{0.4cm}
\nonumber \\ 
\ = & \  
\delta_{\sigma\sigma'}\,
\delta(t-t_2)
\,iG_{\sigma}^{-+}(t_1,t)
\;.
\label{eq:WTmatrix+;-+}
\end{align}

Similarly, the component with  $\mu=+$ and $\nu=-$ can be written 
in the following form,  
\begin{align}
-\Phi_{d,\sigma\sigma'}^{+;+-}(t; t_1, t_2)  = & \ \ 
\theta(t-t_1)
 \left \langle  d_{\sigma}^{}(t_1) \,
\delta n_{d,\sigma'} (t)\,d_{\sigma}^{\dagger} (t_2)  \right \rangle 
\nonumber \\
& +  
\theta(t_1-t)
 \left \langle  
\delta n_{d,\sigma'} (t) \, 
d_{\sigma}^{}(t_1) \,
d_{\sigma}^{\dagger} (t_2) \right \rangle .
\end{align}
The equation of motion of this component is given by
\begin{align}
&
\frac{\partial }{\partial t}\, 
\Phi_{d,\sigma\sigma'}^{+;+-}(t; t_1, t_2) 
+ \Phi_{R,\sigma\sigma'}^{+;+-}(t; t_1, t_2) 
- \Phi_{L,\sigma\sigma'}^{+;+-}(t; t_1, t_2) 
\nonumber \\
 = & \,   
-\delta(t-t_1)
 \left \langle  \,
\left[\,
d_{\sigma}^{}(t) 
\,,\,
\delta n_{d,\sigma'} (t)
\,\right]
d_{\sigma}^{\dagger} (t_2) 
\, \right \rangle 
\rule{0cm}{0.4cm}
\nonumber \\ 
 = & \,   
-\delta_{\sigma\sigma'}\,
\delta(t-t_1)
\,iG_{\sigma}^{+-}(t,t_2)
\,.
\label{eq:WTmatrix+;+-}
\end{align}
Therefore, the equation of motions for the four backward-branch components,
Eqs.\ \eqref{eq:WTmatrix+;++}, 
\eqref{eq:WTmatrix+;--}, 
\eqref{eq:WTmatrix+;-+}, and 
\eqref{eq:WTmatrix+;+-}, 
can also be expressed as the matrix equation 
that is shown in  Eq.\ \eqref{eq:WTmatrix+}.   


\section{Low-energy expansion of the Keldysh vertex functions}
\label{sec:full_vertex_low_energy_form}

In this appendix, we calculate the low-energy expansion of  the Keldysh vertex functions 
$\Gamma_{\sigma\sigma';\sigma'\sigma}^{\mu\mu';\nu'\nu}
(\epsilon,\epsilon';\epsilon',\epsilon)$  
up to linear-order terms with respect to $\epsilon$, $\epsilon'$, $eV$, and $T$, 
 using the  Feynman diagrammatic technique. 
Calculations are carried out for all Keldysh-branch components 
$(\mu,\mu';\nu',\nu)$ 
and internal degrees of freedoms $\sigma, \sigma'$. 
The results are summarized in TABLES \ref{tab:vertex_UU} and \ref{tab:vertex_UD}. 
 It has already been shown  
in Sec.\ \ref{sec:current_conservation_Keldysh_selfenergy_vertex_in_FL_regime}  
 that these results of $\Gamma_{\sigma\sigma';\sigma'\sigma}^{\mu\mu';\nu'\nu}
(\epsilon,\epsilon';\epsilon',\epsilon)$ satisfy the current conservation law 
consistently with the low-energy asymptotic form of 
the  Keldysh self-energies  $\Sigma_{\sigma}^{\nu\mu} (\epsilon)$ in the Fermi-liquid regime,  which are also summarized in TABLE  \ref{tab:self-energy}. 
These  results also explicitly show that all 
the vertex components other than the causal component  
 $\Gamma_{\sigma\sigma';\sigma'\sigma}^{--;--}
(\epsilon,\epsilon';\epsilon',\epsilon)$ 
and its counterpart 
 $\Gamma_{\sigma\sigma';\sigma'\sigma}^{++;++}
(\epsilon,\epsilon';\epsilon',\epsilon)$ 
are pure imaginary in the Fermi-liquid regime. 
Furthermore, the real part of the causal component and the counterpart 
can be deduced up to order $eV$
from the nonequilibrium Ward identity,
as shown in Sec.\  \ref{sec:real_vertex_eV_liniear} and 
TABLE \ref{tab:vertex_UD}.

 We start with a few remarks on the general properties of the 
Keldysh vertex function.
First of all,  the sixteen branch components are linearly dependent,   
 as shown in Eq.\ \eqref{eq:linear_dependency_Keldysh_text} 
and explained more precisely in 
Appendix \ref{sec:linear_dependency_Keldysh}.
\footnote{See also the very recent formulation of the spectral function 
for multipoint correlation functions by  Kugler, Lee, and von Delft 
in Ref.\ \onlinecite{KuglerLeeVonDelft2021}}
The vertex functions also have the symmetrical property 
$\Gamma_{\sigma\sigma';\sigma'\sigma}^{\mu\mu';\nu'\nu}
(\epsilon,\epsilon';\epsilon',\epsilon) 
= 
\Gamma_{\sigma'\sigma;\sigma\sigma'}^{\nu'\nu;\mu\mu'}
(\epsilon',\epsilon;\epsilon,\epsilon')$ under the interchange 
of the variables for incoming and outgoing particles. 
Furthermore,  
the causal vertex function  and its counterpart  
for $\sigma=\sigma'$ identically vanish 
at zero frequencies  $\epsilon=\epsilon'=0$, 
as a result of the Pauli exclusion principle, 
\begin{align} 
\Gamma_{\sigma,\sigma;\sigma,\sigma}^{--;--} (0,0;0,0)=0\,,
\quad  \ 
\Gamma_{\sigma,\sigma;\sigma,\sigma}^{++;++} (0,0;0,0)=0 \,.
\end{align} 
In contrast,  for different levels  $\sigma \neq \sigma'$,  
 the zero-frequency value of  the causal vertex function is finite 
 and  is determined by the off-diagonal susceptibility 
$\chi_{\sigma \sigma'}^{}$, specifically at $eV=T=0$,  
\begin{align} 
 \Gamma_{\sigma\sigma';\sigma'\sigma}^{--;--} (0,0;0,0) 
\,\rho_{d \sigma}^{}  \rho_{d \sigma'}^{} 
\, = &  \   
- \Gamma_{\sigma\sigma';\sigma'\sigma}^{++;++} (0,0;0,0) \, 
\rho_{d \sigma}^{} \rho_{d \sigma'}^{}
\nonumber \\
= &  \   - \chi_{\sigma \sigma'}^{}\,.
\end{align}

In the following, we calculate the imaginary part of  Keldysh vertex 
functions $\Gamma_{\sigma\sigma';\sigma'\sigma}^{\mu\mu';\nu'\nu}
(\epsilon,\epsilon';\epsilon',\epsilon)$  
 up to terms of order   $\epsilon$, $\epsilon'$, $eV$, and $T$.
To this end, we demonstrate how the imaginary part arises first    
in the second order perturbation theory in $U$, 
and then take into account the contributions 
of multiple scattering to all orders in $U$.

\subsection{Perturbation expansion of  the Keldysh vertex function}

\begin{figure}[t]
 \leavevmode
\begin{minipage}{1\linewidth}
 \includegraphics[width=0.28\linewidth]{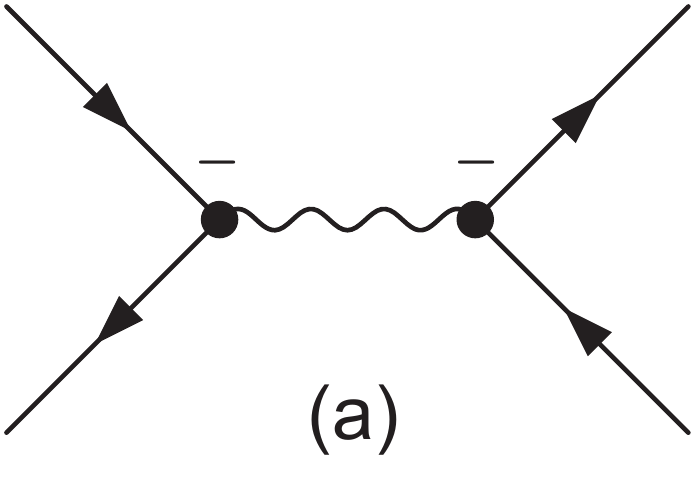}
 \rule{0.02\linewidth}{0cm}
  \includegraphics[width=0.28\linewidth]{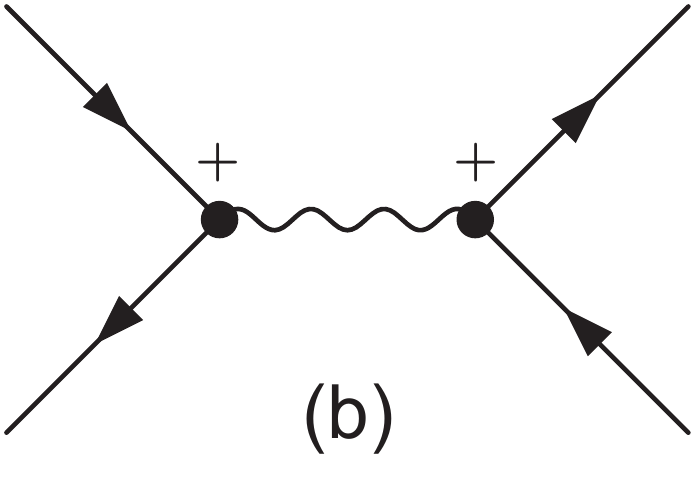}
\rule{0.01\linewidth}{0cm}
\includegraphics[width=0.35\linewidth]{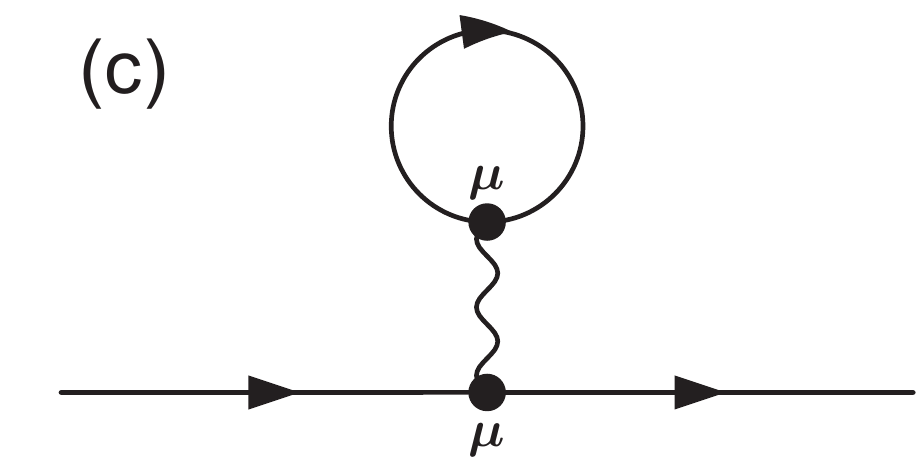}
\end{minipage}
 \caption{
Feynman diagrams for order $U$ contributions, with $\mu = +,\, -\,$  the Keldysh index. 
}
\label{fig:vertex_u1}
\end{figure}

\begin{figure}[t]
 \leavevmode
\begin{minipage}{\linewidth}
\includegraphics[width=0.37\linewidth]{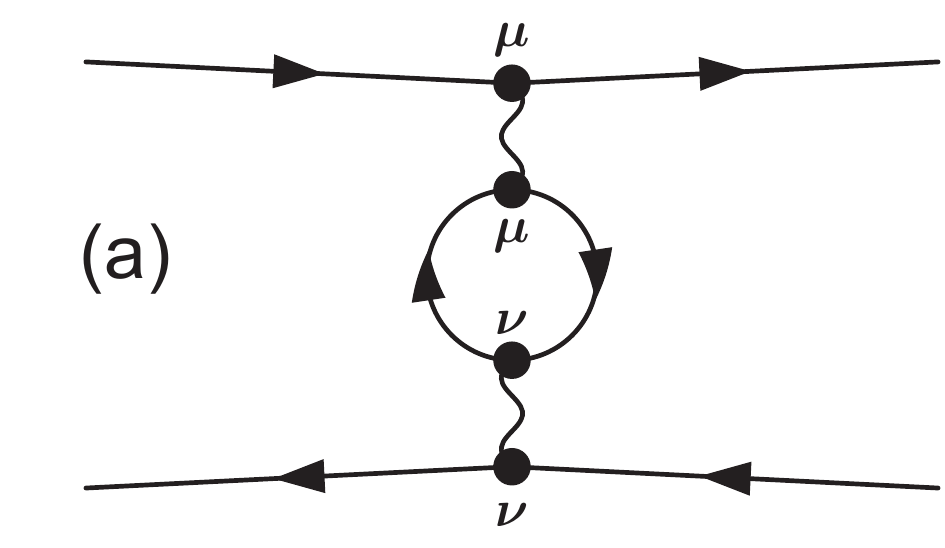}
\rule{0.1\linewidth}{0cm}
\includegraphics[width=0.37\linewidth]{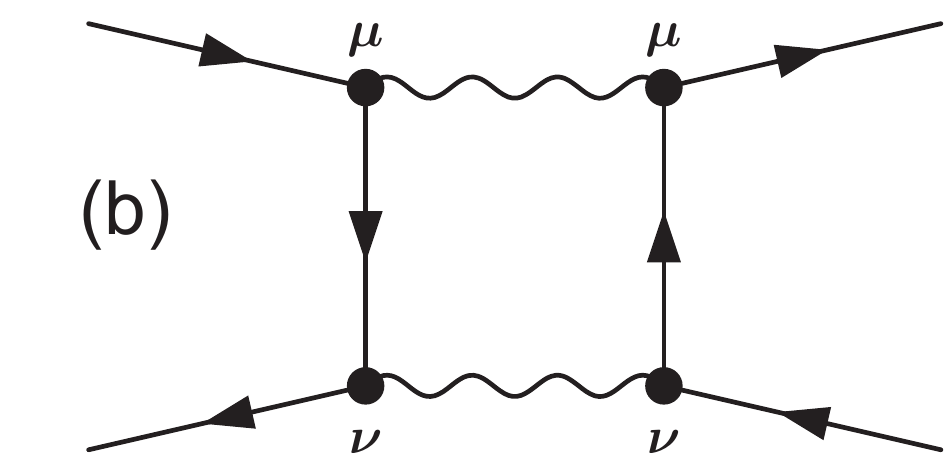}
\\
\includegraphics[width=0.37\linewidth]{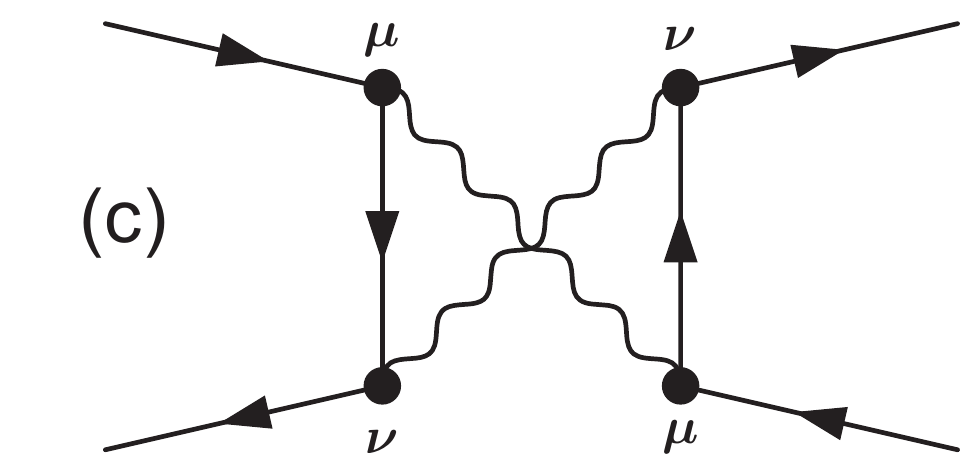}
\rule{0.1\linewidth}{0cm}
\includegraphics[width=0.37\linewidth]{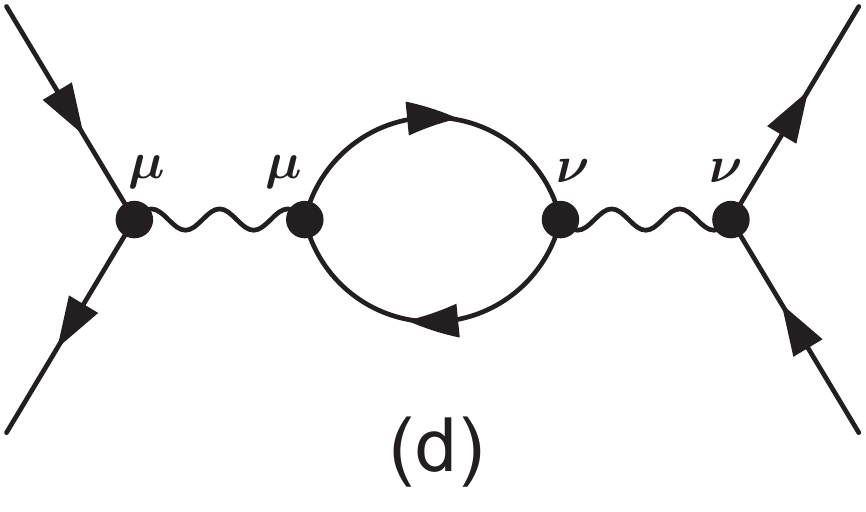}
 \rule{0cm}{2.5cm}
\end{minipage}
 \caption{
Order $U^2$ vertex function 
 $\Gamma_{\sigma\sigma';\sigma'\sigma}^{\mu_1\mu_2;\mu_3,\mu_4}$. 
The diagram (a) is for $\sigma = \sigma'$, 
and (b) and (c) are for $\sigma \neq \sigma'$, 
where $\sigma$ and $\sigma'$ are the level indexes for 
the propagators on the left and right, respectively.  
The diagram (d) also contributes to $\sigma \neq \sigma'$ components 
for multi-level case  $N>2$ as well as to $\sigma = \sigma'$ components.  
}
 \label{fig:vertex_u2}
\end{figure}

Here we briefly summarize  the  Feynman rules for the Keldysh vertex corrections. 
In the perturbation theory with respect to  the Coulomb interactions, 
the sign of an order $U^n$ Feynman diagram of  a vertex component 
is given by 
 \begin{align}
  (-i)^{n}(i)^{2n}  \left(-1\right)^{m} (-1)^{F'} \left\{i\right\} 
\, U^n  =  \,  (-1)^m (-1)^{F'} i^{n+1} U^n  .
\label{eq:Feynman_rule_sign}
\end{align}
Here, $(-i)^n$ and  $(i)^{2n}$ in  the left-hand side 
are associated with $U^n$ which emerges through the time-evolution operator 
and the $2n$ internal Green's functions, respectively.\cite{AGD} 
Another sign factor $(-1)^m$ appears in the Keldysh formalism,  
where  $m$ is the number of $U$'s arising from 
 the time-evolution along the backward branch ($m\leq n$)  
[see, for instance, Figs.\ \ref{fig:vertex_u1} and \ref{fig:vertex_u2}]. 
The factor $(-1)^{F'}$ is determined by the number of fermion loops $F'$,   
which include an additional closed loop 
that may be created when the two external Green's functions with $\sigma'$ 
on the right side of the vertex diagrams are connected with each other  
to construct a three-point correlation function,  
as shown in Fig.\ \ref{fig:3point_diagrams}. 
The remaining  factor  $\{i\}$ in the left-side of Eq.\ \eqref{eq:Feynman_rule_sign} 
corresponds to the coefficient that is assigned in the definitions of 
$\bm{\Phi}_{\gamma,\sigma\sigma'}^{^\alpha}$ and  
$\Gamma_{\sigma\sigma';\sigma'\sigma}^{\mu\mu';\nu'\nu}$,  
described in Eq.\  \eqref{eq:Bethe-Salpeter}.


Figures \ref{fig:vertex_u1} (a) and (b) 
describe the Feynman diagrams for the first-order vertex corrections 
with different levels  $\sigma \neq \sigma'$. 
The sign for each of these two diagrams 
is determined, substituting $n=1$,  $F'=1$, and  (a) $m=0$ and (b)  $m=1$, 
into the right-hand side of Eq.\ \eqref{eq:Feynman_rule_sign}, 
\begin{subequations}
\begin{align}
\Gamma_{\sigma\sigma';\sigma'\sigma}^{(1) --;--} 
\ =&  \ (-1)^0 (-1)^{1}\, i^{1+1}\, U  \ = \   U\;,
\\ 
\Gamma_{\sigma\sigma';\sigma'\sigma}^{(1) ++;++} 
\ =&  \   (-1)^1\,  (-1)^{1}\, i^{1+1}\, U  \ = \  - U\;.
\end{align}
\end{subequations}
Similarly, the order $U$ self-energy, 
described s diagrammatically in  Fig.\  \ref{fig:vertex_u1} (c),   
is given by 
\begin{subequations}
\begin{align}
\!\! \! 
\Sigma^{(1)--}_{\sigma}  =& \   
U \!  \int \! \frac{d\epsilon'}{2 \pi i} \,
e^{i\epsilon'  0^+} \!
\sum_{\sigma'(\neq \sigma)}
G_{\sigma'}^{--} (\epsilon') 
\nonumber \\
 =  &
\int \! \frac{d\epsilon'}{2 \pi i} \,
e^{i\epsilon' 0^+}\!\!
\sum_{\sigma' (\neq \sigma)}
\Gamma_{\sigma\sigma';\sigma'\sigma}^{(1) --;--} 
\,  G_{\sigma'}^{--} (\epsilon') , 
\\
\!\! \!  
\Sigma^{(1)++}_{\sigma}  = & \,  
-U \! \int \! \frac{d\epsilon'}{2 \pi i} \,
e^{-i\epsilon'  0^+}\!
\sum_{\sigma'(\neq \sigma)}
G_{\sigma''}^{++} (\epsilon')
\nonumber \\
= &     
\int \! \frac{d\epsilon'}{2 \pi i} \,
e^{-i\epsilon' 0^+}\! \!
\sum_{\sigma' (\neq \sigma)}
\Gamma_{\sigma\sigma';\sigma'\sigma}^{(1) ++;++} 
\,  G_{\sigma'}^{++} (\epsilon') .
\end{align}
\end{subequations}
Note that these self-energies can also be regarded as the lowest-order terms 
of the skeleton-diagram expansion,  
in which the exact Green's functions $G$ 
instead of the noninteracting ones $G_0$ are assigned to 
the internal lines of the Feynman diagrams.

\subsection{
 $\Gamma_{\sigma\sigma';\sigma'\sigma}^{\nu_1\nu_2;\nu_3\nu_4} 
(\epsilon,\epsilon'; \epsilon',\epsilon)$ 
 for $\sigma=\sigma'$ 
up to linear-order terms in $\epsilon$, $\epsilon'$,  $eV$, and $T$
}

We next consider low-energy asymptotic form of the vertex corrections   
for electrons with  $\sigma = \sigma'$. 
The imaginary part 
 arises first from the order $U^2$ scattering processes 
for $\Gamma_{\sigma\sigma;\sigma\sigma}^{\mu\mu';\nu'\nu}
(\epsilon,\epsilon';\epsilon',\epsilon)$,  
shown in Figs.\ \ref{fig:vertex_u2} (a) and (d).
For each of these two diagrams,
there are six different branch configurations, 
$\Gamma_{\sigma\sigma;\sigma\sigma}^{ --;--}$, 
$\Gamma_{\sigma\sigma;\sigma\sigma}^{ ++;++}$, 
$\Gamma_{\sigma\sigma;\sigma\sigma}^{ --;++}$, 
$\Gamma_{\sigma\sigma;\sigma\sigma}^{ ++;--}$, 
$\Gamma_{\sigma\sigma;\sigma\sigma}^{ -+;+-}$, 
$\Gamma_{\sigma\sigma;\sigma\sigma}^{ +-;-+}$.
In the following, we demonstrate how the imaginary part 
emerges in the second-order skeleton diagrams.
Then, taking into account multiple scatterings to all orders in $U$, 
we calculate the low-energy asymptotic form 
of the full vertex functions 
 up to terms of order $\epsilon$, $\epsilon'$, $eV$, and $T$.

\subsubsection{ 
Low-energy expansion of 
$\Gamma_{\sigma\sigma;\sigma\sigma}^{ --;--}$ 
 \& 
$\Gamma_{\sigma\sigma;\sigma\sigma}^{++;++} $
}

In the second-order perturbation theory in $U$, 
the causal component $(--;--)$ of the vertex function 
with  $\sigma=\sigma'$ is described by    
 the Feynman diagrams shown in Figs.\  \ref{fig:vertex_u2} (a) and (d) 
which have  $F_a'=1$ and $F_d'=2$ closed loops, respectively.
Since $m=0$ and $n=2$ for these processes 
the corresponding vertex functions are given by  
 \begin{subequations}
\begin{align}
&
\!\!\!\!\!\!\!\! 
\Gamma_{\sigma\sigma;\sigma\sigma}^{(2a) --;--} 
(\epsilon,\epsilon'; \epsilon',\epsilon) 
\,= \, 
(-1)^0 i^{2+1} (-1)^{F_a'}
\nonumber \\ 
&  \times \! 
\sum_{\sigma''(\neq \sigma)}
\! U^2 \!
\int \!\frac{d \varepsilon_1}{2\pi} \, 
 G_{\sigma''}^{--}(\varepsilon_1+\epsilon-\epsilon')
\, G_{\sigma''}^{--}(\varepsilon_1) , 
\label{eq:vertex2_----_UpUp_a} 
\\
\nonumber \\
&
\!\!\!\!\!\!\!\! 
\Gamma_{\sigma\sigma;\sigma\sigma}^{(2d) --;--} 
(\epsilon,\epsilon'; \epsilon',\epsilon) 
\, = \,  
(-1)^0 i^{2+1} (-1)^{F_d'}
\nonumber \\
& \times  
\sum_{\sigma''(\neq \sigma)}
\! U^2 \!
\int \!\frac{d \varepsilon_1}{2\pi} \, 
G_{\sigma''}^{--}(\varepsilon_1)\,G_{\sigma''}^{--}(\varepsilon_1) . 
\label{eq:vertex2_----_UpUp_d} 
\end{align}
\label{eq:vertex2_----_UpUp_ad} 
\!\!\!\!\!
\end{subequations}
The sum $\Gamma_{\sigma\sigma;\sigma\sigma}^{(2) --;--}
 \equiv \Gamma_{\sigma\sigma;\sigma\sigma}^{(2a) --;--} 
+\Gamma_{\sigma\sigma;\sigma\sigma}^{(2d) --;--}$ 
 can be written in terms of the particle-hole propagator $X_{\sigma\sigma'}^{--}$, 
defined in Eq.\ \eqref{eq:ph_pp_propagator_def},  
\begin{align}
&
\Gamma_{\sigma\sigma;\sigma\sigma}^{(2) --;--}(\epsilon,\epsilon'; \epsilon',\epsilon) \nonumber \\
& =  
 \sum_{\sigma''(\neq \sigma)}
U^2 
\Bigr[\, 
X_{\sigma''\sigma''}^{--}(\epsilon-\epsilon')
- X_{\sigma''\sigma''}^{--}(0) 
\,\Bigl] .
\label{eq:vertex2_----_UpUp} 
 \end{align}
The corresponding  $(++;++)$ component 
of Figs.\  \ref{fig:vertex_u2} (a) and (d),  
can be obtained, replacing $G_{\sigma''}^{--}$ 
in Eq.\ \eqref{eq:vertex2_----_UpUp_ad} 
by  $G_{\sigma''}^{++}$ and taking $m=2$, as 
\begin{align}
&  \!\! 
\Gamma_{\sigma\sigma;\sigma\sigma}^{(2) ++;++} 
(\epsilon,\epsilon'; \epsilon',\epsilon) 
\nonumber \\
& =   
 \sum_{\sigma''(\neq \sigma)}
U^2 
\Bigl[\, 
X_{\sigma''\sigma''}^{++}(\epsilon-\epsilon') 
- X_{\sigma''\sigma''}^{++}(0) 
\,\Bigr] . 
\end{align}
Therefore, the low-energy behavior of these two components 
are determined by the particle-hole pair propagator 
$X_{\sigma''\sigma''}^{\mu\nu}$.
Furthermore, these results explicitly show that 
$\Gamma_{\sigma\sigma;\sigma\sigma}^{(2) ++;++}  = 
-\left\{\Gamma_{\sigma\sigma;\sigma\sigma}^{(2) --;--} 
\right\}^*$.

To obtain the full vertex function 
 $\Gamma_{\sigma\sigma;\sigma\sigma}^{--;--} 
 (\epsilon,\epsilon'; \epsilon',\epsilon)$,  
all contributions of multiple quasiparticle collision processes are needed 
to be taken into account. 
For the imaginary part,  it can be carried out by replacing bare $U$ 
by the full vertex  $\Gamma_{\sigma\sigma'';\sigma''\sigma}^{--;--}(0,0; 0,0)$  
for $\sigma \neq \sigma''$, defined at  $eV=T=0$.
\cite{AGD,YamadaYosida4,Yoshimori,EliashbergJETP15} 
This is because the damping of a quasiparticle in the low-energy Fermi-liquid regime 
is determined by the scattering processes in which  
 a single particle-hole or particle-particle pair is excited in the intermediate states,
and the scattering matrix element is given by the zero-frequency value of  
the full vertex function that includes all possible higher-order corrections. 
Therefore, the imaginary part of 
 $\Gamma_{\sigma\sigma;\sigma\sigma}^{--;--} 
 (\epsilon,\epsilon'; \epsilon',\epsilon)$  is determined 
by the single particle-hole pair excitation 
shown in Figs. \ref{fig:vertex_singular_general} (a) and (d)  
up to linear-order terms with respect to $\epsilon-\epsilon'$,  $eV$, and $T$, 
\begin{align}
& \!\! \! \!
\mathrm{Im}\, 
\Gamma_{\sigma\sigma;\sigma\sigma}^{--;--} 
(\epsilon,\epsilon'; \epsilon',\epsilon) \, 
\nonumber \\
& \!\! \! \! 
=  \, 
\mathrm{Im} 
\int \!\frac{d \varepsilon_1}{2\pi i} 
\sum_{\sigma''(\neq \sigma)}\! 
\Gamma_{\sigma\sigma'';\sigma''\sigma}^{--;--}(0,0; 0,0) \  
\Gamma_{\sigma''\sigma;\sigma\sigma''}^{--;--}(0,0; 0,0)  
\nonumber \\
& \times 
\biggl[\, 
-G_{\sigma''}^{--}(\varepsilon_1+\epsilon-\epsilon')
\, G_{\sigma''}^{--}(\varepsilon_1)\,
\,+\,
G_{\sigma''}^{--}(\varepsilon_1)\,G_{\sigma''}^{--}(\varepsilon_1)
\,\biggr] 
\nonumber \\ 
& + \  \cdots
 \nonumber \\
=&
\sum_{\sigma''(\neq \sigma)}
\left|\Gamma_{\sigma\sigma'';\sigma''\sigma}^{--;--}(0,0; 0,0) \right|^2
\nonumber \\
&  \qquad \times  
\mathrm{Im}\, 
\biggr[\, 
X_{\sigma''\sigma''}^{--}(\epsilon-\epsilon')
- X_{\sigma''\sigma''}^{--}(0) 
\,\biggl]  \ + \ \cdots
 \nonumber \\
=& \  
\frac{\pi}{\{\rho_{d\sigma}^{}\}^{2}} \sum_{\sigma''(\neq \sigma)}
\chi_{\sigma\sigma''}^{2}
\Bigr[\, 
\mathcal{W}_\mathrm{K}^\mathrm{ph}(\epsilon-\epsilon') 
\,-\,\mathcal{W}_\mathrm{K}^\mathrm{ph}(0) 
\,\Bigl]  \ + \ \cdots \,.
\end{align}
To obtain the last line we have used properties of the 
particle-hole propagator $X_{\sigma\sigma'}^{\mu\nu}$ 
and the collision integral 
$\mathcal{W}_\mathrm{K}^\mathrm{ph}(\omega) 
 \equiv \mathcal{W}_{}^\mathrm{ph}(\omega) + \mathcal{W}_{}^\mathrm{ph}(-\omega)$,  described in Sec.\ \ref{subsec:PH_PP_propagators}
 and TABLE \ref{tab:ph-pp_propagators}.
Note that the susceptibility  for $\sigma'' \neq \sigma$ is given by 
$\chi_{\sigma\sigma''}^{}= -  
\Gamma_{\sigma\sigma'';\sigma''\sigma}^{--;--}(0,0; 0,0) \,
\rho_{d\sigma^{}}
\rho_{d\sigma''^{}}
$.

The real part of the causal vertex function 
for the same level $\sigma=\sigma'$ vanishes 
at low-energies up to linear order terms in $\epsilon$ and $\epsilon'$ 
due to the fermionic  antisymmetrical properties:\cite{ao2017_1_PRL,ao2017_2_PRB}   
 $ \mathrm{Re}\, 
 \Gamma_{\sigma\sigma;\sigma\sigma}^{--;--} 
 (\epsilon,\epsilon'; \epsilon',\epsilon) \,=\,0  + O(\epsilon^2,\epsilon'^2)$.
Therefore, the causal component for $\sigma=\sigma'$ 
 is pure imaginary up to linear order terms with 
respect to $\epsilon-\epsilon'$,  $eV$, and $T$,
\begin{align}
& \!\!\!\! 
\left\{\rho_{d\sigma}^{}\right\}^2\,
\Gamma_{\sigma\sigma;\sigma\sigma}^{--;--} 
(\epsilon,\epsilon'; \epsilon',\epsilon) \ 
\nonumber \\ 
&\!\! 
=   \,
 i \pi \!\! \sum_{\sigma''(\neq \sigma)}
 \chi_{\sigma\sigma''}^{2}
\Bigr[\, 
\mathcal{W}_\mathrm{K}^\mathrm{ph}(\epsilon-\epsilon') 
\,-\,\mathcal{W}_\mathrm{K}^\mathrm{ph}(0) 
\,\Bigl] \, +\cdots.
\end{align}
Correspondingly, the $(++;++)$ component is given by 
$
\Gamma_{\sigma\sigma;\sigma\sigma}^{++;++} 
(\epsilon,\epsilon'; \epsilon',\epsilon) = 
-\left\{
\Gamma_{\sigma\sigma;\sigma\sigma}^{--;--} 
(\epsilon,\epsilon'; \epsilon',\epsilon) \right\}^*$.

\subsubsection{ 
Low-energy expansion of 
$\Gamma_{\sigma\sigma;\sigma\sigma}^{ --;++}$ 
 \& 
$\Gamma_{\sigma\sigma;\sigma\sigma}^{++;--} $
}

We next consider the  $(--;++)$ component, 
for which  order $U^2$ contributions arise from 
the  diagram Fig.\  \ref{fig:vertex_u2} (a).   
In this case,   $\mu=-$,  $\nu=+$, 
and thus  $m=1$,  and  
it has one fermion loop  $F_2'=1$, 
\begin{subequations}
\begin{align}
& 
\!\! \!\! \!\!\!\! 
\Gamma_{\sigma\sigma;\sigma\sigma}^{(2) --;++} 
(\epsilon,\epsilon'; \epsilon',\epsilon) 
\nonumber \\
&=  \  
(-1)^1 i^{2+1} (-1)^{F_2'} 
\nonumber \\
& \qquad 
\times
\sum_{\sigma''(\neq \sigma)} 
\!U^2\!  
\int \!\frac{d\varepsilon_1}{2\pi} \, 
\,G_{\sigma''}^{+-}(\varepsilon_1+\epsilon-\epsilon')
\, G_{\sigma''}^{-+}(\varepsilon_1)
\nonumber \\ 
 &=   \ 
- \sum_{\sigma''(\neq \sigma)}
U^2 X_{\sigma''\sigma''}^{+-}(\epsilon-\epsilon') 
\,.
\end{align}
Similarly, the diagram Fig.\ \ref{fig:vertex_u2} (a) for $\mu=+$ and $\nu=-$ 
describes the  $(++;--)$ component of the order $U^2$ vertex function,  
which takes the form, 
\begin{align}
& 
\!\! \!\! \!\!\!\! 
\Gamma_{\sigma\sigma;\sigma\sigma}^{(2) ++;--} 
(\epsilon,\epsilon'; \epsilon',\epsilon) 
\nonumber \\ 
&=  \  
(-1)^1 i^{2+1} (-1)^{F_2'} 
\nonumber \\
& \qquad 
\times 
\sum_{\sigma''(\neq \sigma)}
U^2 \int \!\frac{d\varepsilon_1}{2\pi} 
\,G_{\sigma''}^{-+}(\varepsilon_1+\epsilon-\epsilon')
\, G_{\sigma''}^{+-}(\varepsilon_1)
\nonumber \\ 
& = \ 
- \sum_{\sigma''(\neq \sigma)}
U^2 X_{\sigma''\sigma''}^{-+}(\epsilon-\epsilon') \,.
\end{align}
\end{subequations}
Note that these two vertex components are pure imaginary.  
Therefore,  the fully renormalized vertex functions 
corresponding to these ones are determined 
by a single particle-hole pair excitation described in the diagram 
 Fig.\  \ref{fig:vertex_singular_general} (a)   
 up to linear order terms 
with respect to $\epsilon-\epsilon'$,  $eV$ and $T$,   
\begin{subequations}
\begin{align}
&\!\!\! 
\Gamma_{\sigma\sigma;\sigma\sigma}^{--;++} 
(\epsilon,\epsilon'; \epsilon',\epsilon) \, 
\nonumber \\
&\!\!
=   
-\int \!\frac{d \varepsilon_1}{2\pi i} 
\sum_{\sigma''(\neq \sigma)}
\Gamma_{\sigma\sigma'';\sigma''\sigma}^{--;--}(0,0; 0,0) 
\,\Gamma_{\sigma''\sigma;\sigma\sigma''}^{++;++}(0,0; 0,0) 
\nonumber \\
& \qquad \qquad \qquad \qquad
\times 
\,G_{\sigma''}^{+-}(\varepsilon_1+\epsilon-\epsilon')
\,  G_{\sigma''}^{-+}(\varepsilon_1) \ + \cdots 
\nonumber \\
& \!\! = 
- \sum_{\sigma''(\neq \sigma)}
 \left|\Gamma_{\sigma\sigma'';\sigma''\sigma}^{--;--}(0,0; 0,0) \right|^2
 X_{\sigma''\sigma''}^{+-}(\epsilon-\epsilon') 
\,+\, \cdots,
\label{eq:vertex_UU_--++}
\end{align}
\begin{align}
%
& \!\! 
\Gamma_{\sigma\sigma;\sigma\sigma}^{++;--} 
(\epsilon,\epsilon'; \epsilon',\epsilon) \, 
\nonumber \\ 
& \!\! =   
- \int \!\frac{d \varepsilon_1}{2\pi i} 
\sum_{\sigma''(\neq \sigma)}
\Gamma_{\sigma\sigma'';\sigma''\sigma}^{++;++}(0,0; 0,0) 
\,\Gamma_{\sigma''\sigma;\sigma\sigma''}^{--;--}(0,0; 0,0) 
\nonumber \\
&  \qquad \qquad \qquad \qquad 
\times
\, G_{\sigma''}^{-+}(\varepsilon_1+\epsilon-\epsilon')
\,  G_{\sigma''}^{+-}(\varepsilon_1) \ + \cdots 
\nonumber \\
& \!\! =   
- \sum_{\sigma''(\neq \sigma)}
 \left|\Gamma_{\sigma\sigma'';\sigma''\sigma}^{--;--}(0,0; 0,0) \right|^2
 X_{\sigma''\sigma''}^{-+}(\epsilon-\epsilon')  \,+\, \cdots .
\label{eq:vertex_UU_++--}
\end{align}
\end{subequations}
These two full vertex components can be rearranged into the symmetric 
and antisymmetric parts,
\begin{align}
& \left\{\rho_{d\sigma}^{}\right\}^2\,
\Bigl[\,
\Gamma_{\sigma\sigma;\sigma\sigma}^{ --;++} 
(\epsilon,\epsilon'; \epsilon',\epsilon) + 
\Gamma_{\sigma\sigma;\sigma\sigma}^{ ++;--} 
(\epsilon,\epsilon'; \epsilon',\epsilon) 
\, \Bigr] 
\nonumber \\
&=  \, 
 -i\, 2\pi 
\sum_{\sigma''(\neq \sigma)}
\chi_{\sigma\sigma''}^{2}
\ \mathcal{W}_\mathrm{K}^\mathrm{ph}(\epsilon-\epsilon') 
\ + \ \cdots , 
\end{align}
\begin{align}
&\left\{\rho_{d\sigma}^{}\right\}^2\,
\Bigl[\,
\Gamma_{\sigma\sigma;\sigma\sigma}^{ --;++} 
(\epsilon,\epsilon'; \epsilon',\epsilon) -  
\Gamma_{\sigma\sigma;\sigma\sigma}^{ ++;--} 
(\epsilon,\epsilon'; \epsilon',\epsilon) 
\, \Bigr] 
\nonumber \\
 &=  \, 
 -i\, 2\pi 
\sum_{\sigma''(\neq \sigma)}
\chi_{\sigma\sigma''}^{2}
\, \big[\,\epsilon-\epsilon' \,\bigr]
\, + \, \cdots\,. 
\end{align}
Note that 
$\mathcal{W}_{}^\mathrm{ph}(\omega) - \mathcal{W}_{}^\mathrm{ph}(-\omega) = \omega$ as mentioned in Sec.\ \ref{subsec:PH_PP_propagators}. 
The explicit expression of $\mathcal{W}_\mathrm{K}^\mathrm{ph}(\omega)$ 
is shown in TABLE \ref{tab:ph-pp_propagators}.

\subsubsection{ 
Low-energy expansion of 
$\Gamma_{\sigma\sigma;\sigma\sigma}^{ -+;+-}$ 
 \& 
$\Gamma_{\sigma\sigma;\sigma\sigma}^{+-;-+} $
}

For $\sigma=\sigma'$ components, 
there are two other  types of order $U^2$ vertex corrections,      
 $\Gamma_{\sigma\sigma;\sigma\sigma}^{ -+;+-}$ 
 and 
 $\Gamma_{\sigma\sigma;\sigma\sigma}^{+-;-+}$,
which emerge from the scattering process illustrated 
in Fig.\  \ref{fig:vertex_u2} (d). 
In this case, the number of loops is $F_d'=2$, 
and these vertex components can be expressed in the following 
form,
\begin{subequations}
\begin{align}
&
\!\! \!\! \!\! 
\Gamma_{\sigma\sigma;\sigma\sigma}^{(2) -+;+-} 
(\epsilon,\epsilon'; \epsilon',\epsilon) \ 
\nonumber \\
=& \ 
(-1)^1 i^{2+1} (-1)^{F_d'}
\sum_{\sigma''(\neq \sigma)}
\! U^2 \! \int \!\frac{d \varepsilon_1}{2\pi} \, 
G_{\sigma''}^{+-}(\varepsilon_1)\,G_{\sigma''}^{-+}(\varepsilon_1)
\nonumber \\
 =&  \  \sum_{\sigma''(\neq \sigma)}
U^2 X_{\sigma''\sigma''}^{+-}(0)  \,,
\end{align}
\begin{align}
& 
\!\! \!\! \!\!
\Gamma_{\sigma\sigma;\sigma\sigma}^{(2) +-;-+} 
(\epsilon,\epsilon'; \epsilon',\epsilon) \ 
\nonumber \\
= & \ 
(-1)^1 i^{2+1}(-1)^{F_d'}
\sum_{\sigma''(\neq \sigma)}
\! U^2 \! \int \!\frac{d \varepsilon_1}{2\pi} 
\, G_{\sigma''}^{-+}(\varepsilon_1)
\, G_{\sigma''}^{+-}(\varepsilon_1)
\nonumber \\
=& \ 
 \sum_{\sigma''(\neq \sigma)}
U^2 X_{\sigma''\sigma''}^{-+}(0)  \,.
\end{align}
\end{subequations}
These components are pure imaginary, 
and the corresponding full vertex functions are determined by 
the multiple scattering processes described in Fig. \ref{fig:vertex_singular_general} (d), 
up to linear order terms in  $eV$, $T$, and frequencies $\epsilon$ and $\epsilon'$,   
\begin{subequations}
\begin{align}
&\Gamma_{\sigma\sigma;\sigma\sigma}^{-+;+-} 
(\epsilon,\epsilon'; \epsilon',\epsilon) \, 
\nonumber \\
& =   
\int \!\frac{d \varepsilon_1}{2\pi i} \, 
\sum_{\sigma''(\neq \sigma)}
\Gamma_{\sigma\sigma'';\sigma''\sigma}^{--;--}(0,0; 0,0) \, 
\Gamma_{\sigma''\sigma;\sigma\sigma''}^{++;++}(0,0; 0,0)  
\nonumber \\
& \qquad \qquad \qquad \quad
\times 
\,G_{\sigma''}^{+-}(\varepsilon_1)\,
 G_{\sigma''}^{-+}(\varepsilon_1) \ + \ \cdots  
 \nonumber \\
&=   
\sum_{\sigma''(\neq \sigma)} 
 \left|\Gamma_{\sigma\sigma'';\sigma''\sigma}^{--;--}(0,0; 0,0) \right|^2
 X_{\sigma''\sigma''}^{+-}(0)  \ + \ \cdots, 
\end{align}
\begin{align}
&\Gamma_{\sigma\sigma;\sigma\sigma}^{+-;-+} 
(\epsilon,\epsilon'; \epsilon',\epsilon) \, 
\nonumber \\
&=   
\int \!\frac{d \varepsilon_1}{2\pi i} \, 
\sum_{\sigma''(\neq \sigma)}
\Gamma_{\sigma\sigma'';\sigma''\sigma}^{++;++}(0,0; 0,0) \, 
\Gamma_{\sigma''\sigma;\sigma\sigma''}^{--;--}(0,0; 0,0)  
\nonumber \\
&  \qquad \qquad \qquad \quad 
\times
\,G_{\sigma''}^{-+}(\varepsilon_1)\,
 G_{\sigma''}^{+-}(\varepsilon_1)
\ + \ \cdots 
 \nonumber \\
&=  
\sum_{\sigma''(\neq \sigma)} 
 \left|\Gamma_{\sigma\sigma'';\sigma''\sigma}^{--;--}(0,0; 0,0) \right|^2
 X_{\sigma''\sigma''}^{-+}(0) 
\ + \ \cdots .
\end{align}
\end{subequations}
Note that 
$X_{\sigma''\sigma''}^{+-}(0) = X_{\sigma''\sigma''}^{-+}(0)$ 
at zero frequency.  
These two vertex components can also be expressed in 
the  symmetrized form,
\begin{align}
&\left\{\rho_{d\sigma}^{}\right\}^2\,
\Bigl[\,
\Gamma_{\sigma\sigma;\sigma\sigma}^{-+;+-} 
(\epsilon,\epsilon'; \epsilon',\epsilon) 
+ \Gamma_{\sigma\sigma;\sigma\sigma}^{+-;-+} 
(\epsilon,\epsilon'; \epsilon',\epsilon)  
\, \Bigr]  
\nonumber \\
& = \,  
i\, 2\pi 
\sum_{\sigma''(\neq \sigma)}
\chi_{\sigma\sigma''}^{2}
\ \mathcal{W}_\mathrm{K}^\mathrm{ph}(0) 
\  + \  \cdots,
\end{align}
\begin{align}
&
\left\{\rho_{d\sigma}^{}\right\}^2\,
\Bigl[\,
\Gamma_{\sigma\sigma;\sigma\sigma}^{-+;+-} 
(\epsilon,\epsilon'; \epsilon',\epsilon) 
-\Gamma_{\sigma\sigma;\sigma\sigma}^{+-;-+} 
(\epsilon,\epsilon'; \epsilon',\epsilon)  
\, \Bigr] 
\nonumber \\
& = \  0 \  + \  O\left((eV)^2, T^2 \right) \,.
\end{align}


\subsection{ 
 $\,\Gamma_{\sigma\sigma';\sigma'\sigma}^{\nu_1\nu_2;\nu_3\nu_4} 
(\epsilon,\epsilon'; \epsilon',\epsilon)$ 
for $\sigma \neq \sigma'$ 
up to linear-order terms in $\epsilon$, $\epsilon'$,  $eV$, and $T$
}

We next consider low-energy behavior of the vertex corrections   
between electrons in different levels $\sigma \neq \sigma'$ 
in the Fermi-liquid regime.  
For the vertex functions of  $\sigma \neq \sigma'$,  
two other Keldysh components 
$\Gamma_{\sigma\sigma';\sigma'\sigma}^{-+;-+}$ and 
$\Gamma_{\sigma\sigma';\sigma'\sigma}^{+-;+-}$  
contribute to the linear order terms 
with respect to  $\epsilon$, $\epsilon'$, $eV$ and $T$,  
in addition to the ones that appear for $\sigma=\sigma'$ 
in the above,  i.e.,
$\Gamma_{\sigma\sigma';\sigma'\sigma}^{ --;--}$, 
$\Gamma_{\sigma\sigma';\sigma'\sigma}^{ ++;++}$, 
$\Gamma_{\sigma\sigma';\sigma'\sigma}^{ --;++}$, 
$\Gamma_{\sigma\sigma';\sigma'\sigma}^{ ++;--}$, 
$\Gamma_{\sigma\sigma';\sigma'\sigma}^{ -+;+-}$, 
$\Gamma_{\sigma\sigma';\sigma'\sigma}^{ +-;-+}$. 
We calculate low-energy expansion of all these components 
in this subsection.

\subsubsection{ 
Low-energy expansion of 
$\Gamma_{\sigma\sigma';\sigma'\sigma}^{ --;--}$ 
 \& 
$\Gamma_{\sigma\sigma';\sigma'\sigma}^{++;++} $
}

For  the causal  $(--;--)$  component, 
 the imaginary part arises from the order $U^2$ scattering processes 
shown in Figs.\  \ref{fig:vertex_u2}  (b),  (c), and (d) 
for  $\sigma \neq \sigma'$.
At low energies, leading-order terms of the imaginary part  
 are determined by a single particle-hole-pair excitations 
in (b) and (d),  and the particle-particle-pair excitations in (c). 
Contributions of each of these three diagrams can be written in the following form, 
taking  $F_b'= F_c'=1$ and  $F_d'=2$, 
\begin{subequations}
\begin{align}
& 
\Gamma_{\sigma\sigma';\sigma'\sigma}^{(2b) --;--} 
(\epsilon,\epsilon'; \epsilon',\epsilon) 
\nonumber \\
& =   
(-1)^0\, i^{2+1} (-1)^{F_b'}\,
U^2 \!  \int \!\frac{d \varepsilon_1}{2\pi} \, 
G_{\sigma}^{--}(\varepsilon_1+\epsilon-\epsilon')\,
G_{\sigma'}^{--}(\varepsilon_1)
\nonumber \\
&=   \, U^2 X_{\sigma\sigma'}^{--} (\epsilon-\epsilon')
\,,
\end{align}
\begin{align}
&\Gamma_{\sigma\sigma';\sigma'\sigma}^{(2c) --;--} 
(\epsilon,\epsilon'; \epsilon',\epsilon) 
\nonumber \\
& =   
(-1)^0\, i^{2+1} (-1)^{F_c'}\,
U^2  \! \int \!\frac{d \varepsilon_1}{2\pi} \, 
G_{\sigma}^{--}(\epsilon+\epsilon' -\varepsilon_1)\,
G_{\sigma'}^{--}(\varepsilon_1)
\nonumber \\
&= \,U^2  Y_{\sigma\sigma'}^{--} (\epsilon+\epsilon')
\,,
\end{align}
\begin{align}
&\Gamma_{\sigma\sigma';\sigma'\sigma}^{(2d) --;--} 
(\epsilon,\epsilon'; \epsilon',\epsilon) 
\nonumber \\
& =   
(-1)^0\, i^{2+1} (-1)^{F_d'}\,
U^2 \! \int \!\frac{d \varepsilon_1}{2\pi} 
\sum_{\sigma''(\neq \sigma, \sigma')} \! 
G_{\sigma''}^{--}(\varepsilon_1)\,
G_{\sigma''}^{--}(\varepsilon_1)
\nonumber  \\
& = \, -\,U^2 
\sum_{\sigma''(\neq \sigma, \sigma')} 
X_{\sigma''\sigma''}^{--} (0) \,.
\end{align}
\label{eq:vertex_U2_UD_causal} \!\!\!
\end{subequations}
Thus, the total  of order $U^2$ causal components is given by 
$\Gamma_{\sigma\sigma';\sigma'\sigma}^{(2) --;--}
 \equiv \Gamma_{\sigma\sigma';\sigma'\sigma}^{(2b) --;--} 
+\Gamma_{\sigma\sigma';\sigma'\sigma}^{(2c) --;--}
+\Gamma_{\sigma\sigma';\sigma'\sigma}^{(2d) --;--}$.  
The contribution of the diagram Fig.\  \ref{fig:vertex_u2} (d)  
appears for $N>2$ as the summation for $\sigma''$ runs over  $N-2>0$ 
possible intermediate states other than $\sigma$ and $\sigma'$.  
Similarly, for the corresponding  $(++;++)$ components, 
order $U^2$ contributions of  the diagrams  Figs.\  \ref{fig:vertex_u2} (b), (c), and (d)  
can be calculated,  replacing $G_{}^{--}$ in 
Eq.\ \eqref{eq:vertex_U2_UD_causal} by
$G_{}^{++}$, as
\begin{align}
&\Gamma_{\sigma\sigma';\sigma'\sigma}^{(2) ++;++} 
(\epsilon,\epsilon'; \epsilon',\epsilon) 
\ = \ - \left\{ \Gamma_{\sigma\sigma';\sigma'\sigma}^{(2)--;--} 
(\epsilon,\epsilon'; \epsilon',\epsilon) \right\}^*
\nonumber \\
& =  U^2 
\Biggl[
X_{\sigma\sigma'}^{++} (\epsilon-\epsilon')
+ Y_{\sigma\sigma'}^{++} (\epsilon+\epsilon')
- \!
\sum_{\sigma''(\neq \sigma, \sigma')} 
X_{\sigma''\sigma''}^{++} (0) 
\Biggr] . 
\end{align}

The imaginary part of  the full vertex function 
that includes multiple scatterings  of all order in $U$ can be calculated 
up to linear order terms with respect to $\epsilon$, $\epsilon'$, $eV$, and $T$, 
from the diagrams show in Figs.\ \ref{fig:vertex_singular_general} (b), (c), and (d). 
The leading-order behaviors are determined by  
 the single particle-hole pair or the particle-particle pair,
 excitation in the intermediate states.  
Therefore,  the imaginary part of the full causal vertex function is given by  
\begin{align}
& 
\mathrm{Im}\,
\Gamma_{\sigma\sigma';\sigma'\sigma}^{(b)--;--} 
(\epsilon,\epsilon'; \epsilon',\epsilon) 
\nonumber  \\
&  
=  \,  
-\mathrm{Im}\,
\int \!\frac{d \varepsilon_1}{2\pi i} \, 
G_{\sigma}^{--}(\varepsilon_1+\epsilon-\epsilon')\,
G_{\sigma'}^{--}(\varepsilon_1)
\nonumber \\
& \qquad \qquad \times 
\Gamma_{\sigma\sigma';\sigma'\sigma}^{--;--}(0,0; 0,0) \, 
\Gamma_{\sigma'\sigma;\sigma\sigma'}^{--;--}(0,0; 0,0) \, 
   +   \cdots 
\nonumber \\
& = \,
\left|\Gamma_{\sigma\sigma';\sigma'\sigma}^{--;--}(0,0; 0,0) \right|^2\, 
\mathrm{Im}\,X_{\sigma\sigma'}^{--} (\epsilon-\epsilon')
  +   \cdots ,
\end{align}
\begin{align}
& 
\mathrm{Im}\,
\Gamma_{\sigma\sigma';\sigma'\sigma}^{(c)--;--} 
(\epsilon,\epsilon'; \epsilon',\epsilon) 
\nonumber  \\
&  
=  \,  
-\mathrm{Im}
\int \!\frac{d \varepsilon_1}{2\pi i} \, 
G_{\sigma}^{--}(\epsilon+\epsilon' -\varepsilon_1)\,
G_{\sigma'}^{--}(\varepsilon_1)
\nonumber \\
& \qquad \qquad \times 
\Gamma_{\sigma\sigma';\sigma'\sigma}^{--;--}(0,0; 0,0) \, 
\Gamma_{\sigma'\sigma;\sigma\sigma'}^{--;--}(0,0; 0,0) \, 
  +   \cdots 
\nonumber \\
&=\,  
\left|\Gamma_{\sigma\sigma';\sigma'\sigma}^{--;--}(0,0; 0,0) \right|^2\, 
\mathrm{Im}\,
 Y_{\sigma\sigma'}^{--} (\epsilon+\epsilon')
  +   \cdots ,
\end{align}
\begin{align}
& 
\mathrm{Im}\,
\Gamma_{\sigma\sigma';\sigma'\sigma}^{(d)--;--} 
(\epsilon,\epsilon'; \epsilon',\epsilon) 
\nonumber  \\
& = 
\mathrm{Im}
\int \!\frac{d \varepsilon_1}{2\pi i} 
\sum_{\sigma''(\neq \sigma, \sigma')} 
G_{\sigma''}^{--}(\varepsilon_1)\,
G_{\sigma''}^{--}(\varepsilon_1)
\\
& \qquad  \times 
\Gamma_{\sigma\sigma'';\sigma''\sigma}^{--;--}(0,0; 0,0) \, 
\Gamma_{\sigma''\sigma';\sigma'\sigma''}^{--;--}(0,0; 0,0) \, 
 +   \cdots 
\nonumber \\
& = 
\,-
\sum_{\sigma''(\neq \sigma, \sigma')} 
\Gamma_{\sigma\sigma'';\sigma''\sigma}^{--;--}(0,0; 0,0) \, 
\Gamma_{\sigma''\sigma';\sigma'\sigma''}^{--;--}(0,0; 0,0) \, 
\nonumber \\
& \quad \times \,
\mathrm{Im}
X_{\sigma''\sigma''}^{--} (0)  
 +   \cdots .
\end{align}
The sum 
$\Gamma_{\sigma\sigma';\sigma'\sigma}^{--;--}  
\equiv \Gamma_{\sigma\sigma';\sigma'\sigma}^{(b)--;--}  
+\Gamma_{\sigma\sigma';\sigma'\sigma}^{(c)--;--}  
+\Gamma_{\sigma\sigma';\sigma'\sigma}^{(d)--;--}  
$ can also be expressed in the following form, 
in terms of the collision integrals,  
\begin{align}
& \mathrm{Im}\,\Gamma_{\sigma\sigma';\sigma'\sigma}^{--;--} 
(\epsilon,\epsilon'; \epsilon',\epsilon) 
\nonumber \\
& 
=   
\frac{\pi}{\rho_{d\sigma}^{} \rho_{d\sigma'}^{}}
\Biggl\{
\chi_{\sigma\sigma'}^{2}
\biggl[ 
\mathcal{W}_\mathrm{K}^\mathrm{ph}(\epsilon-\epsilon') 
- \mathcal{W}_\mathrm{K}^\mathrm{pp}(\epsilon+\epsilon') 
\biggr]
\nonumber \\
 & \qquad \qquad \quad \ \ 
-
\sum_{\sigma''(\neq \sigma, \sigma')} 
\chi_{\sigma\sigma''}^{} \chi_{\sigma'\sigma''}^{}\, 
\mathcal{W}_\mathrm{K}^\mathrm{ph}(0) 
\Biggr\} 
 + \cdots .
\end{align}
The properties of the  PH propagator 
 $X_{\sigma\sigma'}^{\mu\nu}$ and 
PP propagator  $Y_{\sigma\sigma'}^{\mu\nu}$,  
and also the collision integrals 
 $\mathcal{W}_\mathrm{K}^\mathrm{ph}(\omega)$ and
 $\mathcal{W}_\mathrm{K}^\mathrm{pp}(\omega)$, 
are described in Sec.\ \ref{subsec:PH_PP_propagators}.

Note that the causal vertex function for $\sigma \neq \sigma'$
 also has the  $eV$-linear real part  $\mathrm{Re}\, 
\Gamma_{\sigma\sigma';\sigma'\sigma}^{--;--} 
(\epsilon,\epsilon'; \epsilon',\epsilon)$,  
and it is deduced from the Ward identity 
in Sec.\ \ref{sec:real_vertex_eV_liniear} 
 and is shown in  TABLE \ref{tab:vertex_UD}.


\subsubsection{Low-energy expansion of 
$\Gamma_{\sigma\sigma';\sigma'\sigma}^{ --;++}$ 
 \& 
$\Gamma_{\sigma\sigma';\sigma'\sigma}^{++;--} $
}

We next consider  
$\Gamma_{\sigma\sigma';\sigma'\sigma}^{ --;++}$ 
and 
$\Gamma_{\sigma\sigma';\sigma'\sigma}^{++;--}$
which arise first  from order $U^2$  particle-hole ladder diagram 
Fig.\ \ref{fig:vertex_u2} (b).  
The contributions of this process can be calculated,
taking  $F'=1$, as  
\begin{subequations}
\begin{align}
& 
\Gamma_{\sigma\sigma';\sigma'\sigma}^{(2) --;++} 
(\epsilon,\epsilon'; \epsilon',\epsilon)  
\nonumber \\
& = \, 
 (-1)^1\, i^{2+1} (-1)^{F'}
U^2 \!
\int \!\frac{d \varepsilon_1}{2\pi} \, 
G_{\sigma}^{+-}(\varepsilon_1+\epsilon-\epsilon')\,
G_{\sigma'}^{-+}(\varepsilon_1)
\nonumber \\
&= \, 
- U^2 X_{\sigma\sigma'}^{+-}(\epsilon-\epsilon') \,,
\end{align}
\begin{align}
&\Gamma_{\sigma\sigma';\sigma'\sigma}^{(2) ++;--} 
(\epsilon,\epsilon'; \epsilon',\epsilon) 
\nonumber \\
&=\, 
 (-1)^1\, i^{2+1} (-1)^{F'}\,
U^2 \! 
\int \!\frac{d \varepsilon_1}{2\pi} \, 
G_{\sigma}^{-+}(\varepsilon_1+\epsilon-\epsilon')\,
G_{\sigma'}^{+-}(\varepsilon_1)
\nonumber \\
&=\, - U^2 X_{\sigma\sigma'}^{-+}(\epsilon-\epsilon')  \,. 
\end{align}
\end{subequations}
These two components are pure imaginary.
The low-energy asymptotic forms 
of the corresponding full vertex functions    
can be deduced from the diagram  Fig.\ \ref{fig:vertex_singular_general} (b),   
which include  multiple-scattering processes of all orders in $U$,
\begin{subequations}
\begin{align}
&\Gamma_{\sigma\sigma';\sigma'\sigma}^{--;++} 
(\epsilon,\epsilon'; \epsilon',\epsilon) \, 
\nonumber \\
&=  \, 
-\int \!\frac{d \varepsilon_1}{2\pi i} \, 
\Gamma_{\sigma\sigma';\sigma'\sigma}^{--;--}(0,0; 0,0) \, 
\Gamma_{\sigma\sigma';\sigma'\sigma}^{++;++}(0,0; 0,0) \, 
\nonumber \\
& \qquad 
\times
G_{\sigma}^{+-}(\varepsilon_1+\epsilon-\epsilon')\,
 G_{\sigma'}^{-+}(\varepsilon_1) \ + \ \cdots 
 \nonumber \\
&= \,  
- \left|\Gamma_{\sigma\sigma';\sigma'\sigma}^{--;--}(0,0; 0,0) \right|^2 \, 
X_{\sigma\sigma'}^{+-}(\epsilon-\epsilon')  \ + \ \cdots\,,
\label{eq:vertex_UD_--++}
\end{align}
\begin{align}
&\Gamma_{\sigma\sigma';\sigma'\sigma}^{++;--} 
(\epsilon,\epsilon'; \epsilon',\epsilon) \, 
\nonumber \\
& =  \, 
-\int \!\frac{d \varepsilon_1}{2\pi i} \, 
\Gamma_{\sigma\sigma';\sigma'\sigma}^{++;++}(0,0; 0,0) \, 
\Gamma_{\sigma\sigma';\sigma'\sigma}^{--;--}(0,0; 0,0) \, 
\nonumber \\
& \qquad  
\times G_{\sigma}^{-+}(\varepsilon_1+\epsilon-\epsilon')\,
 G_{\sigma'}^{+-}(\varepsilon_1) \ + \ \cdots\,,
 \nonumber \\
&= \, 
- \left|\Gamma_{\sigma\sigma';\sigma'\sigma}^{--;--}(0,0; 0,0) \right|^2 \, 
X_{\sigma\sigma'}^{-+}(\epsilon-\epsilon')  \ + \ \cdots\,,
\label{eq:vertex_UD_++--}
\end{align}
\end{subequations}
Therefore, the symmetrized  and antisymmetrized parts of these two full vertex functions
 can be written, as 
\begin{align}
&\!\!\! 
\rho_{d\sigma}^{}\rho_{d\sigma'}^{}\, 
\Bigl[\,
\Gamma_{\sigma\sigma';\sigma'\sigma}^{--;++} 
(\epsilon,\epsilon'; \epsilon',\epsilon) +
\Gamma_{\sigma\sigma';\sigma'\sigma}^{--;++} 
(\epsilon,\epsilon'; \epsilon',\epsilon) \,\Bigr] 
\nonumber \\
&= \, 
-i\, 2\pi   
\chi_{\sigma\sigma'}^2 \,
\ \mathcal{W}_\mathrm{K}^\mathrm{ph}(\epsilon-\epsilon') 
\ + \ \cdots\,,
\end{align}
\begin{align}
& \!\!\!
\rho_{d\sigma}^{}\rho_{d\sigma'}^{}\, 
\Bigl[\,
\Gamma_{\sigma\sigma';\sigma'\sigma}^{--;++} 
(\epsilon,\epsilon'; \epsilon',\epsilon) -
\Gamma_{\sigma\sigma';\sigma'\sigma}^{--;++} 
(\epsilon,\epsilon'; \epsilon',\epsilon) \,\Bigr] 
\nonumber \\
&=\,  
-i\, 2\pi\,  
\chi_{\sigma\sigma'}^2 
\,  \bigl[\,\epsilon-\epsilon' \, \bigr] \ + \ \cdots \,.
\end{align}
Note that  
$\mathcal{W}_{}^\mathrm{ph}(\omega) - \mathcal{W}_{}^\mathrm{ph}(-\omega) = \omega$, and  
the explicit expression of $\mathcal{W}_\mathrm{K}^\mathrm{ph}(\epsilon-\epsilon')$ 
is shown in TABLE \ref{tab:ph-pp_propagators}.

\subsubsection{ 
Low-energy expansion of 
$\Gamma_{\sigma\sigma';\sigma'\sigma}^{ -+;-+}$ 
 \& 
$\Gamma_{\sigma\sigma';\sigma'\sigma}^{+-;+-} $
}

The following two vertex components, 
$\Gamma_{\sigma\sigma';\sigma'\sigma}^{ -+;-+}$ and 
$\Gamma_{\sigma\sigma';\sigma'\sigma}^{+-;+-} $ 
for $\sigma \neq \sigma$, arise first from
the order $U^2$  diagram Fig.\  \ref{fig:vertex_u2} (c).  
 These vertex functions are determined by intermediate particle-particle-pair excitation 
and are pure imaginary,  which  can be  verified taking  $F'=1$, as 
\begin{subequations}
\begin{align}
& \Gamma_{\sigma\sigma';\sigma'\sigma}^{(2) -+;-+} 
(\epsilon,\epsilon'; \epsilon',\epsilon) \ 
\nonumber \\
&= \,  
 (-1)^1\, i^{2+1} (-1)^{F'}\, U^2 \!
\int \!\frac{d \varepsilon_1}{2\pi} \, 
G_{\sigma}^{+-}(\epsilon+\epsilon' -\varepsilon_1)\,
G_{\sigma'}^{+-}(\varepsilon_1)
\nonumber \\
&= \, 
- U^2 \,Y_{\sigma\sigma'}^{+-}(\epsilon+\epsilon') \,, 
\end{align}
\begin{align}
&\Gamma_{\sigma\sigma';\sigma'\sigma}^{(2) +-;+-} 
(\epsilon,\epsilon'; \epsilon',\epsilon) \ 
\nonumber \\
&=  \,  
 (-1)^1\, i^{2+1} (-1)^{F'}\, U^2 \! 
\int \!\frac{d \varepsilon_1}{2\pi} \, 
G_{\sigma}^{-+}(\epsilon+\epsilon' -\varepsilon_1)\,
G_{\sigma'}^{-+}(\varepsilon_1)
\nonumber \\
&= \, 
- U^2 \,Y_{\sigma\sigma'}^{-+}(\epsilon+\epsilon') 
\,.
\end{align}
\end{subequations}
For these two components, contributions 
of multiple scatterings of all orders in $U$ can also be deduced 
up to  linear-order terms with respect to $|\epsilon-\epsilon'|$, $eV$, and $T$ 
from the diagram Fig.\ \ref{fig:vertex_singular_general} (c),
\begin{subequations}
\begin{align}
&
\Gamma_{\sigma\sigma';\sigma'\sigma}^{-+;-+} 
(\epsilon,\epsilon'; \epsilon',\epsilon) \, 
\nonumber\\
&= \, 
-\int \!\frac{d \varepsilon_1}{2\pi i} \, 
\Gamma_{\sigma\sigma';\sigma'\sigma}^{--;--}(0,0; 0,0) \, 
\Gamma_{\sigma\sigma';\sigma'\sigma}^{++;++}(0,0; 0,0) \, 
\nonumber \\ 
& \qquad \qquad 
\times G_{\sigma}^{+-}(\epsilon+\epsilon'-\varepsilon_1)\,
 G_{\sigma'}^{+-}(\varepsilon_1) \ + \ \cdots
 \nonumber \\
&= \  
-  \left|\Gamma_{\sigma\sigma';\sigma'\sigma}^{--;--}(0,0; 0,0) \right|^2 \,
 Y_{\sigma\sigma'}^{+-}(\epsilon+\epsilon') \ + \ \cdots\,, 
 \label{eq:vertex_UD_-+-+}
\end{align}
\begin{align}
&\Gamma_{\sigma\sigma';\sigma'\sigma}^{+-;+-} 
(\epsilon,\epsilon'; \epsilon',\epsilon) \, 
\nonumber \\
&=  \, 
-\int \!\frac{d \varepsilon_1}{2\pi i} \, 
\Gamma_{\sigma\sigma';\sigma'\sigma}^{++;++}(0,0; 0,0) \, 
\Gamma_{\sigma\sigma';\sigma'\sigma}^{--;--}(0,0; 0,0) \, 
\nonumber \\
& \qquad \qquad 
\times 
G_{\sigma}^{-+}(\epsilon+\epsilon'-\varepsilon_1)\,
 G_{\sigma'}^{-+}(\varepsilon_1)  \ + \ \cdots
 \nonumber \\
&= \,  
-  \left|\Gamma_{\sigma\sigma';\sigma'\sigma}^{--;--}(0,0; 0,0) \right|^2 \,
 Y_{\sigma\sigma'}^{-+}(\epsilon+\epsilon') 
\ + \ \cdots \,. 
 \label{eq:vertex_UD_+-+-}
\end{align}
\end{subequations}
The symmetrized  and antisymmetrized functions of these two 
components can be expressed in the following form,   
\begin{align}
&\!\! 
\rho_{d\sigma}^{}\rho_{d\sigma'}^{}\, 
\Bigl[\,
\Gamma_{\sigma\sigma';\sigma'\sigma}^{-+;-+} 
(\epsilon,\epsilon'; \epsilon',\epsilon) +
\Gamma_{\sigma\sigma';\sigma'\sigma}^{+-;+-} 
(\epsilon,\epsilon'; \epsilon',\epsilon) \,\Bigr]  
\nonumber \\
&=\, 
i\, 2\pi\,  
\chi_{\sigma\sigma'}^2 \, \mathcal{W}_\mathrm{K}^\mathrm{pp}(\epsilon + \epsilon') \ + \ \cdots \,,
\end{align}
\begin{align}
&\!\! 
\rho_{d\sigma}^{}\rho_{d\sigma'}^{}\, 
\Bigl[\,
\Gamma_{\sigma\sigma';\sigma'\sigma}^{-+;-+} 
(\epsilon,\epsilon'; \epsilon',\epsilon) -
\Gamma_{\sigma\sigma';\sigma'\sigma}^{+-;+-} 
(\epsilon,\epsilon'; \epsilon',\epsilon) \,\Bigr]  
\nonumber \\
& =\, 
i\, 2\pi\,  
\chi_{\sigma\sigma'}^2 \ 
\mathcal{W}_\mathrm{dif}^\mathrm{pp}(\epsilon+\epsilon')
\ + \ \cdots \,.
\end{align}
Note that  the bosonic particle-particle collision integral  
$\mathcal{W}_\mathrm{dif}^\mathrm{pp}(\omega) 
\equiv  \mathcal{W}_{}^\mathrm{pp}(\omega) - 
\mathcal{W}_{}^\mathrm{hh}(\omega)$ 
is defined in Eq.\ \eqref{eq:pp_dif}, 
and the explicit expression of $\mathcal{W}_\mathrm{K}^\mathrm{pp}(\omega)$ 
is given in TABLE \ref{tab:ph-pp_propagators}.

\subsubsection{ 
Low-energy expansion of 
$\Gamma_{\sigma\sigma';\sigma'\sigma}^{ -+;+-}$ 
 \& 
$\Gamma_{\sigma\sigma';\sigma'\sigma}^{+-;-+} $
}

For multilevel Anderson impurity with $N>2$,  
vertex corrections for $\sigma' \neq \sigma$ 
 also arise from the diagram (d) of  Fig.\ \ref{fig:vertex_u2}. 
In this case, the intermediate particle-hole pair can be excited 
in $N-2$ different configurations of internal degrees of freedom $\sigma''$ 
as it is required that  $\sigma'' \neq \sigma$ and  $\sigma'' \neq \sigma'$.
Thus, the contributions of the order $U^2$ process are given, 
taking  $F''=2$, by
\begin{subequations}
\begin{align}
&\Gamma_{\sigma\sigma';\sigma'\sigma}^{(2) -+;+-} 
(\epsilon,\epsilon'; \epsilon',\epsilon) \ 
\nonumber \\
&=   \,  
 (-1)^1\, i^{2+1} (-1)^{F''}\,
U^2 \!
\int \!\frac{d \varepsilon_1}{2\pi} \!
\sum_{\sigma''(\neq \sigma, \sigma')} \!  
G_{\sigma''}^{+-}(\varepsilon_1)\,
G_{\sigma''}^{-+}(\varepsilon_1)
 \nonumber \\
 &= \ 
 \sum_{\sigma''(\neq \sigma, \sigma')}
 U^2 X_{\sigma''\sigma''}^{+-}(0) \,,
\end{align}
\begin{align}
& \Gamma_{\sigma\sigma';\sigma'\sigma}^{(2) +-;-+} 
(\epsilon,\epsilon'; \epsilon',\epsilon) \ 
\nonumber \\
& =  \,  
 (-1)^1\, i^{2+1} (-1)^{F''}\,
U^2 \! 
\int \!\frac{d \varepsilon_1}{2\pi} \! 
\sum_{\sigma''(\neq \sigma, \sigma')} \!  
G_{\sigma''}^{-+}(\varepsilon_1)\,
G_{\sigma''}^{+-}(\varepsilon_1)
 \nonumber \\
&=  
 \sum_{\sigma''(\neq \sigma,\sigma')}
U^2 X_{\sigma''\sigma''}^{-+}(0) \,.
\end{align}
\end{subequations}
These contributions  are pure imaginary. 
Thus, similarly to the cases of the other components,  
contributions of the multiple scatterings of all orders in $U$ 
can be deduced from the diagram Fig.\ \ref{fig:vertex_singular_general} (d), 
up to linear order terms with respect to $eV$, $T$, and frequencies 
$\epsilon$ and $\epsilon'$, 
\begin{subequations}
\begin{align}
& \Gamma_{\sigma\sigma';\sigma'\sigma}^{-+;+-} 
(\epsilon,\epsilon'; \epsilon',\epsilon) \, 
\nonumber \\
&=  
\int \!\frac{d \varepsilon_1}{2\pi i} \, 
\sum_{\sigma''(\neq \sigma, \sigma')}
\Gamma_{\sigma\sigma'';\sigma''\sigma}^{--;--}(0,0; 0,0) \, 
\Gamma_{\sigma''\sigma';\sigma'\sigma''}^{++;++}(0,0; 0,0)  
\nonumber \\
& \qquad \qquad \qquad \qquad 
\times 
\,G_{\sigma''}^{+-}(\varepsilon_1)\,
 G_{\sigma''}^{-+}(\varepsilon_1)
 \ + \ \cdots 
 \nonumber \\
&= \  
\sum_{\sigma''(\neq \sigma, \sigma')}
\Gamma_{\sigma\sigma'';\sigma''\sigma}^{--;--}(0,0; 0,0) \, 
\Gamma_{\sigma''\sigma';\sigma'\sigma''}^{--;--}(0,0; 0,0)  \,
\nonumber \\
&\qquad \qquad \quad 
\times
X_{\sigma''\sigma''}^{+-}(0) 
\ + \ \cdots \,,
\end{align}
\begin{align}
& \Gamma_{\sigma\sigma';\sigma'\sigma}^{+-;-+} 
(\epsilon,\epsilon'; \epsilon',\epsilon) \, 
\nonumber \\
&=   
\int \!\frac{d \varepsilon_1}{2\pi i} \, 
\sum_{\sigma''(\neq \sigma, \sigma')}
\Gamma_{\sigma\sigma'';\sigma''\sigma}^{++;++}(0,0; 0,0) \, 
\Gamma_{\sigma''\sigma';\sigma'\sigma''}^{--;--}(0,0; 0,0)  
\nonumber \\
&  \qquad \qquad \qquad  \qquad 
\times
\,G_{\sigma''}^{-+}(\varepsilon_1)\,
 G_{\sigma''}^{+-}(\varepsilon_1)\,
\ + \ \cdots
 \nonumber \\
&= \  
\sum_{\sigma''(\neq \sigma, \sigma')}
\Gamma_{\sigma\sigma'';\sigma''\sigma}^{--;--}(0,0; 0,0) \, 
\Gamma_{\sigma''\sigma';\sigma'\sigma''}^{--;--}(0,0; 0,0)  
\nonumber \\
&  \qquad \qquad \quad 
\times X_{\sigma''\sigma''}^{-+}(0) 
 \ + \ \cdots \,.
\end{align}
\end{subequations}
Furthermore, the symmetrized  and antisymmetrized functions 
 of these two  vertex components can also be written in the form,  
\begin{align}
&\rho_{d\sigma}^{}\,\rho_{d\sigma'}^{}
\Bigl[\,
\Gamma_{\sigma\sigma';\sigma'\sigma}^{-+;+-} 
(\epsilon,\epsilon'; \epsilon',\epsilon) 
+\Gamma_{\sigma\sigma';\sigma'\sigma}^{+-;-+} 
(\epsilon,\epsilon'; \epsilon',\epsilon)  
\,\Bigr]
\nonumber \\
& = \,  
    i\, 2\pi 
\! \sum_{\sigma''(\neq \sigma,\sigma')} \! 
\chi_{\sigma\sigma''}^{}
\chi_{\sigma''\sigma'}^{}
\, \mathcal{W}_\mathrm{K}^\mathrm{ph}(0) 
 \ + \ \cdots \,,
\end{align}
\begin{align}
& \rho_{d\sigma}^{}\,\rho_{d\sigma'}^{}
\Bigl[\,
\Gamma_{\sigma\sigma';\sigma'\sigma}^{-+;+-} 
(\epsilon,\epsilon'; \epsilon',\epsilon) 
- \Gamma_{\sigma\sigma';\sigma'\sigma}^{+-;-+} 
(\epsilon,\epsilon'; \epsilon',\epsilon)  
\,\Bigr]
\nonumber \\
& =  \  0    \ +  O\left( (eV)^2,T^2 \right) .
\end{align}


%

\end{document}